%% file: dissertation.tex
\documentclass[amsthm]{UMassThesis14}
\usepackage{microtype, graphicx, verbatim, tikz, mathrsfs}
\usepackage[shortlabels]{enumitem}
\usepackage[frak=pxtx, scr=esstix]{mathalpha}
\usepackage[defaultlines=2,all]{nowidow}
\usepackage[
    pdfpagelabels,
    hidelinks
]{hyperref}

\usepackage{setspace}
\let\doublespacing\singlespacing
\AtBeginDocument{\singlespacing}

\newtheorem{theorem}{Theorem}[section]
\newtheorem{corollary}{Corollary}[theorem]
\newtheorem{lemma}[theorem]{Lemma}
\newtheorem{definition}{Definition}[section]
\newtheorem*{remark}{Remark}

\newcommand{\B}[1]{\textbf{#1}}
\newcommand{\BT}[1]{\tilde{\textbf{#1}}}
\newcommand{\tr}{{\text{tr}}}
\newcommand{\col}{{\text{col}}}
\newcommand{\row}{{\text{row}}}
\newcommand{\range}{{\text{range}}}
\newcommand{\nullspace}{{\text{null}}}
\newcommand{\rank}{{\text{rank}}}
\newcommand{\linspan}{{\text{span}}}
\newcommand{\var}{{\text{Var}}}
\newcommand{\cone}{{\text{cone}}}
\newcommand{\odotP}{\bar{\odot}}
\newcommand{\diag}[1]{D_{#1}}
\newcommand{\boldpi}{{\boldsymbol{\Pi}}}

\begin{document}

\include{sections/frontmatter}
\include{sections/introduction}
\include{sections/probabilities}
\include{sections/born_identity}
\include{sections/urgleichungen}
\include{sections/classicality}
\include{sections/characterizing}
\include{sections/reconstruction}
\include{sections/backmatter}

\end{document}

%% file: sections/frontmatter.tex
%%%%%%%%%%%%%%%%%%%%%%%%%%%%%%%%%%%%%%%%%%%%%%%%%%%%%%%%%

\title{Redesigning quantum theory}
\author{Matthew B. Weiss} 
\AuthorDegrees{B.A., Brown University\\
M.S., University of Iowa\\
Ph.D., University of Massachusetts Boston}
\DirectedBy{Professor Christopher A. Fuchs}
\ProgramName{Computational Sciences Program} 
\DegreeMonthYear{August}{2026} 
\ThesisOrDissertation{Dissertation}
\DegreeName{Doctor of philosophy} 

\ProgramDirector{%
Daniel Pomerleano, Program Director\\
Computational Sciences Program
}
\DeptChairperson{%
Rahul Kulkarni, Chair\\
Department of Physics
}
\CommitteeMember{%
Christopher A. Fuchs, Distinguished Professor\\
Chairperson of Committee
}
\CommitteeMember{%
Olga Goulko, Assistant Professor\\
Member
}
\CommitteeMember{%
Akira Sone, Assistant Professor\\
Member
}
\CommitteeMember{%
Kourosh Zarringhalam, Professor\\
Member
}
\CommitteeMember{%
Marcus Appleby, PhD\\
University of Sydney\\
Member
}

%%%%%%%%%%%%%%%%%%%%%%%%%%%%%%%%%%%%%%%%%%%%%%%%%%%%%%%%%

\PrintTitlePage
\PrintCopyrightPage
\PrintSignaturetPage

%%%%%%%%%%%%%%%%%%%%%%%%%%%%%%%%%%%%%%%%%%%%%%%%%%%%%%%%%

\begin{UMBAbstract}
QBism understands quantum mechanics to be probability theory supplemented by additional nonclassical coherence conditions. In this dissertation, we develop these nonclassical coherence conditions from first principles, emphasizing the role of a well chosen reference measurement. After treating standard probability on subjective Bayesian lines, we demonstrate an equivalence between the QBist approach and the existing framework of generalized probabilistic theories. We show that the fundamental nonclassical coherence relation may almost always be taken to be a gentle modification of the law of total probability, and give a coherentist account of when an experimental scenario has a classical explanation. Finally, we show that when the reference measurement is chosen to correspond to a complex projective 3-design, the shape of quantum state space is implicit in the probabilities which characterize the reference measurement itself. Thus coherence with this single reference measurement, properly understood, implies coherence with all of finite dimensional quantum mechanics. We then attempt a modest reconstruction of quantum theory along these lines, one practical consequence of which is a method for self-testing complex projective $t$-designs for $t\ge 3$ in a theory-agnostic way.
\end{UMBAbstract}

%%%%%%%%%%%%%%%%%%%%%%%%%%%%%%%%%%%%%%%%%%%%%%%%%%%%%%%%%

\newpage

%%%%%%%%%%%%%%%%%%%%%%%%%%%%%%%%%%%%%%%%%%%%%%%%%%%%%%%%%

\topmatter{\ToCListls{ACKNOWLEDGMENTS}}
\doublespacing

None of this work would have been possible without the extraordinary generosity of Chris Fuchs. Studying with him has been the opportunity of a lifetime, and without his inexhaustible support, I would still be out in the wilderness. Five years ago, he welcomed me into the QBist family, and ever since I've had the benefit of not only his wisdom, but also the strength of the community he has built and so carefully and lovingly maintained over decades. I learned so much of the technical spirit of QBism from conversations with the indispensable John DeBrota and Blake Stacey, on whose shoulders I gratefully stand. From the first day of classes in September 2021, Sachin Gupta has been my most reliable friend and collaborator: so many of our ideas could only take form during those long hours at the whiteboard in our little office. Battling through proofs with Gianluca Cuffaro was an honor, and I'll never forget our time soaking in a hot tub on the roof of a hotel in Anaheim speculating wildly about quantum gravity, or wandering around the labyrinth of MIT with Simone Cepollaro. David Llamas will always have my respect for his commitment to the bit, and for the sensitivity of his bullshit detector. I miss the days when Ghi Coulter-de Wit was a part of the group: we couldn't have made it through that first APS March Meeting without them. I am grateful to Jacques Pienaar for bringing the magic of phenomenology to QBism and for his ability to keep the grants flowing: I picture us walking around Lake V\"axj\"osj\"on with the moon overhead. Any time spent with Marcus Appleby is always a genuine pleasure: his energy and curiosity is infectious---not only did he tirelessly induct us all at UMass Boston into the mysteries of algebraic number theory, but the force of his philosophical conviction is an ongoing inspiration. It was my priviledge as well to spend time with R\"udiger Schack, who makes rigorous Bayesianism look easy.

Thanks to Alex Moll, for having quantum office hours that first semester; to Maxim Olshanii for introducing me to the joys of higher dimensional polytopes; Akira Sone for introducing me to the power of stochastic differential equations; Steve Arnason for teaching me how real physics is done; Kourosh Zarringhalam for making sure my linear algebra was up to snuff; to Rahul Kulkarni for his genuine support and interest. I learned so much from the irreplacable Olgo Goulko, and I miss our trips on the Red Line together. 

My time at UMass Boston would not have been the same without the kindness of Mary Fries and her one-of-kind QBist art. Thanks to Arjun Dhoot for so many wonderful and exciting conversations; to Jack Davis for making Hanover feel like home; to Gino Elia, for keeping us all philosophically honest; to Austin Monaghan, for bravely continuing the work we started; to Chris Sutton, for his absolute genuineness; to Sai Morapakula, for the flowers; to Money Chanalia, I hope we keep running into each other; to Vajra Badha, for keeping our little club together; to Natalia Freitas, for never giving up.

I'd be remiss not to thank Toby, for his teaching me the old ways. I have so much gratitude for Tim and Sally and Julie and Scotty, for remaining in my life all these years. For Kiesha, for understanding what I was trying to do. For Bubbe, for our weekly chats, full of counsel and solicitude. To MMJ, for our emails. In a profound way, everything I have done here has its origin in conversations with Jeremy Silver, when we were just two child philosophers, making the circuit around the Hebrew School playground. For Julian and Anna, for Daniel, for Bilan, Liban, Iman, Bashir, Abo and Hooyo.

Thank you to my Mom and Dad, who have always believed in the ``timespace interchange,'' even when it was most difficult. In a real sense, my Dad taught me how to be a scientist in the most loving way. My mom has always been my most devoted reader, and I don't feel like I understand anything until I can explain it to her. One day, we'll figure out what it all means.

And finally, this dissertation would have been impossible without the support of Ladan, my wife and partner in life. She has truly changed my world for the sweeter. The deepest insights in this work I had beside her, and I am so grateful to have a soulmate on this quantum adventure.

\vspace{1cm}

\emph{This research was supported in part by the National Science Foundation through grants NSF-2210495 and OSI-2328774 as well as through Grant 62424 from the John Templeton Foundation. The opinions expressed in this publication are those of the author and do not necessarily reflect the views of the John Templeton Foundation.}

\vspace{1cm}

\emph{Disclaimer: I must in good faith disclose that several ideas in this dissertation arose in conversation with ChatGPT versions 5.1 to 5.5. ChatGPT was used to check certain proofs, as well as to automate certain algebraic manipulations, but played no role in the writing of the text. I take full responsibility for the content herein.}

%%%%%%%%%%%%%%%%%%%%%%%%%%%%%%%%%%%%%%%%%%%%%%%%%%%%%%%%%

\newpage
\PrintToCLoFLoT
\SetUpMainText

%%%%%%%%%%%%%%%%%%%%%%%%%%%%%%%%%%%%%%%%%%%%%%%%%%%%%%%%%

%% file: sections/introduction.tex
\UMBchapter{Introduction}
\label{ch:introduction}

\section{Foundations}

This dissertation is primarily an exercise in quantum foundations, the ongoing attempt to give a conceptually satisfying account of the lesson that quantum mechanics is trying to teach us about the world and ourselves. It is unapologetically QBist \cite{fuchs2019qbismquantumtheoryheros, fuchsQbismWhereNext2023}. The QBist school of thought holds that quantum mechanics is best understood as a guidebook for an agent trying to gamble on the consequences of their free actions on a world which is just as free to act back. In this way the agent and the world they act upon are locked in a permeable embrace of mutual cocreation. Already subjective Bayesianism understands probability theory to be of a normative character, a way of organizing one's own beliefs to ensure they are as a whole coherent with each other. QBism views quantum theory in exactly the same way, as a tool any agent may use to tend to their mesh of beliefs about the consequences of their own actions precisely in light of nature's vitality. In particular, the key formal move of the QBist approach is to focus attention on \emph{reference measurements}: such measurements have the property that assigning probabilities to their outcomes is mathematically equivalent to assigning a density matrix, the more familiar guise under which a quantum state appears. Quantum states, effects, unitary evolution: they may all be reexpressed in terms of relations between probability assignments. In this way, QBists do not understand quantum theory by analogy with probability theory: they view it as nothing other than probability theory, properly supplemented with new norms which respect nature's lack of ``hidden variables.''

In particular, a key part of respecting this absence is the insistence that the agent ought to use quantum mechanics to organize their beliefs about the consequences \emph{for them} of \emph{their own} actions. To gamble on the consequences for another is as much to deny the openness of the world to mutuality: to treat the consequence for another as settled is to treat it as a hidden variable, an already existing property ready to be trotted out. This personalization leads to certain familiar notions, for consistency's sake, to be recast in at first unusual terms: for example, dynamics is understood in terms of an agent's indifference to whether or not they themselves might perform a particular measurement \cite{debrotaQBismsAccountQuantum2024}. Moreover, just as probability theory is not ``about'' coin flips and dice rolls, but rather helps one organize one's beliefs about coins and dice, so too quantum theory in itself is not ``about'' electrons, hydrogen atoms, or spin glasses; instead it helps the agent organize their beliefs about such phenomena. The beliefs themselves the agent forms in their intercourse with the world, guided by the consequences their world confronts them with. And whereas these consequences are personal to the agent, \emph{any} agent is well advised to use quantum mechanics: its universality attests to the common existential situation we all find ourselves in. But what exactly this situation is, of course, remains still up for debate.

Like others in the quantum foundations community who pursue ``reconstructions'' of quantum mechanics \cite{mullerProbabilisticTheoriesReconstructions2021}, QBists do not wish to take the quantum formalism for granted. We continue to seek a satisfying principled derivation of the nonclassical coherence rules which quantum theory first brought to our attention. In the first part of this dissertation, after developing the usual formalism of probability theory in subjective Bayesian style, I adapt existing arguments in the QBist literature to derive the form of the fundamental nonclassical coherence rule, which in quantum theory corresponds to the Born rule, but which is in reality much more general. The following chapter explores in more detail the mathematical properties of this coherence rule, in particular the nature of the \emph{Born matrix}, which plays a starring role. Ultimately it demonstrates a correspondence between the QBist formalism and the formalism of \emph{generalized probabilistic theories} heavily investigated in the quantum foundations literature \cite{mullerProbabilisticTheoriesReconstructions2021}. The next chapter studies under what circumstances the fundamental nonclassical rule may take an extraordinarily simple form. Indeed, the general QBist strategy in pursuing foundational questions has been to find those circumstances in which the extra normative rules suggested by quantum mechanics may be made to look as close as possible to the familiar rules of standard probability theory. The gap between the two then witnesses the essential difference between classical and nonclassical ways of approaching the world. For this reason, symmetric informationally complete (SIC) reference measurements in quantum mechanics have been an abiding interest for QBists: such measurements have a host of truly remarkable probabilities, not least of which is that they allow one to rewrite the Born rule $P(E|\rho)=\tr(E \rho)$ as
\begin{align}
P(E|\rho)= \sum_i P(E|R_i)\left\{(d+1)P(R_i|\rho) - \frac{d}{n}\right\},
\end{align}
which is as gentle a modification of the law of total probability $\sum_i P(E|R_i)P(R_i|\rho)$ as one could imagine, and which QBists have termed the \emph{Urgleichung}, or fundamental equation. I show that in fact this type of rule is ubiquitous in the landscape of generalized probabilistic theories. Thus the form of this rule is not peculiar to quantum theory, and may even be appealed to in situations where classical explanations are available. But then: what do we even mean by classical? We then turn to the question of when an agent's mesh of beliefs is compatible with a classical explanation in terms of underlying properties which condition the results of measurement, giving a QBist spin on Spekkens's notion of a noncontextual ontological model, grounding it in an agent's indifference to whether or not a reference measurement is performed. Moreover, we recast Bell inequalities and noncontextuality inequalities more generally as providing a lower bound on how close the fundamental nonclassical coherence rule is to the law of total probability. For QBists, violations of these inequalities quantify the cost of the failure to adopt the nonclassical coherence rule itself. 

The final two chapters, however, form the heart and soul of the dissertation. For QBists, the question of \emph{why quantum theory} amounts in some sense to the question: why do states, in the form of probability distributions on reference outcomes, form just \emph{that} particular subset of the probability simplex? For indeed, even though every quantum state corresponds to a probability distribution on the outcomes of a reference measurement, not every conceivable distribution corresponds to a quantum state. Instead, the set of allowed distributions has a very special geometry. Previous attempts at QBist reconstructions of quantum theory have focused their attention on generalizing the image of quantum state space within the probability simplex induced by a SIC-POVM measurement \cite{Appleby_2017}. Such generalizations are called qplexes, and the goal has been to identify a principle by which the qplex corresponding to quantum theory may be picked out among all others. Similar ideas have been applied more generally to measurements corresponding to complex-projective 2-designs, of which SICs are but the simplest example \cite{slomczynskiMorphophoricPOVMsGeneralised2020}. 

Our central innovation is to adopt as a target for generalization not 2-designs, but 3-designs, of contemporary interest for their use in classical shadow estimation procedures \cite{Huang_2020,mao2024magicquditshadowestimation,Kliesch_2021, chen2024nonstabilizernessenhancesthriftyshadow}. We show that (unbiased) 3-design reference measurements allow the shape of quantum state space to be characterized in a strikingly simple way, first in terms of constraints which govern pure state probability distributions, and then more generally in terms of an uncertainty principle, a lower bound on the variance of any of a natural class of observables. In particular, the entire geometry of the state space is encoded in $P(R|R)$, the conditional probability matrix which characterizes the reference measurement itself. In other words, the probabilities which characterize a 3-design  reference measurement characterize the entire theory. The reason is algebraic: the properties of quantum 3-designs imply that the structure coefficients of the Jordan algebra of observables may be extracted from $P(R|R)$ alone. In the end, we find that just as 2-designs make the Born rule look as close as possible to the classical law of total probability, 3-designs make the quantum Jordan product on observables look as close as possible to the classical Jordan product.

In the final chapter, we attempt our own modest reconstruction of finite-dimensional quantum theory. We begin from scratch, with the matrix $P(R|R)$, placing increasingly severe restrictions on its structure, guided by the idea of introducing the gentlest possible modification of the classical rule for multiplying valuations on reference outcomes. In doing so, we retrace the steps in the previous chapter from first principles, motivating a series of assumptions which ultimately imply that the entries of the matrix $P(R|R)$ may be realized as inner products between $d \times d$ Hermitian matrices over $\mathbb{C}$ which constitute a complex projective 3-design. The key step is a ``moment-matching'' assumption. Any valuation on an arbitrary measurement is equivalent to some valuation on the reference measurement: we assume that for some class of measurements, the $m$th moment with respect to the original measurement and the $m$th moment with respect to the reference, calculated according to the nonclassical product, coincide. This leads directly to the terrain of Euclidean Jordan algebras. While our motivations are foundational, the same considerations yield a practical algorithm for self-testing quantum $t$-designs for $t\ge 3$. We show that relations between the probabilities $P(R_i|R_j)$ certify that a Hilbert space representation is possible in a way that other measures of ``designness'' do not: for example, measuring the so-called frame potential and showing that it achieves its minimimum is not sufficient without a guarantee that a Hilbert space exists to begin with. Finally, while we achieve our goal of returning full circle to quantum theory, many deep questions remain regarding both the motivations for our assumptions as well as the geometry they imply, leaving the door open for a great deal of productive work in the future. 

\section{Context}

\subsection{Generalized probabilistic theories}

The last twenty-five years have been a golden age for quantum foundations. The turn of the millennium marked a growing consensus that it ought to be possible to rederive or \emph{reconstruct} quantum mechanics from a few sharp, simple principles, ideally of an information-theoretic character \cite{Berghofer2024, Brassard2005, fuchs2002quantummechanicsquantuminformation, grinbaum2005notionreconstructionquantumtheory}. Today a wide variety of reaxiomatizations of quantum theory are on offer: \cite{Chiribella2011, hardy2001quantumtheoryreasonableaxioms, Hardy2016, Hhn2017, Masanes_2011, Selby2021, vandeWetering:2018kcp, wilceRoyalRoadQuantum2018} give just a taste. This development coincided with the maturation of quantum information theory and the theory of quantum computation. Before this, quantum mechanics had largely been framed in terms of what the theory did \emph{not} allow one to do, e.g., measure position and momentum with arbitrary accuracy at the same time, or from complete information about a whole perfectly predict the behavior of the parts. The shift to the study of quantum communication protocols, quantum cryptography, and quantum algorithms was a shift in emphasis, to what one \emph{can} do with quantum theory, indeed, what would be impossible or difficult to do classically \cite{Nielsen_Chuang_2010}. There is a long prehistory to these ideas, stretching back into the 20th century, but by the end of the 1990s, due to the indefatigable work of a small band of quantum enthusiasts, it is fair to say that a new age of quantum mechanics dawned.

With it, came new ways of thinking about the quantum mechanical formalism as it was handed down by the early pioneers. Whereas Einstein's postulates for special relativity, or the laws of thermodynamics, appeal to physical principles from which the mathematics flows \cite{staceyQuantumTheorySymmetry2019}, quantum mechanics is distinctive in that its axioms, as laid down by von Neumann among others, have an almost entirely mathematical flavor. This, combined with the fact that quantum mechanical predictions are often counterintuitive to the classical mind, has led over the decades to a profusion of interpretations of the theory, from pilot waves \cite{Drr1992} to many worlds \cite{Vaidman_2022}. For most of these interpretations, the strategy is to rewrite the equations in some perspicacious way and then attempt to assign ontological significance to these mathematical expressions. From the beginning, however, another paradigm has been pursued, which recognizes that the comparison between classical and quantum mechanics can only take one so far. This paradigm instead tries to \emph{generalize} quantum mechanics, to place quantum mechanics in a landscape of possible theories, and by comparison throw its distinctiveness into new light.  Already in the 1930's, Wigner, Jordan, and von Neumann generalized the algebraic structure of quantum mechanical observables in hopes of understanding how essential matrix representations are to the theory, in the process classifying the so-called Euclidean Jordan algebras \cite{farautAnalysisSymmetricCones1994,Jordan1934-cm}, of which we will have much more to say. von Neumann and Birkhoff went on to pursue generalizations of Boolean logic under the heading of ``quantum logic'' \cite{Birkhoff1936}. Later Mackey, Davies, Ludwig, and others began instead from the \emph{convex structure} of quantum theory \cite{hartkamperFoundationsQuantumMechanics1974, Davies1970, ludwigAxiomaticBasisQuantum1985, mackey2004mathematical, Randall1978}, a theme that would be taken up in the 21st century under the heading of \emph{generalized probabilistic theories}.

The study of \emph{generalized probabilistic theories} \cite{BARNUM20113, barrettInformationProcessingGeneralized2007, Janotta_2014, lami2018nonclassicalcorrelationsquantummechanics, mullerProbabilisticTheoriesReconstructions2021, PLAVALA20231}, or GPTs, starts from an observation about what classical probability theory and quantum theory have in common: the state space is a convex subset of a vector space, and the effect space is a convex set of nonnegative linear functionals on that subset. It is easier to understand what this means with an example. The most familiar GPT is classical probability theory as it is used to formulate a theory of measurement, that is, a theory which attempts to understand observed behavior in terms of the unobserved. Given a set of mutually exclusive and exhaustive events $\{\lambda_i\}$, the familiar law of total probability tells us that we may write the probability of the outcome $E$ of a measurement as
\begin{align}
P(E)=\sum_i P(E|\lambda_i)P(\lambda_i).
\end{align}
Gathering up the probabilities $P(\lambda_i)$ into a column vector $|\sigma)$, and gathering up the conditional probabilities $P(E|\lambda_i)$ into a row vector  $(E|$, we can write the outcome probability as a linear functional acting on a vector $P(E) = (E|\sigma)$. In the language of GPTs, we would call $|\sigma)$ a \emph{state} and $(E|$ an \emph{effect}. Here the state has an interpretation as a state of uncertainty about which event $\lambda_i$ occurred, or in language borrowed from quantum foundations, which value a ``hidden variable'' takes. Another way it is sometimes put is that in classical theory, an observed outcome is always grounded in some underlying properties of a system which preexist the measurement. But not all GPTs are classical! Whereas the states of classical probability theory form a simplex, in general the states of a GPT may form an essentially arbitrary convex subset of a vector space, while the effects are a set of linear functionals which give valid probabilities when they act on states.

Indeed, it turns out quantum mechanics fits exactly this pattern. The Born rule tells us that the probability of an outcome $E$ can be written
\begin{align}
P(E) = \tr(E\sigma),
\end{align}
where on the right hand side of the equation, $E$ is a positive semidefinite matrix representing an effect, and $\sigma$ is a positive semidefinite matrix with trace 1, representing a state. Don't let the fact that these are matrices fool you: one may equivalently think of $\sigma$ as a vector, and $\tr(E\cdot)$ as a linear functional, just as in classical probability theory. In this sense, for both classical probability theory and quantum mechanics, the structure for calculating probabilities is the same\footnote{Although only in the first case can one interpret this structure as the ``law of total probability.''}. The convex sets involved, however, are quite different: for example, the space of states of the simplest quantum system, the qubit, may be identified with the surface and interior of a \emph{sphere}\footnote{The situation is much more complicated in higher dimensions.} in three dimensions. 

Thus far we've encountered a simplex state space, and a spherical state space. But why not any other essentially arbitrary convex shape? For instance, one could develop an analogous theory e.g., where the state space is a square, an icosahedron, or an ice-cream cone, as long as the shape is convex. The corresponding effect space must consist of linear functionals that give valid probabilities on those states, and one may proceed to study the properties of the resulting, generally nonclassical, theory---what it has in common with classical or quantum  theory, and how it differs in perhaps surprising ways \cite{Aubrun2022, barnumCloningBroadcastingGeneric2006, barnum2008teleportationgeneralprobabilistictheories, Garner_2017, Plvala2016}. In this way, GPTs offer a  minimal, but expressive framework for exploring possible theories. Atop this loose structure, one may introduce some well motivated postulates governing the behavior of the theory \cite{Barnum2014, Heinosaari2019, Krumm2019, PhysRevLett.108.130401, Wright2021}, and study how these postulates constrain the geometry of the state and effect spaces. Alternatively, one may start with a geometry \cite{Kolangatt2025} and study the resulting behavior of the theory.  In particular, over the years, GPT researchers have constructed many ``toy'' theories which challenge naive notions about what is the ``essence'' of quantum mechanics. 

For example, it is by now well known that both experiment and quantum mechanics violate Bell inequalities, and this is evidence against the applicability of so-called local realist models in physics \cite{Scheidl2010}. In the classic CHSH scenario  as usually described \cite{PhysRevLett.23.880}, two parties, traditionally called Alice and Bob, are distributed one qubit each (which might be realized as the spins of two electrons, or the polarizations of two photons) which are entangled in the ``singlet'' state. After being separated sufficiently that no signal could pass between them in the course of the experiment, Alice chooses to perform measurement $A_1$ or $A_2$ and Bob chooses to perform measurement $B_1$ or $B_2$. The correlations between the outcomes of these measurements according to quantum theory are in general inconsistent with the idea that the two qubits have some prearranged way of responding to measurements. In other words, if one rejects any kind of ``spooky action at a distance,'' one must reject the supposition that the outcomes of measurements reveal preexisting properties of the qubits. The CHSH inequality itself is a constraint on probability assignments compatible with classical assumptions, and quantum mechanics violates this inequality. But one could ask: is this the maximum conceivable violation in \emph{any} theory, compatible with locality? The answer, perhaps surprisingly, turns out to be no. A GPT affectionately known as \emph{boxworld} achieves the maximal violation compatible with the assumption that Alice and Bob's systems cannot signal each other: it consists in a composite GPT where the individual state spaces are not a simplex, nor a sphere, but instead a simple square \cite{barrettInformationProcessingGeneralized2007, mullerProbabilisticTheoriesReconstructions2021}.

In the same vein, in light of entanglement, quantum theory is often described as \emph{holistic}. As Schr\"odinger put it, one may find that the ``best possible knowledge of a whole does not include best possible knowledge of its parts---and that is what keeps coming back to haunt us'' \cite{cat}. One may ask, however: are there theories which are strictly \emph{more holistic} than quantum theory? Again, the answer is yes \cite{Hardy2011}. It turns out that, entanglement notwithstanding, the quantum state of a composite can be reconstructed solely from separate measurements on its parts, including their correlations (and assuming the same experiment can be run many times). This property is called \emph{local tomography}. One may consider a variant of quantum theory which is defined over the real numbers instead of the complex numbers\footnote{This provides, in fact, an example of a Euclidean Jordan algebra.}. The simplest such system, sometimes called a \emph{rebit}, has a state space corresponding to a disk. It turns out such a theory is \emph{bilocally tomographic}: to reconstruct the state of a composite one must perform measurements not only on the parts, but on \emph{pairs} of parts. One may further generalize this idea to construct theories which are $n$-locally tomographic \cite{Daki2014}, all of which are in this sense strictly more holistic than standard quantum theory.

These examples challenge qualitative assessments of what quantum mechanics is really about, whether by constructing classical models that reproduce certain aspects of the theory, or by proposing nonclassical models which exemplify yet more dramatically some notion taken to be properly quantum. The study of GPTs allows one to see ``around'' the givenness of the quantum formalism, and the hope in quantum foundations is that these new vantage points will bring into view ever more incisive ways of clarifying \emph{why quantum theory} among the widest range of potential theories.

 \subsection{QBism}

QBism \cite{fuchsQbismWhereNext2023, fuchs2019qbismquantumtheoryheros} is a research program in the foundations of quantum mechanics which begins by taking a stand on the nature of probabilities. The reader may be familiar with the fact that the history of probability theory is largely the story of the uneasy coexistence of two camps: on the one hand, the frequentists who hold that probabilities represent limiting frequencies (the proportion of times an event would occur given an infinite number of trials), and on the other hand, the Bayesians who hold that probabilities represent degrees of belief \cite{hackingEmergenceProbabilityPhilosophical2006}. QBism takes a \emph{subjective Bayesian} view of probabilities, following the great Italian probabilist Bruno de Finetti \cite{PhysRevA.104.022207, RevModPhys.85.1693}. Remarkably, de Finetti was able to \emph{derive} the laws of probability\footnote{ That probabilities should lie between 0 and 1; that $P(A \text{ or } B)=P(A) + P(B)$ for mutually exclusive events; that $P(A \text{ and } B)=P(A|B)P(B)$, and so on.} by defining the probability $P(E)$ as the price at which a gambler would be willing to buy or sell a ticket which pays \$1 if $E$ should occur, and demanding that no sequence of buying and selling tickets leads to a sure loss. In light of this, the laws of probability may be understood entirely in terms of ensuring a gambler's behavior is consistent: the rules protect the gambler from falling into a situation where they are \emph{guaranteed} to lose money, regardless of what events actually occur. In this sense, we may say that the laws of probability are \emph{normative ideals}: they are prescriptions for how one \emph{should} behave, supposing one has a stake in success. For example, the law of total probability $P(A)=\sum_i P(A|B_i)P(B_i)$, where $\{B_i\}$ is a set of mutually exclusive and exhaustive events, is a \emph{normative criterion} by which one may check consistency between the probability assignments which are implicated in it.  In short, for followers of de Finetti, probability theory is \emph{advice} by which an agent may regulate their behavior.
 
 QBism understands quantum theory in precisely the same way. Quantum theory should not be regarded as a \emph{description of physical reality}, but rather as a set of consistency constraints on one's probability assignments in light of nature's vitality \cite{staceyQuantumTheorySymmetry2019} as witnessed by the failure of classical models. This view is made the more convincing by rewriting the theory with respect to a \emph{reference measurement}. But what is a reference measurement? Suppose one is confronted with a cloud of particles. Classically, this cloud may be characterized by the positions and momenta of all the particles that make it up. If we consider the outcome $E$ of some measurement on the system, the law of total probability would urge on us the norm
 \begin{align}
 P(E)=\sum_{\{x, p\}} P(E|\{x, p\})P(\{x,p\}),	
 \end{align}
which relates $P(E)$ to one's conditional probabilities for $E$ given a particular configuration of positions and momenta as well as one's probability distribution over these positions and momenta themselves. What justifies this? First, classically, any other property of the cloud of particles supervenes on, or is grounded in, the positions and momenta. In this sense, a measurement of the positions and momenta serves as a reference measurement. At the same time, classically, it is a matter of indifference whether one actually performs this reference measurement or not. One may appeal to the law of total probability either way, since classically, we suppose that each particle ``has'' a position and ``is'' moving in a particular direction at a particular speed: the reference measurement simply reveals what is the case. But quantum mechanics forces us to question this assumption, that we may describe objects as having definite properties independent of measurement.

Indeed, nonclassicality of the form familiar from quantum theory arises by dropping the assumption that a system may be characterized by definite properties independent of measurement, while retaining the assumption that reference measurements exist. To dramatize this, we may then compare two scenarios. In the first, a reference measurement with outcomes $\{R_i\}$ is performed on a system after which an arbitrary measurement with an outcome $E$ is made. By the law of total probability,
\begin{align}
P_1(E) = \sum_i P(E|R_i)P(R_i).
\end{align}
In the second scenario, the measurement with outcome $E$ is made directly. Here the reference outcomes $\{R_i\}$ remain hypothetical. We will see, however, that assuming $\{R_i\}$ is a reference measurement implies that we may nevertheless appeal to a new nonclassical norm, which is in fact a \emph{deformation} of the law of total probability,
\begin{align}
\label{2}
P_2(E) &= \sum_{ij} P(E|R_i)\Phi_{ij}P(R_j).
\end{align}
There is no inconsistency since the law of total probability gives us no guidance should the intermediate events turn out to \emph{not happen at all}! Nevertheless, the second equation provides us a new consistency constraint to compensate for the loss. In fact this equation is nothing other than the Born rule, if you like, in disguise.

In short, QBism views quantum theory as a kind of ``user's manual'' \cite{fuchs2019qbismquantumtheoryheros} for anyone interested in gambling on a world which we may say, in light of the failure of classical models, is undergoing ceaseless creation.  From this point of view, QBism does not give up the ideal of a rational agent: rather, nonclassicality brings with it extra coherence relations which a gambler ought to impose on their probability assignments.  In a striking inversion, endowing the world with a fundamental openness translates into more restrictive constraints on the agent who wants to gamble well upon it. In fact, the same interpretation can be applied to any nonclassical GPT, although the particular constraints will differ: for the QBist, any GPT is just a convenient way for an agent to take stock of their probability assignments and evaluate them for consistency.

Philosophically, then, QBism takes the freedom of the interested agent as fundamental: upon acting on their world, the agent gets a free response as the world pushes back. This freedom, this vitality, this creative capacity, are the message of nonclassicality. Going still further, for QBists, the free response of nature is taken to be an \emph{experience} of the agent \cite{fuchsQbismWhereNext2023}, although this should not be understood too narrowly: indeed, the mutual cocreation of agent and world suggests that the traditional boundaries between subject and object must flex and bend. By taking the experience of the agent seriously, by positing that quantum theory is in fact a first person theory\footnote{As is any user's manual.}, QBism is able to resolve to some degree the interpretative difficulties e.g., raised by entanglement. In the CHSH scenario, rather than viewing Alice and Bob from a third person ``objective'' point of view from which vantage the correlations between the results of their measurements imply some kind of ``spooky \emph{context-dependence} at a distance,'' QBism urges one for example, to take the point of view of some third person, Charlie, who will perform a measurement on Alice, eliciting her result, and a measurement on Bob, eliciting his result, and who wants to gamble on the outcomes. The correlations in a Bell experiment are precisely correlations for Charlie, for whom the results of measurement are created, locally, on the spot.

More broadly, for QBists, nonclassicality urges us to grapple with a notion of experience which is not so much localized in someone's brain, but which is instead found at the threshold as self and other freely negotiate a boundary. This idea has led QBists to associate with phenomenologists and enactivists for whom experience and action are not separate \cite{berghofer2023phenomenology, Gefter:2024jbq}. What is real is on the surface: the world presents itself to us as invitations to act; cognition itself is embodied, embedded into an environment that cocreates it, enacted not only through neurons firing, but also through everything an organism does, so that we may regard mind itself to be extended out into what was traditionally called the world \cite{Rowlands2010-pz}. This too is part of the message of nonclassicality, the failure of classical models to adequately provide good advice. 

As we have observed, from a technical point of view, the QBist reconstruction project begins from the observation that distributions on reference outcomes ought to be restricted in some way. In GPT language, the valid reference distributions ought to live in the special subset of the probability simplex which corresponds to the image of the GPT state space induced by the choice of reference measurement. Wandering out of this subset will generally yield negative probabilities. The question for QBists has therefore been: how can one motivate the particular shape of quantum state space within the probability simplex on the outcomes of a well-chosen reference measurement? As we noted earlier, this led to the introduction of the notion of a \emph{qplex}, a generalization of the image of quantum state space within the simplex induced by a symmetric informationally complete (SIC) measurement. Just as the landscape of GPTs allows one to place quantum theory in a broader context, so too the landscape of qplexes allows the particular features of quantum theory to be brought to the fore. \cite{Appleby_2017} showed how one can pick out the qplex corresponding to quantum theory on grounds of symmetry. In this dissertation, however, we take a different approach, instead generalizing the image of quantum state space induced by a so-called complex projective 3-design reference measurement \cite{waldronIntroductionFiniteTight2018}. 

Complex projective or quantum state $t$-designs are ensembles of states whose $t$th moment reproduces the $t$th moment of quantum state space as a whole, and are widely used in quantum information theory. In particular, the structure of 3-designs is intimately tied to the Jordan algebraic structure of quantum theory \cite{obst2024wignerstheoremstabilizerstates}, and so we will make much use of the pioneering work of Jordan, von Neumann and Wigner in formalizing and classifying the Euclidean Jordan algebras \cite{wilceRoyalRoadQuantum2018, Jordan1934-cm}.   Jordan began by noting that quantum observables form a commutative, but not associative algebra under the symmetric product $A \odot B = (AB+BA)/2$. Generalizing the essential properties of this product, he gave birth to the notion of a Jordan algebra, and already by 1934, von Neumann, Wigner, and Jordan demonstrated that any so-called Euclidean Jordan algebra must be a composite of simple Jordan algebras, corresponding essentially to quantum mechanics over different number fields. Later the work of Koecher and Vinberg \cite{kriegMinnesotaNotesJordan1999} showed that such Jordan algebras are in correspondence with  \emph{symmetric cones}, giving them an entirely geometric interpretation \cite{farautAnalysisSymmetricCones1994}. Work on Jordan algebras continues to this day: they are of interest to pure mathematicians, as well as those working in convex optimization \cite{alizadehIntroductionFormallyReal2012}, not to mention those working in quantum foundations \cite{wilceRoyalRoadQuantum2018, barnumStronglySymmetricSpectral2019}. In particular, we show that for QBists, the import of Euclidean Jordan algebras lies in the interplay between valuations on the outcomes of arbitrary measurements and their equivalent valuations on reference outcomes. If one minimally deforms the classical rule for multiplying valuations in a way analogous to how one deforms the classical law of total probability into the Born rule, then demanding that there exists a class of measurements for which the $m$th moment of a reference valuation with respect to the nonclassical product reproduces the $m$th moment of the original valuation, one is led almost inevitably to a Euclidean Jordan algebra. By giving a QBist motivation for the Jordan algebraic structure of quantum theory, we hope to open up a new chapter in the QBist reconstruction project.

\section{Chapter summaries}

With this background established, we offer detailed chapter summaries to give a sense of the general flow of argument in the dissertation.

\vspace{0.5cm}

\B{Chapter \ref{ch:probabilities}: Probabilities}. This chapter lays the conceptual and formal groundwork for everything that follows. We begin in subjective Bayesian style by presuming that there is an agent who is trying to make better decisions. In accordance with the philosophy of QBism, we regard the agent as free to act on a world beyond them while the world has the freedom to return back consequences beyond the agent's control. We first derive the familiar laws of probability theory in the style of de Finetti \cite{ProbabilitiesBettingOdds}, interpreting probabilities as bets, and demanding that no sequence of bets leads the agent to a sure loss. This demand implies that the gambler's bets must satisfy certain standards of consistency, which turn out to be just the rules of probability theory. We also discuss the use of van Fraassen's reflection principle to make sense of how an agent should think of their probability assignments across time \cite{vanfraassenBeliefWill1984}. We extend the analysis to random variables more generally \cite{definettiTheoryProbabilityCritical2017, definettiForesightItsLogical1992}, and show that valuations on events form an algebra. We then discuss how to treat gambles on the outcomes of several experiments, and introduce the de Finetti representation theorem which gives subjective Bayesian meaning to the idea of ``estimating unknown probabilities,'' e.g., inferring the probability that a coin will come up heads or tails from the results of flipping the coin many times. If a probability is the fair price the gambler assigns to a ticket, how can a probability be ``unknown''? The theorem provides an answer: if the joint distribution on many coin flips is so-called exchangeable, it may be expressed in terms of a underlying distribution on the bias of the coin, and so the gambler may act \emph{as if} there were an unknown probability they are trying to estimate. Next, we emphasize how the law of total probability $P(E)= \sum_i P(E|R_i)P(R_i)$ holds precisely because the events $\{R_i\}$ are mutually exclusive and exhaustive so that the gambler believes that exactly one of them will occur. What if, however, the events $\{R_i\}$ are called off? Is there any coherence condition which yet relates the probabilities $P(E)$, $P(E|R_i)$ and $P(R_i)$? In the special case that these events can be related to another set of mutually exclusive and exhaustive events $\{\lambda_i\}$ of a certain kind, then all is not lost. We show that we may then rewrite the law of total probability in terms of a matrix $\Phi$, giving us our first taste of a formal structure which will be thoroughly investigated in this work. We then discuss the idea of a reference measurement, another central theme, at first grounding it in classical notions. We make the central nonclassical assumption:  we drop the idea that a system is characterized by definite properties independent of measurement, but retain the idea of a reference measurement. We develop a series of definitions for system, state, and effect, grounded solely in the relations between the probabilities that the gambler assigns to the outcomes of measurement, so that a notion of state and effect spaces with convex geometry emerges. We then come to the centerpiece of this chapter. We compare two scenarios. In the first, a measurement $A$ is performed, followed by the reference measurement, followed by a measurement $B$; in the second, $A$ is performed and then $B$ directly. We prove that under very general assumptions, the probabilities the agent assigns in the two scenarios may be related by a nonclassical coherence condition: this is our core representation theorem. In the first scenario, the agent ought to use the familiar law of total probability $P(B|R, A) = P(B|R)P(R|A)$ whereas in the second they ought to use $P(B|A) = P(B|R)\Phi P(R|A)$, where we have employed here a matrix notation which we will frequently use throughout the text. In other words, the nonclassical coherence rule takes the form of a deformation of the law of total probability by the presence of the Born matrix $\Phi$, where $\Phi$ satisfies $P(R|R)\Phi P(R|R)= P(R|R)$: here $P(R|R)$ is the conditional probability matrix which characterizes the reference measurement itself. We then briefly discuss how to incorporate dynamics into this formalism, in QBist spirit appealing to van Fraassen's reflection principle to argue that assigning a particular dynamics is equivalent to the agent's being indifferent to whether an intermediate measurement is performed. Finally we treat briefly the subject of when the gambler might be led to carve subsystems out of a whole.

\B{Chapter \ref{ch:born_identity}: The Born Identity}. Having established the fundamental nonclassical coherence condition in the previous chapter, we turn to discussing the mathematical properties of the Born matrix $\Phi$, identifying it as a particular kind of generalized inverse of $P(R|R)$, namely, a $\{1\}$-inverse \cite{GeneralizedInverses2003a}. Indeed, previous accounts of the Born matrix focused their attention on the case that $P(R|R)$ is invertible; an important innovation of the present account is a systematic treatment in the case that $P(R|R)$ is singular. We show that the fundamental coherence condition in fact restricts states $P(R|\rho)$ to live in the column space of $P(R|R)$ and effects $P(E|R)$ to live in the row space of $P(R|R)$. We then introduce a central tool of analysis: the rank factorization of a matrix, which allows us to represent states and effects as $r$-dimensional vectors, where $r=\rank(P(R|R))$. The $\{1\}$-inverse condition, that $P(R|R)\Phi P(R|R)= P(R|R)$ can be reinterpreted as a resolution of the identity $\B{S}\Phi \B{R}=I_r$, where the rows of $\B{R}$ are compact representatives of the reference effects and the columns of $\B{S}$ are compact representatives of the reference states. This representation makes clear that in fact our framework is essentially equivalent to the framework of generalized probabilistic theories \cite{mullerProbabilisticTheoriesReconstructions2021}. Our derivation of the nonclassical coherence condition $P(E|\rho) = P(E|R)\Phi P(R|\rho)$ amounts to a proof of a simplified version of Ludwig's embedding theorem \cite{lami2018nonclassicalcorrelationsquantummechanics}, which ensures that any GPT has a vector space representation. We discuss the two core examples of GPTs: classical probability theory and quantum mechanics. We then develop the key ideas and constructions of frame theory, showing that if one starts with a GPT, then our reference measurement formalism arises as a ``double-sided'' frame representation. Indeed, we define more rigorously what a reference measurement is from the GPT point of view, introducing the notion of informational completeness and its ability to separate states or effects. We reproduce a classic proof that reference measurements whose effects form a basis, that is, minimal informationally complete (MIC) measurements, exist in any unrestricted GPT (which allows all mathematically possible states and effects), and explain how the de Finetti representation theorem can be adapted to a wide class of GPTs\footnote{There is an interesting analogy here. The quantum de Finetti theorem was originally proven in the 1970's \cite{hudsonLocallyNormalSymmetric1976}, but a much more straightforward and conceptually satisfying proof was provided in \cite{cavesUnknownQuantumStates2002} by introducing a reference measurement and appealing to the classical de Finetti representation theorem. The proof for GPTs follows the same pattern. Similarly, Ludwig proved his original theorem in \cite{ludwigAxiomaticBasisQuantum1985, lamiNonclassicalCorrelationsQuantum2018} building on his work from the 1970's. The introduction of a reference measurement makes the finite dimensional version of the proof straightforward.} to make sense of the tomographic procedure of ``estimating unknown states.''

\B{Chapter \ref{ch:urgleichungen}: Die Urgleichungen}\footnote{Material in this chapter is based on \cite{weiss2024depolarizingreferencedevicesgeneralized}.}. In this chapter, we discuss the conditions under which the Born matrix may be chosen to take a very simple form, $\Phi = \alpha I + (1-\alpha)wu^\dagger$ so that $P(E|\rho)=\sum_i P(E|R_i)\big\{\alpha P(R_i|\rho) + (1-\alpha)P(R_i|\mu)\big\}$, where $|\mu)$ is a distinguished state which plays the role of the state of complete uncertainty and $J$ is the matrix of all 1's. We call it the \emph{Protourgleichung}. We first show that any GPT can be brought into a standard ``Bloch'' form, introducing a block matrix notation which facilitates the proofs of this chapter. Our first theorem is that in any unrestricted GPT, given a reference measurement, one may always construct a set of reference states which allows $\Phi$ to take the abovementioned form: we conclude therefore the Protourgleichung is ubiquitous. An alternative characterization of such reference measurements may be given in terms of the corresponding channel operator: $\Phi$ may take Protourgleichung form iff the channel operator is depolarizing. In this case, the matrix $P(R|R)$ acts simply on vectors in its column space, depolarizing them in just such a way as to invert the action of the Protourgleichung $\Phi$. We then adapt the recently introduced notion of morphophoricity \cite{slomczynskiMorphophoricPOVMsGeneralised2020, szymusiakCanQBismExist2025} to our setting, in fact extending the definition to include weighted morphophoric measurements. Such measurements imply that the reference measurement maps the GPT state space into the probability simplex in a shape-preserving way. We show that weighted morphophoric measurements always exist in an unrestricted GPT, and that for a MIC, the choice of reference states which give a depolarizing channel is fixed up to scale/sign. A consequence: the effects of any unbiased morphophoric MIC measurement must form a regular simplex, that is, a symmetric informationally complete measurement, and we give an example from quantum theory. Continuing to connect the result to quantum theory, and elaborating on the famous result of Andrew Scott \cite{scottTightInformationallyComplete2006, slomczynskiMorphophoricPOVMsGeneralised2020}, we provide an alternative proof that unweighted morphophoric quantum reference measurements whose effects are proportional to pure states must correspond to complex projective 2-designs. Finally, we turn to a quantitative measure of how much the law of total probability is deformed, $\lVert I - \Phi \rVert$, with respect to any unitarily invariant norm. This measure has been discussed primarily for MIC measurements in quantum theory \cite{debrotaSymmetricInformationallyComplete2020}. We discuss the challenges that come from defining this measure for reference measurements which form overcomplete frames. In particular, we show that there are many choices of $\Phi$ matrix which act identically on the relevant subspace $\col(P)$ as a Protourgleichung but which have different values for the LTP deformation $\lVert I - \Phi \rVert$. Moreover, we show by the method of Lagrange multipliers that for the Frobenius norm, the unique Born matrix which minimizes the LTP deformation over the choice of $\Phi$ matrix does not necessarily act as a Protourgleichung in general: we discuss the special case that it does. We thus conclude with a word of caution about the general applicability of this measure.

\B{Chapter \ref{ch:classicality}: (No) return to classicality}. When does a set of probabilities across different experiments have a classical explanation? As we saw in chapter \ref{ch:probabilities}, merely being able to write a coherence relation in terms of a Born matrix is not enough. In order to answer this question, we adapt Spekkens's notion of a noncontextual ontological model \cite{schmidCharacterizationNoncontextualityFramework2021, schmidStructureTheoremGeneralizednoncontextual2024, shahandehUnifiedLinearAlgebraic2025} to the QBist way of thought, emphasizing that the key notion of classicality is the indifference of the gambler to whether a reference measurement is performed or not. We discuss the notion of contextuality from a linear algebraic point of view, relating it to the notion of a rank factorization, and following recent work \cite{shahandehUnifiedLinearAlgebraic2025}, show how noncontextual classical models may be understood in terms of nonnegative matrix factorizations. We then give a QBist spin on the equivalent notion of a simplex embedding, a linear embedding of states and effects into the probability simplex and the dual space of response functions in a probability preserving way. In particular, we show how the noncontextuality of an ontological model of a scenario which includes a reference measurement $\{R_i\}$ is equivalent to the assumption that $\{R_i\}$ forms a reference for the ultimate classical reference measurement $\{\lambda_i\}$. We discuss the computational details of calculating simplex embeddings, in terms of alternating projections, as well as a more sophisticated scheme using linear programming supplemented with the ability to switch from the vertex to the halfspace representation of a convex set or cone \cite{selbyLinearProgramTesting2024}. While the result is not original, we show how the construction may be understood in terms of a variant of the Born matrix, and in a special circumstance the Born matrix itself. In such a case the question turns on whether a nonnegative Born matrix can be found. We then adapt an old result \cite{tamGeometricTreatmentGeneralized1981} characterizing the conditions under which $\Phi$ may be taken to be a stochastic matrix. We then turn to the subject of Bell inequalities and more generally noncontextuality inequalities. For the QBist, the key point is the gambler is well advised to use the nonclassical coherence condition $P(E|R) = P(E|R)\Phi P(R|\rho)$. Attempts to build so-called Bell local models simply defy this rule despite the fact that performing a reference measurement or not makes a difference. After recalling the general setup of Bell inequalities, focusing on the CHSH scenario, we show that in fact Bell inequality violations may be reinterpreted as providing lower bounds on $\lVert I - \Phi \rVert$. In this way, such violations may be understood as quantifying the cost of abandoning the fundamental nonclassical coherence condition. We show that the same analysis may be applied to noncontextuality inequalities more generally. 

\B{Chapter \ref{ch:characterizing}: Characterizing quantum state space with a single quantum measurement\footnote{Material in this chapter is based on \cite{weissCharacterizingQuantumStatespace2025}.}}. So far we have discussed nonclassicality in very general terms, and in some sense that's all we need: given enough experimentation, adherence to the fundamental nonclassical coherence rule is enough to guarantee consistency with e.g., quantum mechanics. But we desire something more: to characterize the state and effect spaces of quantum theory specifically in coherentist terms. In this chapter, we show that remarkably the state space of $d$-dimensional quantum theory can be derived from studying the behavior of a single reference measurement---if the reference measurement corresponds to a complex projective 3-design. In this privileged case, not only does each quantum state correspond to a probability distribution over the outcomes of a single measurement, but also the probability distributions which correspond to quantum states can be elegantly characterized as those which respect a generalized uncertainty principle. The latter takes the form of a lower bound on the variance of a natural class of observables as measured by the reference. We give simple equations which pure state probability distributions must satisfy, and contextualize these results by showing how 3-designs allow the structure coefficients of the Jordan algebra of observables to be extracted from the probabilities which characterize the reference measurement itself. In fact, just as 2-designs make the Born rule appear as a gentle modification of the law of total probability, 3-designs make the quantum Jordan product appear as a gentle modification of the classical Jordan product on valuations. Taken as a whole, this is in line with the QBist view that quantum theory can be primarily understood as a set of normative constraints on probability assignments on reference outcomes which reflect nature's lack of hidden variables. Moreover, our result further cements the significance of 3-designs, already of interest e.g., for their use in classical shadow estimation protocols, in quantum information science. In an appendix, we also discuss how unitary maps corresponding to symmetries of the state space manifest in Jordan algebraic terms.

\B{Chapter \ref{ch:reconstruction}: Reconstruction}: This chapter forms a pair with the previous. There, we showed how 3-design reference measurements give rise to a remarkably simple characterization of quantum mechanics in probabilistic terms. The central insight is that just as 2-designs make the Born rule look as close as possible to the law of total probability, 3-designs make the quantum rule for multiplying valuations on the reference measurement---the Jordan product on $d\times d$ Hermitian matrices over $\mathbb{C}$---look as close to the classical rule as possible, the latter being the Hadamard or elementwise product on valuations. Indeed, this is the Jordan product on $\mathbb{R} \oplus \dots \oplus \mathbb{R}$. In this chapter, we start from scratch and attempt to climb our way back to quantum mechanics, and a 3-design reference measurement in particular, by placing ever more severe restrictions on $P(R|R)$, the conditional probability matrix which characterizes the reference measurement itself. On the one hand, this is an exercise in quantum foundations in the spirit of the QBist reconstruction of quantum mechanics inspired by SIC-POVMs; on the other hand, it provides a practical means of self-testing (unbiased) quantum $t$-designs for $t\ge 3$, since the results of the chapter show precisely what restrictions on $P(R|R)$ guarantee a Hilbert space representation. We begin by assuming that $P\equiv P(R|R)$ takes a simple form: symmetric, with a constant along the diagonal. The assumption that $\Phi = \alpha I + (1-\alpha)J/n$ is a $\{1\}$-inverse of $P$ restricts the eigenstructure of $P$ and implies that $P$ depolarizes vectors in its column space. We then discuss how valuations on arbitrary measurements can be transferred to equivalent valuations on the reference measurement. We prove that any deformation of the classical rule for multiplying valuations which satisfies a handful of desiderata motivated by the idea that we seek the gentlest possible modification of the classical rule must take the form $x \odot y = \gamma (x \circ y) + (1-\gamma)(\overline{y} x + \overline{x} y - (x \cdot y)u)/n$ for some constant $\gamma$ and where $u$ is the vector of all 1's. The hope is that on the one hand, the classical product $x \circ y$, the entrywise product of valuations, has an equivalent valuation on some measurement, as does the nonclassical product of valuations $x \odot y$. The difference between them, we suppose, is that $x \odot y$ ought not to depend on the choice of reference measurement. Defining an inner product which reproduces the nonclassical probability rule, and demanding that $\odot$ be consistent with this inner product, fixes the value of $\gamma$ in terms of $\alpha$: moreover, it shows that the algebra defined by the nonclassical product is so-called formally real. We then demand that for some class of measurements, the $m$th power of a reference valuation with respect to our product reproduces the $m$th moment of the original measurement's valuation. To make this coherent, we must assume that the product is power associative, so that the $m$th power has an unambiguous meaning. By a classic result of Jordan, von Neumann, and Wigner \cite{Jordan1934-cm}, imposing power associativity (along with our other assumptions) is equivalent to imposing that the nonclassical product satisfies the so-called Jordan identity. This provides a means of implementing power associativity in terms of constraints on the matrix entries of $P(R|R)$ itself, which we work out in terms of the third order polarization of our product. Now that the algebra of valuations must be a Euclidean Jordan algebra, we can appeal to many powerful results. First, we lift our product on valuations to a product on probability distributions and response functions, recasting the Jordan identity in those terms. We then discuss how state and effect spaces arise from a Euclidean Jordan algebra in terms of the self-dual cone of squares with respect to the Jordan product. In fact, this leads to a means of characterizing the state space in terms of an uncertainty principle: a distribution $P(R|\rho)$ is valid iff it satisfies this uncertainty principle, and this constraint can be reformulated in terms of the postive semidefiniteness of a matrix $\mathcal{L}_{P(R|\rho)}$, which in fact is the linear operator which performs the (lifted) Jordan product itself. Further appealing to self-duality, we rescale our product by a self-duality constant $\kappa$ into a form convenient for discussing the idempotents of the Jordan algebra. We show how many key linear algebraic results port to the Jordan algebraic setting. In fact, we make good on our original ambition: the spectral theorem for Euclidean Jordan algebras guarantees that any reference valuation is equivalent to a valuation on a measurement whose effects form mutually orthogonal idempotents. We give both vector and scalar characterizations of the primitive idempotents, as well as for their orthogonality, and show that our assumptions imply that the reference states themselves must be primitive. Finally, after explaining how to check whether $P(R|R)$ gives rise to a simple Euclidean Jordan algebra, we appeal to the famous classification of the latter to nail down quantum theory: the rank of $P(R|R)$ must be the square of the maximum number of mutually orthogonal primitive idempotents. This fixes the values of $\kappa$ and $\alpha$, and we then show that our assumptions imply that the reference measurement forms a complex projective 3-design. We show how the above considerations, while motivated by foundational concerns, amount to an algorithm for self-testing complex projective $t$-designs for $t \ge 3$: indeed, the constraints that we put on $P(R|R)$ guarantee that the states and effects have a Hilbert space representation, and once this representation is secured, one can rely on the minimization of the frame potential to certify higher order designs. Finally, recapitulating our results, we give an overview of what questions remain open about our construction, and speculate about future directions for research into the subject.

\vspace{0.5cm}

\emph{An open-source python library, \texttt{redesigning}, implementing many of the constructions developed throughout this dissertation has been made available \cite{matthewweissHeyredhatRedesigningV12026}}.

%% file: sections/probabilities.tex
\UMBchapter{Probabilities}
\label{ch:probabilities}

\section{Introduction}

We begin from the presumption that there is an agent---You---who wants to make better decisions. The agent is free to act on a world beyond them, but the world returns back consequences beyond the control of the agent: that same freedom the agent has, the world has too. How should the agent organize their behavior in light of this mutual freedom? We begin by developing standard probability theory along subjective Bayesian lines. Then, after defining the notion of reference measurement, state, and effect, we prove a fundamental representation theorem which narrows down the form of the fundamental nonclassical coherence condition necessary to respect nature's openness. 

\section{de Finetti style derivation of probability theory}

In the 1930's, the great Italian probabilist Bruno de Finetti taught us that if we interpret probabilities as bets, the laws of probability can be derived from the simple demand that a gambler ought to agree only to buy or sell tickets in such a way that avoids a guaranteed loss \cite{definettiTheoryProbabilityCritical2017, definettiForesightItsLogical1992, platoCreatingModernProbability1994}. Of course, you might lose any given bet: your horse simply may not come in. But you might find yourself making a series of transactions such that \emph{regardless of what happens}, you end up losing money. This is what probability theory helps you prevent from happening by teaching you how your probability assignments ought to mesh together. In this sense, the laws of probability are not laws in a Newtonian sense, laws which ``govern,'' but are instead essentially good advice, norms which one ought to follow to the extent that one has a stake in success. This is the starting point of subjective Bayesianism \cite{jeffreySubjectiveProbabilityReal2004}, and in what follows we hew closely to the account given in \cite{ProbabilitiesBettingOdds}.

\begin{definition}[Probability]
	The probability $P(E)$ of an event $E$ is the price at which a gambler is willing to buy or sell a lottery ticket promising ``Pay \$1 if $E$.'' 
\end{definition}

\noindent Although this definition may seem arbitrary, it is not hard to show that we may take such a ticket to be fundamental. The following proof is illustrative of the general style of argument, which proceeds by appealing to the equivalence of holding different ensembles of tickets.

\begin{theorem}
If a gambler is willing to buy or sell a ticket ``Pay \$1 if $E$'' at $P(E)$, then they ought to be willing to buy or sell a ticket ``Pay \$$x$ if $E$'' at $xP(E)$.
\end{theorem}
\begin{proof}
\label{arbitrary_payout}
Suppose $x=\frac{p}{q}$, a rational number, and consider the following three tickets,
\begin{enumerate}[label=(\arabic*)]
\item ``Pay \$1 if $E$''
\item ``Pay \$$\frac{1}{q}$ if $E$''
\item ``Pay \$$\frac{p}{q}$ if $E$.''	
\end{enumerate}
Clearly, holding $q$ tickets of type (2) is equivalent to holding a single ticket of type (1). The gambler therefore ought to be willing to buy or sell at the same price for both: $P_1(E)=qP_2(E)$, so that $P_2(E)=\frac{1}{q}P_1(E)$. But holding $p$ tickets of type (2) is equivalent to holding a single ticket of type (3): $P_3(E)=pP_2(E)$. Thus  $P_3(E)=\frac{p}{q}P_1(E)$. By the idealization of continuity, the gambler ought to be willing to buy or sell a ticket ``Pay \$$x$ if $E$'' at $xP(E)$ for any real number $x$.
\end{proof}

\noindent We now proceed to derive the basic laws of probability.
\begin{theorem}
A probability $P(E)$ ought to satisfy $0 \leq P(E) \leq 1$.	
\end{theorem}
\begin{proof}
	On the one hand, suppose the gambler is willing to sell a ticket ``Pay \$1 if $E$'' for a price $P(E)<0$. Equivalently, they are willing to pay someone to take it off their hands, which leads to a sure loss. We therefore require $P(E)\ge0$. On the other hand, suppose the gambler is willing to buy such a ticket for a price $P(E)>1$. Equivalently, they are willing to pay more for the ticket than it will ever pay off, guaranteeing a loss: so we require $P(E)\leq 1$. 
\end{proof}
\begin{remark}
If a gambler is certain that an event $E$ will occur,  they ought to assign $P(E)=1$, since otherwise they must expect to lose money. Conversely, if a gambler sets $P(E)=1$, then they must expect to lose money unless they are certain that $E$ will occur.	
\end{remark}
\begin{theorem}
	$P(A \text{ or } B)=P(A)+P(B)$, for mutually exclusive events $A, B$.
\end{theorem}
\begin{proof}
A gambler judges events $\{A_i\}$ to be \emph{mutually exclusive} if they believe only one of them will occur. Consider the following three tickets:
\begin{enumerate}[label=(\arabic*)]
\item ``Pay \$1 if $A$ or $B$''
\item ``Pay \$1 if $A$''
\item ``Pay \$1 if $B$.''	
\end{enumerate}
Suppose  the gambler buys $(1)$ and sells $(2)$ and $(3)$. From buying $(1)$, if $A$ or $B$ occurs, the gambler  makes $1-P(A \text{ or } B)$.  Suppose $A$ occurs. From selling $(2)$, the gambler makes $P(A)-1$. From selling $(3)$, since only one of $A$ or $B$ can occur and $B$ did not occur,  the gambler makes $P(B)$, for a total of $P(A)+P(B)-1$. Moreover, this is just the same in the case that $B$ occurs.  To avoid a sure loss, we require
\begin{align}
\big(1-P(A \text{ or } B)\big)+\big(P(A)+P(B)-1\big)\ge0	\Longrightarrow P(A \text{ or } B)\leq P(A)+P(B).
\end{align}
Alternatively, suppose the gambler sells $(1)$ and buys $(2)$ and $(3)$. From selling $(1)$, if $A$ or $B$ occurs, the gambler makes $P(A \text{ or } B)-1$, and from buying $(2)$ and $(3)$, the gambler makes $1-P(A)-P(B)$. To avoid a sure loss, then
\begin{align}
\big(P(A \text{ or } B)-1\big) + \big(1- P(A) - P(B)\big) \ge 0 \Longrightarrow P(A \text{ or } B)\ge P(A)+P(B).	
\end{align}
We conclude that the gambler ought to set $P(A \text{ or } B)=P(A)+P(B)$.
\end{proof}
\begin{remark}
Clearly the above proof can be extended to any finite number of mutually exclusive events. Under the idealization that the gambler can buy or sell a countably infinite number of tickets whose payout depends on a countable infinity of mutually exclusive events, one arrives at the countable additivity of probabilities.	This however was no small matter of controversy for de Finetti himself \cite{platoCreatingModernProbability1994}.
\end{remark}
\begin{theorem}
	$P(A \text{ and } B)=P(A|B)P(B)$ for events $A, B$.
\end{theorem}
\begin{proof}
Consider the following three tickets:
\begin{enumerate}[label=(\arabic*)]
\item``Pay \$1 if $A$ and $B$; pay \$$P(A|B)$ if not $B$''
\item ``Pay \$1 if $A$ and $B$''
\item ``Pay \$$P(A|B)$ if not $B$.''
\end{enumerate}
Let $P(A|B)$ be the price at which the gambler is willing to buy or sell (1), so that the gambler will get a refund if $B$ does not occur. Holding the ticket at that price means they will win \$1 if $A$ occurs given that $B$ occurs, and so we can think of $P(A|B)$ as a conditional probability. Meanwhile, let $P(A \text{ and } B)$ be the price the gambler is willing to buy or sell $(2)$, and for (3), notice that by Theorem \ref{arbitrary_payout}, the gambler ought to set price $P(A|B)P(\text{not B})$. Clearly, holding ticket $(1)$ is equivalent to holding $(2)$ and $(3)$, and so they should have the same price:
\begin{align}
P(A|B) &= P(A \text{ and } B)+P(A|B)P(\text{not }B)\\
&=	P(A \text{ and } B)+P(A|B)\big(1-P(B)\big)\\
&=P(A \text{ and } B)+P(A|B)-P(A|B)P(B),
\end{align}
from which we conclude $P(A \text{ and }B)=P(A|B)P(B)$.
\end{proof}

\begin{remark}
If the gambler judges that $A$ does not depend on $B$, then $P(A|B) = P(A)$, and $P(A \text{ and } B) = P(A)P(B)$.
\end{remark}

\begin{corollary}[Bayes's Rule]
	Since $P(A \text{ and } B) = P(A|B)P(B) = P(B|A)P(A)$, if $P(A)\neq 0$, 
	\begin{align}
	P(B|A) &= \frac{P(A|B)P(B)}{P(A)}.
	\end{align}
\end{corollary}

\noindent With these theorems established, we have already erected essentially the whole structure of probability theory as usually formalized by Kolmogorov's axioms.  We now prove a useful lemma, the law of total probability (LTP). 

\begin{lemma}
	$P(A)=\sum_i P(A|B_i)P(B_i)$ for an event $A$ and a collection of mutually exclusive and exhaustive events $\{B_i\}_{i=1}^n$.
\end{lemma}
\begin{proof}
A gambler judges events $\{B_i\}$ to be \emph{exhaustive} if they believe at least one of them must occur. From the relationship between joint and conditional probabilities and from the additivity of probabilities, we have
\begin{align}
	\sum_i P(A|B_i)P(B_i)&=\sum_i P(A \text{ and } B_i)\\
	&=P\big((A \text{ and } B_1) \text{ or }  \dots \text{ or } (A \text{ and } B_n)\big)\\
	&=P\big(A \text{ and }(B_1 \text{ or } \dots \text{ or } B_n)\big).
\end{align}
The event ($B_1$ or $\dots$ or $B_n$) is just the event that some $B_i$ occurs. But since the gambler believes $\{B_i\}$ is a set of mutually exclusive and exhaustive events, they are certain that exactly one of the $B_i$'s will occur. Thus they ought to buy or sell a ticket ``Pay \$1 if $A$ and ($B_1$ or $\dots$ or $B_n$)'' at the same price as a ticket ``Pay \$1 if $A$.'' We conclude that $P(A)=\sum_i P(A|B_i)P(B_i)$.
\end{proof}

\begin{corollary}[Bayes's Rule: LTP version] Given an event $A$ such that $P(A)\neq 0$, and a collection of mutually exclusive and exhaustive events $\{B_i\}$,
\begin{align}
P(B_i|A)=\frac{P(A|B_i)P(B_i)}{P(A)}=\frac{P(A|B_i)P(B_i)}{\sum_j P(A|B_j)P(B_j)}.
\end{align}
\end{corollary}

\section{Reflection}

Bayes's rule tells us that a gambler ought to set $P(A|B) = P(A, B)/P(B)$. Here $P(A|B)$ expresses the fair price the gambler has assigned at the present moment (call it $t=0$), that is, before the events $A$ and $B$ have been resolved. But what about later, say at some time $t = \tau$? Suppose event $B$ befalls the gambler. Must they set $P_\tau(A)= P_0(A|B)$, where we have tagged probabilities with subscripts denoting the time? This is the standard Bayesian update rule, but in fact, without further assuptions, there is no coherence argument which would compel the gambler to do so \cite{hackingSlightlyMoreRealistic1967, debrota2024quantumdynamicshappenspaper}. Nevertheless, one may argue from coherence that a gambler's probability assignments at different times ought to be constrained in a particular way, summarized by van Fraassen's \emph{reflection principle} \cite{vanfraassenBeliefWill1984, goldsteinPrevisionPrevision1983, shaferSubjectiveInterpretationConditional1983, fuchsBayesianConditioningReflection2012}. We follow closely the exposition in \cite{debrota2024quantumdynamicshappenspaper}. Crucially, in order to formulate the reflection principle we must consider the gambler's beliefs now \emph{about their own future probability assignments}. For example, $P_0(P_\tau(A) = q)$ signifies how strongly the gambler believes now (at $t=0$) that at time $t=\tau$ they will assign probability $q$ to event $A$. The reflection principle constrains the probabilities the gambler assigns now, given their beliefs about how they will assign probabilities in the future.

\begin{theorem}[Reflection principle]
Let $P_0(A)$ be the probability the gambler assigns to event $A$ at $t=0$, and let $P_\tau(A)$ be the probability the gambler assigns to event $A$ at $t=\tau$. Suppose that $P_{0}(P_{\tau}(A) = q) = 1$. Then coherence demands that $P_0(A)=P_\tau(A)=q$.
\end{theorem}
\begin{proof}
Suppose at $t=0$, the gambler sets $P_{0}(P_{\tau}(A) = q) = 1$: that is, the gambler at $t=0$ is certain that at time $\tau$ they will assign probability $q$ to event $E$. This means that if $q< P_0(A)$, then at $t=0$ the gambler is willing to buy a ticket at price $\$ P_0(A) $ even as they also believe that later they will be willing to sell it for the lower price $\$q$. Similarly, if $q > P_0(A)$, then at $t=0$, the gambler is willing to sell a ticket at $\$P_0(A)$ even as they also believe that later they will be willing to buy it for a higher price $\$q$. Either way, unless $P_0(A) = P_\tau(A) = q$, the gambler faces a sure loss.
\end{proof}

\begin{theorem}[Reflection principle: conditional version]
Let $P_0(A|P_\tau (A) = q)$ be the fair price the gambler assigns at $t=0$ to a conditional ticket on $A$ which gives a refund if at $t=\tau$, the gambler assigns $P_\tau (A) \neq q$. We assume that $P_0(P_\tau(A)=q) > 0$ so that the conditional probability is well defined. Coherence then demands that
\begin{align}
P_0(A|P_\tau (A) = q)=q.
\end{align}
\end{theorem}
\begin{proof}
We want to show that $P_0(A|P_\tau (A) = q)=q$. $P_0(A|P_\tau (A) = q)$ represents the fair price at which the gambler would buy or sell a ticket ``Pay $\$1$ if $A$ and $P_\tau (A) = q$; pay $P_0(A|P_\tau (A) = q)$ if  $P_\tau (A) \neq q$.'' In other words, $P_0(A|P_\tau (A) = q)$ is the fair price the gambler assigns for a conditional ticket on $A$, whose cost is refunded unless $P_\tau (A) = q$. In particular, the gambler is willing at $t=0$ to buy such a ticket at price $P_0(A|P_\tau (A) = q)$. Suppose $P_0(A|P_\tau (A) = q) > q$. If in the end $P_\tau (A) = q$, this means that at time $\tau$, the gambler is willing to sell the ticket ``Pay $\$1$ if $A$'' at price $q$. But if they buy the original ticket for $P_0(A|P_\tau (A) = q)$ and later sell the $A$-ticket at $q$, they lose $P_0(A|P_\tau (A) = q)-q$. Of course, if in the end $P_\tau (A) \neq q$, they get a refund: thus conditional on $P_\tau(A) = q$, setting $P_0(A|P_\tau (A) = q) > q$ exposes the gambler to a sure loss. (Notice that this is independent of whether $A$ itself occurs.) By the same token, the gambler is willing at $t=0$ to sell the conditional ticket at price $P_0(A|P_\tau (A) = q)$. Suppose $P_0(A|P_\tau (A) = q) < q$. If in the end $P_\tau (A) = q$, the gambler is willing at time $\tau$ to buy the ticket ``Pay $\$1$ if $A$'' at price $q$. Since $P_0(A|P_\tau (A) = q) < q$, the gambler certainly loses $q-P_0(A|P_\tau (A) = q)$ conditional on $P_\tau (A) = q$. Thus coherence demands that $P_0(A|P_\tau (A) = q)=q$.
\end{proof}

\begin{corollary}
By reflection, $P_0(A|P_\tau(A)=q) = q$, so that the law of total probability demands
\begin{align}
P_0(A) &= \sum_i P_0(A|P_\tau(A)=q_i)P_0(P_\tau(A)=q_i)= \sum_i q_i P_0(P_\tau(A)=q_i).
\end{align}
\end{corollary}
\begin{remark}
In the present, the gambler may contemplate their future beliefs about event $A$. There are several possibilities: they might assign probabilities $q_1, q_2, \dots$ to event $A$, e.g., given their other beliefs about what might happen between now and then. In particular, still in the present, they assign probability $P_0(P_\tau(A)=q_1)$ that in the future they will judge $q_1$ to be a fair price for an $A$ ticket, $P_0(P_\tau(A)=q_2)$ that they will judge $q_2$ as a fair price for an $A$ ticket, and so on. In order for their present belief about $A$ to be coherent with their beliefs now about their future beliefs, the gambler ought to adopt $P_0(A)=\sum_i q_i P_0(P_\tau(A)=q_i)$: the probability they assign now to $A$ should be a mixture of the probabilities they might assign in the future, weighted by the probability that they will make that assignment.
\end{remark}

\section{Sample spaces and all that}

In our arguments, we have implicitly made use of the idea that an agent may assign probabilities not just to isolated events, but to also more general propositions involving events, e.g., to the event  that $A$ \emph{and} $B$ occur, or the event that $A$ \emph{or} $B$ occurs, or event  that $A$ does \emph{not} occur at all. We can formalize this logic using a \emph{Boolean algebra}.  In particular, following de Finetti \cite{definettiForesightItsLogical1992}, we may consider a notion of ``constituent'' or ``atomic'' events into which arbitrary events can be refined. Given a set of $n$ events $\{E\}$, we may consider the $2^n$ possible conjunctions of these $n$ events and their negations. For example, for two events $A, B$, we may consider the events
\begin{align*}
	&(1) \  A \text{ and } B && (3) \ \text{not } A \text{ and } B\\
	&(2) \ A \text{ and not } B && (4) \ \text{not } A \text{ and not B}.
\end{align*}
If we consider any pair of these atomic events, they must differ in at least one event. For example, $(1)$ requires that $B$, but $(2)$ requires that not $B$; $(1)$ requires that $A$, but $(3)$ requires that not $A$, and so forth. Thus the $2^n$ atomic events so constructed are all mutually exclusive. Moreover, if we consider the disjunction of all the atomic events that implicate one of the original events, we recover that original event. For example, if we consider the disjunction $(1)$ or $(2)$, we have
\begin{align}
(A \text{ and } B) \text{ or } (A \text{ and not } B)= A \text{ and } (B \text{ or not } B) = A.
\end{align} 
Thus given any set of events, we may resolve them into a set of mutually exclusive atomic events $\Omega$. For $n$ events, we require no more than $2^n$ atomic events, but possibly less since we may remove atomic events which are equivalent or impossible. More general events may then be identified with subsets of $\Omega$, logical disjunction (or) with set union ($\cup$), logical conjunction (and) with set intersection $(\cap)$, and logical negation with set complement $(\neg)$. The empty subset may be regarded as an event the agent regards as impossible, and the full set $\Omega$ may be regarded as an event the agent regards as necessary.

From these considerations, we can see that de Finetti probability theory is equivalent to Kolmogorov's axiomatic definition \cite{kolmogorovFoundationsTheoryProbability2018} in terms of probability spaces.

\begin{definition}[Probability space]
A finitely additive probability space consists of a \emph{sample space} $\Omega$, the set of atomic events; an \emph{event space} $E$, the set of all subsets (the powerset) of $\Omega$; and a \emph{probability measure} $P$, which assigns to each event in the event space a probability. The probability measure must satisfy $\forall A \in E: P(A) \ge 0$, $P(\Omega)=1$, and $P(\bigcup_{i=1}^N A_i) = \sum_{i=1}^N P(A_i)$ for a finite family of disjoint events $A_i$. One can work more generally with event spaces forming a nontrivial $\sigma$-algebra, and impose that the final axiom holds for any countable sequence of disjoint events: this is a countably additive probability space.
\end{definition}

Using set theoretic notation, we can write a somewhat more compact proof of the law of total probability (LTP). Let $A=\{A_0, A_1, \dots\}$ be a partition of the sample space $\Omega$ into disjoint subsets whose union is $\Omega$. Since $A$ is therefore a mutually exclusive and exhaustive set of events, we have for any other event $B$
\begin{align}
P(B) &= P(B \cap \Omega) = P( (B \cap A_0) \cup (B \cap A_1) \cup \dots )\\
&=\sum_i P(B \cap A_i) = \sum_i P(B|A_i)P(A_i).
\end{align}
In particular, if we consider the ultimate partition given by $\Omega$ itself, we have
\begin{align}
\label{classique}
P(B) = \sum_{\omega \in \Omega} \delta_{\omega \in B} P(\omega),
\end{align}
that is, the conditional probabilities in the LTP are just so-called indicator functions.

\subsection{Measures}

For de Finetti, countable additivity was not forced upon the gambler by coherence alone: assuming it is a mathematical convenience which however does not exhaust all the possibilities open to the gambler \cite{platoCreatingModernProbability1994}. But let us see what it buys us. Consider for example, the case where the sample space $\Omega$ is uncountably infinite: let $\Omega$ be the real line $\mathbb{R}$. In the Kolmogorovian measure theoretic treatment of probabilities, one doesn't assign probabilities to any arbitrary subsets of $\mathbb{R}$, but specifically to \emph{measurable subsets}. For example, one might choose the Borel $\sigma$-algebra on $\mathbb{R}$ which is generated by the open intervals of $\mathbb{R}$, and which is the smallest collection of subsets of $\mathbb{R}$ which contains the open intervals, is closed under complements, countable unions, and countable intersections. 

A \emph{probability measure} on the $\sigma$-algebra assigns probabilities to events. Under certain regularity conditions, a probability measure $dP$ can be expressed in terms of a probability density $p(x)$ so that one may express the probability of falling within an interval as
\begin{align}
P(x\in [a,b]) = \int_{[a,b]}dP = \int_a^b p(x) dx
\end{align}
where $p(x)\ge 0$ and integrates over the line to 1. For such continuous distributions, individual points have probability 0: there is no contradiction here since an interval is an uncountable union of points. This same procedure described above can be generalized to more exotic sample spaces. In this dissertation, however, we will almost entirely focus on discrete events, e.g., measurements with a finite number of outcomes. 

\section{Random variables}

Intuitively, a random quantity $X$ is a quantity that could take different values $\{x_i\}$ conditional on this or that event occurring: it is a valuation on outcomes. In betting terms, we can regard $X$ as a ticket or contract such that if event $E_i$ occurs, one gains (or owes) some amount $x_i$. (We will always refer to this as a gain, with the understanding that a negative gain means a loss.) In his \emph{Theory of Probability} \cite{definettiTheoryProbabilityCritical2017}, de Finetti defines the \emph{prevision}, or fair price, $P(X)$ of a random quantity $X$ in the following way: a gambler picks $P(X)$ with the understanding that it commits them to accepting any bet with gain $c(X-P(X))$ where $c$ is arbitrary and ``at the choice of an opponent.'' 

To understand the logic, suppose $c=1$. Then $X-P(X)$ is just the gain or loss the gambler realizes after having paid $P(X)$ for a ticket, $X$ now taking one of its values $x_i$. $c=2$ corresponds to the case of the gambler having purchased two such tickets. If $c=-1$, then $P(X)-X$ is the gain or loss the gambler realizes after selling a ticket at price $P(X)$. The point is that $c$ can be chosen adversarially: in declaring $P(X)$, the gambler has specified the \emph{certain gain} $P(X)$ that they consider equivalent to the \emph{uncertain gain} $X$. Moreover, we say $P(X)$ is the \emph{fair price} the gambler has assigned to $X$: the gambler is willing to take either side of a gamble at that rate. 

de Finetti's coherence principle amounts to a single normative injunction: ``It is assumed that You do not wish to lay down bets which will with \emph{certainty} result in a loss for You'' \cite{definettiTheoryProbabilityCritical2017}. A set of previsions is coherent, then, if there is no combination of bets whose gains are all uniformly negative. In particular, given a collection of random quantities $X_i$, coherence demands that we must choose previsions $P(X_i)$ such that there is no linear combination
\begin{align}
\sum_i c_i (X_i - P(X_i)) < 0.
\end{align}
Otherwise, if the gambler announces previsions $\{P(X_i)\}$, an adversary could induce them to buy or sell tickets in amounts $\{c_i\}$, forcing upon the gambler a certain loss. That is, the adversary could pick $\{c_i\}$ such that $\sum_i c_i X_i <  \sum_i c_i P(X_i)$: no matter what values are realized for each $X_i$, the gambler has overpaid. 

For example, consider the random quantities $X, Y,$ and $X+Y$, and the linear combination
\begin{align}
\big((X+Y) - P(X+Y) \big) - \big(X - P(X)\big) - \big(Y - P(Y)\big) &= P(X) + P(Y) - P(X+Y) \ge 0.
\end{align}
By the same token we can consider the linear combination
\begin{align}
-\big((X+Y) - P(X+Y) \big) + \big(X - P(X)\big) + \big(Y - P(Y)\big) &= P(X+Y) - P(X) - P(Y) \ge 0.
\end{align}
Putting these two inequalities together, we find that coherence demands $P(X+Y) = P(X) + P(Y)$. In fact, it would have sufficed to observe that the initial linear combination is a constant \emph{regardless} of the actual payoffs of $X$, $Y$, and $X+Y$. If it were a negative constant, the gambler would always lose money; if it were a positive constant, flipping the signs in the linear combination, swapping buying for selling, one would end up again with a negative constant. Thus the only constant that is coherent is $0$. Similarly, consider random quantities $X$ and $aX$ for some real constant $a$, and the linear combination
\begin{align}
\big(aX - P(aX)\big) - a\big(X - P(X)\big) &= aP(X)-P(aX) \ge0.
\end{align}
Again this is a constant regardless of outcome, and so the only possibility is that $P(aX) = aP(X)$. Thus in general, previsions behave linearly
\begin{align}
P\left(\sum_i a_i X_i \right) = \sum_i a_i P(X_i).
\end{align}

With this in hand, suppose we have a ticket $X$ such that one gains $x_i$ conditional on event $E_i$ occurring. At the same time, we can consider tickets $E_i$ which pay $\$1$ if a particular event $E_i$ occurs, and 0 otherwise. (We conflate the symbol for the ticket and the event.) As we have seen, $P(E_i)$ is just the probability of the event $E_i$. Since we can decompose the random quantity as $X=\sum_i x_i E_i$, by the linearity of previsions,
\begin{align}
P(X) = P\left(\sum_i x_i E_i\right)=\sum_i x_i P(E_i).
\end{align}
We conclude: the fair price for a random quantity $X$ is just what we normally think of as the expectation value, denoted $\mathbb{E}[X]$ or $\langle X\rangle$. But here the interpretation is that $\langle X\rangle$ is the certain quantity which, if it were swapped for the uncertain quantity $X$, the gambler would be indifferent to the exchange. Similarly, the variance,
\begin{align}
\var[X]= P\big((X-P(X))^2\big)= P(X^2)- P(X)^2,
\end{align}
may be interpreted as the fair price of a ticket that pays more when your stated prevision of $X$ is badly off. Finally, by the homogeneity of previsions, e.g., $P(aX)= aP(X)$, we may contemplate previsions of random quantities which cannot in themselves be interpreted as gains or losses of money. We simply pick a conversion factor $a$ such that $aX$ \emph{is} a monetary value: then let $P(X) = (1/a)P(aX)$.

\subsection{The algebra of random variables}

Appealing to the set of atomic events $\Omega$, we can more formally define a random variable $X$ to be a map from $\Omega$ to a set of valuations. In other words, a random variable assigns a valuation to atomic events. We focus here on the particular case of real valued random variables, $X: \Omega \rightarrow \mathbb{R}$, which we now show form an algebra. Indeed, suppose now we have two random variables $G:\Omega \rightarrow \mathbb{R}$ and $H: \Omega \rightarrow \mathbb{R}$. We can add them as $(H+G)(\omega) = H(\omega) + G(\omega)$; we can multiply them by scalars as $(\lambda H)(\omega) = \lambda H(\omega)$; and finally we can multiply them as $(HG)(\omega) = H(\omega)G(\omega)$. This latter operation may be viewed as the pointwise or Hadamard product on $\mathbb{R}^n$, where $|\Omega|=n$. (As usual, we assume $n$ is finite, but really it need not be.)

Let $A=\{A_0, A_1, \dots\}$ be a partition of $\Omega$. We say a random variable $X$ is $A$-measurable if $X$ is constant on each block of $A$: $\forall \omega \in A_i: X(\omega) = x_i$. The intuition is that $X$ assigns a particular payoff depending only upon which composite event $A_i$ occurs. We then have $P(X=x_i) = \sum_{\omega \in A_i} P(\omega)$, and 
\begin{align}
\langle X\rangle = \sum_{\omega\in \Omega } X(\omega) P(\omega) = \sum_i x_i \sum_{\omega \in A_i} P(\omega) = \sum_i x_i P(X=x_i) .
\end{align}
Moreover, given $X$ which is $A$-measurable and $Y$ which is $B$-measurable, we have
\begin{align}
\label{product}
\langle XY\rangle &= \sum_{\omega \in \Omega} X(\omega)Y(\omega) P(\omega) = \sum_{ij} x_i y_j \sum_{\omega \in A_i \cap B_j} P(\omega)\\
&=\sum_{ij} x_i y_j P(X=x_i, Y=y_j).
\end{align}
Here $P(X=x_i, Y=y_j)$ is shorthand for $ P(X=x_i \text{ and } Y=y_j)$.
From this we can see that the familiar formula for the expectation value of the product of two random variables $\langle X Y\rangle$ can be derived from the pointwise product of valuations on the underlying sample space.

\section{From composites to the de Finetti representation}

In the preceding, we used the fact that $X$ and $Y$ correspond to alternative coarse grainings $\{A_i\}$ and $\{B_i\}$  of the same underlying atomic events. For example, consider rolling a single die. The atomic events $\Omega$ correspond to the different outcomes of the die roll, and e.g., a random variable $Z$ might map ``1 pip shows'' $\rightarrow 1$, ``2 pips show'' $\rightarrow 2$ and so on. For convenience, we might therefore write the sample space $\Omega=\{1,2,3,4,5,6\}$\footnote{From this point of view, any semantic meaning applied to the atomic events may be recovered from the random variables that implicate them.}. We might then consider the partition $A_0=\{2,4,6\}, A_1=\{1,3,5\}$ and call the event corresponding to the former ``even pips,'' and the event corresponding to the latter ``odd pips.'' We could distinguish them with an $A$-measurable random variable, for example, which takes $+1$ for $\omega \in A_0$ and $-1$ for $\omega \in A_1$. We could also consider the partition $B_0=\{1,2,3\}, B_1=\{4,5,6\}$, and distinguish them with an $B$-measurable random variable which takes $+1$ on $B_0$ and $-1$ on $B_1$. We could then contemplate the probability that the pip is both even and in the first half, that is, $P(A = +1, B=+1)=P(2)$: this turns out to be an atomic event.

If however we'd like to gamble on two rolls of the die, we will need to expand our sample space. We can take the Cartesian product of sample spaces $\Omega \times \Omega = \{(1,1), (1,2), \dots, (2, 1), (2,2), \dots, (6, 1), (6, 2), \dots\}$. Now the atomic events correspond to getting particular pairs of rolls. Going in reverse, define random variables \(D_1(i,j)=i\) and \(D_2(i,j)=j\). Then \(D_1=i\) means the event that the first die comes up $i$ regardless of the second and corresponds to the subset \(\{(i,j):j=1,\dots,6\}\)---and similarly for $D_2$. Then
\begin{align}
P(\text{first die shows 1}) = P(D_1 = 1) = \sum_j P(D_1=1, D_2=j) = \sum_j P((1,j)).
\end{align}
In the special case that the two rolls are independent, we have $P(D_1=i, D_2=j)=P(D_1=i)P(D_2=j)$. But how can we capture the intuition that we are rolling ``the same die'' in each trial? One strong assumption is that the rolls are not just independent but identically distributed, so that
\begin{align}
P(D_1=i, D_2=j, D_3=k, \dots) = P(D=i)P(D=j)P(D=k)\dots,
\end{align}
where a single distribution $P(D=\cdot)$ is used for every roll. But where does that single distribution come from? In particular, how can we justify the idea that precisely by rolling the dice many times, we may sharpen our judgement about what probabilities to assign to each outcome in \emph{any} roll of the die? de Finetti recognized that the more basic judgement is that when we assign joint probabilities to repeated trials the order of the rolls should not matter. This led to one of his great accomplishments: the de Finetti representation theorem, which shows that an infinitely exchangeable joint distribution over many trials can be represented as a mixture of independent and identically distributed trials. In particular, it justifies acting ``as if'' there is an unknown single roll distribution which may be sharpened by repeated rolling. The ``as if'' matters: for a subjective Bayesian, there are no such thing as an ``unknown distribution,'' as probabilities are by definition beliefs on the part of the gambler.

Suppose the gambler assigns a joint probability distribution to the outcomes of $n$ trials, e.g., $n$ rolls of the dice. Further suppose that the gambler is indifferent to the order of the die rolls, and could imagine rolling the die indefinitely. In this case, the gambler ought to assign a joint probability distribution which is (infinitely) \emph{exchangeable}. What does this mean? A distribution is symmetric (finitely exchangeable) if it is invariant under permutations of its arguments, corresponding to the different trials. A distribution is (infinitely) exchangeable if it is symmetric and extendible to a symmetric distribution on arbitrarily many random trials: formally, the distribution over $n$ trials can be obtained from a distribution over $n+m$ trials by marginalizing over the last $m$ trials\footnote{We note that exchangeability is a weaker notion that the independence of trials. A famous example of a distribution which is the former but not the latter is provided by Polya's urn model. We also note that one may prove finite de Finetti theorems where only finite exchangeability is demanded: the representation then proves to be approximate with a certain quantifiable error.}. de Finetti proves the following theorem \cite{platoCreatingModernProbability1994, cavesUnknownQuantumStates2002}, which is a special case of a broader result:

\begin{theorem}[discrete classical de Finetti representation theorem]

Suppose \(P(x_1,\dots,x_n)\) is the \(n\)-trial marginal of an infinitely exchangeable sequence of \(k\)-valued random variables $X_i$ where $x_i$ denotes which of the $k$ values $X_i$ takes. Then there exists a unique probability measure \(P(\mathbf p)\) on \(\Delta_k\) such that
\begin{align}
P(x_1, x_2, \dots, x_n) &= \int_{\Delta_k} P(x_1|\B{p})\dots P(x_n|\B{p})P(\B{p})d\B{p}\\
&= \int_{\Delta_k} p_1^{n_1}p_2^{n_2}\dots p_k^{n_k} P(\B{p})d\B{p},
\end{align}
where $\B{p} = (p_1, \dots, p_k)$; $\Delta_k$ is the probability simplex over $k$ outcomes; $P(\B{p})$ is a probability density function on the simplex; and $n_j$ is the number of times the outcome $j\in \{1,\dots, k\}$ occurred. Note that $P(x_i|\B{p})$ just pulls out the probability $p_i$ corresponding to the outcome denoted by $x_i$. 
\end{theorem}

Upon obtaining outcomes on $m$ trials, denoting them $T_m$, one may use Bayes's rule to pass from the prior $P(\B{p})$ to the posterior 
\begin{align}
P(\B{p}|T_m) = \frac{P(T_m|\B{p})P(\B{p})}{P(T_m)} = \frac{P(T_m|\B{p})P(\B{p})}{\int_{\Delta_k} P(T_m|\B{p})P(\B{p})d\B{p}},
\end{align}
where $P(T_m|\B{p}) =p_1^{n_1}\dots p_k^{n_k}$, where the $\{n_j\}$ refer to the number of times the $j$th outcome actually occurred. This justifies the usual idea that to determine the ``unknown probability distribution'' over the outcomes of e.g., a roll of the dice, one collects the frequency data over many trials. But for de Finetti, this is an ``as if'': probabilities do not exist in nature, unknown and waiting to be known: probabilities are judgements of an individual gambler. de Finetti therefore begins with the gambler who has assigned joint probabilities over many trials, who is committed to an exchangeable distribution, and who can then write their joint distribution in terms of a probability distribution over which probabilities to assign to one of the six pips coming up. As data comes in from the trials, the theorem justifies the gambler in using Bayes's rule to update their probabilities over probabilities, $P(\B{p})$, using the outcome frequencies. In principle, as the number of trials increases, $P(\B{p})$ may become highly peaked around a particular distribution, a single assignment of probabilities to each of the $k$ outcomes of a single trial, e.g., each roll of the dice. The point is that even if several gamblers start with different priors, assuming there are no pathologies (assignments of zero probabilities) if they update on the same data for long enough, they will converge on the same distribution $P(\B{p})$. Thus the theorem is really about how through repeated experiments, different gamblers ought to come to agreement in their beliefs.

\section{Calling the whole thing off}

The law of total probability $P(E)=\sum_i P(E|R_i)P(R_i)$ urges on the gambler a particular relationship between the probabilities they assign to events $E$ and $\{R_i\}$. The key assumption is that exactly one of the $\{R_i\}$'s will in fact occur. Suppose however that after assigning probabilities $P(E|R_i)$ and $P(R_i)$, the gambler learns that no $R_i$ will occur after all. Can we derive any coherence condition that $P(E)$, $P(E|R_i)$, and $P(R_i)$ must satisfy---or are all bets off? 

In general, yes, all bets are off. But it may be that the gambler can relate events $E$ and $\{R_j\}$ to another set of mutually exclusive and exhaustive events $\{\lambda_k\}$. We assume that the $\lambda_k$'s mediate the relevance of the $R_j$'s to $E$ in the sense that $P(E|R_j, \lambda_k) = P(E|\lambda_k)$. Indeed, this implies
\begin{align}
P(E|R_j) &= \sum_k P(E, \lambda_k|R_j) = \sum_k P(E|\lambda_k,R_j) P(\lambda_k|R_j) = \sum_k P(E|\lambda_k)P(\lambda_k|R_j).
\end{align}
Now since the events $\{\lambda_k\}$ are mutually exclusive and exhaustive, if it turns out that no $R_j$ will occur, there remains a norm that the gambler can appeal to, namely
\begin{align}
	\label{LTP_AC}
P(E) = \sum_k P(E|\lambda_k)P(\lambda_k).
\end{align}
Under a special circumstance, we can take an alternative perspective on this same norm. First, we note that according to Bayes' rule,
\begin{align}
P(\lambda_k|R_j)=\frac{P(R_j|\lambda_k)P(\lambda_k)	}{\sum_i P(R_j|\lambda_i)P(\lambda_i)}.
\end{align}
We then ask: is it possible to find a matrix $\Phi$ which satisfies 
\begin{align}
\sum_{kl}P(\lambda_j |R_k)\Phi_{kl}P(R_l|\lambda_m)	=\delta_{jm}?
\end{align}
In more compact linear algebraic notation, we require
\begin{align}
\label{prob_identity}
P(\lambda|R)\Phi P(R|\lambda) &= I,	
\end{align}
where $P(\lambda|R), P(R|\lambda)$ denote conditional probability matrices. (We shall also use $P(R), P(\lambda)$ as a compact notation for the corresponding probability vectors.)

When can one find such a $\Phi$? We may appeal to some linear algebra. Let $P(\lambda|R) \in \mathbb{R}^{r \times m}$ and $P(R|\lambda)\in \mathbb{R}^{m \times r}$ so that $I$ is the $r\times r$ identity matrix. We will later show in Theorem (\ref{phi-existence}) that such a $\Phi$ can be found iff $P(\lambda|R), P(R|\lambda)$ are both of rank $r$. Explicitly, for example, we could take $\Phi = P(\lambda|R)^+ P(R|\lambda)^+$ where $A^+$ denotes the Moore-Penrose pseudoinverse of $A$. Now if it is possible to find such a $\Phi$, we say that gambler believes that the events $\{R_i\}$ constitute a \emph{reference} for events $\{\lambda_i\}$. To appreciate this, we may insert the resolution of the identity provided by Eq. (\ref{prob_identity}) into Eq.  (\ref{LTP_AC}),
\begin{align}
P(E)&=P(E|\lambda)P(\lambda) = P(E|\lambda)\Big[P(\lambda|R)\Phi P(R|\lambda)\Big]P(\lambda)=P(E|R)\Phi P(R),
\end{align}
which looks like the law of total probability save for the presence of this matrix $\Phi$. 

To recapitulate, if the gambler believes that some $R_i$ will occur, in order to be coherent, they ought to follow a norm: the law of total probability,
\begin{align}
	P_1(E) = \sum_{i}P(E|R_i)P(R_i).
\end{align}
If instead it turns out some event $R_i$ will \emph{not} occur, but nevertheless the gambler can relate the events $\{R_i\}$ to another set of mutually exclusive and exhaustive events $\{\lambda_i\}$ which they believe \emph{will} occur, and moreover they judge that the events $\{R_i\}$ constitute a reference for the events $\{\lambda_i\}$ in the linear algebraic sense we have described, then the gambler may appeal to a \emph{deformation} of the law of total probability,
\begin{align}
\label{p2}
	P_2(E)&=\sum_{ij}P(E|R_i)\Phi_{ij}P(R_j),
\end{align}
 where $\Phi$ satisfies $P(\lambda |R)\Phi P(R|\lambda)	=I$. Despite the fact that the events $\{R_i\}$ are counterfactual in the second scenario, nevertheless if the agent believes that the events $\{R_i\}$ form a reference for the events $\{\lambda_i\}$, they can reuse the probabilities they would assign in the first scenario to deal with the second.

It may seem that there is something contradictory here. In general the matrix $\Phi$ may contain negative entries in defiance of our earlier demonstration that assigning negative probabilities is incoherent! Of course, the inconsistency is merely apparent: after all, we have just shown that adopting $P_2(E)=P(E|R)\Phi P(R)$ is equivalent to adopting $P_2(E)=P(E|\lambda)P(\lambda)$. Furthermore, recall that we derived the conditions for coherence on the assumption that one of the events $R_i$ would really happen: in the derivation of the law of total probability, it is assumed that $\{R_i\}$ constitute a mutually exclusive and exhaustive set of events. But if that is no longer the case, then the law of total probability no longer applies.

We can also give a de Finetti style justification for Eq. (\ref{p2}). Using the resolution of the identity from Eq. (\ref{prob_identity}), we have in particular 
\begin{align}
P(\lambda) &= \Big[P(\lambda|R)\Phi P(R|\lambda)\Big] P(\lambda)=P(\lambda|R)\Phi P(R).
\end{align}
In other words, since the gambler judges $\{R_i\}$ to be a reference for $\{\lambda_i\}$, we can write the probabilities $P(\lambda_i)$ as
\begin{align}
P(\lambda_i) = \sum_j x_j^{(i)} P(R_j),
\end{align}
where $ x_j^{(i)}=\sum_{k}P(\lambda_i|R_k)\Phi_{kj}$. We can then compare on the one hand, tickets $(1_i)$, ``Pay \$1 if $\lambda_i$''; and on the other hand, the bundle of tickets $(2_i)$,
\begin{itemize}
\item``Pay \$$x_1^{(i)}$ if $R_1$''
\item \ \ \vdots
\item ``Pay \$$x_m^{(i)}$ if $R_m$.''
\end{itemize}
The price at which the gambler ought to buy or sell $(1_i)$ is $P(\lambda_i)$, while the price at which the gambler ought to buy or sell the bundle of tickets $(2_i)$ is $ \sum_j x^{(i)}_j P(R_j)$. Because $\{R_i\}$ is a reference for $\{\lambda_i\}$, the fair price that the gambler assigns to $(1_i)$ and $(2_i)$ are the same. Thus the gambler ought to be willing to exchange the one for the other. If the gambler suspects that one of the $R_i$'s will not occur after all, they ought to aggregate all their $\{R_i\}$-related tickets into a bundle of the form $(2_i)$, which they can convincingly market as equivalent to $(1_i)$. Having converted their $(2_i)$'s into $(1_i)$'s, should the $\{R_i\}$'s in fact be called off (nothing is certain), their fortunes will be secured.

\subsection{Coulda, woulda, shoulda}

The possibility of reselling $R$-tickets notwithstanding, one might still worry: if adopting $P_2(E)=P(E|R)\Phi P(R)$ is equivalent to adopting $P_2(E)=P(E|\lambda)P(\lambda)$, what is the usefulness of the former, if the latter is available? Suppose in fact that the gambler believes some $\{\lambda_i\}$ will occur, but they do not believe they can assign informed probabilities to them. In that case, $P_2(E)=P(E|\lambda)P(\lambda)$ cannot offer them much advice, nor apparently can $P_2(E)=P(E|R)\Phi P(R)$, since we defined $\Phi$ to satisfy $P(\lambda|R)\Phi P(R|\lambda) = I$, whereas we have stipulated the gambler feels they cannot in good conscience assign good conditional probabilities to these events.

Suppose however that there is (yet) another set of mutually exclusive and exhaustive events $\{R^\prime_i\}$ to which the gambler \emph{has} assigned probabilities, and which are presumed related to the events $\{\lambda_i\}$. Even though the gambler is not informed enough to assign specific probabilities $P(R^\prime|\lambda)$ or $P(\lambda|R)$, if they nevertheless believed that if they \emph{did} assign such probabilities, that $P(R^\prime|\lambda)$ and $P(\lambda|R)$ would both be rank $r$, that is, that there \emph{would} exist a $\Phi$ such that
\begin{align}
P(\lambda|R)\Phi P(R^\prime |\lambda) = I,
\end{align} 
then multiplying by $P(R^\prime|\lambda)$ on the left and $P(\lambda|R)$ on the right, we find that this $\Phi$ must satisfy
\begin{align}
\label{early-1-inverse}
 P(R^\prime |R) \Phi P(R^\prime | R) = P(R^\prime |R).
\end{align}
In fact, this is the defining equation for a $\{1\}$-inverse of $P(R^\prime |R)$, and as we will see, it turns out that \emph{every} matrix has at least one $\{1\}$-inverse. In other words, if the gambler believes (a) that $\{R_i\}$ and $\{R^\prime_i\}$ form references for $\{\lambda_i\}$ and (b) that they have assigned probabilities $P(R^\prime |R)$, then they can appeal to 
\begin{align}
P_2(E) = P(E|\lambda)\Big[P(\lambda|R)\Phi P(R^\prime|\lambda)\Big]P(\lambda) = P(E|R)\Phi P(R^\prime), 
\end{align}
even if they have not actually assigned probabilities $P(\lambda|R)$ or $P(R^\prime|\lambda)$ since $\Phi$ may now be defined as any $\{1\}$-inverse of $P(R^\prime |R)$. We will later explore the equivalence between these two definitions: the key is that $P(R^\prime|R)$ is itself rank $r$.

In practice, it may be that we can in some sense identify the events $\{R_i\}$ and $\{R_i^\prime\}$. For example, the events $\{R_i\}$ might refer to an initial measurement on a system, after which ``the same measurement'' is repeated giving outcome $\{R_i^\prime\}$. To identify the two means that all things being equal (e.g., if the order of the measurements were reversed) the gambler would assign the same probabilities in both instances, $P(R_i)=P(R_i^\prime)$. In such a circumstance, we can write, as before,
\begin{align}
	P_2(E)=P(E|R)\Phi P(R),
\end{align}
where $P(R |R)\Phi P(R |R)=P(R |R)$. Crucially, neither expression depends on any probabilities assigned to $\{\lambda_i\}$: instead we rely upon the gambler's judgements $P(R_i|R_j)$, and their judgement that $\{R_i\}$ forms a reference for the events $\{\lambda_i\}$.

Nevertheless, in the absence of reasons to assign probabilities to the events $\{\lambda_i\}$, why should the gambler believe that the events $\{R_i\}$ form a reference for them? They may well have legitimate, independent reasons for believing so. But even if they don't, as we have observed, a $\{1\}$-inverse $\Phi$ exists for \emph{any} matrix so that regardless of whether $\{R_i\}$ forms a reference, we can always calculate $P_2(E)=P(E|R)\Phi P(R)$ for $\Phi$ any $\{1\}$-inverse of $P(R|R)$ in the case that the events $\{R_i\}$ are called off. On the one hand, the gambler may hope that the $\{R_i\}$ form at least a partial reference for $\{\lambda_i\}$, so that $P_2(E)=P(E|R)\Phi P(R) $ would still have some advice to offer, albeit unreliable advice: think of $P(R|\lambda)$ having an approximate left inverse. On the other hand, if the gambler has independent reasons for assigning $P_2(E)$ directly, to the extent that $P_2(E)=P(E|R)\Phi P(R) $ is in fact satisfied by their mesh of beliefs, the gambler is the more justified in saying that $\{R_i\}$ is a good reference after all. Of course, just because  the gambler assigns $P_2(E)\neq P(E|R)\Phi P(R)$ does not imply that $\{R_i\}$ is \emph{not} a good reference: it might also be that the gambler simply ought to adjust their assignments $P_2(E)$, $P(E|R_j)$, and $P(R_i)$. In what follows, we will be able to treat this sort of question more systematically.

\section{Pulling out the rug: nonclassicality}

\subsection{Hidden variables}

So far we have been a bit fast and loose with our language, talking about events and outcomes which might occur or might be called off, like a horse race on a stormy day. There is nothing in itself wrong with this: there are many brands of subjective Bayesianism to which the development of probability theory we have given here would be acceptable. For example, in interpreting the construction of the previous section, we might gloss the events $\{\lambda_i\}$ as referring to whether or not something has this or that underlying property, which may not be accessible to observation, even as it conditions the results of observation. In other words, in the language of quantum foundations, we could consider the $\lambda_i$'s to be \emph{hidden variables}. The assumption that $P(R|R) = P(R|\lambda)P(\lambda|R)$ is in fact a nontrivial constraint, even if the $\lambda_i$'s are ``hidden.'' This is because we assume $\{R_i\}$ forms a reference for $\{\lambda_i\}$, which linear algebraically means that all three matrices must have rank $r$: $P(R|R)\in \mathbb{R}^{n\times n}_r$, $P(R|\lambda)\in \mathbb{R}^{n\times r}_r$ and $P(\lambda|R)\in \mathbb{R}^{r\times n}_r$. Put another way, the assumption  is that $P(R|R)$, a stochastic matrix, has a \emph{rank factorization} into the product of two stochastic matrices of the same rank as $P(R|R)$. We will return to this subject in Chapter \ref{ch:classicality}. For now we simply observe that such a factorization cannot be found for just any stochastic matrix! The assumption that some $\{\lambda_i\}$ really occurs is thus a strong constraint on the probabilities $P(R|R)$ the gambler might assigns.

In a similar vein, we may interpret the measurement $\{R_i\}$ as disturbing these hidden variables, even as it forms a reference for them. On the one hand, 
\begin{align}
P(E|R)P(R) = P(E|\lambda) P(\lambda|R) P(R|\lambda)P(\lambda),
\end{align}
shows that $P(\lambda|R) P(R|\lambda)$ encodes the disturbance to the hidden variables as a result of the measurement, while
\begin{align}
P(E|R)\Phi P(R) = P(E|\lambda)\Big[P(\lambda|R)\Phi P(R|\lambda)\Big] P(\lambda) = P(E|\lambda)P(\lambda)
\end{align}
allows us to appreciate that the role of $\Phi$ is precisely to take this disturbance into account and undo it.

We will return to this theme in a later chapter when we discuss notions of classicality. As discussed in the introduction, however, in light of quantum mechanics, the assumption that there are ``hidden variables'' of this sort is impossible in general to maintain. We therefore begin from an entirely different premise. The agent or gambler identifies something external to themselves as a system, external in the sense of beyond their control. They freely act upon that system, and their action has a consequence: the system is as free to respond to the agent's grasp as the agent was free to reach out. In particular, there is no presumption that the consequences are conditioned by some underlying properties of the system. From this point of view, measurements are nothing other than actions the agent takes on their world, and outcomes or events are nothing other than the consequences of these acts. So going forward, we will admit no events which are not consequences, or outcomes, of an act.

\subsection{Classical reference measurements}

Before we begin, however, let us review classical probability theory once more, specifically from the point of view of measurement theory. Having designated a part of the world as \emph{the system}, one supposes that the system is ``in'' one of some number of states defined by some conjunction of properties, states which are mutually exclusive and exhaust all possibilities. We may denote them $\{\lambda_i\}$. For example, in a classical Newtonian system, each $\lambda_i$ might correspond to the positions and momenta of some number of particles. This set may be finite or infinite: for simplicity of exposition, let us for now assume there are some finite number $n$ of underlying states. Out of ignorance, one may assign probabilities $P(\lambda_i)$ to these underlying possible states of the system. A measurement with outcomes $\{E_i\}$ is formalized as a set of conditional probabilities $P(E_i|\lambda_j)$: given that the system is in the state $\lambda_j$, what probabilities ought one to assign to the different outcomes of the measurement? To say that $\{\lambda_i\}$ form a mutually exclusive and exhaustive set means that we believe exactly one of them is the case. Then we can invoke the law of total probability,
\begin{align}
P(E_i) = \sum_j P(E_i|\lambda_j)P(\lambda_j).
\end{align}
By a reductionist principle, the $\{\lambda_i\}$ ought to be identified with the set $\Omega$ of atomic events upon which more general events $\{E_i\}$ supervene: each $E_i$ is identified with a subset of $\Omega$ with the stipulation that subsets $\{E_i\}$ partition $\Omega$ into disjoint subsets. From this point of view, the different measurements one may perform upon the system correspond to different partitions $\{E_i\}$ of $\Omega$, different ways of carving up $\Omega$ into mutually disjoint subsets. It is assumed that exactly one of the $\lambda_i$'s is true: and the outcome of any measurement is simply whichever outcome corresponds to a subset containing $\lambda_i$, the true state. But most measurements are coarse instruments for determining this $\lambda_i$. 

Thus in classical probability theory there is an ultimate \emph{reference measurement}, the one which distinguishes all the atomic events $\{\lambda_i\}$. For example, in classical physics, one may ideally read off the positions and momenta of a cloud of point particles without disturbing them in any way so that $P(\lambda|\lambda)=I$. In the formalism of the previous section, we have
\begin{align}
P_1(E) &= P(E|\lambda) P(\lambda) && P_2(E) = P(E|\lambda) \Phi P(\lambda),
\end{align}
where $\Phi$ satisfies $P(\lambda|\lambda) \Phi P(\lambda|\lambda) = P(\lambda|\lambda)$. But $P(\lambda|\lambda)= I$ and so $\Phi=I$, and $P_2(E)=P_1(E)$. In other words, classically, one need not necessarily distinguish between the properties of the system and the outcomes of a reference measurement since the latter just reads off the former.

But what if we drop the assumption that a system is characterized by some definite properties independent of measurement? This may seem contradictory: how can a system be characterized at all if not by its properties? The resolution is that one may nevertheless retain the idea of a \emph{reference measurement} and the $\Phi$-formalism we have already developed can provide a series of norms to which a gambler can appeal.

\subsection{System, reference, state, effect}

\begin{definition}[System]
A \emph{system} is identified by the agent as precisely something beyond their control but which they can nevertheless act upon. 
\end{definition}

\begin{definition}[Reference measurement]
Let $\rho_1$ and $\rho_2$ be two outcomes of an agent's actions on a system. $R$ is a reference measurement with outcomes $\{R_i\}$ iff
\begin{align}
\forall i: P(R_i|\rho_1) = P(R_i|\rho_2) \Longleftrightarrow \forall E: P(E|\rho_1) = P(E|\rho_2).
\end{align}
In other words, if the agent assigns the same reference probabilities conditional on outcome $\rho_1$ or $\rho_2$, then they ought to assign the same probabilities to \emph{any} outcome $E$ conditional on outcome $\rho_1$ or $\rho_2$.
\end{definition}

\begin{definition}[State]
A \emph{state} is an equivalence class of measurement outcomes conditional upon all of which a gambler would assign the same probabilities to a subsequent reference measurement--and so to all measurements.
\end{definition}
\begin{remark}
In the definition of a reference measurement, $\rho_1$ and $\rho_2$ would belong to the equivalence class $\rho$. The gambler calls $\rho$ a \emph{state} since for them each outcome in the equivalence class implies the same ``state of expectation'' with respect to the reference measurement. We will then write simply $P(E|\rho)$: in practice $\rho$ will be implemented by some particular member of the equivalence class, and we will often interchangeably refer to $\rho$ and $P(R|\rho)$ as a state. Indeed, if the gambler has a particular ``state of expectation'' with respect to the reference measurement, it must be the result of some constellation of circumstances, the consequences of many actions. Considering these as a whole, this single collective consequence must lie by definition in the equivalence class.
\end{remark}
\begin{remark}
From the definition, a reference measurement \emph{separates} distinct states: conversely, if two states yield all the same probabilities with respect to reference measurement, they are really one state after all. Although the outcomes in the equivalence class might be genuinely different experiences for the agent, the gambler believes these differences make no difference at all to their intercourse with the system.
\end{remark}

\begin{remark}
A \emph{state preparation} e.g., of $\rho$, is an idealization: the agent takes an action among whose consequences is a desired member of the equivalence class $\rho$, and to which consequence the agent assigns probability 1.
\end{remark}

\begin{definition}[Preparatory measurement]
A \emph{preparatory measurement} is any measurement with outcomes $\{\rho_i\}$ such that the gambler assigns
\begin{align}
\forall i, j: P(R_i|\rho_j) = P(R_i|\rho_j, \text{any other consequences}).
\end{align}
In other words, the gambler would assign the same reference probabilities conditional on outcome $\rho_i$ regardless of any other consequences: for the gambler, the outcome $\rho_i$ is all that matters as far as the system is concerned.
\end{definition}

\begin{definition}[Reference states]
Let $E_1$ and $E_2$ be two different outcomes, not necessarily of the same measurement. $\{R_i\}$ is a set of \emph{reference states} iff
\begin{align}
\forall i: P(E_1|R_i) = P(E_2|R_i) \Longleftrightarrow \forall \rho: P(E_1|\rho) = P(E_2|\rho).
\end{align}
In other words, if the agent assigns the same probabilities to outcomes $E_1$ and $E_2$ conditional on the reference states, then they ought to assign the same probabilities to outcomes $E_1$ and $E_2$ conditional on  \emph{any} state. 
\end{definition}

\begin{definition}[Effect]
An \emph{effect} is an equivalence class of measurement outcomes to which the gambler would assign the same probabilities conditional on a set of reference states---and thus they would assign the same probabilities conditional on any state.
\end{definition}

\begin{remark}
If $E_1$ and $E_2$ lead to the same ``effect'' conditional on reference outcomes, they ought to lead to the same ``effect'' conditional on any outcome. Thus we call $E$ an \emph{effect}, and talk of measuring $E$ as a short hand for some member of the equivalence class of which $E_1$ and $E_2$ are a part. A set of reference states \emph{separate} distinct effects: conversely, if two effects yield all the same probabilities with respect to the reference states, they must be the same effect after all. We will often interchangably refer to $E$ and $P(E|R)$ as an effect.
\end{remark}

\begin{remark}
We assume that the reference measurement is a preparatory measurement, and that the reference outcomes are reference states. Thus conditioning on a reference outcome leaves subsequent measurement outcomes independent of any other consequences, and the same reference measurement separates states and effects.
\end{remark}

\subsection{State space}

Given a reference measurement, let us now consider the set $\mathscr{S}$ of probability distributions $\{P(R|\rho)\}$ which the gambler may consistently assign to its outcomes. Equivalently $\mathscr{S}$ is the set of states $\{\rho\}$, and we shall call the set $\mathscr{S}$ the \emph{state space}. Suppose the agent performs a preparatory measurement with outcomes $\{\pi_i\}$ followed by the reference measurement. By the law of total probability,
\begin{align}
P(R|\rho) = \sum_i P(R|\pi_i)P(\pi_i|\rho).
\end{align}
Since the preparatory measurement can be chosen arbitrarily, we conclude that $\mathscr{S}$ is a \emph{convex set}: if $\{P(R|\pi_i)\}$ is a set of distributions the gambler is willing to assign to the reference measurement, then the gambler ought to also be willing to assign $P(R|\rho)=\sum_i P(R|\pi_i)p_i$ for probabilities $p_i$ summing to unity\footnote{Alternatively, as in \cite{staceyQuantumTheorySymmetry2019}, one may appeal to the reflection principle. We saw that $P_0(A) = \sum_i q_i P_0(P_\tau(A) = q_i)$. In particular, 
\begin{align}
P_0(R) = \sum_j P(R) P_0(P_\tau(R) = P(R)) = \sum_j P(R) p_j,
\end{align}
for probabilities $p_j$. Here the convexity of the state space comes from fact that the gambler ought to adopt now mixtures of state assignments they might make in the future.}.
\begin{definition}[Convex set]
A \emph{convex set} $S$ is a subset of a vector space such that $\forall x, y\in S, p \in [0,1]: px + (1-p)y \in S$. 
\end{definition}

For convenience, we recall here some elementary facts \cite{rockafellarConvexAnalysis1970} about convex sets. An \emph{extreme point} of a convex set is any point which cannot be written as a nontrivial convex combination of different points in $S$: $x$ is extreme whenever $x = py + (1-p)z$ for $y,z\in S$ and $p\in (0,1)$ implies $y=z=x$. By the Heine-Borel theorem, in $\mathbb{R}^n$ a set $S$ is \emph{compact} if and only if it is \emph{closed} and \emph{bounded}. A set is \emph{closed} if it includes its boundary: if we have a sequence of points $\lim_{n\rightarrow \infty} x_n = x$, then the limit point $x\in S$. A set is \emph{bounded} if it can be enclosed in a sphere of finite radius. We note that continuous functions on compact sets are bounded and attain their minimum and maximum values \cite{macauleyNovelProofHeineBorel2008,rudinPrinciplesMathematicalAnalysis1976}. By the Krein-Milman theorem, a compact convex set is the \emph{convex hull} of its extreme points \cite{rudinFunctionalAnalysis1991}, and given a set $X$, its convex hull is
\begin{align}
\text{conv}(X) = \left\{\sum_i p_i x_i : x_i \in X, p_i\ge 0, \sum_i p_i =1 \right\}.
\end{align}
Since $\mathscr{S}\subseteq \Delta_n$, where $\Delta_n$ is the probability simplex on $n$ outcomes, the state space is clearly bounded, and since we define $\mathscr{S}$ in terms of the convex hull of the distributions the gambler is willing to assign to the reference measurement, we may take $\mathscr{S}$ to be closed, and therefore compact. Finally, it is often convenient to work with unnormalized probability vectors, including even the 0 vector, giving us what we shall call the \emph{state cone} $\hat{\mathscr{S}}$, a proper convex cone.

\begin{definition}[Convex cone]
A \emph{cone} $C$ is a subset of a vector space $V$ such that if $x\in C$, then $\lambda x \in C$ for $\lambda >0$. A cone is \emph{convex} if for any $x,y\in C$ and $\lambda, \mu>0$, $\lambda x + \mu y \in C$. A convex cone is \emph{pointed} if $C \cap (-C) = \{0\}$ \cite{farautAnalysisSymmetricCones1994}: geometrically this means $C$ contains no full line through the origin.  A convex cone is \emph{proper} if it is pointed and closed. A convex cone is \emph{spanning} if $C - C = V$.
\end{definition}
We recall some more useful facts. A ray of a cone $\{\lambda x: \lambda \ge 0\}$ generated by a point $x\in X$ is \emph{extremal} whenever $x=y+z$ for $y,z\in C$ implies there exist $\alpha,\beta\ge 0$ such that $y = \alpha x$ and $z=\beta x$. The \emph{conic hull} of a set $X$ is 
\begin{align}
\text{cone}(X) = \left\{\sum_i \lambda_i x_i : x_i \in X, \lambda_i \ge 0\right\},
\end{align}
and finally, a closed pointed finite dimensional convex cone is the conic hull of its extreme rays \cite{rockafellarConvexAnalysis1970}. A \emph{base} of a cone is any nonempty convex set $B$ within the cone not containing the origin such that every nonzero point in the cone can be written as $x=\lambda b$ for $\lambda > 0$ and $b\in B$. We conclude that the state space itself is the base of the state cone, picked out by the hyperplane on which $\sum_i P(R_i|\rho) = 1$.

\subsection{Effect space}

Let us now consider the set $\mathscr{E}$ of response functions $\{P(E|R)\}$ which the gambler may consistently assign conditional on the outcomes of the reference measurement. Equivalently, $\mathscr{E}$ is the set of effects $\{E\}$, and we shall call $\mathscr{E}$ the \emph{effect space}. Now on the one hand, the gambler may always coarse grain outcomes of a measurement,
\begin{align}
P(E_1 \cup E_2|R) = P(E_1|R) +  P(E_2|R),
\end{align}
treating two outcomes as equivalent. On the other hand, the gambler may always fine grain outcomes of a measurement e.g., by performing another measurement conditional on an outcome of the first, which the gambler believes only depends upon that outcome, and treating two step measurement as a single measurement. Clearly,
\begin{align}
P(E_1|R) = \sum_i P(E_{i}^{(1)}|E_1)P(E_1|R),
\end{align}
since $  \sum_i P(E_{i}^{(1)}|E_1)=1$. By alternating fine graining and coarse graining, from a single measurement, we may obtain arbitrary combinations $\sum_i P(E_i|R)p_i$ for $p_i \in [0,1]$ but where $\{p_i\}$ need not sum to 1. For example:

\[
\begin{tikzpicture}[
    every node/.style={inner sep=2pt, outer sep=2pt},
    line/.style={thick}
]

% Top row
\node (a1) at (0,0) {$P(E_1| R)$};
\node (a2) at (4,0) {$P(E_2| R)$};
\node (a3) at (8,0) {$P(E_3| R)$};

% Middle row
\node (m1) at (6.5,-1.5) {$p\,P(E_3|R)$};
\node (m2) at (10,-1.5) {$(1-p)\,P(E_3|R)$};

% Bottom row
\node (b1) at (0,-3) {$P(E_1| R)$};
\node (b2) at (5,-3) {$P(E_2| R)+p\,P(E_3| R)$};
\node (b3) at (10,-3) {$(1-p)\,P(E_3| R)$};

% Edges
\draw[line] (a1) -- (b1);
\draw[line] (a2) -- (b2);
\draw[line] (a3) -- (m1);
\draw[line] (a3) -- (m2);
\draw[line] (m1) -- (b2);
\draw[line] (m2) -- (b3);

\end{tikzpicture}
\]

As before when we extended the state space to a state cone, and then restricted ourselves to the base of normalized states, let us consider for mathematical convenience an \emph{effect cone} $\hat{\mathscr{E}}$, where we allow any conic combination of effects. In particular, two effects play a special role. On the one hand, let $0$ denote the equivalence class of outcomes the gambler judges to be impossible. Then $P(0|R)=0$ must be the zero vector. On the other hand, let $1$ denote the equivalence class of outcomes the gambler judges to be necessary. Then $P(1|R)= (1,\dots,1)$ must be the row vector of all 1's. Since $\hat{\mathscr{E}}$ is a subset of the nonnegative orthant and contains the zero vector $P(0|R)$, the effect cone must be a pointed convex cone.

But we can say more. Given any outcome $E$, we may always consider the outcome that $E$ does not occur, that is, the complement of $E$, with probability $P(\neg E) = 1-P(E)$. Thus $\forall E: P(\neg E|R) = P(1|R) - P(E|R) \in \mathscr{E}$: the ``opposite effect'' $P(\neg E|R)$ must also be an effect. Now given any proper convex cone $C$, we can define a partial order\footnote{Recall that a partial order satisfies reflexivity, antisymmetry, and transitivity.} $x \leq y$ iff $y-x \in C$. In particular, the effect cone defines a partial order on effects, and 
\begin{align}
P(\neg E |R) = P(1|R)-P(E|R) \in \hat{\mathscr{E}} \Longleftrightarrow P(E |R) \leq P(1|R).
\end{align}
Moreover, $P(E|R)- P(0|R) \in \hat{\mathscr{E}}$ so that in terms of the partial ordering induced by the effect cone
\begin{align}
\forall E: P(0|R) \leq P(E|R) \leq P(1|R).
\end{align}
We thus recover the idea that we may take nonnegative combinations of effects only insofar as the sum can appear in a measurement, whose effects must sum to $P(1|R)=(1,\dots, 1)$. The effect space $\mathscr{E}$ is thus itself a compact convex set, a subset of the hypercube $\Delta_n^*$ of response functions, and the intersection of two cones, one opening upwards from $P(0|R)$ and the other opening downwards from $P(1|R)$.

The picture that has thus emerged is that normalized states $P(R|\rho)$ live in the state cone $\hat{\mathscr{S}}$, a proper closed convex cone, and in particular, on the base picked out by $P(1|R)=(1,\dots, 1)$. All this is taking place within the probability simplex $\Delta_n$ on $n$ outcomes. Meanwhile, the effects $P(E|R)$ live in the effect cone $\hat{\mathscr{E}}$, also a proper closed convex cone. Effects in the effect space itself satisfy $P(0|R) \leq P(E|R) \leq P(1|R)$ with respect to the partial ordering defined by the effect cone, and all this takes place within the space of response functions on $n$ outcomes $\Delta_n^*$.

\subsection{Nonclassical consistency}

\begin{figure}[t]
\centering
\includegraphics[scale=0.5]{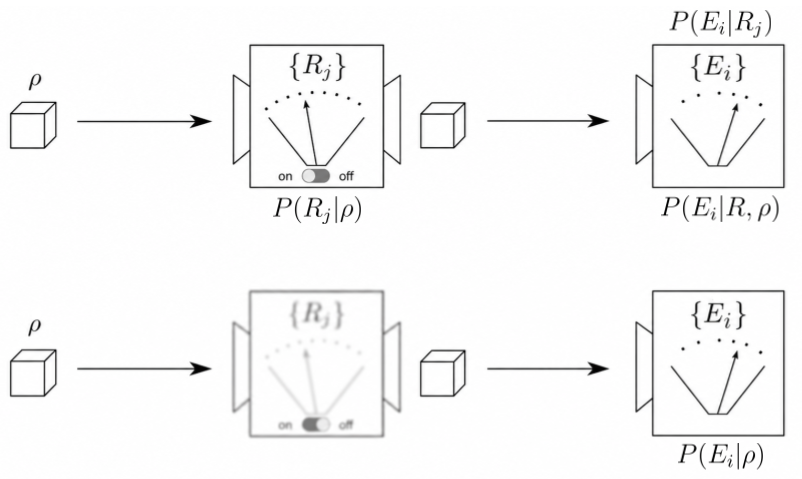}
\caption{Two scenarios \cite{fuchsQbismWhereNext2023}.}
\label{fig:sky_ground}
\end{figure}

Let us now consider the two scenarios depicted schematically in Fig. \!\ref{fig:sky_ground} in terms of ``measuring devices.'' The box in the middle is a reference device, which can be turned on or off. In the first scenario, the reference device is ``on'' signifying that after the outcome $\rho$ of a preparatory measurement, the agent performs a reference measurement $\{R_i\}$ followed by an arbitrary measurement $\{E_i\}$. Since the outcomes of both measurements really occur, a gambler betting on their outcomes is urged to adopt the law of total probability,
\begin{align}
P(E_i|R, \rho) = \sum_j P(E_i|R_j)P(R_j|\rho),
\end{align}
to check the consistency of their assignments. In the second scenario, after the preparatory measurement yields outcome $\rho$, the agent measures $\{E_i\}$ directly: the reference measurement is ``turned off.'' We ask again: is there any norm that we can appeal to that constrains $P(E_i|\rho)$ in terms of $P(R_j|\rho)$ and $P(E_i|R_j)$? There ought to be, precisely because $\{R_i\}$ is a reference measurement: by definition, the probabilities $P(R_i|\rho)$ fully characterize the state $\rho$, and the conditional probabilities $P(E_i|R_j)$ fully characterize the effects $\{E_i\}$. Thus $P(E_i|\rho)$ ought to be \emph{some} function of these two distributions,
\begin{align}
P(E_i|\rho) = \mathcal{F}\big( P(E_i|R), P(R|\rho)\big).
\end{align}
We now show that $\mathcal{F}(\cdot, \cdot)$ must take a particular form. Our argument is indebted to the one given in \cite{PhysRevA.104.022207} which itself was inspired by \cite{Wright2021, buschQuantumStatesGeneralized2003, Caves2003GleasonTypeDO}.

\begin{theorem}
\label{thm:form-of-Phi}
Let $\{R_i\}_{i=1}^n$ be a reference measurement. The function $\mathcal{F}(\cdot, \cdot)$ in $P(E|\rho) = \mathcal{F}\big( P(E|R), P(R|\rho)\big)$ which expresses the coherence condition the gambler ought to impose on their probability assignments when the reference measurement remains counterfactual, is bilinear. In particular,
\begin{align}
P(E|\rho) = P(E|R)\Phi P(R|\rho),
\end{align} 
where $\Phi$ is any matrix satisfying $P(R|R)\Phi P(R|R)= P(R|R)$, $P(R|R)$ being the conditional probability matrix which characterizes the reference measurement itself.
\end{theorem}
\begin{proof}
Suppose that before the reference measurement, the gambler performs a preparatory measurement $\{\rho_i\}$. If conditional on $\rho_i$, the gambler would assign $P(R|\rho_i)$ to the reference measurement, then they ought also to be willing to assign $P(R|\rho)=\sum_i P(R|\rho_i)P(\rho_i)$ to the reference measurement. The same applies should the gambler perform some arbitrary measurement with outcome $E$ instead: $P(E|\rho) = \sum_i P(E|\rho_i)P(\rho_i)$. From
\begin{align}
P(E|\rho) &= \mathcal{F}\big( P(E|R), P(R|\rho) \big) = \mathcal{F}\big( P(E|R), \sum_i P(R|\rho_i)P(\rho_i) \big)\\
&= \sum_i P(E|\rho_i)P(\rho_i) = \sum_i \mathcal{F}\big( P(E|R), P(R|\rho_i)\big) P(\rho_i),
\end{align}
we conclude that $f(x) = \mathcal{F}(P(E|R), x)$ for some fixed $P(E|R)$ preserves convex combinations on the state space $\mathscr{S}$,
\begin{align}
	\label{f0}
f\big( px + (1-p) y\big)  = pf(x) + (1-p)f(y) && p\in [0,1] && x,y\in \mathscr{S}.
\end{align}
We shall now extend $f$ to a series of functions defined on progressively more general domains. First, we define $f_1$ not just on the state space $\mathscr{S}$ but also the state cone $\hat{\mathscr{S}}$. Now in general, any vector $x$ in a cone can be written uniquely as $\lambda y$ for $\lambda >0$ and $y$ in a base of the cone: the only exception to uniqueness is when $x=0$. In our case, since the base of the state cone is picked out by the normalization constraint on probability vectors, any $x\in \hat{\mathscr{S}}$ can be written $x=\overline{x}y$ for $y\in \mathscr{S}$ itself, where $\overline{x}=\sum_i x_i$.  We thus define $f_1(x) = \overline{x}f(x/\overline{x})$. Clearly this agrees with $f$ on $\mathscr{S}$. Further, we assume that $f_1(0) = P(E|0)=\mathcal{F}\big(P(E|R), P(R|0)) =\mathcal{F}\big(P(E|R), 0)=  0$ so that $f_1$ is centered since any event conditioned on the impossible event ought to be assigned zero probability. 

From the definition $f_1(x) = \overline{x}f(x/\overline{x})$, we have for $t>0$,
\begin{align}
f_1(tx) = \overline{tx}f\left(\frac{tx}{\overline{tx}}\right) = t \overline{x} f\left(\frac{x}{\overline{x}}\right) = t f_1(x),
\end{align}
so that in fact $f_1$ is homogeneous for $t \ge 0$. Moreover, $f_1$ is also additive on  $\hat{\mathscr{S}}$:
\begin{align}
f_1(x+y) &= \overline{x+y}f\left(\frac{x+y}{\overline{x+y}}\right)= (\overline{x} +\overline{y})f\left(\frac{\overline{x}}{\overline{x}+\overline{y}}\frac{x}{\overline{x}} + \frac{\overline{y}}{\overline{x}+\overline{y}}\frac{y}{\overline{y}} \right) \\
&= (\overline{x}+\overline{y})\left[ \frac{\overline{x}}{\overline{x}+\overline{y}} f\left(\frac{x}{\overline{x}}\right) + \frac{\overline{y}}{\overline{x}+\overline{y}} f\left(\frac{y}{\overline{y}}\right) \right]=\overline{x}f\left(\frac{x}{\overline{x}}\right) + \overline{y} f\left(\frac{y}{\overline{y}}\right) \\
&= f_1(x) + f_1(y).
\end{align}
 We now extend the domain to include $\hat{\mathscr{S}} - \hat{\mathscr{S}}=\linspan(\mathscr{S})$. By definition, any vector $x \in \linspan(\mathscr{S})$ can be written $x = x^+ - x^-$ for $x^+, x^- \in \hat{\mathscr{S}}$. Thus let $\forall x \in \linspan(\mathscr{S}): f_2(x) = f_1(x^+) - f_1(x^-)$. Clearly, for vectors in the cone, $f_2$ agrees with $f_1$. In fact, we could use \emph{any} decomposition of $x$ into the difference of two vectors in $\hat{\mathscr{S}}$. Let $x=x^+ - x^-=y^+ - y^-$ be two such decompositions. Then $x^+ + y^- = y^+ + x^- \in \hat{\mathscr{S}}$ so that
 \begin{align}
  f_1(x^+ + y^-) &= f_1(y^+ + x^-)\\
   f_1(x^+) + f_1(y^-) &= f_1(y^+) + f_1(x^-)\\
   f_1(x^+) - f_1(x^-) &= f_1(y^+) - f_1(y^-)\\
   f_2(x^+ - x^-) &= f_2(y^+ - y^-) = f_2(x).
 \end{align} 
It is straightforward to see that $f_2$ is also homogeneous for $t\ge 0$ and centered. It is also additive on $\hat{\mathscr{S}} - \hat{\mathscr{S}}$. Let $x = x^+ - x^-$ and $z = z^+ - z^-$. Since $x+z = (x^+ + z^+) - (x^- + z^-)$, 
\begin{align}
f_2(x+z) &= f_1(x^+ + z^+) - f_1(x^- + z^-) = f_1(x^+) + f_1(z^+) - f_1(x^-) - f_1(z^-) = f_2(x)+ f_2(z).
\end{align}
Since $f_2$ is additive on $\hat{\mathscr{S}} - \hat{\mathscr{S}}$, we have $f_2(0) = f_2(x + (-x)) = f_2(x) + f_2(-x) = 0$ showing that $f_2(-x) = - f_2(x)$. Let $t = -s < 0$. Then $f_2(tx) =f_2(-sx) = -f_2(sx)= -s f_2(x) = t f_2(x)$ so that $f_2$ is in fact homogeneous over all of $\mathbb{R}$. We have thus established that $f_2$ is homogeneous over $\mathbb{R}$ and additive on $\hat{\mathscr{S}} - \hat{\mathscr{S}}$: in other words, it is linear on $\linspan(\mathscr{S})$. Finally, we extend our function (not uniquely) to a linear function on the entirety of the $n$-dimensional vector space $V$. Let $\{b_i\}$ be an orthonormal basis for 
$\linspan(\mathscr{S})$ and $\{c_i\}$ be an orthonormal basis for its complement: together they form a basis for $V$. Let $f_3(b_i) = f_2(b_i)$ while $f_3(c_i)=a_i $ may be arbitrary, e.g., 0. We may decompose any vector $x= x_b + x_c$, where $x_b$ lies in $\linspan(\mathscr{S})$ and $x_c$ lies in its complement. Since $f_3$ is assumed linear,
\begin{align}
	f_3(x) &= f_3(x_b + x_c) = f_3(x_b) + f_3(x_c) = \sum_i (b_i \cdot x_b) f_3(b_i) + \sum_i (c_i \cdot x_c) f_3(c_i) \\
	&= \sum_i (b_i \cdot x_b) f_2(b_i) + \sum_i (c_i \cdot x_c) a_i.
\end{align}
We have thus constructed a function $f_3(x)$ such that when $x$ is fully supported on $\linspan(\mathscr{S})$, it agrees with $f_2(x)$; and when $x$ is in the state cone $\hat{\mathscr{S}}$, $f_2(x)$ agrees with $f_1(x)$; and when $x$ is in the state space $\mathscr{S}$, $f_1(x)$ agrees with our original function $f(x)=\mathcal{F}(P(E|R), x)$. Since $f_3$ is linear on the entire vector space, and its restriction to $\mathscr{S}$ agrees with $f$, we conclude that $f$ itself is linear on its domain. Thus $\mathcal{F}(\cdot, \cdot)$ is linear in its second argument.

Let us now consider the first argument. Since we may always coarse grain measurements, letting $E_{12}$ be the event that $E_1$ or $E_2$ befalls the agent, we must have
\begin{align}
P(E_{12}|\rho) &= \mathcal{F}\big(P(E_{12}|R), P(R|\rho) \big) = \mathcal{F}\big( P(E_1|R) + P(E_2|R), P(R|\rho)\big)\\
&= P(E_1|\rho) + P(E_2|\rho) = \mathcal{F}\big( P(E_1|R), P(R|\rho)\big) +  \mathcal{F}\big( P(E_2|R), P(R|\rho)\big),
\end{align}
so that $\mathcal{F}(\cdot, \cdot)$ is additive in its first argument on the effect space $\mathscr{E}$. Notice that we assume that $E_1$ and and $E_2$ are part of the same measurement so that $P(E_1|R) + P(E_2|R) \leq P(1|R)$. But we may also fine grain measurements, performing a subsequent measurement conditional on the outcome of the first which the gambler believes depends only upon the outcome of the first. Letting $E_k$ denote the outcome of the first measurement and $F_j$ the outcome of the second, we have $P(E_{jk}|\rho)=P(F_j, E_k|\rho) = P(F_j|E_k)P(E_k|\rho)$ where by writing $P(E_{jk}|\rho)$ we have treated the two measurements as one. Thus
\begin{align}
P(E_{jk}|\rho) &= \mathcal{F}\big(P(E_{jk}|R), P(R|\rho) \big) = \mathcal{F}\big( P(F_j|E_k)P(E_k|R), P(R|\rho)\big)\\
&= P(F_j|E_k)P(E_k|\rho) = P(F_j|E_k) \mathcal{F}\big( P(E_k|R), P(R|\rho)\big),
\end{align}
so that $\mathcal{F}(\cdot, \cdot)$ is homogeneous for $p \in [0,1]$. Let $g(x) = \mathcal{F}(x, P(R|\rho))$ for some fixed $P(R|\rho)$: we have shown that $g$ is additive on the effect space $\mathcal{E}$ and homogeneous for $p\in [0,1]$, where we recall that the effect space is the subset of the effect cone satisfying $P(0|R) \leq P(E|R) \leq P(1|R)$.

Now any vector in the effect cone can be written not uniquely $x = \lambda_x \tilde{x}$ for $\lambda_x > 0$ and $\tilde{x} \in \mathscr{E}$. We define $g_1(x)$ on the entire effect cone by taking $g_1(x) = \lambda_x g(\tilde{x})$.  In fact it doesn't matter which decomposition of $x$ we choose. Let $x=\lambda_x \tilde{x} = \lambda_x^\prime \tilde{x}^\prime$ and define $g_1(x)$ in terms of the former and $g_1^\prime(x)$ in terms of the latter. We have $\tilde{x}^\prime = (\lambda_x/\lambda_x^\prime) \tilde{x}$, and without loss of generality we may assume $\lambda_x < \lambda_x^\prime$. Then $g(\tilde{x}^\prime ) = g\left((\lambda_x/\lambda_x^\prime)\tilde{x}\right)= (\lambda_x /\lambda_x^\prime) g(\tilde{x}) $ so that $\lambda_x^\prime g(\tilde{x}^\prime ) = \lambda_x g(\tilde{x})$, and thus $g_1^\prime(x) = g_1(x)$. We now show that the extension is homogeneous for $t\ge 0$. Let $t>1$. If $x=\lambda_x \tilde{x}$ for $\tilde{x}\in \mathscr{E}$, then $tx =t \lambda_x \tilde{x}=\lambda_{tx} \tilde{x}$, so that we may take $\lambda_{tx} = t \lambda_x$. Since $1/t \in [0,1]$, we have $g_1(tx) = \lambda_{tx} g(tx/\lambda_{tx}) = t \lambda_x g(x/\lambda_x) = t g_1(x)$ 
so that $g_1$ is homogeneous for $t\ge 0$. Finally, we show that the extension is additive on $\hat{\mathscr{E}}$. Let $x+y=\lambda_x \tilde{x} + \lambda_y \tilde{y} $ for $\tilde{x}, \tilde{y} \in \mathscr{E}$ and $\lambda_x,\lambda_y >0$. Let 
\begin{align}
\tilde{z} = \frac{\lambda_x}{\lambda_x + \lambda_y}\tilde{x} + \frac{\lambda_y}{\lambda_x + \lambda_y}\tilde{y} \in \mathscr{E},
\end{align}
so that $z = (\lambda_x + \lambda_y)\tilde{z} = x+y$. Then
\begin{align}
g(\tilde{z}) &=  \frac{\lambda_x}{\lambda_{x}+\lambda_y}g(\tilde{x}) + \frac{\lambda_y}{\lambda_{x}+\lambda_y}g(\tilde{y}),
\end{align}
so that $(\lambda_x + \lambda_y)g(\tilde{z}) = \lambda_x g(\tilde{x}) + \lambda_y g(\tilde{y})$ or $g_1(x+y)= g_1(x) + g_1(y)$. We conclude $g_1$ is also additive on $\hat{\mathscr{E}}$. We now extend the domain to include $\hat{\mathscr{E}} - \hat{\mathscr{E}} = \linspan(\mathscr{E})$. Again we decompose $x=x^+ - x^-$ for $x^+, x^-\in \hat{\mathscr{E}}$ and define $g_2(x) = g_1(x^+) - g_1(x^-)$. By the same argument we used in examining the state dependence, $g_2$ must be homogeneous over $\mathbb{R}$ and additive on $\hat{\mathscr{E}} - \hat{\mathscr{E}}$, that is, linear on $\linspan(\mathscr{E})$. Finally, as before, we may then extend $g_2$ to a linear function $g_3$ defined on the whole vector space which agrees with $g_2$ on $\linspan(\mathscr{E})$. Since $g_3$ is linear, and the restriction of $g_3$ to the effect space $\mathscr{E}$ agrees with $g$, we conclude that $\mathcal{F}(\cdot, \cdot)$ is linear in its first argument.

Putting these two results together, we conclude that $\mathcal{F}(\cdot, \cdot)$ is a bilinear function, and therefore can be represented as a matrix $\Phi$,
\begin{align}
P(E|\rho) = \mathcal{F}\big(P(E|R), P(R|\rho)\big) = P(E|R) \Phi P(R|\rho).
\end{align}
Suppose that the arbitrary measurement $\{E_i\}$ is in fact $\{R_i\}$, the reference measurement itself. For consistency, we require
\begin{align}
P(R|\rho) = P(R|R)\Phi P(R|\rho),
\end{align}
and specializing to the reference states, we require
\begin{align}
P(R|R) = P(R|R)\Phi P(R|R).
\end{align}
But this is the defining equation of a $\{1\}$-inverse of the conditional probability matrix $P(R|R)$, which characterizes the reference states in terms of the reference measurement, or equivalently, the reference measurement in terms of reference states.
\end{proof}

\begin{remark}
In fact, $P(R|\rho) = P(R|R)\Phi P(R|\rho)$ and $P(E|R) = P(E|R)\Phi P(R|R)$ imply that any $P(R|\rho)$ must lie in the column space of $P(R|R)$ and any $P(E|R)$ must lie in the row space of $P(R|R)$. We will discuss this in more detail in the next chapter. Only if these column and row space conditions are met will $P(R|\rho) = P(E|R)\Phi P(R|\rho)$ give identical results for any choice of $\Phi$. 
\end{remark}

We conclude that in our two scenarios, a gambler ought to check their probability judgements against the following two norms,
\begin{align}
P(E|R, \rho) = P(E|R)P(R|\rho) && P(E|\rho) = P(E|R)\Phi P(R|\rho),
\end{align}
where $\Phi$ is any $\{1\}$-inverse of the conditional probability matrix $P(R|R)$ which characterizes the reference measurement itself. In the first scenario, the reference measurement is performed in between the preparatory measurement (with outcome $\rho$) and the final measurement. $P(E|R, \rho)$ is no different than in the classical case: it is given simply by the law of total probability. In the second scenario, the reference measurement is counterfactual. But since the reference probabilities characterize the states and effects completely, there must be some way of expressing the probability $P(E|\rho)$ in terms of them. We have shown that the right normative rule is given by a deformation of the law of total probability by the interposition of the Born matrix $\Phi$. 

In this way, we can supplement standard probability theory to handle situations where systems cannot be defined by properties independent of measurement but which \emph{can} nevertheless be fully characterized by the probabilities an agent assigns to a reference measurement, in the sense that if two preparations yield the same probabilities on the reference, they must yield the same probabilities on all measurements. In this way, we weaken the classical notion of a system, without sacrificing the idea that an agent is served well by imposing on themselves a notion of consistency. In a sense, the outcomes of the reference play the role of the atomic sample space. But now other events, e.g., the outcomes of other measurements, are not simple stochastic coarse grainings or subsets of the atomic events, as we saw in Eq. \!(\ref{classique}), $P(E) = \sum_{\omega \in \Omega} \delta_{\omega \in E} P(\omega)$. Instead, we have \emph{quasistochastic} regraining as in $ P(E|\rho) = P(E|R)\Phi P(R|\rho)$. Why ``regraining''? Because the measurement $\{E_i\}$ may have not just fewer, but even more outcomes than the reference! This is quite unlike the classical case.

\subsection{$\Phi$-duality}

Now that we have established our nonclassical coherence rule $P(E|\rho) = P(E|R)\Phi P(R|\rho)$ as well as the basic geometry of the state and effect cones, we can define a notion of $\Phi$-duality. Call $\hat{\mathscr{S}}^{\star}$, the cone $\Phi$-dual to the state cone, in the following sense,
\begin{align}
\hat{\mathscr{S}}^{\star} = \{y \in V^* | \ \forall x \in \hat{\mathscr{S}}: y \Phi x \ge 0\}.
\end{align}
Clearly, $\hat{\mathscr{E}} \subseteq \hat{\mathscr{S}}^{\star}$. Notice that this is not the usual notion of duality for cones due to the presence of the Born matrix $\Phi$ \cite{farautAnalysisSymmetricCones1994}. Similarly, we may consider the cone $\Phi$-dual to the effect cone, 
\begin{align}
\hat{\mathscr{E}}^{\star} = \{ x \in V | \ \forall y \in \hat{\mathscr{E}}: y \Phi x \ge 0\},
\end{align}
so that so that $\hat{\mathscr{S}} \subseteq \hat{\mathscr{E}}^{\star}$.

\section{``Dynamics''}

So far we have focused on the case of preparatory measurements $\{\rho_i\}$ such that $P(R|\rho_i) = P(R|\rho_i, \text{any other events})$: the agent believes only the outcome $\rho_i$ is relevant as far as the reference measurement is concerned. We have focused on this case because the reference measurement itself is taken to be preparatory, and we have only considered the probabilities for some final measurement $\{E_i\}$. But  the gambler may update their state assignment conditional on the result of a measurement in a more general way. For example, consider the following sequence of events: the gambler performs a preparatory measurement with outcome $\rho$, followed by a measurement $\{E_i\}$, which is then followed by a reference measurement $\{R_i\}$. We can consider the latter two to form a composite measurement with outcomes $\{E_i \text{ and } R_j\}$. We can then compare this to the situation where after obtaining outcome $\rho$, the gambler first performs the reference measurement $\{R_i\}$, followed by $\{E_i\}$, followed by $\{R_i\}$ again. We may relate the two situations, as we have seen, using the Born matrix. We have
\begin{align}
P(R_i, E_j | \rho) &= \sum_{lm} P(R_i, E_j|R_l)\Phi_{lm} P(R_m|\rho),
\end{align}
so that
\begin{align}
	\label{conditional-evolution}
P(R_i|E_j,\rho) &= \frac{1}{P(E_j|\rho)}\sum_{lm} P(R_i, E_j|R_l)\Phi_{lm} P(R_m|\rho).
\end{align}

Suppose now that we are interested in how the gambler should update their state assignment in time, or better said, how the gambler ought to harmonize their beliefs now about the present and their beliefs now about the future. In order to formalize this, we again invoke the reflection principle. We saw earlier that $P_0(E) = \sum_i q_i P_0(P_\tau(E)=q_i)$, where $P_0(E)$ is the probability the gambler assigns at $t=0$ to $E$ and $P_0(P_\tau(E)=q_i)$ is the probability the gambler assigns at $t=0$ to the possibility that they will assign probability $q_i$ to event $E$ at time $t=\tau$. Now suppose the agent plans to undertake a sequence of measurements: first, a preparatory measurement with outcome $\rho$ at $t_0$; at $t_1$, a measurement with outcomes $\{E_i\}$; and at $t_2$, a measurements with outcomes $\{F_i\}$. Let $\tau$ be a time after the measurement at $t_1$ but before the measurement at $t_2$. By the law of total probability,
\begin{align}
P_0(F_j) &= \sum_i P_0\Big(F_j\Big|P_\tau(F_j) = P(F_j|E_i, \rho)\Big)P_0\Big(P_\tau(F_j)=P(F_j|E_i, \rho)\Big).
\end{align}
On the one hand, by reflection, 
\begin{align}
 P_0\Big(F_j\Big|P_\tau(F_j) = P(F_j|E_i, \rho)\Big) = P(F_j|E_i, \rho).
\end{align}
On the other hand, 
\begin{align}
P_0\Big(P_\tau(F_j)=P(F_j|E_i, \rho)\Big) = P(E_i|\rho),
\end{align}
since now at $t_0$ the agent naturally believes that when $t=\tau$ rolls around they will assign $P(F_j|E_i,\rho)$ to getting outcome $F_j$ conditional to $E_i$ with probability $P(E_i|\rho)$. Putting these two statements together, and expressing all probabilities with respect to the reference measurement, 
\begin{align}
P_0(F_j) &= \sum_i P(F_j|E_i, \rho)P(E_i|\rho)\\
&= \sum_i \big\{P(F_j|R)\Phi P(R|E_i, \rho) \big\} P(E_i|\rho)\\
&= \sum_i P(F_j|R)\Phi \left\{\frac{1}{P(E_i|\rho)} P(R, E_i|R)\Phi P(R|\rho) \right\} P(E_i|\rho)\\
&= P(F_j|R)\Phi \left\{\sum_i P(R, E_i|R)\right\} \Phi P(R|\rho),
\end{align}
where in the second to last line we've used Eq. (\ref{conditional-evolution}). Thus reflection implies that at $t_0$, the agent should assign a probability $P_0(F_j)$ which is obtained from the initial state $P(R|\rho)$ by averaging over the updating maps associated to each outcome $E_i$ of the intermediate measurement. 

Let us now suppose that the agent is in fact indifferent to whether they perform the intermediate measurement $\{E_i\}$ or not. Indifference means that they would assign the same probabilities at $t=\tau$, just before the measurement at $t_2$, regardless of whether they end up performing the measurement or not. But this just means that they should assign 
\begin{align}
P_0(F_j) &= P(F_j|R)\Phi \left\{\sum_i P(R, E_i|R)\right\} \Phi P(R|\rho),
\end{align}
even if the $\{E_i\}$ measurement is not performed. Because of the agent's indifference, even if they don't perform the measurement, they should update their state $P(R|\rho)$ using $\sum_i P(R, E_i|R)$. This is principle of dynamics in our framework which takes an agent's actions and their consequences to be primary: to determine the dynamics, that is, how the gambler ought to update their state assignment in time \emph{in the absence of measurement}, the gambler ought to identify precisely those measurements which according to their mesh of probabilities they are indifferent to. Because of this indifference, they ought to use the same update rule even in the measurement's absence. 

We note that this reasoning implies that the set of possible dynamics $\{P(R, E|R)\}$ is a subset of substochastic matrices: this subset must be convex, and in fact a subset of a convex cone satisfying $0 \leq P(R, E|R) \leq J/n$, where $0$ denotes the matrix of all 0's, $J$ the matrix of all 1's, and the inequality is understood in the conic sense. Further, let us consider the special case that the gambler performs a measurement with a single certain outcome, and then updates their state. What could this mean? Following \cite{debrotaQBismsAccountQuantum2024}, we take such an action to correspond to carving the system out of its background, the consequence being ``the sense of the objects existence for the agent.'' Indifference to this kind of measurement means that even in the absence of the system being present to the agent, the agent ought to update their state according to some $P(R,1|R)\equiv P_\chi(R|R)$, which we note need not be expressible as a mixture: it may correspond to an extreme ray of the cone. Given two such stochastic matrices $P_\chi(R|R)$ and $P_{\chi^\prime}(R|R)$, it may be that 
\begin{align}
P_{\chi}(R|R)\Phi P_{\chi^\prime}(R|R)\Phi = P_{\chi^\prime}(R|R)\Phi P_\chi(R|R)\Phi = P(R|R)\Phi,
\end{align}
in which case we say that the two are inverses of each other, and $P_\chi(R|R)$ along with its inverse correspond to symmetries of the state space.

\section{Decomposition}

Finally, let us discuss briefly what it might mean for the gambler to carve out more than one nonclassical system? Certainly, the gambler may introduce a reference measurement for each system they are free to act upon. Let us at first suppose the gambler assigns probabilities independently to the measurements on the different systems. Clearly, for two systems $A$ and $B$ treated independently, since $P(E^{(A)},E^{(B)}|\rho^{(A)}, \rho^{(B)}) = P(E^{(A)}|\rho^{(A)}) \otimes P(E^{(B)}|\rho^{(B)})$, where $\otimes$ denotes the tensor product, we have
\begin{align}
&P(E^{(A)},E^{(B)}|\rho^{(A)}, \rho^{(B)})\\
&=\Big\{P(E^{(A)}|R^{(A)}) \otimes P(E^{(B)}|R^{(B)})\Big\} \Big\{ \Phi^{(A)} \otimes \Phi^{(B)} \Big\} \Big\{P(R^{(A)}|\rho^{(A)}) \otimes P(R^{(B)}|\rho^{(B)})\Big\},\nonumber
\end{align}
where $\{R_i^{(A)}\}$ and $\{R_i^{(B)}\}$ But what if we assign more general distributions to the reference measurements? In particular, what if we allow for actions and consequences that implicate the whole collective, not just separate measurements on the parts? For the moment, let us assume that if $\{R_i^{(A)}\}$ is a reference for $A$ and $\{R_i^{(B)}\}$ is a reference for $B$, then $\{R_i^{(A)} \cap R_j^{(B)}\}$ is a reference for $A$ and $B$. Since $P(R^{(A)}, R^{(B)}| R^{(A)}, R^{(B)}) = P(R^{(A)}|R^{(A)}) \otimes P(R^{(B)}|R^{(B)})$, we may take $\Phi = \Phi_A \otimes \Phi_B$  as this will be a $\{1\}$ inverse of $P(R^{(A)}, R^{(B)}| R^{(A)}, R^{(B)})$. Thus more generally,
\begin{align}
P(E|\rho) = P(E|R^{(A)}, R^{(B)})\Big\{\Phi^{(A)} \otimes \Phi^{(B)}\Big\}P(R^{(A)}, R^{(B)}|\rho),
\end{align}
where $E$ and $\rho$ denote outcomes of measurements on the collective, which as a special case may reduce to the conjunction of outcomes of measurements on the two systems, $E=E^{(A)}_i \cap E^{(B)}_j$. 

Now given the state $\mathscr{S}^{(A)}$ and effect $\mathscr{E}^{(A)}$ spaces of $A$, and the same for $B$, what can we say about the state and effect spaces of the collective? Since a state space must be convex, the smallest collective state space $\mathscr{S}^{(AB)}$ would consist solely of mixtures of separate state assignments for system $A$ and system $B$,
\begin{align}
\mathscr{S}^{(AB)}_{\min} &= \Big\{P(R^{(A)}, R^{(B)}|\rho) = \sum_i p_i P(R^{(A)}|\rho_i^{(A)}) \otimes P(R^{(B)}|\rho_i^{(B)}) \Big\} .
\end{align}
We call such states \emph{separable}. In contrast, the largest possible collective state space consistent with the gambler's beliefs about $A$ and $B$ would consist of those assignments to $\{R_i^{(A)} \cap R_j^{(B)}\}$ which imply valid probabilities on all effects in $\mathscr{E}_A$ and in $\mathscr{E}_B$. 
\begin{align}
\mathscr{S}^{(AB)}_{\max} &= \Big\{ P(R^{(A)}, R^{(B)}|\rho) \Big| \forall E^{(A)} \in \mathscr{E}^{(A)}, E^{(B)} \in \mathscr{E}^{(B)}: \nonumber \\
&0 \leq \big( P(E^{(A)}|R^{(A)}) \otimes P(E^{(B)}|R^{(B)}) \big)\big(\Phi^{(A)} \otimes \Phi^{(B)}\big)P(R^{(A)}, R^{(B)}|\rho) \leq 1\Big\}.
\end{align}
Any state in $\mathscr{S}^{(AB)}_{\max}$ which is not a part of $\mathscr{S}^{(AB)}_{\min}$, that is, which is not separable, we call \emph{entangled}. Symmetrically, the smallest collective effect space would consist of conic combinations of effects on $A$ and effects on $B$ less than $P(1|R^{(A)}, R^{(B)})$, while the largest would consist of effects $P(E|R^{(A)}, R^{(B)})$ which satisfy $0\leq P(E|R^{(A)}, R^{(B)})\{\Phi^{(A)} \otimes \Phi^{(B)}\}\{P(R^{(A)}|\rho^{(A)}) \otimes P(R^{(B)}|\rho^{(B)})\} \leq 1$ for all $P(R^{(A)}|\rho^{(A)}) \in \mathscr{S}^{(A)}$ and $P(R^{(B)}|\rho^{(B)}) \in \mathscr{S}^{(B)}$. So coherence with the gambler's assignments to independent measurements on $A$ and $B$ demands that the collective state and effect spaces satisfy
\begin{align}
\mathscr{S}^{(AB)}_{\min} \subseteq \mathscr{S}^{(AB)} \subseteq \mathscr{S}^{(AB)}_{\max} && \mathscr{E}^{(AB)}_{\min} \subseteq \mathscr{E}^{(AB)} \subseteq \mathscr{E}^{(AB)}_{\max}.
\end{align}
Of course, we may phrase all this in terms of the state and effect cones as well---and say a little more, in fact. It will be helpful, however, to establish the following lemma first.

\begin{lemma}[$\Phi$-duality reverses inclusion]
Suppose cone $K^{\star}$ is $\Phi$-dual to cone $K$, and $L^{\star}$ is $\Phi$-dual to cone $L$, and that $K \subseteq L$. It follows that $L^{\star} \subseteq K^{\star}$.
\end{lemma}
\begin{proof}
	By definition,
\begin{align}
K^{\star} &= \big\{ y \ \big| \ \forall x\in K: y \Phi x \ge 0 \big\} && L^{\star} = \big\{ y \ \big|\ \forall x\in L: y \Phi x \ge 0 \big\}.
\end{align}
Since $K\subseteq L$, if $y$ satisfies $\forall x\in L: y \Phi x \ge 0$, then certainly $\forall x\in K \subseteq L: y \Phi x \ge 0$. But this means that if $y \in L^{\star}$, then $y \in K^{\star}$. We conclude $L^{\star}\subseteq K^{\star}$.
\end{proof}

On the one hand, we have already established the following cone sandwiches: $\hat{\mathscr{S}}_{AB}^{\min} \subseteq \hat{\mathscr{S}}_{AB} \subseteq \hat{\mathscr{S}}_{AB}^{\max}$ and $\hat{\mathscr{E}}_{AB}^{\min} \subseteq \hat{\mathscr{E}}_{AB} \subseteq \hat{\mathscr{E}}_{AB}^{\max}$. But also $\hat{\mathscr{S}}_{AB} \subseteq \hat{\mathscr{E}}_{AB}^*$ and $\hat{\mathscr{E}}_{AB} \subseteq \hat{\mathscr{S}}_{AB}^*$: any state and effect cones must be subsets of the set of mathematically possible states (and effects, respectively). Finally, we note that since
\begin{align}
\hat{\mathscr{E}}_{AB}^{\min} = \cone\{ a \otimes b, a\in \hat{\mathscr{E}}_A, b \in \hat{\mathscr{E}}_B\},
\end{align}
we have
\begin{align}
\left(\hat{\mathscr{E}}_{AB}^{\min}\right)^* = \big\{ x \ \big| \ (a\otimes b)\Phi x \ge 0, a \in \hat{\mathscr{E}}_A, b\in \hat{\mathscr{E}}_B\big\} = \hat{\mathscr{S}}_{AB}^{\max},
\end{align}
and similarly $(\hat{\mathscr{S}}_{AB}^{\min})^* = \hat{\mathscr{E}}_{AB}^{\max}$. Putting all this together, we have the following inclusions:
\begin{align}
\hat{\mathscr{S}}_{AB}^{\min} &\subseteq \hat{\mathscr{S}}_{AB} \subseteq \hat{\mathscr{E}}_{AB}^* \subseteq (\hat{\mathscr{E}}_{AB}^{\min})^*=\hat{\mathscr{S}}_{AB}^{\max}\\
\hat{\mathscr{E}}_{AB}^{\min} &\subseteq \hat{\mathscr{E}}_{AB} \subseteq \hat{\mathscr{S}}_{AB}^* \subseteq (\hat{\mathscr{S}}_{AB}^{\min})^*=\hat{\mathscr{E}}_{AB}^{\max}.
\end{align}

These mathematical niceties aside, let us return to our assumption that $\{R_i^{(A)} \cap R_j^{(B)}\}$ forms a reference for the collective of $A$ and $B$: in the quantum foundations literature, this assumption is called \emph{local tomography}. In general, there is no reason to think this ought to be satisfied. It may be that there is strictly more vitality in the systems together than apart. Whenever multiple systems are involved, therefore, in general, one must take care to identify reference measurements at each level of collectivity, taking into account reference measurements $\{R_i^{(AB)}\}$ upon the collective as a whole. Indeed, it is ideal to start from the utmost whole, and decompose the whole into parts, rather than build up the whole from the parts. On what principle may one decompose a whole? Precisely in a principle of indifference, that as regards the gambler's judgements about measurements in some set $A$, they are indifferent to what measurements in some other set $B$ are performed. This amounts to saying the gambler has carved out two systems which may be addressed separately. Formally, this amounts to
\begin{align}
a, a^\prime \in A; b\in B: \sum_i P(a_i, b_j|\rho) = \sum_i P(a_i^\prime, b_j|\rho) =P(b_j|\rho)\\
 a \in A; b, b^\prime \in  B: \sum_j P(a_i, b_j|\rho) = \sum_j P(a_i, b_j^\prime|\rho) = P(a_i|\rho).
\end{align}
In other words, the marginal probability $P(b|\rho)$ the gambler assigns to a measurement in $B$ is independent of the choice of which measurement the gambler might also make in $A$: indeed, this is a prerequisite for considering a marginal state of $B$ separate from $A$ at all---that is, considering the whole as consisting of parts to begin with. In the quantum foundations literature, these are usually called the ``no signalling'' conditions. From there, the gambler may consider what comes of restricting themselves to measurements in $\mathcal{A}$, identifying a reference measurement, and building up a coherent mesh of probabilities---and do the same for measurements in $\mathcal{B}$. Then given e.g., $\mathscr{S}^{(A)}$ and $\mathscr{S}^{(B)}$ they may identify which composite state space $\mathscr{S}^{(AB)}$ recovers (at least part of) the original state space $\mathscr{S}$ of the whole.

Going in the opposite direction, supposing we can identify reference measurements $\{R_i^{(A)}\}$ on system $A$ and $\{R_i^{(B)}\}$ on system $B$, then we can characterize a collective with a joint distribution $P(R_i^{(A)}, R_j^{(B)}|\rho)$. Marginalizing over either measurement, we obtain $P(R^{(A)}|\rho)$ and $P(R^{(B)}|\rho)$, and since each is assumed to be a reference measurement, we have
\begin{align}
P(E^{(A)}|\rho) &= P(E^{(A)}|R^{(A)})\Phi^{(A)}P(R^{(A)}|\rho)\\
 P(E^{(B)}|\rho) &= P(E^{(B)}|R^{(B)})\Phi^{(B)}P(R^{(B)}|\rho) .
\end{align}
At the same time, 
\begin{align}
P(E^{(A)}, E^{(B)}|\rho) = \big\{P(E^{(A)}|R^{(A)}) \otimes P(E^{(B)}|R^{(B)})\big\}\big\{\Phi^{(A)}\otimes \Phi^{(B)}\big\} P(R^{(A)}, R^{(B)}|\rho),
\end{align}
whose marginals clearly agree so that ``no signalling'' is satisfied.

\section{Conclusion}

In this chapter, we have developed the essentials of probability theory from a subjective Bayesian point of view, emphasizing throughout the coherence of an individual agent's mesh of beliefs. Taking up the QBist point of view, that the agent, their own actions, and the consequences of those actions for them ought to be taken as primary, we introduced a notion of system, state, and effect that formalizes a particular notion of nonclassicality. Central to this notion is the idea of a reference measurement. We showed that our assumptions imply a fundamental nonclassical norm which the gambler ought to abide by which supplements the familiar law of total probability. The formalism we have developed here is expansive enough to aid a gambler grappling with an open-ended, creative nature whose unfolding the agent themselves is implicated in. In some sense, this chapter contains everything such a gambler really needs. In the next chapter, however, we will bring an extra, linear algebraic layer of sophistication to the formalism, indeed, showing its equivalence to the standard framework of \emph{generalized probabilistic theories}, of which quantum theory is an example. 

%% file: sections/born_identity.tex
\UMBchapter{The Born Identity}
\label{ch:born_identity}

\section{Introduction}

Armed with the fundamental nonclassical coherence condition, expressed in terms of the Born matrix $\Phi$, we now turn to studying its properties linear algebraically. In the course of doing so, we show that the conditional probability matrix $P\equiv P(R|R)$ which characterizes the reference measurement defines two fundamental subspaces: consistency requires that distributions on reference outcomes $P(R|\rho)$ must live in $\col(P)$ and response functions $P(E|R)$, as well as valuations on reference outcomes, ought to live in $\row(P)$. Introducing bases for these spaces, we pass to a compact representation of states and effects which shows the essential equivalence between the QBist framework and the framework of generalized probabilistic theories or GPTs. In the other direction, we show how given a GPT, one can develop a ``double sided'' frame representation which reproduces the nonclassical coherence condition we began with. Relying on already existing results in the literature, we define what properties a measurement must have in a GPT to be a proper reference measurement, showing that generally speaking such measurements always exist, and moreover, allow one to prove a nonclassical de Finetti representation theorem.

\section{The Born matrix}

As we have seen, having identified a reference measurement, the gambler may consider two scenarios. In the first scenario, the gambler performs a preparatory measurement, followed by a reference measurement, followed by some arbitrary measurement, in which case they may invoke the familiar classical coherence condition, the law of total probability,
\begin{align}
P(E|R, \rho) = P(E|R)P(R|\rho).
\end{align}
In the second scenario, they skip the reference measurement. Despite the fact that in the latter case, the reference measurement is hypothetical, the gambler may appeal to a nonclassical coherence condition, whose form we derived in the last chapter to be
\begin{align}
P(E|\rho) = P(E|R)\Phi P(R|\rho),
\end{align}
where $\Phi$ satisfies $P(R|R)\Phi P(R|R) = P(R|R)$. Here $P(R|R)$ is the conditional probability matrix which characterizes the reference measurement itself. In fact, this is the defining equation of a $\{1\}$-inverse of $P(R|R)$, and we shall now study this linear algebraic object more closely. 

We first establish that \emph{every} matrix over $\mathbb{R}$ has a $\{1\}$-inverse. In fact, our proof holds over $\mathbb{C}$, and the result itself holds over any field $\mathbb{F}$, as we will show later. We begin with this proof, however, which relies on the singular value decomposition, in order to build an intuition as to how one might calculate one numerically.

\begin{theorem}[Existence and form of $\{1\}$-inverse]
\label{existence-of-1-inverse}
Every matrix $M \in \mathbb{C}^{m \times n}$ has a $\{1\}$-inverse, i.e., a matrix $M^{(1)}$ such that
\begin{align}
MM^{(1)}M=M.	
\end{align}
\end{theorem}
\begin{proof} Every matrix has a singular value decomposition $M=U\Sigma V^\dagger$. Let
\begin{align}
M^{(1)} = 
V 
\left(\begin{array}{c|c}
\sigma^{-1}& A\\ \hline
B & C \\ 
\end{array}\right)
U^\dagger	,
\end{align}
where $\sigma$ is a diagonal matrix of the $r=\rank(M)$ non-zero singular values, and where the blocks $A, B, C$ are completely arbitrary. Observe that
\begin{align}
M M^{(1)}M&= \left[U 
\left(\begin{array}{c|c}
\sigma& 0\\ \hline
0 & 0 \\ 
\end{array}\right)
V^\dagger\right]
\left[V 
\left(\begin{array}{c|c}
\sigma^{-1}& A\\ \hline
B & C \\ 
\end{array}\right)
U^\dagger	
\right]\left[  U 
\left(\begin{array}{c|c}
\sigma& 0\\ \hline
0 & 0 \\ 
\end{array}\right)
V^\dagger	 \right]\\
&=U\left(\begin{array}{c|c}
I&  \sigma A\\ \hline
0 & 0 \\ 
\end{array}\right)\left(\begin{array}{c|c}
\sigma& 0\\ \hline
0 & 0 \\ 
\end{array}\right)V^\dagger =U\left(\begin{array}{c|c}
\sigma&  0\\ \hline
0 & 0 \\ 
\end{array}\right)V^\dagger=M,
\end{align}
showing that $M^{(1)}$ is a $\{1\}$-inverse. Conversely, suppose $MM^{(1)}M=M$, that is,
\begin{align}
\left[U	\left(\begin{array}{c|c}
\sigma& 0\\ \hline
0 & 0 \\ 
\end{array}\right)V^\dagger \right]M^{(1)} \left[U	\left(\begin{array}{c|c}
\sigma& 0\\ \hline
0 & 0 \\ 
\end{array}\right)V^\dagger\right] &= U	\left(\begin{array}{c|c}
\sigma& 0\\ \hline
0 & 0 \\ 
\end{array}\right)V^\dagger.
\end{align}
Multiplying on the left by $U^\dagger$ and on the right by $V$ yields
\begin{align}
\left(\begin{array}{c|c}
\sigma & 0\\ \hline
0 & 0 \\ 
\end{array}\right)V^\dagger M^{(1)} U	\left(\begin{array}{c|c}
\sigma& 0\\ \hline
0 & 0 \\ 
\end{array}\right) &=\left(\begin{array}{c|c}
\sigma & 0\\ \hline
0 & 0 \\ 
\end{array}\right).
\end{align}
\noindent This gives us one constraint: writing the blocks of $V^\dagger M^{(1)} U$ as $\left(\begin{array}{c|c} X& A \\ \hline  B & C \end{array}\right)$, we see that $\sigma X \sigma = \sigma$, which implies $X=\sigma^{-1}$. Then multiplying $V^\dagger M^{(1)} U$ by $V$ from the left and $U$ from the right, we arrive at $M^{(1)} = V 
\left(\begin{array}{c|c}
\sigma^{-1}& A\\ \hline
B & C \\ 
\end{array}\right)
U^\dagger$, as desired.
\end{proof}

\begin{remark}
Suppose that $M$ is in fact invertible. Then $M^{-1}\left[MM^{(1)}M\right]M^{-1}=M^{-1}MM^{-1}$ implies that $M^{(1)}=M^{-1}$, the usual inverse. In the case that $A=B=C=0$, $M^{{(1)}}= M^+$, the Moore-Penrose pseudoinverse.
\end{remark}

In fact, there is a whole zoo of generalized inverses $X$ for singular matrices $M$ which are classified according to a scheme due to Penrose. Each species is specified by a subset of the following list of axioms they satisfy \cite{GeneralizedInverses2003a}:
\begin{align*}
&(1)	\ \ MXM=M\\
&(2) \ \ XMX=X\\
&(3) \ \  (MX)^\dagger = MX\\
&(4) \ \ (XM)^\dagger = XM\\
&(5) \ \ MX = XM\\
&(5^k) \ \ M^k X =XM^k\\
&(6^k) \ \ MX^k = X^kM
\end{align*}
Notice that all of these are satisfied by the true matrix inverse. The only restriction on a Born matrix per se is that it is a $\{1\}$-inverse: but it may satisfy other axioms. For instance, the Moore-Penrose pseudoinverse is the unique $\{1,2,3,4\}$-inverse. Later we will consider reasons one might prefer one choice over another. As a taste, we offer one proof in the interim. One might wish $W(R|\rho) = \Phi P(R|\rho)$ to be a \emph{quasiprobability vector}, that is, not necessarily nonnegative, but nevertheless summing to 1. We can guarantee this if the Born matrix $\Phi$ is itself quasistochastic, that is, satisfying $u^\dagger \Phi = u^\dagger$ where $u=(1,\dots,1)^\dagger$.
\begin{lemma}
   \label{taken_quasistochastic}
Let $\Phi$ be the $\{1\}$-inverse of a stochastic matrix $P$. Then $\Phi$ can be chosen to be quasistochastic, so that it takes probability vectors to quasiprobability vectors.
\end{lemma}
\begin{proof}
Let $u=(1,\dots, 1)^\dagger$, and pick any vector $v$ such that $u^\dagger v = 1$ and any $\{1\}$-inverse $\Phi_0$ not necessarily quasistochastic. Then let
\begin{align}
\Phi = \Phi_0 + v(u^\dagger - u^\dagger \Phi_0).
\end{align}
On the one hand, $u^\dagger \Phi = u^\dagger$, so that $\Phi$ is quasistochastic. Using the stochasticity of $P$ and  $\{1\}$-inverse axiom, we have $u^\dagger \Phi_0 P = u^\dagger P \Phi_0 P = u^\dagger P = u^\dagger$, and so
\begin{align}
P \Phi P = P\Phi_0 P + Pv(u^\dagger - u^\dagger \Phi_0)P = P + Pv(u^\dagger - u^\dagger) = P.
\end{align} 
\end{proof}

\section{Subspaces from coherence}

We derived that $\Phi$ ought to be a $\{1\}$-inverse of $P(R|R)$ by considering the special case that both the preparatory and final measurements are the reference measurement itself: $P(R|R) = P(R|R)\Phi P(R|R)$. But we can actually say a bit more. By the same logic, for any state $\rho$, we have by coherence with the reference measurement,
\begin{align}
P(R|\rho) = P(R|R)\Phi P(R|\rho).
\end{align}
In fact, this puts a strong restriction on the gambler's assignments $P(R|\rho)$. To see this, we must understand the action of $P(R|R)\Phi$. 

\begin{lemma}
\label{1-inverse-proj-col}
Let $M^{(1)}$ be a $\{1\}$-inverse of $M$. Then $M M^{(1)}$ projects onto the range of $M$, that is, the column space of $M$ \cite{GeneralizedInverses2003a}.
\end{lemma}
\begin{proof}
Let $\Pi = M M^{(1)}$. Then by axiom (1), $\Pi^2 = M M^{(1)}M M^{(1)} = M^{(1)}M = \Pi$: $\Pi$ is idempotent, and so a (not necessarily orthogonal) projector. On the one hand, $\range(\Pi) = \range(M M^{(1)}) \subseteq \range(M)$. On the other hand, for $x\in \range(M)$, we can write $x = My$: $x$ is some linear combination of the columns of $M$. Then $\Pi x = \Pi M y = M M^{(1)} M y = M y = x$ so that $\Pi$ fixes all $x \in \range(M)$, from which we conclude $\range(\Pi) = \range(M)$.
\end{proof}

\begin{corollary}
   Recognizing that $P(R|R) \Phi$ is a projector onto $\range(P(R|R))$, from $P(R|\rho) = P(R|R)\Phi P(R|\rho)$ we arrive at another coherence condition: \emph{the set $\mathscr{S}$ of probability distributions on the reference measurement must lie in $\range(P(R|R))$}. Equivalently, the agent should only assign distributions $P(R|\rho)$ to the reference measurement which are linear combinations of the columns of $P(R|R)$.
\end{corollary}

Similarly, for any effect $E$, we have by coherence with the reference measurement,
\begin{align}
P(E|R) = P(E|R) \Phi P(R|R),
\end{align}
which leads us to ask about the action of $\Phi P(R|R)$ on row vectors.

\begin{lemma}
Let $M^{(1)}$ be a $\{1\}$-inverse of $M$. Then $M^{(1)} M$ projects row vectors onto the rowspace of $M$.
\end{lemma}
\begin{proof}
$(M^{(1)} M)^2 = M^{(1)} M M^{(1)} M = M^{(1)} M$ so $M^{(1)} M$ is a projector. On the one hand, if $(x M^{(1)})M = x$, then $x \in \row(M)$, since $x$ is a linear combination of the rows of $M$. Conversely, if $x = y M$ for some row vector $y$, then $x M^{(1)} M = y M M^{(1)} M = yM = x$.
\end{proof}

\begin{corollary}
   We thus arrive at another coherence condition: since $P(E|R) = P(E|R) \Phi P(R|R)$, \emph{the effect space $\mathscr{E}$ of conditional probability distributions $P(E|R)$ must lie in $\row(P(R|R))$}.
   
   In particular, if $P(R|R)$ is symmetric, then, identifying vectors and covectors, both states and effects must live in the same subspace.
\end{corollary}

\begin{lemma}
Let $P(E|R)\in \row(P(R|R)$ and $P(R|\rho) \in \col(P(R|R))$ and suppose $\Phi$ and $\Phi^\prime$ are both $\{1\}$-inverses of $P(R|R)$. Then 
\begin{align}
P(E|R)\Phi P(R|\rho) = P(E|R)\Phi^\prime P(R|\rho).
\end{align}
\end{lemma}
\begin{proof}
Since $P(E|R)\in \row(P(R|R))$, we can write $P(E|R) = a^\dagger P(R|R)$ for some $a$; since $P(R|\rho) \in \col(P(R|R))$, we can write $P(R|\rho) = P(R|R)b$ for some $b$. Then
\begin{align}
P(E|R)\Phi P(R|\rho) = a^\dagger P \Phi P b  = a^\dagger P b = a^\dagger P \Phi^\prime P b = P(E|R)\Phi^\prime P(R|\rho).
\end{align}
\end{proof}

\subsection{A word on observables}

Classically, we defined a random variable as a valuation e.g., $X:\Omega \rightarrow \mathbb{R}$ on the atomic events, or in the language of measurements, on the outcomes of the finest grained reference measurement. In the nonclassical setting, the role of $\Omega$ is played by the outcomes of a reference measurement, but these do not correspond to ``atomic events'' whose coarse grainings give all possible composite events: this difference is witnessed by the presence of the Born matrix $\Phi$. Nevertheless, we may still take a random variable to be $X: \{R_i\}\rightarrow \mathbb{R}$, a valuation of reference outcomes. To see this, denote by $\mathfrak{x}$ a row vector of valuations on a measurement with outcomes ${X_i}$. We may always rewrite it as a vector of valuations on the reference measurement,
\begin{align}
x = \mathfrak{x} P(X|R) \Phi,
\end{align}
so that $\langle \mathfrak{x}\rangle_\rho = \mathfrak{x} P(X|\rho) = \mathfrak{x} P(X|R)\Phi P(R|\rho) = x P(R|\rho) = \langle x \rangle_\rho$. Indeed, since $\langle \mathfrak{x}\rangle_\rho = \langle x\rangle_\rho$, the agent should value them at the same price.

Moreover, since $P(R|\rho)\in \col(P)$, any component of $x$ orthogonal to $\col(P)$ will not contribute to any expectation values. Indeed, if we write $x = x^{(\parallel)} + x^{(\perp)}$, then
\begin{align}
x P(R|\rho) &=  (x^{(\parallel)} + x^{(\perp)})P(R|\rho) =x^{(\parallel)}P(|\rho),
\end{align}
so that the gambler ought to assign the same price regardless. Thus without loss of generality, the gambler need only consider valuations on reference outcomes which lie in $\col(P)$. At the same time, if $x = \mathfrak{x} P(X|R) \Phi$, then since $P(X_i|R)\in \row(P)$, $x \in \row(P)\Phi$. We will often take $\Phi$ to be invertible, so that $x\in \row(P)$ itself. If $P=P^T$, then such an $x$ will automatically be in $\col(P)$.

\section{Rank factorization}

For convenience let $P\equiv P(R|R)$. Let $n$ be the number of outcomes of the reference measurement and let $r=\rank(P)=\dim(\col(P))=\dim(\row(P))$. We have just shown that our effects $P(E|R)$ live in $\row(P)$ while our states $P(R|\rho)$ live in $\col(P)$. Thus rather than work in the $n$-dimensional space of probability vectors, we may work directly in the $r$-dimensional spaces in which the states and effects live. To this end, we may introduce bases for the row and column spaces of $P$, and one nice way of thinking about this is in terms of a \emph{rank factorization} of $P(R|R)$.

\begin{theorem}
\label{full-rank-factorization-exists}
Suppose $P\in \mathbb{F}^{m\times n}_r$. There always exists a full rank factorization $P=\B{E}\B{S}$ where $\B{E}\in \mathbb{F}^{m\times r}_r$ and $\B{S}\in \mathbb{F}^{r \times n}_r$. Here $r$ is the rank of $P$.
\end{theorem}
\begin{proof}
Since $P$ has rank $r$, it has $r$ linearly independent columns. Let column vectors $e_1, \dots, e_r$ be a basis for the column space of $P$, and arrange them into an $m\times r$ matrix
\begin{align}
\B{E} = \begin{pmatrix} e_1 &\vline & \dots & \vline & e_r \end{pmatrix}.
\end{align}
By construction, every column $p_1, \dots, p_n$ of $P$ is a linear combination of the columns of $\B{E}$. Arrange the expansion coefficients for each column of $P$ into column vectors themselves $s_1, \dots, s_n$. Let
\begin{align}
\B{S} &= 	\begin{pmatrix} s_1 &\vline & \dots & \vline & s_n \end{pmatrix},	
\end{align}
which is an $r \times n$ matrix. Then $P = \B{ES}$ as desired.
\end{proof}
\begin{remark}
We could have equally well begun with a basis for the row space of $P$, constructing $\B{S}$ first, and then $\B{E}$. A straightforward way of obtaining a full rank factorization is by means of the singular value decomposition,
\begin{align}
P &= U\Sigma V^{\dagger} = \left[\begin{matrix} U_1 & U_2 \end{matrix}\right] \left[ \begin{matrix} \Sigma_r & 0 \\ 0 & 0 \end{matrix} \right]\left[ \begin{matrix} V_1^{\dagger} \\ V_2^{\dagger} \end{matrix}\right] =  U_1(\Sigma_r V_1^{\dagger}) = \B{ES},
\end{align}
where $r$ is the rank of $P$. 
\end{remark}
In particular, let $P(R|R)=\B{RS}$ be a rank decomposition of the conditional probability matrix which characterizes the reference measurement itself. If we write
\begin{align}
\B{R} = \sum_i |i)(R_i| = \begin{pmatrix} (R_1| \\ \vdots \\ (R_n| \end{pmatrix} && \B{S} = \sum_j |S_j)(j| = \begin{pmatrix} |S_1) & \cdots & |S_n) \end{pmatrix},
\end{align}
we may consider the rows of $\B{R}$ to be a compact representation of the effects $(R_i|$ and the columns of $\B{S}$ to be a compact representation of the states $|S_i)$ since by construction
\begin{align}
P(R_i|R_j) = (R_i|S_j).
\end{align}
In fact, this will lead to an alternative perspective on the Born matrix itself. To attain it, we first reprove a standard lemma: we have seen that any matrix has a $\{1\}$-inverse, but if a matrix has full row rank or full column rank, we can say even more. 
\begin{lemma}
\label{left-right-1-inverses}
Let $M\in \mathbb{F}^{r \times n}_r$: then $M^{(1)}$ is a right inverse of $M$. Similarly, let $M\in \mathbb{F}^{n \times r}_r$: then $M^{(1)}$ is a left inverse of $M$ \cite{GeneralizedInverses2003a}. 
\end{lemma}
\begin{proof}
If $M$ has full row rank, its $n$ columns span $\mathbb{F}^r$. Thus we can express basis vectors $e_i$ in $\mathbb{F}^r$ as linear combinations of the columns of $M$: $e_i = M r_i$ for some coefficient vectors $r_i$. But then arrange these $r_i$'s into the columns of a matrix $R$. Clearly, $I=MR$, so $M$ has a right inverse. But then multiply the $\{1\}$-inverse axiom by $R$: $M M^{(1)}M R = MR \Longrightarrow  M M^{(1)} = I$, as desired. On the other hand, if $M$ has full column rank, its $n$ rows span $\mathbb{F}^r$, so we can write the basis covectors $e_i = l_i M$, which amounts to $I= LM$, so $M$ has a left inverse. Then $L M M^{(1)}M  = LM \Longrightarrow  M^{(1)} M = I$, as desired.
\end{proof}
\begin{corollary}
\label{moore-penrose-left-right}
The Moore-Penrose pseudoinverse \cite{GeneralizedInverses2003a} may be calculated, in the case that $\B{R}$ has full column rank and in the case that $\B{S}$ has full row rank respectively, as
\begin{align}
	\B{R}^L = (\B{R}^\dagger \B{R})^{-1}\B{R}^\dagger && \B{S}^R = \B{S}^\dagger(\B{S}\B{S}^\dagger)^{-1}
\end{align}
\end{corollary}
\begin{proof}
We first must establish the invertibility $\B{R}^\dagger \B{R}$ and $\B{S}\B{S}^\dagger$. Suppose  $\B{R}^\dagger \B{R}$ were not invertible. Then $\B{R}^\dagger \B{R}x=0$ for some non-zero vector $x$. But then $x^\dagger\B{R}^\dagger \B{R}x=||\B{R}x||^2=0$ which can only be true if $\B{R}x$ is the zero vector. But then $\B{R}$ can't have a left inverse. We conclude that $\B{R}^\dagger \B{R}$ is invertible after all. The argument is the same for the invertibility $\B{S}\B{S}^\dagger$. Then clearly
$\B{R}^L\B{R}=\left[(\B{R}^\dagger \B{R})^{-1}\B{R}^\dagger \right]\B{R} =I= \B{S}\left[ \B{S}^\dagger(\B{S}\B{S}^\dagger)^{-1} \right]=\B{S}\B{S}^R$.
\end{proof}

But once we have one left or right inverse, we actually can obtain them all. More generally, given any $\{1\}$-inverse, we have the following parameterization of the space of all $\{1\}$-inverses\footnote{We observe in passing that all these constructions are now independent of the choice of scalars.}.

\begin{lemma}
\label{all-1-inverses}
Suppose $M^{(1)}$ is a $\{1\}$-inverse of $M \in \mathbb{F}^{n\times m}$ so that $MM^{(1)}M=M$. Then any other $\{1\}$-inverse can be obtained via
\begin{align}
\label{all-1-inverses-eq}
\tilde{M}^{(1)} = M^{(1)} + A - M^{(1)} M A M M^{(1)}.	
\end{align}
where $A \in \mathbb{F}^{m\times n}$ is completely arbitrary  \cite{GeneralizedInverses2003a, campbellGeneralizedInversesLinear2009, liGeneralFrameDecompositions1995, piziakMatrixTheory2007}.
\end{lemma}
\begin{proof}
On the one hand,
\begin{align}
M \tilde{M}^{(1)} M &= M\big(M^{(1)}  + A - M^{(1)}  M A M M^{(1)} \big)M \\ 
&= M M^{(1)}  M+ M A M- (M M^{(1)}  M) A (M M^{(1)} M) \\
&= M.
\end{align}
Conversely, suppose $\tilde{M}^{(1)} $ is a 1-inverse. We can obtain $\tilde{M}^{(1)} $ from $M^{(1)}$ by taking $A = \tilde{M}^{(1)}  - M^{(1)} $. Then the RHS of Eq. (\ref{all-1-inverses-eq}) becomes
\begin{align}
&= M^{(1)}  + (\tilde{M}^{(1)} - M^{(1)} )- M^{(1)}  M (\tilde{M}^{(1)} - M^{(1)} )MM^{(1)} \\
&=\tilde{M}^{(1)}  -M^{(1)}  (M \tilde{M}^{(1)} M)M^{(1)}  +M^{(1)}  (MM^{(1)}  M)M^{(1)}  \\
&= \tilde{M}^{(1)} .
\end{align}
\end{proof}

Let $P(R|\rho)$ be a state: by coherence, it must live in $\col(P)$. Let $P=\B{RS}$ be a rank decomposition of $P$. By construction, the columns of $\B{R}$ form a basis for $\col(P)$, and so we must have $P(R|\rho) = \B{R}|\rho)$ for some $|\rho)$. Similarly, by coherence, any effect $P(E|R)$ must live in $\row(P)$, but since the rows of $\B{S}$ form a basis for $\row(P)$, we must have $P(E|R) = (E|\B{S}$ for some $(E|$. We then come to the following lemma.

\begin{lemma}[The Born identity]
   \label{born_id_lemma}
Let $P=\B{R}\B{S}$ be full rank factorization such that $\B{R} \in \mathbb{F}^{n \times r}_r$ and  $\B{S} \in \mathbb{F}^{r\times n}_r$. Then 
\begin{align}
P \Phi P = P \Longleftrightarrow \B{S} \Phi \B{R} = I.
\end{align}
\end{lemma}
\begin{proof}
Clearly, acting on $\B{S} \Phi \B{R} = I$ from the left with $\B{R}$ and from the right with $\B{S}$ yields $P \Phi P = P$. Conversely, by Lemma (\ref{left-right-1-inverses}), $\B{R}$ has a left inverse $\B{R}^L$ and $\B{S}$ has a right inverse $\B{S}^R$. Acting from the left with $\B{R}^L$ and from the right with $\B{S}^R$  on $P\Phi P = P$, we conclude $\B{R}^L(\B{RS})\Phi(\B{RS})\B{S}^R = \B{R}^L(\B{RS})\B{S}^R$, that is, $\B{S}\Phi \B{R} = I$.
\end{proof}
Immediately, we have
\begin{align}
P(E|\rho) &= P(E|R)\Phi P(R|\rho) = (E|\B{S} \Phi \B{R} |\rho) = (E|\rho).
\end{align}
We conclude that for \emph{any} effect $E$ and state $\rho$, we may write $P(E|\rho) = (E|\rho)$. Going in the other direction, we have explicitly
\begin{align}
|\rho)= \B{R}^L P(R|\rho),
\end{align}
where $\B{R}^L$ is some left inverse of $\B{R}$ and
\begin{align}
(E| = P(E|R)\B{S}^R,
\end{align}
where $\B{S}^R$ is some right inverse of $\B{S}$. Putting them together, we have
\begin{align}
(E|\rho) = P(E|R)\B{S}^R\B{R}^L P(R|\rho).
\end{align}
Indeed, $P(\B{S}^R\B{R}^L)P = \B{RS}(\B{S}^R\B{R}^L)\B{RS}=P$ shows that we may always take $\Phi=\B{S}^R\B{R}^L$ as a Born matrix, a $\{1\}$-inverse of $P$. If we do so, however, such a Born matrix enjoys another generalized inverse axiom, namely,
\begin{align}
\Phi P \Phi = \B{S}^R \B{R}^L (\B{R}\B{S}) \B{S}^R \B{R}^L =  \B{S}^R \B{R}^L = \Phi.
\end{align}
In other words, such a $\Phi$ is a $\{1,2\}$-inverse. We note but do not prove the following theorem of Bjerhammar \cite{GeneralizedInverses2003a},
\begin{theorem}
   \label{Bjerhammar}
Let $X$ be a $\{1\}$-inverse of a matrix $M$. Then $X$ is a $\{1,2\}$ inverse if and only if $\rank(X) = \rank(M)$.
\end{theorem}

In this way, coherence with the reference implies that we can express $P(E|\rho)$ as the action of a covector $(E|=P(E|R)\B{S}^R$ on a vector $|\rho)=\B{R}^L P(R|\rho)$ in an $r=\rank(P)$ dimensional vector space. As such we will often refer to these objects as states and effects as well. Moreover, consider conditional probability matrices $P(R|E,R)$ which capture how the gambler ought to update their state assignments conditional on a subsequent measurement result $E$: $P(R|E, \rho)=P(E|\rho)^{-1}P(R|E,R)\Phi P(R|\rho)$. By coherence the row space of $P(R|E,R)$ ought to be $\subseteq \row(P)$ and the column space of such a matrix ought to be $\subseteq \col(P)$. Thus we may construct $T=\B{R}^L P(R|E,R) \B{S}^R$ which acts natively on states $|\rho)$ or effects $(E|$. 

Finally, let us make a brief comment about normalization. Suppose we have a measurement with outcomes $\{E_i\}$. The gambler assigns to each outcome an effect $P(E_i|R)$ so that together they form the rows of a conditional probability matrix $P(E|R)$. To guarantee that probabilities sum to 1, such a matrix ought to be column stochastic: equivalently, $\sum_i P(E_i|R) = P(1|R)= (1,\dots, 1)$. We see then that the effects in a measurement must sum to the effect $P(1|R)$ which assigns probability 1 to all states. Conceptually, this is like saying the gambler believes one of the outcomes must certainly occur; mathematically, it guarantees that the probabilities $P(E|\rho) = P(E|R)\Phi P(R|\rho)$ sum to 1. Picking a rank decomposition $P(R|R) = \B{RS}$, we have
\begin{align}
(1| = P(1|R)\B{S}^R && \forall \rho: (1|\rho) = 1 && \sum_i (E_i| = (1|,
\end{align}
where the latter holds for any measurement.

Of course, the representation of states, effects, and update maps so obtained is not at all unique, as the following theorem, whose proof we reproduce here, shows.

\begin{theorem}
Any two full rank factorizations $P=\B{R}_1\B{S}_1=\B{R}_2\B{S}_2$ are related by some invertible matrix $T$ so that $\B{R}_2=\B{R}_1T^{-1}$ and $\B{S}_2 = T\B{S}_1$ \cite{piziakFullRankFactorization1999}.
\end{theorem}
\begin{proof}
Since $\B{R}_2$ has full column rank, it has a left inverse; since $\B{S}_2$ has full row rank, it has a right inverse. Consider the two matrices $\B{R}_2^L \B{R}_1$ and $\B{S}_1 \B{S}_2^R$. We have $(\B{R}_2^L \B{R}_1)(\B{S}_1 \B{S}_2^R)=\B{R}_2^LP\B{S}_2^R=\B{R}_2^L \B{R}_2 \B{S}_2 \B{S}_2^R=I_r$.  Moreover,
\begin{align}
r = \rank(\B{S}_2) = \rank(\B{R}_2^L\B{R}_2\B{S}_2)= \rank(\B{R}_2^L\B{R}_1\B{S}_1) = \rank(\B{R}_2^L \B{R}_1),
\end{align}
where the last follows from the fact that $\B{S}_1$ has full row rank. Similarly,
\begin{align}
r = \rank(\B{R}_2) = \rank(\B{R}_2\B{S}_2\B{S}_2^R)=\rank(\B{R}_1\B{S}_1\B{S}_2^R)=	\rank(\B{S}_1\B{S}_2^R),
\end{align}
where the last follows from the fact that $\B{R}_1$ has full column rank. Thus $\B{R}_2^L \B{R}_1$ and $\B{S}_1 \B{S}_2^R$ are both full rank, and in fact they must be inverses of each other. Let $T=\B{R}_2^L \B{R}_1$ and $T^{-1} = \B{S}_1 \B{S}_2^R$. We then have
\begin{align}
\B{R}_1 T^{-1}=\B{R}_1 [\B{S}_1 \B{S}_2^R]=\B{R}_2\B{S}_2\B{S}_2^R=\B{R}_2 && T	\B{S}_1 = [\B{R}_2^L \B{R}_1]\B{S}_1 = \B{R}_2^L \B{R}_2\B{S}_2 = \B{S}_2,
\end{align}
as desired. 
\end{proof}
\begin{remark}
 Our freedom in choosing states $|\rho)$ and effects $(E|$ amounts to a choice of invertible $T$. But if a $\Phi$ matrix works for one choice, it works for them all: $\B{S}^\prime \Phi \B{R}^\prime = T\B{S}\Phi \B{R} T^{-1} = T T^{-1}=I$.	
\end{remark}

\section{Generalized probabilistic theories}

We defined a \emph{state} to be an equivalence class of consequences $\rho = \{\rho_i\}$ such that
\begin{align}
\forall j, k: P(R|\rho_j) = P(R|\rho_k),
\end{align}
that is, after any measurement with an outcome in the equivalence class, the gambler would assign the same probabilities to a subsequent reference measurement. A state $\rho$ may be identified on the one hand with a particular member of the equivalence class of outcomes, e.g., which is actually realized, and on the other hand, with a unique state of expectation about the outcomes of the reference measurement $P(R|\rho)$. Similarly, a \emph{effect} is an equivalence class of measurement outcomes $E = \{E_i\}$ such that
\begin{align}
\forall j,k: P(E_j|R) = P(E_k|R),
\end{align}
that is, conditional on a reference outcome, the gambler would assign the same probabilities to any outcome in the equivalence class upon a subsequent measurement. We may identify an effect $E$ with a particular member of the equivalence class, and at the same time the conditional distribution $P(E|R)$. Finally, we defined the \emph{reference measurement} itself to be a preparatory measurement such that 
\begin{align}
 P(R|\rho) = P(R|\rho^\prime ) &\Longleftrightarrow \forall E \in \mathscr{E}: P(E|\rho) = P(E|\rho^\prime) \Longleftrightarrow \rho \simeq \rho^\prime\\
P(E|R) = P(E^\prime|R) &\Longleftrightarrow \forall \rho \in \mathscr{S}: P(E|\rho) = P(E^\prime|\rho) \Longleftrightarrow E \simeq E^\prime,
\end{align}
that is, if two states are equivalent with respect to the reference measurement, they are equivalent with respect to any measurement, and similarly for effects. 

This framework established, we derived a norm a gambler may appeal to in the case that the reference measurement remains hypothetical,
\begin{align}
   P(E|\rho) = \mathcal{F}(P(E|R), P(R|\rho)) = P(E|R)\Phi P(R|\rho).
\end{align}
Moreover, we developed a geometrical picture of the states and effects. The state space $\mathscr{S}$ ought to be a convex set since we can always consider convex combinations of reference distributions: allowing for arbitrary rescalings, we arrive at the state cone $\hat{\mathscr{S}}$, recovering $\mathscr{S}$ as a base cut by the effect $P(1|R)=(1,\dots,1)$ which ensures normalization of probability vectors. Similarly, by contemplating coarse and fine graining of effects, we consider another convex set: the effect space $\mathscr{E}$, and its associated cone $\hat{\mathscr{E}}$. The effect space itself must satisfy $P(0|R) \leq P(E|R) \leq P(1|R)$ with respect to the ordering induced by the effect cone: this comes from the fact that if $P(E|R)$ is an effect then so must $P(\neg E|R) = P(1|R)-P(E|R)$. We also introduced a notion of $\Phi$-duality, so that e.g., $\hat{\mathscr{S}}^\star$ is the cone $\Phi$-dual to the state cone, and consists of all mathematically possible effects. None of these cones were assumed to be spanning of the $n$ dimensional vector space $\mathscr{V}$ in which the probability vectors live. In fact, we have just shown that $\mathscr{S} \subseteq \col(P(R|R))$ and $\mathscr{E} \subseteq \row(P(R|R))$.

Introducing bases for the column and row spaces of $P(R|R)$, we were able to rewrite 
\begin{align}
P(E|\rho) = (E|\rho),
\end{align}
where now we work in an $r=\rank(P(R|R))$ dimensional space $V$. In this setting, $P(E|\rho)$ is given by a linear functional which we may identify with the effect acting directly on a vector which we may identify with the state. Measurements become sets of effects $\{(E_i|\}$ which sum to the image of $P(1|R)$, that is, $\sum_i (E_i| = (1|$, such that $\forall \rho: (1|\rho)=1$. State updates conditional on measurement outcomes can be represented by linear operators $T$ acting on this same space.  In this representation, the state and effect cones become by construction \emph{generating cones}, that is, they satisfy $C-C=V$. Moreover, the effect cone is now a subset of the dual of the state cone in the usual sense of convex duality \cite{rockafellarConvexAnalysis1970}, i.e., $C^* = \{ y \in V^*: \forall x \in C: (y|x) \ge 0\}$, and similarly for the state cone. The ``cone sandwiches'' we exhibited to better understand possible collectives may also be reexpressed in terms of standard convex duality.

In fact, what we have arrived at is a \emph{convex operational theory} or \emph{generalized probabilistic theory} (GPT) \cite{mullerProbabilisticTheoriesReconstructions2021, barrettInformationProcessingGeneralized2007}, and in essence, we have proven a version of Ludwig's embedding theorem for finite dimensional GPTs \cite{lamiNonclassicalCorrelationsQuantum2018, ludwigAxiomaticBasisQuantum1985}.  For comparison, we reproduce a somewhat informal statement of that theorem as summarized in \cite{ludovicolamiGeneralProbabilisticTheories2018}.
\begin{theorem}[Finite dimensional Ludwig's embedding theorem, informal]
   \label{ludwig}
Let $\mathscr{S}$ be the set of states, and $\mathscr{E}$ be the set of effects. A \emph{theory} is a function $f:\mathscr{E} \times \mathscr{S} \rightarrow [0,1]$ which takes a state and an effect and returns a probability. Let $f$ satisfy the following axioms:
\begin{enumerate}
\item The function $f$ separates states and effects: if $\rho_1, \rho_2\in \mathscr{S}$ satisfy $\forall E \in \mathscr{E} : f(E, \rho_1) = f(E, \rho_2)$, then $\rho_1=\rho_2$. Similarly, if $E_1, E_2 \in \mathscr{E}$ satisfy $\forall \rho \in \mathscr{S}: f(E_1,\rho) = f(E_2,\rho)$, then $E_1=E_2$.
\item There is a trivial effect $u$ such that $\forall \rho \in \mathscr{S}: f(u, \rho)=1$, and for every $E\in \mathscr{E}$ there is an opposite effect $\neg E$ such that $\forall \rho \in \mathscr{S}: f(\neg E, \rho) = 1- f(E, \rho)$.
\item Probabilistic mixtures of states (respectively, effects) are again states (respectively, effects). For all states $\rho_1, \rho_2\in \mathscr{S}$ and probabilities $p\in[0,1]$, there exists a state $\tau$ such that $\forall E \in \mathscr{E}: f(E, \tau) = p f(E,\rho_1) + (1-p)f(E, \rho_2)$, and similarly for effects.
\end{enumerate}
Then there exists a vector space $V$ such that 
\begin{itemize}
\item $\mathscr{S} \subseteq V, \mathscr{E} \subseteq V^*$ are convex sets with $0, u \in \mathscr{E}$ and $\mathscr{E} = u - \mathscr{E}$.
\item $\mathbb{R}^+ \cdot \mathscr{S} = \hat{\mathscr{S}}$ is a cone, $\mathbb{R}^+ \cdot \mathscr{E} \subseteq \hat{\mathscr{S}}^*=\{y \in V^*: \forall x \in \hat{\mathscr{S}}, y(x)\ge0\} \subseteq V^*$ is the corresponding dual cone, and the set of states satisfies $\mathscr{S} = \{x\in \hat{\mathscr{S}}: e(x) = 1\}$. 
\item The generalized Born rule holds: $f(E, \rho) = E(\rho)$.
\end{itemize}
Here $V^*$ is the dual of the vector space $V$, the space of all linear functionals acting on $V$. We call $e$ the normalization functional.
\end{theorem}

\begin{remark}
The essential difference in our development is that we begin from the assumption of a reference measurement, and admit conditioning only on the consequences of an agent's own actions. The basic properties of the state and effect spaces for us derive from demanding coherence with the reference measurement.
\end{remark}

\begin{remark}
We merely assume that the states are a \emph{subset} of the cone dual to the effects, and the effects are a \emph{subset} of the cone dual to the states. In the language of GPTs, we do not assume the ``no-restriction hypothesis.'' 
\end{remark}

\begin{remark}
Another subtlety is that in the standard GPT presentation,
one takes states to live in an ordered real vector space $V$ and effects to live in the dual space of linear functionals $V^*$ so that $P(E|\rho) = E(\rho)$, where $E$ is a linear functional acting on a state $\rho$---without necessarily equipping $V$ with an inner product. But every finite dimensional vector space over $\mathbb{R}$ or $\mathbb{C}$ may be turned into an inner product space simply by choosing a basis and defining an inner product on the coordinates (identifying $V$ with $\mathbb{R}^n$ or $\mathbb{C}^n$). The finite dimensional Riesz representation theorem \cite{axlerLinearAlgebraDone2024} guarantees that one may identify each linear functional $E\in V^*$ with a unique vector $E \in V$ such that $P(E|\rho) = E(\rho) = \langle E, \rho\rangle = (E|\rho)$. For us, the inner product is inherited from the probability representation, and thus in essence we already work in the framework of ``geometric GPT's'' introduced in \cite{szymusiakCanQBismExist2025}.
\end{remark}

\subsection{Examples}
Classical probability theory is an example of a GPT: the state space is the space of probability distributions on $n$ outcomes, i.e., the probability simplex $\Delta_n$, and the effect space is the hypercube $\Delta_n^*$ dual to the simplex, the space of response functions. In other words, $P(E|\rho) = \sum_i P(E|\lambda_i)P(\lambda_i|\rho)$. The normalization functional is $(1|=(1,\dots, 1)$; measurements correspond to (column) stochastic matrices; and general transformations correspond to (sub)stochastic matrices.

Quantum theory is also a GPT. The state space of quantum theory is the space of $d \times d$ density matrices, i.e., positive semidefinite Hermitian matrices $\rho \ge 0$ with $\tr(\rho)=1$, which tells us that the normalization functional is $\tr(\cdot)$. The effect space consists of positive semidefinite Hermitian matrices $0 \leq E \leq I$ with no restriction on their trace. Measurements in quantum theory correspond to \emph{positive operator valued measures} (POVMs), which map events to positive semidefinite operators on Hilbert space or, more simply, which are collections of effects such that $\sum_i E_i=I.$  Both the state space and the effect space consist of PSD matrices: quantum theory is what is known as a \emph{self-dual} theory, for which effects can always be rescaled into states. Transformations correspond to completely positive trace preserving (CPTP) maps, of which unitary transformations, which preserve pure states, are a special case. Quantum theory provides an example of a Euclidean Jordan algebra, about which we will have much to say in the sequel.

Notice that we can consider quantum states and effects as elements of vector spaces over $\mathbb{R}$: \text{Herm}$(d)$ is a real vector space of dimension $d^2$. For example, for a qubit, the state space corresponds to the closed unit ball in $\mathbb{R}^3$, which we can see by expanding in the orthogonal basis provided by the Pauli matrices
\begin{align}
\rho &= \frac{1}{2}\Big(I + \tr(\sigma_x\rho) \sigma_x + \tr(\sigma_y\rho)  \sigma_y + \tr(\sigma_z\rho)  \sigma_z\Big).
\end{align}
In this representation, the normalization functional is $(1|=(1,0,0,0)$, and we have 
\begin{align}
 (E|=\frac{1}{2}\begin{pmatrix} \tr(E) & \tr(\sigma_x E) & \tr(\sigma_y E) &\tr(\sigma_z E)\end{pmatrix} && |\rho) = \begin{pmatrix} 	1\\ \tr(\sigma_x \rho)\\\tr(\sigma_y \rho)\\ \tr(\sigma_z \rho) \end{pmatrix}
\end{align}
so that
\begin{align}
P(E)= \tr(E\rho) = (E|\rho).
\end{align}
This same ``Bloch sphere'' construction can be generalized to any dimension $d$ by employing, for example, the generalized Gell-Mann matrices, which are similarly Hermitian, traceless, and orthogonal---and we will take recourse to the analogous representation in nonquantum theories. 

More generally, one can consider constructing GPTs from first principles, with more exotic state and effect spaces, e.g., theories with square state spaces, state spaces corresponding to higher dimensional spheres,  state spaces shaped like ice cream cones, and deduce many illuminating theorems relating their geometries to a diverse set of operationally grounded axioms.

\section{Frame theory}

Let us now work in the opposite direction, seeing how we can begin from the GPT framework and arrive at the Born matrix formalism. The bridge is \emph{frame theory} \cite{waldronIntroductionFiniteTight2018}.
\begin{definition}[Frame]
A \emph{frame} is a set of vectors $\{|\phi_i)\}$ in an inner product space $V$ over a field $\mathbb{F}$ such that there exist constants $0<a\leq b < \infty$ such that
\begin{align}
\forall x \in V: a(x|x) \leq \sum_i |(\phi_i| x)|^2 \leq b (x|x).
\end{align}
\end{definition}
\begin{remark}
 If all the frame vectors have equal norm, this is called a \emph{equal norm}, \emph{unweighted}, or \emph{unbiased} frame.
\end{remark}
\noindent One may begin from this definition, but it is easier to motivate it in the following way. Suppose $V$ to be finite dimensional, with dimension $r$. Then let $\mathcal{A}\in \mathbb{F}^{n \times r}_r$ be an \emph{analysis operator},
\begin{align}
\mathcal{A} = \sum_i |i)(\phi_i|:
\end{align}
$\mathcal{A}$ analyzes a vector $|x) \in V$ in terms of the frame vectors,
\begin{align}
|y) = \mathcal{A}|x) = \sum_i (\phi_i|x) |i).
\end{align}
Since $\mathcal{A}$ has full column rank, it has a left inverse. Another way of saying this is that the frame vectors span $V$. A \emph{synthesis operator} $\mathcal{S}$ is any choice of left inverse for $\mathcal{A}$. Writing it
\begin{align}
\mathcal{S} = \sum_i |\tilde{\phi}_i)(i|,
\end{align}
where $\{|\tilde{\phi}_i)\}$ are called \emph{dual vectors}, we have the resolution of the identity
\begin{align}
I = \mathcal{S}\mathcal{A} = \sum_i |\tilde{\phi}_i)(\phi_i|.
\end{align}
Notice that we may rewrite
\begin{align}
\sum_i |(\phi_i| x)|^2  = \sum_i (x|\phi_i)(\phi_i|x) = (x | \left[\sum_i |\phi_i)(\phi_i| \right]|x) = (x | F |x),
\end{align}
where 
\begin{align}
F=\mathcal{A}^\dagger \mathcal{A} = \sum_i |\phi_i)(\phi_i|
\end{align}
 is called the \emph{frame operator}, which is clearly not only Hermitian but also positive semidefinite. The frame condition then reads
\begin{align}
\forall |x) \in V: a (x|x) \leq (x|F|x) \leq b (x|x),
\end{align}
so that $a,b$ are the smallest and largest eigenvalues of $F$, respectively. As we saw in Corollary (\ref{moore-penrose-left-right}), $F$ must be invertible. Suppose it were not. Then for some $|x)$, $(x|F|x) = (x|\mathcal{A}^\dagger \mathcal{A}|x)=\lVert \mathcal{A}|x) \rVert^2 = 0$ which can only be true if $ \mathcal{A}|x)$ is the zero vector. But then $\mathcal{A}$ couldn't have a left inverse, a contradiction. Conversely, if $F$ is invertible, then the corresponding analysis operator has a left inverse, and the frame condition is satisfied. Of course, one can consider frames for \emph{subspaces} of a vector space, in which case $F$ must be invertible on that subspace, and the restriction of $\mathcal{A}$ to that subspace must have a left inverse.

A \emph{tight frame} is one for which $a=b$, implying
\begin{align}
I = \frac{1}{a}F = \sum_i \frac{1}{a} |\phi_i)(\phi_i|,
\end{align}
that is, the dual vectors are simply the frame vectors rescaled. When $a=1$, the tight frame is known as a Parseval frame. We note that
\begin{align}
G = \mathcal{A}\mathcal{A}^\dagger = \sum_{ij} (\phi_i|\phi_j) |i)(j|
\end{align}
is the Gram matrix of the frame. Calculating the Gram matrix is a useful way of checking whether the frame vectors really span the space: they will if and only if the rank of $G$ is $r$, the dimension of $V$. To see this, recall the following familiar lemma:
\begin{lemma}
Let $A = XX^\dagger$ and $B=X^\dagger X$. Then $A$ and $B$ have the same non-zero eigenvalues. 
\end{lemma}
\begin{proof}
$Av = \lambda v \Longrightarrow XX^\dagger v = \lambda v \Longrightarrow X^\dagger X (X^\dagger v) = \lambda (X^\dagger v)\Longrightarrow B v^\prime = \lambda v^\prime$. Thus if $\lambda, v$ is an eigenvalue/eigenvector pair of $A=XX^\dagger$, then $\lambda, X^\dagger v$ is an eigenvalue/eigenvector pair of $B=X^\dagger X$.
\end{proof}
\begin{corollary}
   For a Parseval frame, the $r\times r$ frame operator $F= \mathcal{A}^\dagger \mathcal{A} = I$ has $r$ eigenvalues equal to 1. Thus the $n\times n$ Gram matrix $G=\mathcal{A} \mathcal{A}^{\dagger}$ also has $r$ eigenvalues equal to 1, and is consequently a rank-$r$ projector. More generally, since the frame operator is invertible iff $\{|\phi_i)\}$ form a frame, and $F$ and $G$ share the same nonzero spectrum, then the Gram matrix will be rank-$r$ if and only if the $n$ frame vectors span the $r$-dimensional space $V$. 
\end{corollary}

Finally, call
\begin{align}
\mathcal{F}(\{|\phi_i)\}) = \tr(F^2) = \sum_{ij} |(\phi_i|\phi_j)|^2
\end{align}
the \emph{frame potential}. The frame potential allows one to characterize tight frames as the following lemma shows \cite{waldronIntroductionFiniteTight2018}.
\begin{lemma}
   \label{frame_pot_lower_bound}
Let $\{|\phi_i)\}$ be a frame for a vector space of dimension $r$ with frame operator $F=\sum_i |\phi_i)(\phi_i|$. The frame potential satisfies
\begin{align}
\mathcal{F}(\{|\phi_i)\}) \ge \frac{1}{r}\tr(F)^2
\end{align}
with equality iff $\{|\phi_i)\}$ forms a tight frame.
\end{lemma}
\begin{proof}
 Since $F$ is positive semidefinite and invertible, it will have $r$ nonnegative eigenvalues. Recalling
\begin{align}
\tr(F) = \sum_{k=1}^r \lambda_k && \tr(F^2) = \sum_{k=1}^r \lambda_k^2,
\end{align}
and applying the Cauchy-Schwartz inequality, 
\begin{align}
\left(\sum_{k=1}^r \lambda_k \cdot 1 \right)^2 \leq \left(\sum_{k=1}^r \lambda_k^2\right)\left(\sum_{k=1}^r 1^2\right),
\end{align}
we conclude
\begin{align}
\tr(F^2) \ge \frac{1}{r}\tr(F)^2.
\end{align}
Equality is given when the vectors $(\lambda_1, \dots, \lambda_r)=a(1,\dots, 1)$, that is, they are proportional to each other so that all the eigenvalues are the same.  We conclude that $\tr(F) = ra$, and
\begin{align}
F = \frac{\tr(F)}{r} I.
\end{align}
Thus the frame must be tight with frame constant $a=\tr(F)/r$. 
\end{proof}
\begin{corollary}
   In the special case that the frame vectors are normalized, $\tr(F) = \sum_i (\phi_i|\phi_i) = n $ so that the lower bound becomes $\mathcal{F}(\{|\phi_i)\}) \ge n^2/r$.
\end{corollary}

\subsection{Dual frames}
Taking the adjoint of the resolution of the identity
\begin{align}
I = \mathcal{A}^\dagger \mathcal{S}^\dagger = \sum_i |\phi_i)(\tilde{\phi}_i|,
\end{align}
we see that
\begin{align}
\mathcal{S}^\dagger = \sum_i |i)(\tilde{\phi}_i|
\end{align}
may be viewed as the analysis operator of a \emph{dual frame}, with frame vectors $\{ |\tilde{\phi}_i)\}$,  dual frame operator $\tilde{F} = \mathcal{S}\mathcal{S}^\dagger = \sum_i |\tilde{\phi}_i)(\tilde{\phi}_i|$, and with a choice of synthesis operator,
\begin{align}
\mathcal{A}^\dagger = \sum_i |\phi_i)(i|.
\end{align}
For any frame, there is a \emph{canonical dual frame}. One way of characterizing it, useful to us, is the following.
\begin{lemma}
   \label{dual_analysis_range}
The range of the dual analysis operator $\mathcal{S}^\dagger$ to be the same as the range of the original analysis operator $\mathcal{A}$ iff $\mathcal{S} = \mathcal{A}^+$, the Moore-Penrose pseudoinverse.
\end{lemma}
\begin{proof}
If $\col(\mathcal{S}^\dagger) = \col(\mathcal{A})$, then we can write
\begin{align}
\mathcal{S}^\dagger = \mathcal{A}X,
\end{align}
for some $X$. But then $I = \mathcal{S}\mathcal{A} = X^\dagger \mathcal{A}^\dagger \mathcal{A} = X^\dagger F$. Since $F$ is invertible, this means that $X^\dagger = F^{-1} = X$, since $F$ is Hermitian. We conclude that canonical synthesis operator is given by
\begin{align}
\mathcal{S}_{\text{can}} = F^{-1}\mathcal{A}^\dagger = (\mathcal{A}^\dagger \mathcal{A})^{-1}\mathcal{A}^\dagger = \mathcal{A}^+,
\end{align}
that is, the canonical synthesis operator is Moore-Penrose pseudoinverse $\mathcal{A}^+$ of the analysis operator. Conversely, if $\mathcal{S}=\mathcal{A}^+ = F^{-1}A^\dagger$, then $\mathcal{S}^\dagger = \mathcal{A}F^{-1}$ so that $\col(\mathcal{S}^\dagger) = \col(\mathcal{A} F^{-1}) = \col(\mathcal{A})$ since $F^{-1}$ is invertible\footnote{We can understand this as a basic property of the pseudoinverse itself. To see it, recall that $X=U \Sigma V^\dagger$ is the singular value decomposition of $X$, then $X^+ = V \Sigma^+ U^\dagger$, where $\Sigma^+$ is the diagonal matrix of the reciprocals of the nonzero singular values. Since the first $r=\rank(X)$ columns of $U$ form a basis for $\col(X)$ and the first $r$ rows of $V^\dagger$ form a basis for $\row(X)$, we have $\range(X^+) = \range(X^\dagger)$. If $X$ is Hermitian, then $\range(X^+)=\range(X)$.}.
\end{proof}
\noindent By the above argument, the canonical dual frame operator $\tilde{F}=F^{-1}$, and so the canonical dual frame satisfies
\begin{align}
\forall |x) \in V: \frac{1}{b} (x|x) \leq (x|F^{-1}|x) \leq \frac{1}{a} (x|x).
\end{align}

Notice that since $\mathcal{S}$ is a left inverse of $\mathcal{A}$, $\mathcal{A}\mathcal{S}\mathcal{A} = \mathcal{A}$: we can equivalently say that $\mathcal{S}$ is any $\{1\}$-inverse of $\mathcal{A}$. From Lemma \ref{1-inverse-proj-col}, we know therefore that $\mathcal{A}\mathcal{S}$ is a projector onto $\range(\mathcal{A})$. We may therefore understand this now familiar fact in a new way: we take a vector in the $n$-dimensional space where the frame representation lives, project it into the original $r$-dimensional space $V$, and then return back to the $n$-dimensional space. Since in general a frame is overcomplete, there may be many different $n$-dimensional vectors which project to the same $r$-dimensional vector. But there is a special choice: that given by the analysis operator itself, that is, the one that lives in $\range(\mathcal{A})$ and $\Pi = \mathcal{A}\mathcal{S}$ is the projector onto that subspace. We note in the special case that $\mathcal{S}=\mathcal{A}^+$, we have $\Pi = \mathcal{A}\mathcal{A}^+ = \mathcal{A} (\mathcal{A}^\dagger \mathcal{A})^{-1} \mathcal{A}^\dagger = \Pi^\dagger$: the projection is in fact an orthogonal projection.

Finally, as in Lemma (\ref{all-1-inverses}), notice that given one choice of synthesis operator, we can write any other choice of synthesis operator 
\begin{align}
\mathcal{S} = \mathcal{S}_0 + X(I - \Pi),
\end{align}
since $\mathcal{S}\mathcal{A} =  \mathcal{S}_0\mathcal{A} + X(I - \Pi)\mathcal{A} = I$, where $X$ is completely arbitrary, as $(I-\Pi)\mathcal{A}=0$.

\subsection{Frame coordinates and dual coordinates}

Given a frame with analysis and synthesis operators
\begin{align}
\mathcal{A} = \sum_i |i)(\phi_i| && \mathcal{S} = \sum_i |\tilde{\phi}_i)(i|,
\end{align}
and a choice of dual frame with analysis and synthesis operators
\begin{align}
\mathcal{S}^\dagger = \sum_i |i)(\tilde{\phi}_i| && \mathcal{A}^\dagger = \sum_i |\phi_i)(i|,
\end{align}
we have two resolutions of the identity $\mathcal{S}\mathcal{A}=I=\mathcal{A}^\dagger \mathcal{S}^\dagger$, providing two representations of a vector $|x)$,
\begin{align}
|x) = \mathcal{S}\mathcal{A}|x) = \sum_i (\phi_i|x) |\tilde{\phi}_i) = \mathcal{A}^\dagger \mathcal{S}^\dagger|x) = \sum_i (\tilde{\phi}_i|x) |\phi_i).
\end{align}
Let $|y)$ denote the vector with components $(\phi_i|x)$: we will call this the \emph{frame representation} in terms of \emph{regular coordinates}. Meanwhile, let $|\tilde{y})$ denote the vector with components $(\tilde{\phi}_i|x)$: the \emph{dual frame representation} in terms of \emph{dual coordinates}.

Suppose we seek a \emph{transfer matrix} $\mathcal{T}$ which takes us between the two representations. Unless the frame forms a basis, there will be multiple choices of dual frame, and moreover, for a given choice of dual frame, there will be multiple choices of transfer matrices $\mathcal{T}$. As you might suspect, $\mathcal{T}$ must be a $\{1\}$-inverse of the Gram matrix $G$, satisfying $G\mathcal{T} G = G$. Since $G=\mathcal{A}\mathcal{A}^\dagger$ and $\mathcal{A}$ and $\mathcal{A}^\dagger$ have left and right inverses respectively, this implies $\mathcal{A}^\dagger \mathcal{T}\mathcal{A} =I$. Then
\begin{align}
\mathcal{A}^\dagger \mathcal{T}\mathcal{A}|x) = \mathcal{A}^\dagger \mathcal{T}|y) = \mathcal{A}^\dagger |\tilde{y}^\prime )=|x),
\end{align}
that is, the dual synthesis operator acts on $|\tilde{y}^\prime )$ to give us back $|x)$. In this sense, $\mathcal{T}$ delivers perfectly good dual coordinates, but they need not lie in the range of the dual analysis operator $\mathcal{S^\dagger}$. If we select the canonical dual frame, however, then the range of $\mathcal{S^\dagger}$ is the same as the range of $\mathcal{A}$. Moreover, $\range(G) \subseteq \range(\mathcal{A})$, and taking $\mathcal{T}= G^+$, $\range(\mathcal{T}) = \range(G)$ so that $\mathcal{T}|y) = |\tilde{y})$ for $|\tilde{y}) =\mathcal{S}^\dagger |x)$. Finally, $G|\tilde{y}) = \mathcal{A} \mathcal{A}^\dagger \mathcal{T} \mathcal{A}|x) = \mathcal{A}|x)=|y)$ takes us back to dual coordinates to regular frame coordinates.

\subsection{Double sided frames}

Suppose we introduce two interlocking frames, one for vectors and another for covectors: we shall call such a construction a \emph{double sided frame}, a notion implicit in the literature, but original to us. In the context of a GPT, this would be like introducing a frame for states consisting of reference effects $\{(R_i|\}$, and a frame for effects consisting of reference states $\{|S_i)\}$. In other words, the frame vectors (or covectors) are in fact states and effects in the GPT. If the GPT dimension is $r$, then we demand that the reference states and effects both span the $r$-dimensional space: and in this context, we call them \emph{informationally} or \emph{tomographically} complete. (We will have more to say on this notion in the following section.) Of course,  we would also like the reference effects to form a legitimate measurement, that is, $\sum_i (R_i| = (1|$. 

In other words, let
\begin{align}
\B{R} = \sum_i |i)(R_i| && \B{S} = \sum_i |S_i)(i|
\end{align}
be the analysis operators for states and effects, respectively. If the reference effects are spanning, then $\B{R}$ will have a left inverse, $\B{R}^L\B{R}=I$ and similarly if the reference states are spanning, then $\B{S}$ will have a right inverse $\B{S}\B{S}^R= I$. Putting these two resolutions of the identity together, we have
\begin{align}
\B{S}(\B{S}^R \B{R}^L)\B{R}  = \B{S} \Phi \B{R} = I.
\end{align}
Indeed, this is a necessary and sufficient condition for such a $\Phi$ to exist:
\begin{theorem}
   \label{phi-existence}
Let $A \in \mathbb{F}^{r \times m}_r$ and $B \in \mathbb{F}^{n \times r}$. Then there exists a $\Phi\in \mathbb{F}^{m\times n}$ such that $A \Phi B = I$ iff $\rank(A) = \rank(B)=r$.
\end{theorem}
\begin{proof}
If $A \Phi B = I$, then from
\begin{align}
r &= \rank(I) = \rank(A \Phi B) \leq \rank(A) \leq r\\
r &= \rank(I) = \rank(A \Phi B) \leq \rank(B) \leq r,
\end{align}
we conclude that $\rank(A) = \rank(B)=r$.  Conversely, if $\rank(A)=r$, then $A$ has a right inverse $A^R$, and if $\rank(B)=r$, then $B$ has a left inverse $B^L$. Then we may take $\Phi = A^R B^L$: in fact, by Theorem (\ref{Bjerhammar}) such a $\Phi$ is a $\{1,2\}$-inverse.
\end{proof}
\noindent Now if we have any $\Phi$ satisfying $\B{S}\Phi \B{R} = I$, then clearly, $\BT{S}\equiv \B{R}^L=\B{S}\Phi$ is a valid synthesis operator for $\B{R}$, and $\BT{R} \equiv \B{S}^R=\Phi \B{R}$ is a valid synthesis operator for $\B{S}$: in this sense, the two frames are interlocking. Thus if we start from $\B{S}\Phi \B{R} = I$, we can derive $\BT{R}$ and $\BT{S}$; conversely if we start from $\BT{S}$ and $\BT{R}$, we can obtain a Born matrix $\Phi = \BT{R}\BT{S}$. Of course, this is not the only choice of $\Phi$. As we have seen, satisfying $\B{S}\Phi \B{R} = I$ is equivalent to $P\Phi P = P$, where $P=\B{R}\B{S}$: that is, $\Phi$ must be a $\{1\}$-inverse of $P$, and unless $P$ is invertible, there is no unique choice. But however the twin synthesis operators are obtained, we may define dual effects and dual states as the rows and columns respectively of
\begin{align}
\BT{S} = \sum_i |\tilde{S}_i)(i| && \BT{R} = \sum_i |i)(\tilde{R}_i|.
\end{align}
Our two resolutions of the identity 
\begin{align}
\BT{S}\B{R} = I = \B{S}\BT{R}
\end{align}
then yield
\begin{align}
|\rho) &= \BT{S}\B{R}|\rho) = \sum_i |\tilde{S}_i)(R_i|\rho) = \sum_i P(R_i|\rho) \ |\tilde{S}_i)\\
|\rho) &= \B{S}\BT{R}|\rho) = \sum_i |S_i)(\tilde{R}_i|\rho) = \sum_i W(R_i|\rho) \ |S_i) = \sum_{ij} \Phi_{ij} P(R_j|\rho) |S_i),
\end{align}
where we write $W(R_i|\rho)$ to signify that these quantities need not be probabilities.

As usual, the Moore-Penrose pseudoinverse plays a special role in picking out a $\Phi$. Let $F_S$ and $F_R$ be the frame operators for our two interlocking frames. In the canonical case, we have $\B{S}^R = \B{S}^\dagger F_S^{-1} = \Phi \B{R}$ and $\B{R}^L = F_R^{-1}\B{R}^\dagger =  \B{S}\Phi$, and  the choice of $\Phi=\B{S}^R\B{R}^L = \B{S}^\dagger F_S^{-1} F_R^{-1}\B{R}^\dagger$ tells us that in fact $\Phi = P^+$.
\begin{theorem}
If $M=AB$ is a rank factorization, $M^+ = B^+ A^+ = B^\dagger (BB^\dagger)^{-1} (A^\dagger A)^{-1} A^\dagger$ \cite{GeneralizedInverses2003a}.
\end{theorem}

Finally, just as $P=\B{RS} = \sum_{ij} (R_i|S_j) |i)(j|$,  we may define the channel operator 
\begin{align}
C = \B{S}\B{R} = \sum_i |S_i)(R_i|,
\end{align}
so that $(E|C|\rho) = \sum_i (E|S_i)(R_i|\rho) = P(E|R)P(R|\rho)=P(E|R, \rho)$. If $P$ is the double sided frame analogue of a Gram matrix, then $C$ is the double sided frame analogue of the frame operator. In fact, $C$ picks out yet another choice of Born matrix, namely $\Phi = \B{R}C^{-2}\B{S}$. 
\begin{theorem}
Let $M=AB$ be a rank factorization and $BA$ is invertible. Then $M^\sharp = B (BA)^{-2} A$ is a $\{1,2,5\}$-inverse \cite{GeneralizedInverses2003a}, or \emph{group inverse}.
\end{theorem}

\section{Informational completeness}

\subsection{Separation}

The essence of a GPT is a state space and effect space which separate each other. This means that states are uniquely determined by the probabilities they induce on all the effects, and the effects are uniquely determined by the probabilities they induce on all the states. In other words, a GPT is a model which is entirely grounded in probability assignments: it contains no extraneous structure beyond the probabilities an agent might assign to all possible measurements on a system. This property of ``separation'' may be checked in a purely linear algebraic way, as the following theorem shows.

\begin{theorem}[Adapted from \cite{shahandehUnifiedLinearAlgebraic2025}]
\label{separation-theorem}
Let $\mathscr{S}$ and $\mathscr{E}$ be state and effect spaces. The following two statements are equivalent:
\begin{enumerate}
\item $\mathscr{S}$ and $\mathscr{E}$ separate each other.
\item $\mathscr{E}$ separates vectors in $\linspan(\mathscr{S})$ and $\mathscr{S}$ separates vectors in $\linspan(\mathscr{E})$.
\end{enumerate}
\end{theorem}
\begin{proof}
 $(2)\Longrightarrow (1)$ is immediate since $\mathscr{S} \subseteq \linspan(\mathscr{S})$ and $\mathscr{E} \subseteq \linspan(\mathscr{E})$. As for $(1)\Longrightarrow (2)$: by the assumption,
 \begin{align}
 \forall \pi \in \mathscr{S}: (E|\pi) = (E^\prime |\pi) &\Longleftrightarrow (E| = (E^\prime| \\
 \forall E \in \mathscr{E}: (E|\pi) = (E|\pi^\prime) &\Longleftrightarrow |\pi) = |\pi^\prime).
 \end{align}
Now suppose on the contrary that $\mathscr{E}$ did \emph{not} separate $\linspan(\mathscr{S})$. This would mean that
\begin{align}
\exists |x), |y) \in \linspan(\mathscr{S}) \text{ such that } \forall E \in \mathscr{E}: (E|x) = (E|y) \text{ but } |x) \neq |y).
\end{align}
On the one hand, $\forall E \in \mathscr{E}: (E|\big(|x)-|y)\big) = 0$ tells us in particular that $(1|\big(|x)- |y)\big)=0$, where $\forall \pi \in \mathscr{S}: (1|\pi) =1$. On the other hand, $|x)-|y) \in \linspan(\mathscr{S})$ so we can express it as a linear combination of states $|x)-|y)  = \sum_i \alpha_i |\pi_i)$ yielding
\begin{align}
(1|\big(|x)- |y)\big)= \sum_i \alpha_i (1|\pi_i) = \sum_i \alpha_i= 0.
\end{align}
Separating out the nonnegative from the negative components in the linear combination, we have $\sum_{\alpha_i \ge 0} \alpha_i = \sum_{\alpha_i < 0} |\alpha_i| = N$ for some $N$. Thus  
\begin{align}
|\sigma^+) = \sum_{\alpha_i \ge 0} \frac{\alpha_i}{N} |\pi_i) && |\sigma^-) =  \sum_{\alpha_i < 0} \frac{|\alpha_i|}{N} |\pi_i) 
\end{align}
are both convex combinations of states and so states themselves. Therefore
\begin{align}
\forall E \in \mathscr{E}: \frac{1}{N}(E|\big( |x) - |y) \big) &=  \frac{1}{N}(E|\sum_i \alpha_i |\pi_i) =  (E| \left\{ \sum_{\alpha_i \ge 0} \frac{\alpha_i}{N} |\pi_i) - \sum_{\alpha_i < 0} \frac{|\alpha_i|}{N} |\pi_i)\right\}\\
&= \big(E|(|\sigma^+) - |\sigma^-\big) = 0.
\end{align}
But since the effects separate the states, $\forall E \in \mathscr{E}: (E|\sigma^+) = (E|\sigma^-)$ implies that $|\sigma^+) = |\sigma^-)$. But $|\sigma^+) - |\sigma^-) = \frac{1}{N}\big(|x)-|y)\big)$ this implies that $|x)=|y)$, contradicting our original assumption. We conclude $\mathscr{E}$ separates $\linspan(\mathscr{S})$ after all.
 
 We use a different proof strategy to show that $\mathscr{S}$ must separate $\linspan(\mathscr{E})$. Again assuming the contrary means that
 \begin{align}
 \exists (x|, (y| \in \linspan(\mathscr{E}) \text{ such that } \forall \pi \in \mathscr{S}: (x|\pi) = (y|\pi) \text{ but } (x| \neq (y|.
 \end{align} 
In other words, $\forall \pi \in \mathscr{S}: \big((x|-(y|)|\pi) = 0$. Now suppose we pick an effect $(e_0|$ in the \emph{relative interior} \cite{boydConvexOptimization2004} of the effect space. Recall that a point is in the relative interior of a set $S$ just when there exists an $\epsilon > 0$ such that the intersection of a ball of radius $\epsilon$ centered on $x$ with the affine hull of $S$ is contained in the set: the relative interior is appropriate for talking about the interior of a set which is confined e.g., to a subspace. Since the effect space contains the $0$ vector, in fact, its affine hull is the $\linspan(\mathscr{E})$\footnote{Recall that $\text{aff}(S) = \left\{\sum_i \alpha_i x_i, x_i \in S, \sum_i \alpha_i = 1\right\}$. If $0\in S$, then we can extend $y= \sum_i \alpha_i x_i + (1-\overline{\alpha})0$ where $\overline{\alpha} = \sum_i \alpha_i$. Then $y$ is written as an affine combination. In the other direction, an affine combination is already a linear combination.}.  Then for sufficiently small $\epsilon > 0 $, let
\begin{align}
(e| = (e_0| + \epsilon \big((x|-(y|)\big).
\end{align}
Since $(e_0|$ is in the relative interior, then $(e|$ will be a valid effect. But 
\begin{align}
\forall \pi: (e|\pi) = (e_0|\pi) + \epsilon \big((x|-(y|)\big)|\pi) = (e_0|\pi),
\end{align}
showing that the $\pi$'s can't distinguish $(e_0|$ from $(e|$. But this contradicts our assumption: we conclude that $\mathscr{S}$ can separate $\linspan(\mathscr{E})$ after all.
\end{proof}

\begin{corollary}
   Let $\{(R_i|\}$ be a set of effects whose span is $\linspan(\mathscr{E})$ and $\{|R_i)\}$ be a set of states whose span is $\linspan(\mathscr{S})$. Then the probabilities $P(R|\rho)$ suffice to characterize any state $\rho$ and the probabilities $P(E|R)$ suffice to characterize any effect $E$. We call $\{(R_i|\}$ and $\{|R_i)\}$ \emph{informationally complete}, and we call a preparatory measurement characterized by the conditional probability matrix $P(R|R)$ a \emph{reference measurement}.
\end{corollary}

In our earlier development, we took the existence of a reference measurement as an assumption. If we start instead with an arbitrary GPT, is there guaranteed to be a reference measurement? We reproduce here a proof given in \cite{barnumCloningBroadcastingGeneric2006,barrettFinettiTheoremTest2009} that informationally complete measurements exist in any GPT which allows all mathematically possible effects, that is, which satisfies the no-restriction hypothesis. In fact, it shows that minimal IC (MIC) measurements always exist, where $n=r$: the number of outcomes is precisely the dimension of the GPT. The effects of a MIC therefore form a basis, and so $\Phi = P(R|R)^{-1}$.
\begin{theorem}
   \label{mics-always-exist}
There exists a MIC in any GPT satisfying the no-restriction hypothesis.
\end{theorem}
\begin{proof}
Pick an arbitrary basis $\{(b_i|\}_{i=1}^r$. We have $\sum_i (b_i| = (s|$, whatever $(s|$ may be, but we can always find some invertible matrix $V$ which takes $(s|V = (1|$ so that the covectors $(b^\prime_i|=(b_i|V$ sum to $(1|$. Now let $m$ be the minimal value attained by any of the $(b^\prime_i|$'s on any of the states $\rho$ of the GPT, that is, $m = \min_{i, \rho \in \mathscr{S}} (b^\prime_i|\rho)$. Because the state space is assumed compact, the minimum can be attained. In finite dimensions, every linear functional is continuous, and so the image of $f(\rho) = (E|\rho): \mathscr{S}\rightarrow \mathbb{R}$ must be itself compact. Thus the minimum is some finite value.  Clearly, $(b^{\prime\prime}_i| = (b^\prime_i| - m(1|$ satisfies
\begin{align}
\forall i, \rho \in \mathscr{S}: (b^{\prime\prime}_i| \rho) = (b^\prime_i|\rho) - m \ge \min_{i, \rho \in \mathscr{S}} (b^\prime_i|\rho) - m\ge 0.
\end{align}
Moreover, $\sum_i (b^{\prime\prime}_i| = (1-rm)(1|$ so that $\sum_i (b^{\prime\prime}_i|\rho) = 1-rm > 0$, since each term is individually nonnegative, and they are assumed to not all be identically 0. Thus let
\begin{align}
(R_i| = \frac{1}{1-rm}\Big( (b_i|V - m (1| \Big).
\end{align}
$(R_i|$ must still span the space, and since there are $r$ elements, it must form a basis. Moreover, the $(R_i|$ are nonnegative on all states, and sum to $(1|$, and so represent a logically possible measurement.
\end{proof}

\begin{remark}
\emph{Any} measurement is informationally complete for \emph{some} set of states and effects. 
\end{remark}

\begin{remark}
In particular, one may always construct an equiangular measurement in no-restriction GPT. One simply begins with a regular $(r-1)$-simplex with $r$ vertices. Then as in the theorem, one shrinks the simplex until the effects corresponding to the vertices (obtained by adding in a component proportional to $(1|$) are nonnegative on all states in the GPT. Such effects must form a linearly independent basis.
\end{remark}

\begin{remark}
In \emph{any} GPT, not just no-restriction GPTs, reference measurements exist. At worst, one may simply consider all allowed measurements: the effects of each individually sum to $(1|$ so that they may be rescaled to form one single measurement. Since GPT effects must separate the states, this will be a reference measurement.
\end{remark}

\section{de Finetti for GPT's}

Using informationally complete measurements, \cite{barrettFinettiTheoremTest2009} proves a de Finetti representation theorem for a wide class of GPT's\footnote{In fact, they work in the essentially equivalent framework of ``test spaces.''}. Analogous to the classical de Finetti representation theorem, this theorem gives a subjective Bayesian justification for when one can act \emph{as if} repeated reference measurements are estimating ``unknown states.''

\begin{theorem}
Suppose $|\rho^n) \in V^{\otimes n}$ is exchangeable, meaning that it is
\begin{enumerate}[(a)]
\item   invariant under permutations of the $n$ systems;
\item  $n$-fold nonsignalling, that is, nonsignalling across every bipartite split: this means that the marginal probability of obtaining an outcome on system $A$ does not depend on what measurement is done on $B$;
\item extendable, so that $|\rho^n)$ may be obtained as the marginal of a state $|\rho^{N})$ for any $N>n$,
\end{enumerate}
then $|\rho^n)$ may be uniquely expressed 
\begin{align}
|\rho^n) = \int_{\mathcal{S}} d\mu(\rho) |\rho)^{\otimes n},
\end{align}
where $\mu$ is a unique probability measure on the single system state space $\mathscr{S}$ which is independent of $n$.
\end{theorem}

The authors adopt the proof strategy of \cite{cavesUnknownQuantumStates2002} in the setting of quantum mechanics: they introduce an informationally complete reference measurement to express $|\rho^n)$ as a joint probability distribution on $n$ copies of the reference measurement, and then appeal to the classical de Finetti theorem. A crucial assumption, then, is that performing reference measurements separately on each part of the collective is equivalent to performing a reference measurement on the whole: this assumption is usually called \emph{local tomography}. A famous example where local tomography fails is quantum theory over the real numbers. Finally, we note that \cite{fuchsFinettiRepresentationTheorem2004} provides an analogous de Finetti theorem justifying the ``estimation of unknown quantum processes'': the proof strategy makes use of the Choi representation theorem and so may be extended to GPTs which admit a similar duality between states and transformations \cite{chiribellaProbabilisticTheoriesPurification2010}.

\section{Conclusion}

In this chapter, we have explored the properties of the Born matrix $\Phi$, seeing that it must form a $\{1\}$-inverse of $P(R|R)$, the conditional probability matrix that characterizes the reference measurement. Further pushing the notion of nonclassical coherence allowed us to derive fundamental subspace restrictions for states, effects, and valuations. Using the technique of rank factorization, we then showed that the QBist framework is essentially equivalent to that of the framework of generalized probabilistic theories, allowing results of the latter to be imported into the former and vice versa. We now turn to the question of what is the simplest form the fundamental nonclassical coherence condition $P(E|\rho)= P(E|R)\Phi P(R|\rho)$ can take, showing that in GPT-land, it may almost always be expressed as a particularly gentle modification of the law of total probability.

%% file: sections/urgleichungen.tex
\UMBchapter{Die Urgleichungen}
\label{ch:urgleichungen}

\section{Introduction}

As we have seen, on very general principles, we may derive a nonclassical coherence condition
\begin{align}
P(E|\rho) = P(E|R)\Phi P(R|\rho),
\end{align}
where $\Phi$ is any $\{1\}$-inverse of $P(R|R)$, the conditional probability matrix that characterizes the reference measurement. In particular, in quantum theory, if the reference measurement is taken to be a SIC-POVM \cite{RevModPhys.85.1693}, we arrive at the QBist \emph{Urgleichung}, or fundamental equation, 
\begin{align}
P(E|\rho ) = \sum_i P(E|R_i)\left\{(d+1)P(R_i|\rho) - \frac{1}{d}\right\},
\end{align}
which looks quite close to the law of total probability: the only difference is that one rescales and shifts the reference probabilities $P(R|\rho)$. Now this precise expression is unique to a SIC, but a SIC is an example of a so-called quantum state 2-design, and any measurement forming a 2-design will enjoy a similar expression \cite{slomczynskiMorphophoricPOVMsGeneralised2020}. More generally, we can ask: in any GPT, when it is possible to choose a reference measurement which yields a \emph{Protourgleichung},
\begin{align}
P(E|\rho) = \sum_i P(E|R_i)\Big\{ \alpha P(R_i|\rho) + (1-\alpha)P(R_i|\mu)\Big\},
\end{align}
where $\mu$ represents a state of complete uncertainty? If $\alpha = d+1$ and $P(R|\mu) = 1/d^2$, as it does for a SIC, we recover the Urgleichung itself. We now show that in fact such expressions are ubiquitous in the landscape of GPTs: in a GPT which allows all mathematically possible states and effects, a set of reference states may be chosen so that any reference measurement gives rise to a \emph{Protourgleichung}. An equivalent condition can be stated in terms of the channel operator associated with the reference measurement: it must be depolarizing, and we show that this places restrictions on the action of $P(R|R)$. We then explore in our terms the recently introduced notion of morphophoricity \cite{slomczynskiMorphophoricPOVMsGeneralised2020, szymusiakCanQBismExist2025}, the idea that a reference measurement maps a state space into the probability simplex in a shape preserving way, providing an alternative proof that such measurements correspond to the tight IC-POVMs pioneered by Andrew Scott \cite{scottTightInformationallyComplete2006}. Finally, we examine critically a measure of nonclassicality found in the QBist literature, $\lVert I - \Phi\rVert$ with respect to any unitarily invariant norm \cite{debrotaSymmetricInformationallyComplete2020}, showing that in the overcomplete case, this measure is unable to capture the essence of the \emph{Protourgleichung} in a univocal way.

\section{Bloch form}

In what follows, it will prove useful to bring our states and effects into a standard \emph{Bloch (block) form}. First we would like the normalization functional, corresponding to $P(1|R)$, to be represented as $(1|=(1,0,\dots,0)$ so that states can be represented as
\begin{align}
|\rho) =  \left(\begin{array}{c}  
1  \\
\hline
\rho_0
\end{array}\right).
\end{align}
We would also like to distinguish a particular state $|\mu)$ to represent the ``state of complete uncertainty.'' In some GPTs, there is a unique choice of such a state. For example, in classical probability theory, such a state would correspond to $(1/n, \dots, 1/n)$. Notice that this is the unique state which is invariant under all the symmetries of the state space which in this case are permutations of the simplex vertices. In quantum theory, such a state would correspond to $I/d$: similarly, this is the unique state invariant under the \emph{unitary} symmetries of the state space. More generally, given a GPT state space, we may consider a state  $|\mu)$ which is invariant under the reversible transformations which preserve the state space, and adopt this as the ``state of complete uncertainty'' \cite{Janotta_2014,szymusiakCanQBismExist2025}. But in general, there may not be a single unique such state \cite{scandoloInformationtheoreticFoundationsThermodynamics2019}. Moreover, in classical and quantum theory, several ideas coincide: a state which is invariant under state space symmetries, the geometric center of the state space, the maximal entropy state. In terms of our reference measurement formalism, we might be inclined to pick a state corresponding to an eigenvector with eigenvalue 1 of $P(R|R)$: indeed, since $P(R|R)$ is column stochastic, it must have at least one such eigenvector which is a probability distribution \cite{hornMatrixAnalysis1985}: but in general, it will not have a unique such eigenvector. Thus the choice of $|\mu)$ depends essentially on the nature of the GPT. Of course, the simplest solution is to take the uniform average of all the extremal states. For our purposes, however, it suffices merely to distinguish \emph{some} state $|\mu)$ in the interior of the state space, and we leave open the question of how to choose it. We will then work in a representation such that $|\mu)=(1,0, \dots, 0)^\dagger$. 

\begin{lemma}
Any finite-dimensional GPT can be brought, non-uniquely, into Bloch form: for a choice of normalized state $|\mu)$, one may choose coordinates such that
\begin{align}
(1|=\begin{pmatrix} 1 & 0 & \dots & 0 \end{pmatrix}
\qquad
|\mu)=
\begin{pmatrix}
1\\
0\\
\vdots\\
0
\end{pmatrix},
\end{align}
where $(1|$ is the normalization functional.
\end{lemma}
\begin{proof}
Let $V$ be the $r$ dimensional real vector space spanned by the states of the GPT, and let $(1|\in V^*$ be the unit effect such that $\forall |\rho) \in \mathscr{S}: (1|\rho)=1$. Choose a normalized state $|\mu)$. Every vector $|v) \in V$ decomposes uniquely as
\begin{align}
|v) = (1|v)|\mu) + |v_0),
\end{align}
where $|v_0) = |v) - (1|v)|\mu)$ and $(1|v_0)=0$. We can thus split $V = \mathbb{R}|\mu) \oplus V_0$ where we shall call $V_0 = \{ |x) \in V: (1|x)=0\}$ the ``traceless subspace.'' Picking a basis $\{ |b_1), \dots, |b_{r-1})\}$ for $V_0$, we may define an invertible matrix
\begin{align}
B=\begin{pmatrix}
|\mu) & |b_1) & \cdots & |b_{r-1})
\end{pmatrix}.
\end{align}
Then $(E|\rho) = (E| B B^{-1} |\rho) = (E^\prime|\rho^\prime)$ shifts us into the new coordinate system in a probability preserving way. In particular, since $B(1,0,\dots)^\dagger = |\mu)$, 
\begin{align}
|\mu^\prime) = B^{-1}|\mu) = \begin{pmatrix} 1 \\ 0 \\ \vdots \end{pmatrix}.
\end{align}
At the same time, since $(1|\mu)=1$ and $(1|b_i)=0$, 
\begin{align}
(1^\prime| = (1|B = (1,0,\dots),
\end{align}
as desired. The choice of block representation is not unique, since it depends on the chosen normalized state $|\mu)$ and on the choice of basis for $V_0$.
\end{proof}

In what follows therefore without loss of generality, we will assume that $(1|=(1,0,\dots)$ and $|\mu)=(1,0,\dots)^\dagger$ so that effects and states may be expressed
\begin{align}
(E| = 	\left(\begin{array}{c|c}w & E_0 \end{array}\right) && |\rho) =  \left(\begin{array}{c}  
1  \\
\hline
\rho_0
\end{array}\right),
\end{align}
where $w = P(E|\mu)$. Moreover, a measurement may be expressed as a block matrix whose rows are effects,
\begin{align}
\B{E}=
\left(\begin{array}{c|ccc}  
& & & \\
w & & \B{E}_0 \\
& & &\\
\end{array}\right),
\end{align}
where $w$ is a column vector of weights satisfying $\sum_{i=1}^n w_i = 1$ and the columns of $\B{E}_0 \in \mathbb{R}^{n \times (r-1)}$ sum to 0. Thus $\sum_{i=1}^n (E_i| = (1|$, or more compactly, $u^\dagger \B{E} = (1|$ where $u^\dagger = (1,\dots, 1)$. Interpreting $|\mu)$ as the state of complete ignorance, the weights $w_i$ can be understood as the \emph{biases} of the measurement $ P(E_i|\mu) = w_i$: even in the state of complete ignorance, the gambler might nevertheless assign nonuniform probabilities to the outcomes of a measurement signifying its bias.

\section{Ubiquitous Urgleichung}

An operator $\B{R}$ representing a measurement can be thought of as a linear map from the state space into the probability simplex, $P(R|\rho) = \B{R}|\rho)$. Expressing it in block form makes clear that it can alternatively be understood as an \emph{affine transformation} from the traceless subspace instead,
\begin{align}
\label{p_affine}
P(R|\rho) &= \B{R}|\rho) = 	\left(\begin{array}{c|ccc}  
& & & \\
w & & \B{R}_0 \\
& & &\\
\end{array}\right) \left(\begin{array}{c}  
1  \\
\hline
\rho_0
\end{array}\right)= w + \B{R}_0 \rho_0. 
\end{align}
Assuming that $\{R_i\}$ is an informationally complete measurement, if we were to demand that the quasiprobabilities $W(R|\rho)=\BT{R}|\rho)$ be close as possible to $P(R|\rho) = \B{R}|\rho)$, we could do no better than to assume that
\begin{align}
\label{w_affine}
W(R|\rho) = w+ \alpha \B{R}_0 \rho_0,
\end{align}
for some choice of a single parameter $\alpha$. Comparing Eq. (\ref{p_affine}) and Eq. (\ref{w_affine}), we have
\begin{align}
W(R|\rho) &= \alpha(P(R|\rho)-w) + w= \alpha P(R|\rho) + (1-\alpha)P(R|\mu), 
\end{align}
since $w=P(R|\mu)$. Recalling that $P(E|\rho)=P(E|R)\Phi P(R|\rho) = P(E|R)W(R|\rho)$, we obtain what we shall call the \emph{Protourgleichung},
\begin{align}
P(E|\rho) &= \sum_{jk}P(E|R_j)\Phi_{jk} P(R_k|\rho)\\
&=	\sum_j P(E|R_j)\Big\{\alpha P(R_j|\rho) + (1-\alpha)P(R_j|\mu)\Big\},
\end{align}
that is, we may take $\Phi = \alpha I + (1-\alpha)w u^\dagger$. Can this always be achieved? We now establish the following theorem:
\begin{theorem}
   \label{universal-urgleichung}
Given an informationally complete measurement with $n$ outcomes in a no-restriction GPT of dimension $r$ with a distinguished interior state $|\mu)$, one can always choose a set of reference states which furnish a quasiprobability representation $W(R|\rho)=\alpha P(R|\rho) + (1-\alpha)P(R|\mu)$, for some parameter $|\alpha|\ge 1$.
\end{theorem}
\begin{proof}
Let 
\begin{align}
\B{S} = \left(\begin{array}{ccc}  
1 & \dots & 1 \\
\hline
& \\
& \B{S}_0 \\
& &\\
\end{array}\right)	&& 
\BT{R}=
\left(\begin{array}{c|ccc}  
& & & \\
w & & \alpha\B{R}_0 \\
& & &\\
\end{array}\right).
\end{align}
The Born identity requires that $\B{S}\BT{R}=I$. We have
\begin{align}
 \left(\begin{array}{ccc}  
1 & \dots & 1 \\
\hline
& \\
& \B{S}_0 \\
& &\\
\end{array}\right)\left(\begin{array}{c|ccc}  
& & & \\
w & & \alpha\B{R}_0 \\
& & &\\
\end{array}\right) &= \begin{pmatrix}u^\dagger w & \alpha u^\dagger \B{R}_0 \\
\B{S}_0w & \alpha \B{S}_0\B{R}_0
 \end{pmatrix}= \begin{pmatrix}1& 0 \\
\B{S}_0w & \alpha \B{S}_0\B{R}_0
 \end{pmatrix},
\end{align}
where the last follows from $u^\dagger w =1$ and $u^\dagger \B{R}_0 = 0$. To satisfy the Born identity, then, our states $\B{S}$ must satisfy
\begin{align}
	\label{protourg_const}
\B{S}_0w=0 && \alpha\B{S}_0\B{R}_0 = I_{r-1}.	
\end{align}
The measurement matrix itself is
\begin{align}
\B{R}=
\left(\begin{array}{c|ccc}  
& & & \\
w & & \B{R}_0 \\
& & &\\
\end{array}\right).
\end{align}
By informational completeness, it must have (not uniquely) a left inverse $\B{R}^L$, which we may express in block form as
\begin{align}
\B{R}^L = \begin{pmatrix} \ell \\ \B{R}_0^L \end{pmatrix}. 
\end{align}
Indeed, 
\begin{align}
\B{R}^L \B{R}= \begin{pmatrix} \ell \\ \B{R}_0^L  \end{pmatrix} \begin{pmatrix} w & \B{R}_0 \end{pmatrix} = \begin{pmatrix} \ell w & \ell \B{R}_0 \\ \B{R}_0^L w & \B{R}_0^L \B{R}_0\end{pmatrix} = \begin{pmatrix} 1 & 0 \\ 0 & I_{r-1} \end{pmatrix},
\end{align}
so that $\B{R}_0^L w = 0$ and $\B{R}_0^L \B{R}_0=I_{r-1}$. We conclude that if we take $\B{S}_0 \propto \B{R}_0^L$, we can satisfy the two constraints in Eq. \ref{protourg_const}. The key is to choose the constant $\alpha$ so that $|S_i)$, the columns of $\B{S}$, are valid states. To this end, let $\B{S}_0 = \alpha^{-1}\B{R}_0^L$ where $\alpha$ has been chosen so that
\begin{align}
\forall i, E \in \mathscr{E}: (E|S_i) = 	\left(\begin{array}{c|c}  
E_1 & E_0
\end{array}\right)\left(\begin{array}{c}  
1 \\
\hline\\[-10pt]
\frac{1}{\alpha}(R^L_0)_i \\
\end{array}\right) &= E_1 + \frac{1}{\alpha}E_0 \cdot (R^L_0)_i \ge 0,
\end{align}
where $(R^L_0)_i$ denotes the $i$'th column of $\B{R}_0^L$. Now $E_1 = P(E|\mu) > 0$ since we chose $|\mu)$ to be an interior state. Thus we can ensure $\forall i, E \in \mathscr{E}: (E|S_i) \ge 0$ by choosing $|\alpha| \ge 1$ sufficiently large. That such an $\alpha$ can always be found follows from the same considerations as in Theorem \ref{mics-always-exist}: the set of effects is assumed to be compact, and in finite dimensions, linear functionals are all continuous, so that $f(E) = (E|S_i)$ sends the effect space to a compact subset of $\mathbb{R}$ on which the function attains its minimum and maximum. Moreover, because we work in a no-restriction GPT, $\mathscr{S} = \{\rho \ | \ \forall E \in \mathscr{E}: (E|\rho) \ge 0\}$.

In short, we select some left inverse $\B{R}^L$ of the measurement matrix $\B{R}$. The latter $r-1$ rows of $\B{R}^L$ form a left inverse $\B{R}_0^L$ of $\B{R}_0$ which additionally satisfies $\B{R}_0^L w = 0$. We then take $\B{S}_0 = \frac{1}{\alpha}\B{R}_0^L$ where $|\alpha|$ is chosen sufficiently large that $|S_i)$ are valid states. Thus $\B{S}_0w = 0$ and $\alpha \B{S}_0 \B{R}_0 = I_{r-1}$, as desired, and since $\B{S}\BT{R}=I$, $\B{S}$ has a right inverse, and so $r$ linearly independent rows: the states $|S_i)$ are informationally complete. Finally, since $w=P(R|\mu)$, we have $\BT{R}|\rho) = \alpha P(R|\rho) + (1-\alpha)P(R|\mu)$ as desired. 
\end{proof}

\begin{remark}
The condition $\B{S}_0w=0$ amounts to $\B{S}w = |1)$: the weighted average of the reference states is the state of complete uncertainty. In a self-dual theory, and for an unbiased measurement, this implies that the reference states themselves correspond to a measurement. 
\end{remark}

Another perspective on this construction is provided by considering the channel operator corresponding to reference measurement $C = \sum_{i=1}^n |S_i)(R_i|$. Notice that $(E|C|\rho) = P(E|R)P(R|\rho)$. We can understand this in reflection terms: the gambler is now contemplating their future where they expect first to perform the reference measurement followed by an arbitrary final measurement $\{E_i\}$. What state should they use now to assign probabilities to the final measurement: precisely, $\mathcal{C}|\rho) = \sum_i P(R_i|\rho) |S_i)$. 

\begin{corollary}
   \label{depolarizing-corollary}
	A reference measurement furnishes a Protourgleichung with quasiprobabilities $W(R|\rho)=\alpha P(R|\rho) + (1-\alpha)P(R|\mu)$ iff its channel operator is depolarizing with parameter $1/\alpha$. 
\end{corollary}
\begin{proof}
Working out the form of the channel operator under the assumptions of Theorem (\ref{universal-urgleichung}) yields
\begin{align}
 \B{S}\B{R} &=\left(\begin{array}{ccc}  
1 & \dots & 1 \\
\hline
& \\
& \B{S}_0\\
& &\\
\end{array}\right)\left(\begin{array}{c|ccc}  
& & & \\
w & & \B{R}_0 \\
& & &\\
\end{array}\right)\\
&= \begin{pmatrix}u^\dagger w & u^\dagger \B{R}_0 \\ \B{S}_0 w & \B{S}_0 \B{R}_0 \end{pmatrix}= \begin{pmatrix}1 & 0 \\ 0 & \frac{1}{\alpha}I_{r-1}\end{pmatrix}\\
&=\frac{1}{\alpha}I+\left(1-\frac{1}{\alpha}\right) |1)(\mu|,
\end{align}
implying the reference measurement acts as a depolarizing channel with parameter $1/\alpha$,
\begin{align}
|\rho) \mapsto \frac{1}{\alpha}|\rho)+\left(1-\frac{1}{\alpha}\right)|\mu).
\end{align}
The argument can be read in reverse so that the assumption that the channel is depolarizing and informationally complete implies a quasiprobability representation of the desired form.
\end{proof}

\begin{remark}
It may be useful to see step by step ``why'' the Protourgleichung works. We begin by inserting the Born identity into the expression for $P(E|\rho)$,
\begin{align}
P(E|\rho) &= \B{E}|\rho)	= \B{E}\left[\B{S}\Phi \B{R}\right] |\rho).
\end{align}
If $\Phi$ acts as a Protourgleichung, we have
\begin{align}
&=\B{E}\B{S}\Big[\alpha \B{R}|\rho)+(1-\alpha)P(R|\mu)\Big]=\B{E}\Big[\alpha\B{SR}|\rho) +(1-\alpha)\B{S}P(R|\mu)\Big].
\end{align}
But we know that $\B{SR}$ is just the channel operator $C$, which is depolarizing with parameter $1/\alpha$, and $\B{S}P(R|\mu)=|\mu)$. The effect, then, is simply to balance depolarization with repolarization,
\begin{align}
&=\B{E}\Bigg[\alpha\left[ \frac{1}{\alpha}|\rho)+\left(1-\frac{1}{\alpha}\right)|\mu)\right] +(1-\alpha)|\mu)\Bigg]\\
&=\B{E}\Big[|\rho)- (1-\alpha)|\mu) +(1-\alpha)|\mu)\Big]\\
&= \B{E}|\rho).
\end{align}
\end{remark}

\begin{remark}
In the case that $\alpha=1$, $W(R|\rho)=P(R|\rho)$: the Protourgleichung collapses back to the law of total probability. Since $\BT{R}=\B{R}$, the channel operator is simply the identity $C=\B{SR}=I$, which implies that $P^2=\B{R}\big[ \B{SR}\big] \B{S}=\B{RS}=P$: the conditional probability matrix which characterizes the reference measurement is a projector. Moreover, notice that
\begin{align}
	\forall E, \rho: (E|\rho) = \Big[(E|\B{S}\Big]\Big[\B{R}|\rho)\Big]=P(E|R)  P(R|\rho).
\end{align}
 Thus $\B{S}$ provides a linear map which embeds the effect space into the hypercube dual to the probability simplex, and $\B{R}$ provides a linear map which embeds the state space into the probability simplex itself in such a way that all probabilities $P(E|\rho)$ are preserved. We have therefore constructed a \emph{simplex embedding}, or equivalently a noncontextual ontological model \cite{schmidCharacterizationNoncontextualityFramework2021, selbyLinearProgramTesting2024}: in short, if we can take $\alpha=1$, then our GPT must be equivalent to a classical theory. In the sequel, we shall discuss the possibility of classical representations more thoroughly.
\end{remark}

For a depolarizing reference measurement constructed according to the above prescription, the conditional probability matrix $P(R|R)$ which characterizes the reference measurement takes the general form,
\begin{align}
P(R|R) &= \B{RS} = 		\left(\begin{array}{c|ccc}  
& & & \\
w & & \B{R}_0 \\
& & &\\
\end{array}\right)\left(\begin{array}{ccc}  
1 & \dots & 1 \\
\hline
& \\
& \B{S}_0 \\
& &\\
\end{array}\right)\\
&= wu^\dagger + \B{R}_0 \B{S}_0.
\end{align}
We then have the following theorem:
\begin{theorem}
\label{equiangular}
The conditional probability matrix $P\equiv P(R|R)$ which characterizes a depolarizing reference measurement acts on vectors $y\in \text{col}(P)$ in its column space as
\begin{align}
Py &= \frac{1}{\alpha} y + \left(1- \frac{1}{\alpha}\right) \overline{y} w,
\end{align}
where $w=P(R|\mu)$ and $\overline{y}$ denotes the sum of the components of $y$. If the reference measurement is a MIC, then in fact
\begin{align}
P(R|R) = \frac{1}{\alpha}I + \left(1- \frac{1}{\alpha}\right) w u^\dagger,
\end{align}
and if moreover the reference measurement is unbiased, then $P(R|R) = a I + b J$, where $J$ is the matrix of all 1's with $a=1/\alpha$ and $b= (1-1/\alpha)/r$, that is, $P(R|R)$ is compound symmetric.
\end{theorem}
\begin{proof}
From $P = w u^\dagger + \B{R}_0 \B{S}_0$ and using the fact that $\B{S}_0 w = 0$, we see that $Pw=w$. Moreover, since $P$ is column stochastic, $u^\dagger P = u^\dagger$. By construction $P \Phi P = P$ where $\Phi = \alpha I + (1-\alpha)w u^\dagger$, or $\alpha P^2 + (1-\alpha)w u^\dagger = P$. Now suppose $y = Px \in \col(P)$. Noticing that $w u^\dagger Px = w u^\dagger x$, we have for $y \in \text{col}(P)$,
\begin{align}
\alpha P y + (1-\alpha)w u^\dagger y = y,
\end{align}
or
\begin{align}
Py &= \frac{1}{\alpha} y + \left(1- \frac{1}{\alpha}\right) \overline{y} w.
\end{align}
Now if the reference measurement is a MIC then $P(R|R)$ is an $r \times r$ invertible matrix and $\col(P) = V$, the entire vector space so that 
\begin{align}
P(R|R) = \frac{1}{\alpha}I + \left(1- \frac{1}{\alpha}\right) w u^\dagger.
\end{align}
If the measurement is unbiased, then $w=u/n$, and $P(R|R) = a I + bJ$ where $a= 1/\alpha$ and $b=(1-a)/r$.
\end{proof}
\begin{remark}
Supposing $y\in \text{col}(P)$ is a probability vector $P(R|\rho)$, we have
\begin{align}
P(R|R)P(R|\rho) = \frac{1}{\alpha}P(R|\rho) + \left(1- \frac{1}{\alpha}\right)P(R|\mu).
 \end{align}
We could have proven this equally well starting from the structure of the channel operator.
\end{remark}

\begin{lemma}
\label{depolarizing-P-spectrum}
Let $\{R_i\}$ be a depolarizing reference measurement such that $\Phi = \alpha I + (1-\alpha)w u^\dagger$. Then $P(R|R)$, the conditional probability matrix which characterizes it must have 1 eigenvalue equal to 1, $r-1$ eigenvalues equal to $1/\alpha$, and the rest 0.
\end{lemma}
\begin{proof}
Recall $P(R|R)=wu^\dagger +\B{R}_0 \B{S}_0$, and that: $u^\dagger w =1$, $u^\dagger \B{R}_0=0$, $\B{S}_0 w = 0$, and $\B{S}_0\B{R}_0=\alpha^{-1}I_{r-1}$. We first note that
\begin{align}
P(R|R)w = w u^\dagger w + \B{R}_0\B{S}_0w = w,
\end{align}
so that $w$ is an eigenvector with eigenvalue 1. Meanwhile, for any vector $x$,
\begin{align}
P(R|R)\B{R}_0 x = w u^\dagger \B{R}_0x + \B{R}_0\B{S}_0 \B{R}_0x = \alpha^{-1} \B{R}_0x,
\end{align}
so that every vector in $\col(\B{R}_0)$ is an eigenvector with eigenvalue $\alpha^{-1}$. Since $u^\dagger w = 1$, but $u^\dagger \B{R}_0 = 0$, we conclude that $w \notin \col(\B{R}_0)$. We have so far accounted for $r$ linearly independent eigenvectors. Finally,
\begin{align}
\rank(P(R|R)) \leq \rank(wu^\dagger) + \rank(\B{R}_0\B{S}_0)\leq 1 + (r-1) = r,
\end{align}
so that the remaining eigenvalues are zero.
\end{proof}

\section{Morphophoricity}

The key to the construction of a depolarizing reference measurement is the demand that
\begin{align}
\alpha \B{S}_0\B{R}_0 = I_{r-1},
\end{align}
so that $\B{S}_0 = \frac{1}{\alpha}\B{R}_0^{L}$ where $\alpha$ is chosen so that the resulting states are nonnegative on all effects in the GPT.  Suppose however that we further demand that the reference states take the particular form
\begin{align}
\B{S} = 	\left(\begin{array}{ccc}  
1 & \dots & 1 \\
\hline
& \\
& c\B{R}_0^\dagger W^{-1} \\
& &\\
\end{array}\right),
\end{align}
where $W^{-1}$ is the diagonal matrix with the reciprocals of the weights $1/w_i$ along the diagonal. If the constant $c$  were equal to 1,  this would amount to taking $|S_i) = (1/w_i)|E_i)$, that is, to taking the reference states proportional to effects. This is always possible in a self-dual GPT: otherwise, employing the same trick as before, one may choose the constant $c$ to sufficiently depolarize the prospective states until they are nonnegative on all effects. Either way, we'll call such a reference measurement a \emph{parallel update} reference measurement. By the same reasoning as in Corollary (\ref{depolarizing-corollary}), if such a measurement is to furnish a depolarizing channel, we must have
\begin{align}
   \label{traceless-morpho-condition}
\alpha c\B{R}_0^\dagger W^{-1}\B{R}_0 = I_{r-1}	,
\end{align}
or  $\B{S}_0=c\B{R}_0^\dagger W^{-1} = \frac{1}{\alpha}\B{R}_0^L$, and since $c$ is fixed by the demand that $c\B{R}_0^\dagger W^{-1}$ furnish valid states, this fixes the value of $\alpha$. Finally, yet another way of saying it is that we require that the columns of $\B{R}_0$  be orthonormal with respect to the metric provided by $\mathcal{M}=\alpha cW^{-1}$.

If this is possible, we have the following interesting consequence. Recalling that $P(R|\rho)=\B{R}|\rho)=w+\B{R}_0 \rho_0$, we have
\begin{align}
\big{\lVert}P(R|\rho) - P(R|\sigma) \big{\lVert}^2_{\mathcal{M}} &= \big{\lVert}w+ \B{R}_0 \rho_0 - w-\B{R}_0 \sigma_0 \big{\lVert}^2_{\mathcal{M}}\\
&=\big{\lVert}\B{R}_0 (\rho_0 - \sigma_0)\big{\lVert}^2_\mathcal{M}\\
&= (\rho_0 - \sigma_0)^\dagger\B{R}_0^\dagger  \mathcal{M} \B{R}_0(\rho_0 - \sigma_0)\\
&= (\rho_0 - \sigma_0)^\dagger\alpha c\B{R}_0^\dagger W^{-1}\B{R}_0(\rho_0 - \sigma_0)\\
&= \big{\lVert}\rho_0-\sigma_0 \big{\lVert}^2.
\end{align}
In other words, up to the diagonal metric $\mathcal{M}$, the measurement map from the traceless subspace into the probability simplex preserves Euclidean distances. We'll call such a measurement \emph{weighted morphophoric}. In particular, for an unbiased measurement, we have $\B{R}_0^\dagger \B{R}_0 = \frac{1}{\alpha cn}I$ so that the traceless part of the effects form a tight frame for the traceless subspace (the subspace orthogonal to $|\mu)$), and 
\begin{align}
	\big{\lVert}P(R)_\rho - P(R)_\sigma \big{\lVert}^2&= \frac{1}{\alpha cn} \big{\lVert}\rho_0-\sigma_0 \big{\lVert}^2,
\end{align}
that is, the measurement map is a \emph{similarity} with respect to the Euclidean metric on both the probability simplex and the traceless subspace. Such a measurement is properly \emph{morphophoric}.

\begin{remark}
``Morphophoric'' in Greek means \emph{form-bearing}: indeed, a morphophoric measurement embeds a state space into the probability simplex in a shape preserving way.	The terminology was introduced in the quantum mechanical setting in \cite{slomczynskiMorphophoricPOVMsGeneralised2020}. A rigorous mathematical treatment of the subject in the more general setting of ``geometric GPTs'' equipped with inner products can be found in \cite{szymusiakCanQBismExist2025}. Morphophoric measurements in general yield Urgleichung-like expressions which are somewhat more sophisticated; we confine our interest here to the very simplest deformations of the law of total probability which correspond to quasiprobability representations of the form $W(R|\rho)=\alpha P(R|\rho) + (1-\alpha)P(R|\mu)$. This motivates our introduction of the notion of weighted morphophoricity, that is, the demand that $\B{R}_0^\dagger W^{-1}\B{R}_0 \propto I$ in order to handle biased measurements on equal footing with unbiased.
\end{remark}

\begin{lemma}
Weighted morphophoric measurements exist in any no-restriction GPT of dimension $r$ with a distinguished interior state $|\mu)$, for any number of outcomes $n\ge r$ and arbitrary choice of biases $w_i >0$ and $\sum_i w_i=1$.
\end{lemma}
\begin{proof}
   We seek $\B{R}, \B{S}$ of the form
   \begin{align}
\B{R}=	\left(\begin{array}{c|ccc}  
& & & \\
w & & \B{R}_0 \\
& & &\\
\end{array}\right) && \B{S}=\left(\begin{array}{ccc}  
1 & \dots & 1 \\
\hline
& \\
& c\B{R}_0^\dagger W^{-1} \\
& &\\
\end{array}\right).
\end{align}
Pick $r-1$ orthonormal vectors in the subspace orthogonal to $W^{1/2}u$, and use them to build an $n \times (r-1)$ matrix $\B{K}_0$ such that $\B{K}_0^\dagger W^{1/2}u = 0$. Then let $\B{R}_0 = (\alpha c)^{-1/2}W^{1/2}\B{K}_0$. Clearly, $u^\dagger \B{R}_0 = 0$ and moreover, $\alpha c \B{R}_0^\dagger W^{-1} \B{R}_0 = \B{K}_0^\dagger \B{K}_0 = I_{r-1}$. As in Theorem \ref{universal-urgleichung}, by taking $|\alpha|$ sufficiently large with $c=1$, the traceless part of $(R_i|$ can be made arbitrarily small and so since $w_i>0$, the $(R_i|$'s can be made positive on the compact state space. This is enough to ensure they are valid effects since we work in a no-restriction GPT: $\mathscr{E} = \{E \ | \ \forall \rho \in \mathscr{S}: (E|\rho) \in [0,1]\}$. Similarly, since $|\mu)$ is interior, vectors sufficiently close to $|\mu)$ will be valid states: thus we can take $|c|$ sufficiently large that $|S_i)$ are positive on the compact effect space. Since $\B{R}_0$ has rank $r-1$, the effects (and states) span the full $r$-dimensional space, giving informational completeness.
\end{proof}

Suppose we have a weighted morphophoric measurement. We have shown that if we consider the (properly depolarized) states proportional to the reference effects, they furnish a depolarizing channel. But we are not limited to that particular choice of reference states: as we have also seen, any $\B{S}$ built from a left inverse of $\B{R}_0$ will suffice. There is a special case, however.

\begin{theorem}
For a weighted morphophoric MIC reference measurement, the choice of reference states which furnish a depolarizing channel is fixed up to sign.
\end{theorem}
\begin{proof}
   From Theorem (\ref{universal-urgleichung}), given a choice of reference effects, we require of the reference states that they satisfy $\beta \B{S}_0\B{R}_0 = I_{r-1}	$ in order for the resulting channel to be depolarizing. At the same time, morphophoricity demands Eq. (\ref{traceless-morpho-condition}), that is, $\alpha c \B{R}_0^\dagger W^{-1}\B{R}_0 = I_{r-1}$.  Hence
   \begin{align}
(\alpha c \B{R}_0^\dagger W^{-1}	-\beta\B{S}_0)\B{R}_0 = M\B{R}_0 = 0.
\end{align}
 For a MIC, $\B{R}_0^\dagger  \in \mathbb{C}^{(r-1)\times r}_{r-1}$: by the rank nullity theorem, therefore, $\dim(\nullspace(\B{R}_0^\dagger ))=1$. Because the reference effects sum to $(1|$, we have $\B{R}_0^\dagger u = 0$, and so $\nullspace(\B{R}_0^\dagger) = \linspan(u)$. Meanwhile, denoting the $i$th row of $M$ by $m_i^\dagger$, we have $\B{R}_0^\dagger m_i = 0$: that is, each of the $m_i$ must lie in $\nullspace(\B{R}_0^\dagger) = \linspan(u)$. Thus every row of $M$ must be proportional to $u^\dagger$ so that we can write $M=xu^\dagger$ for some $x \in \mathbb{F}^{r-1}$. From Theorem (\ref{universal-urgleichung}), we also have $\B{S}_0 w=0$ and $\B{R}_0^\dagger W^{-1} w = \B{R}_0^\dagger  u = 0$. Thus $Mw = 0$. But also $Mw = x u^\dagger w = 0$. But since the weights must sum to 1, we conclude that $x=0$, and so $M=0$, and that $\alpha c \B{R}_0^\dagger W^{-1}	=\beta\B{S}_0 $. For a weighted morphophoric MIC, the choice of depolarizing reference states is essentially fixed up to $c$: we assume one takes $|c|$ as small as possible. Note, however, that we can take $\B{S}_0 =-c\B{R}_0^\dagger W^{-1}$ and $\beta=-\alpha$: hence the reference states are fixed up to sign.
\end{proof}

\begin{corollary}
Suppose we have an unbiased morphophoric MIC reference measurement. Its depolarizing reference states are fixed up to sign to be proportional to effects (modulo $c$). By Theorem \ref{equiangular}, the conditional probability matrix which characterizes the reference measurement must be compound symmetric, that is, of the form $P=a I + bJ$. Thus any unbiased morphophoric MIC measurement is a SIC, that is, a \emph{symmetric informationally complete} measurement, forming a regular simplex in the effect space. Note, however, that such measurements need not be extremal, that is, lying on the boundary of the effect space. As we have observed, it is easy to see that nonextremal SICs exist in any GPT: one may simply embed a sufficiently small $(r-1)$-simplex (with $r$ vertices) in the effect space such that the effects are non-negative on all states.
\end{corollary}

\subsection{An example from quantum theory}

 Consider the following extremal symmetric informationally complete measurement (SIC-POVM) in $d=2$ quantum theory. The effects can be written in terms of Pauli matrices,
\begin{align}
R_0 = \frac{1}{4}\left(I + \frac{1}{\sqrt{3}}(\sigma_x + \sigma_y + \sigma_z)\right) && R_2 &= \frac{1}{4}\left(I + \frac{1}{\sqrt{3}}(-\sigma_x - \sigma_y + \sigma_z)\right)	\\
R_1 = \frac{1}{4}\left(I + \frac{1}{\sqrt{3}}(\sigma_x - \sigma_y - \sigma_z)\right) && R_3&= \frac{1}{4}\left(I + \frac{1}{\sqrt{3}}(-\sigma_x + \sigma_y - \sigma_z)\right),
\end{align}
or more compactly in the Bloch representation,
\begin{align}
\B{R} = \left(\begin{array}{c|ccc} 
 \frac{1}{4}u & \B{R}_0
 \end{array}\right)	&& \B{R}_0 = \frac{1}{4\sqrt{3}}\begin{pmatrix} 1 & 1 & 1 \\
 1 & -1 & -1\\
-1 & -1 & 1 \\
-1 & 1 & -1  \end{pmatrix}.
\end{align}
Since quantum theory is self-dual, we can choose states proportional to effects $\B{S} = \left(\begin{array}{@{}c@{}} u^\dagger \\ \hline 4\B{R}_0^\dagger \end{array}\right)$, in which case $\B{R}_0^\dagger W^{-1}\B{R}_0 = 4\B{R}_0^\dagger \B{R}_0= (1/3)I$ so that $\alpha=3$ (and $c=1$). Then $\B{SR}$ corresponds to a depolarizing channel with parameter $1/3$,
\begin{align}
|\rho) \mapsto \frac{1}{3}|\rho) + \frac{2}{3}|\mu),
\end{align}
and we have the Urgleichung
\begin{align}
P(E|\rho) = \sum_i P(E|R_i)\left[3P(R_i|\rho)-\frac{1}{2}\right].
\end{align}
Alternatively, however, we could choose the states \emph{antipodal} to the aforementioned, which also form a SIC, that is, we could take $\B{S}^\prime = \left(\begin{array}{@{}c@{}} u^\dagger \\ \hline -4\B{R}_0^\dagger \end{array}\right)$, in which case  $\B{S}_0=-\B{R}_0^\dagger W^{-1}=-4\B{R}_0^\dagger $ and $\alpha=-3$. Then 
\begin{align}
|\rho) \mapsto -\frac{1}{3}|\rho) + \frac{4}{3}|\mu)
\end{align}
with $P(E|\rho)=\sum_i P(E|R_i) \left[-3P(R_i|\rho)+1\right]$. In quantum theory,  complete positivity for a depolarizing channel demands that $-1/(d^2-1) \leq  1/\alpha \leq 1$ \cite{debrotaLudersChannelsExistence2019}. Indeed, when $d=2$, the lower bound becomes $-1/3$, which this latter channel saturates. Thus for both choices of reference states, we have a valid depolarizing channel. It is worth remarking, however, that the latter choice also leads to an inversion of the state space. For an arbitrary GPT, it may or may not be the case that the channel with an inverted depolarization parameter is a valid channel.

\subsection{Tight IC-POVMs}

Thanks to the pioneering work of Andrew Scott, it has long been known that in the case of quantum theory, measurements which give rise to tight frames for the traceless subspace (called tight IC-POVMs) correspond to so-called complex projective 2-designs \cite{scottTightInformationallyComplete2006}. We will have more to say about $t$-designs in the sequel: here we confine ourselves to giving an alternative proof using the formalism we have developed in this chapter.

\begin{theorem}
In quantum theory over $\mathbb{C}$, weighted morphophoric reference measurements whose effects are extremal correspond to weighted complex projective 2-designs.
\end{theorem}
\begin{proof}
For a depolarizing reference measurement, we have
\begin{align}
C = \B{SR}=\sum_{i=1}^n |S_i)(R_i|=	\frac{1}{\alpha}I+\left(1-\frac{1}{\alpha}\right)|\mu)(1|
\end{align}
Given that we are working with quantum theory over $\mathbb{C}$, we can interpret this expression  in terms of vectorized operators, which will be elements of $\mathbb{C}^{d^2}$ \cite{rungtaUniversalStateInversion2001}. Let $|A) = (A \otimes I )\sum_{i=1}^d |i,i\rangle$ be the (row) vectorization of $A$. It consists of the rows of $A$ laid end to end to form a column vector. Then
\begin{align}
(A|B) &= \Bigg[ \sum_{i=1}^d \langle i,i| (A^{\dagger} \otimes I) \Bigg]		 \Bigg[ (B\otimes I) \sum_{j=1}^d |j,j\rangle \Bigg] \\
&= \sum_{i,j} \langle i|A^{\dagger} B|j \rangle \otimes \langle i|j \rangle= \sum_i \langle i | A^{\dagger} B |i \rangle= \tr(A^{\dagger}B).
\end{align}
 In other words, $(A|B)$ reproduces the Hilbert-Schmidt inner product $\tr(A^\dagger B)$ on operators. Note that in this case $(A|$ really does mean the conjugate transpose of $|A)$. 
 
In this formalism, the vectorized identity operator $(I|\equiv (1|$ is the normalization functional, while $\frac{1}{d}|I)=|\mu)$ corresponds to the maximally mixed state. Thus 
\begin{align}
\label{quantumC}
\sum_{i=1}^n |S_i)(R_i|= \frac{1}{\alpha}I_{d^2}	+ \left(1-\frac{1}{\alpha}\right)\frac{1}{d}|I_d)(I_d|.
\end{align}
Taking the trace of the right-hand side of Eq. \!(\ref{quantumC}), we find
\begin{align}
\sum_{i=1}^n \tr(R_i S_i) &= \frac{1}{\alpha}(d^2) +\left(1-\frac{1}{\alpha}\right)\frac{1}{d}(d)=\frac{1}{\alpha}(d^2-1)+1.
\end{align}
Quantum theory over $\mathbb{C}$ is a self-dual theory, and so we can always take reference states directly proportional to effects: thus let $R_i = \tr(R_i)S_i$---this along with the depolarizing assumption amounts to weighted morphophoricity. Moreover, let us assume that our states are pure so that $\tr(S_i^2)=1$. Then taking the trace of the left hand side of Eq. \!(\ref{quantumC}), we find 
\begin{align}
	\sum_{i=1}^n \tr(R_i S_i) &= \sum_{i=1}^n \tr(R_i)\tr(S_iS_i) = \sum_{i=1}^n \tr(R_i)=\tr\left(\sum_{i=1}^n R_i\right)=\tr(I)=d.
\end{align}
Equating these two expressions fixes the value of $\alpha$:
\begin{align}
	\frac{1}{\alpha}(d^2-1)+1 &= d \Longrightarrow
	\alpha = d+1.
\end{align}
Thus our channel operator becomes
\begin{align}
C=\sum_{i=1}^n \tr(R_i)|S_i)(S_i| &= \frac{1}{d+1}I_{d^2}	+ \left(1-\frac{1}{d+1}\right)\frac{1}{d}|I_d)(I_d|\\
&= \frac{1}{d+1}\Big(I_{d^2}+|I_d)(I_d|\Big).
\end{align}
Normalizing so that the trace of both sides is 1, we find
\begin{align}
\label{todesign}
	\sum_{i=1}^n w_i|S_i)(S_i| = \frac{1}{d(d+1)}\Big(I_{d^2} + |I_d)(I_d|\Big),
\end{align}
where $w_i = \tr(R_i)/d$ so that $\sum_i w_i = 1$.  
 Considering the right hand side of Eq.\! (\ref{todesign}), notice that we can rewrite it as
 \begin{align}
	I_{d^2} + |I_d)(I_d| &= I_d \otimes I_d + \sum_{jk}|j,j\rangle\langle k, k|.
\end{align}
Taking the partial transpose of the second tensor factor and multiplying by $\frac{1}{2}$, we find:
\begin{align}
	 \frac{1}{2}\Big(I_d \otimes I_d + \sum_{jk}|j,k\rangle\langle k, j| \Big)=  \frac{1}{2}\Big(I_d \otimes I_d + \text{SWAP}\Big)= \Pi_{\text{sym}^2},
\end{align}
where $\Pi_{\text{sym}^2}$ is the projector onto the permutation symmetric subspace of two tensor factors. Considering the left hand side of Eq. \!(\ref{todesign}), we must also take the transpose of the second tensor factor. It is a useful lemma that for a pure state $S = |\psi\rangle\langle \psi|$, we have $|S)(S| = S \otimes S^T$. We conclude
\begin{align}
\sum_{i=1}^n w_i S_i \otimes S_i = \frac{2}{d(d+1)}\Pi_{\text{sym}^2}.	
\end{align}
Meanwhile, Schur-Weyl duality tells us that the integral over tensor powers of pure states is
\begin{align}
\int |\psi\rangle\langle\psi|^{\otimes t} d\psi = \frac{\Pi_{\text{sym}^t}}{\tr(\Pi_{\text{sym}^t})} = \frac{1}{\binom{t+d-1}{d-1}}\Pi_{\text{sym}^t},
\end{align}
 which in the case of $t=2$, delivers us
 \begin{align}
 	\int |\psi\rangle\langle\psi|^{\otimes 2} d\psi = \frac{2}{d(d+1)}\Pi_{\text{sym}^2}=\sum_{i=1}^n w_i S_i \otimes S_i.
 \end{align}
Indeed, this is the defining equation for a weighted complex projective 2-design, showing that such designs coincide precisely with pure weighted morphophoric reference measurements \cite{scottTightInformationallyComplete2006} in quantum theory.
\end{proof}

\begin{corollary}
	For a pure weighted morphophoric reference measurement in quantum theory, we have $\alpha=d+1$ so that $W(R|\rho)=\alpha P(R|\rho) + (1-\alpha)P(R|\mu)$ becomes
	\begin{align}
	W(R|\rho) = (d+1)P(R|\rho) -d P(R|\mu).
	\end{align}
 For an unbiased measurement:
 	\begin{align}
	W(R|\rho) = (d+1)P(R|\rho) -\frac{d}{n}u.
	\end{align}
For a MIC, with $n=d^2$:
 	\begin{align}
	W(R|\rho) = (d+1)P(R|\rho) -\frac{1}{d}u.
	\end{align}
It has long been known that the unbiased complex projective 2-designs with the minimal number of elements correspond to pure SICs \cite{renesSymmetricInformationallyComplete2004} with $P(R_i|R_j)= \frac{d\delta_{ij} +1}{d(d+1)}$, if indeed SICs exist in any dimension \cite{applebySICsAlgebraicNumber2017, bengtssonSICsExplanations2020}. We then arrive at the Urgleichung \cite{fuchsQbismWhereNext2023},
\begin{align}
P(E|\rho) = \sum_{j=1}^{d^2} P(E|R_j)\left[ (d+1)P(R_j|\rho) - \frac{1}{d}\right].
\end{align}
\end{corollary}

\begin{remark}
In appendix \ref{realqm}, we give the analogous result for quantum mechanics over $\mathbb{R}$.
\end{remark}

\section{The LTP Deformation}

As we have seen, the only difference between the law of total probability and our nonclassical coherence condition is the presence of the Born matrix $\Phi$, a $\{1\}$-inverse of $P(R|R)$, the conditional probability matrix which characterizes the reference measurement itself. Inspired by this, \cite{debrotaSymmetricInformationallyComplete2020} introduced the distance between the identity and $\Phi$ with respect to any unitarily invariant matrix norm\footnote{For example, a Schatten $p$-norm $||A||_p	= \left(\sum_i \sigma_i^p\right)^{\frac{1}{p}}$ where the $\sigma_i$'s are the singular values of $A$ \cite{hornMatrixAnalysis1985}.} as a measure of the nonclassicality of the reference measurement,
\begin{align}
\mathcal{D}(\Phi) =\lVert I - \Phi \rVert.
\end{align}
We shall call it here the \emph{LTP deformation}. By minimizing this quantity over all allowed reference measurements with a certain number of outcomes, one obtains the \emph{irreducible LTP deformation}, a property of the underlying theory itself (e.g., $d$ dimensional quantum theory over $\mathbb{C}$), and which is meant to be a measure of nonclassicality on  grounds that classically one may always pick a reference measurement for which $\Phi=I$, simply reading off an objective property of the system without disturbing it. The conclusion of \cite{debrotaSymmetricInformationallyComplete2020} was among all minimal informationally complete (MIC) measurements (with $d^2$ outcomes), parallel update SIC reference measurements achieve the irreducible LTP deformation in all dimensions---if they exist.
 
Now for a MIC, the reference states and effects must form bases, and we have $\Phi = P(R|R)^{-1}$ univocally. But if we want to calculate $\mathcal{D}(\Phi)$ for ``overcomplete'' reference measurements, we must make a choice of Born matrix. We therefore introduce the \emph{minimal LTP deformation}
 \begin{align}
 	\mathcal{D}_{\text{min}} = \min_{\Phi}||I - \Phi|| \quad \text{s.t.} \quad P\Phi P = P,
 \end{align}
which is a constrained convex optimization problem. In particular, the Frobenius norm squared is strictly convex, and since the feasible set is itself convex, the problem in this case has a unique solution \cite{boydConvexOptimization2004}. One can further obtain the irreducible LTP deformation by minimizing over both reference measurements with a certain number of outcomes as well as over $\Phi$ itself.

But this quantity may be misleading. On the one hand, we have seen that for any depolarizing reference measurement we may take $\Phi_1 = \alpha I + (1-\alpha)w u^\dagger$. This is arguably the simplest Born matrix in algebraic terms. It gives an LTP deformation of e.g.,
\begin{align}
\lVert I - \Phi_1 \rVert_F = |1-\alpha|\sqrt{n(1+\lVert w \rVert^2) - 2}.
\end{align}
which in the unbiased case becomes $\lVert I - \Phi_1 \rVert_F = |1-\alpha|\sqrt{n-1}$. But in general this will \emph{not} be the $\Phi$ matrix which gives the smallest possible LTP deformation for the given measurement. In other words, in the non-MIC case, there may be a divergence between algebraic simplicity and matrix distance. For example, there are other $\Phi$ matrices that act as a Protourgleichung, that is, which act just like $\Phi = \alpha I + (1-\alpha)w u^\dagger$ on the relevant subspace, but which may have different $\mathcal{D}(\Phi)$ values. For example, putting two forms of the Born identity together $\B{S}\BT{R}=\BT{S}\B{R}=I$ yields $\B{S}\BT{R}\BT{S}\B{R}=I$ so that we may take
\begin{align}
\Phi_2 &= \BT{R}\BT{S} = 
\left(\begin{array}{c|ccc}  
& & & \\
w & & \alpha\B{R}_0 \\
& & &\\
\end{array}\right)\left(\begin{array}{ccc}  
1 & \dots & 1 \\
\hline
& \\
& \alpha\B{S}_0\\
& &\\
\end{array}\right)\\
&= wu^\dagger +\alpha^2 \B{R}_0 \B{S}_0.
\end{align}
This Born matrix indeed acts as a Protourgleichung for a depolarizing reference measurement. Since $P(R|\rho) = \B{R}|\rho) = w + \B{R}_0 \rho_0$, we have
\begin{align}
	\Phi_2 P(R|\rho) &= \left[ wu^\dagger +\alpha^2 \B{R}_0 \B{S}_0\right][w + \B{R}_0 \rho_0]\\
	&= w+\alpha^2 \B{R}_0 \B{S}_0 \B{R}_0 \rho_0\\
	&= w+\alpha \B{R}_0 \rho_0\\
	&= \alpha P(R|\rho) + (1-\alpha)P(R|\mu),
\end{align}
where we have used the fact that $\alpha \B{S}_0 \B{R}_0 = I$. If we suppose that this Born matrix is unitarily diagonalizable, that is, a normal matrix, then we can easily calculate its LTP deformation. By the same argument as in Lemma \ref{depolarizing-P-spectrum}, the Born matrix $\Phi=wu^\dagger +\alpha^2 \B{R}_0 \B{S}_0$ will have 1 eigenvalue equal to 1, $r-1$ eigenvalues equal to $\alpha$, and the rest 0. Then $\lVert I - \Phi_2 \rVert = \lVert I - VDV^\dagger \rVert = \lVert V(I - D)V^\dagger \rVert = \lVert I - D\rVert$, where we have used the unitary invariance of the norm, and $D$ is the diagonal matrix with the aforementioned eigenvalues along its diagonal. We conclude
\begin{align}
\lVert I - \Phi_2\rVert_F &= \left\lVert \text{diag}\left(
0,\,
\overbrace{1-\alpha,\ldots,1-\alpha}^{r-1},\,
\overbrace{-1,\ldots,-1}^{n-r}
\right)\right\rVert_F \\
&= \sqrt{(r-1)(1-\alpha)^2 + (n-r)}.
\end{align}
But suppose we add $\Pi_\perp$, the projector onto the complement of $\col(P)$, so that $\Phi_3 = wu^\dagger +  \alpha^2 \B{R}_0 \B{S}_0 + \Pi_\perp$. This will make no difference in the action of $\Phi_3$ since $\Pi_\perp$ gives 0 on any vector in $\col(P)$. But crucially, now
\begin{align}
\lVert I - \Phi_3\rVert_F &=  \left\lVert \text{diag}\left(
0,\,
\overbrace{1-\alpha,\ldots,1-\alpha}^{r-1},\,
\overbrace{0,\dots,0}^{n-r}
\right)\right\rVert_F  \\
&= |1-\alpha|\sqrt{r-1}.
\end{align}
In fact, we can get a more explicit form for this Born matrix. Using $P = w u^\dagger + \B{R}_0 \B{S}_0$ and $\Phi_1 = \alpha I + (1-\alpha)w u^\dagger$, we have for the projector onto $\col(P)$, $\Pi = P \Phi = w u^\dagger + \alpha \B{R}_0 \B{S}_0$.  Then 
\begin{align}
	\label{turns-out-min}
\Phi_3 = wu^\dagger +  \alpha^2 \B{R}_0 \B{S}_0 + \Pi_\perp = I + \alpha(\alpha -1)\B{R}_0 \B{S}_0.
\end{align}

The point, of course, is that all three choices of $\Phi$ matrices behave the same on the subspace $\col(P(R|R))$, even if they differ elsewhere. The $\Phi_1$ matrix we considered is arguably the simplest: it depends only on the fact that the reference measurement is depolarizing without being sensitive to the details of the measurement itself. $\Phi_2$ and $\Phi_3$ act exactly the same as the first on the relevant subspace: the second is confined entirely to that subspace, while the third is able to achieve the smallest LTP deformation of the three by adding support judiciously onto the complement.

Now since $P(R|R)\Phi$ is a projector onto $\col(P(R|R))$, we could define an alternative measure, 
\begin{align}
\mathcal{D}_{\text{sub}}(\Phi) = \lVert (I - \Phi)\Pi\rVert,
\end{align}
where $\Pi$ is the orthogonal projector onto $\col(P)$. Any $\Phi$ matrices which act identically on the subspace to which probability vectors are in any case constrained would be given identical scores by this measure: thus this measure would judge the three $\Phi$ matrices we considered above to have the same LTP deformation. But then again, is this what we really want? If we can make a $\Phi$ matrix simpler in an algebraic sense by adding support on the complement of $\col(P)$, why shouldn't we? We conclude that measure one uses depends on the goal: for non-MIC reference measurements, there is no one univocal measure of LTP deformation.

We further dramatize this issue by proving the following theorem. Using the method of Lagrange multipliers, we can solve for the $\Phi$ matrix which achieves $ \min_{\Phi}||I - \Phi|| \text{ s.t. } P\Phi P = P$ with respect to the Schatten 2-norm (the Frobenius norm) analytically, at least for unbiased reference measurements. 
 
\begin{theorem}
The unique Born matrix which minimizes $||I - \Phi||_F$ for an unbiased depolarizing reference measurement is
 	\begin{align}
 	\Phi =I + \left(1-\frac{1}{\alpha}\right)\B{S}_0^\dagger (\B{S}_0\B{S}_0^\dagger)^{-1}(\B{R}_0^\dagger\B{R}_0)^{-1}\B{R}_0^\dagger.	
 	\end{align}
\end{theorem}
\begin{proof}
Define a Lagrangian
\begin{align}
\mathscr{L}(\Phi, \Lambda) &= ||I-\Phi||_2^F + \tr\Big(\Lambda^T(P\Phi P-P)\Big)\\
&= n - 2\tr(\Phi) + ||\Phi||^2 + \tr(\Lambda^T P\Phi P) - \tr(\Lambda^T P)
\end{align}
where $\Lambda$ is a matrix of Lagrange multipliers.
Recalling that
\begin{align}
\frac{\partial \tr(X)}{\partial X} = I &&	\frac{\partial ||X||^2}{\partial X} = 2X && \frac{\partial \tr(AX)}{\partial X} = A^T, 
\end{align}
we differentiate with respect to $\Phi$ and demand that the result vanishes.
\begin{align}
\frac{\partial\mathscr{L}(\Phi, \Lambda) }{\partial \Phi}	&= -2I+2\Phi + (P\Lambda^TP)^T=0
\end{align}
yields
\begin{align}
\Phi = I - \frac{1}{2}	P^T\Lambda P^T,
\end{align}
where, considering that $P\Phi P = P$, $\Lambda$ must satisfy
\begin{align}
\label{lambda}
P^2-P=\frac{1}{2}PP^T \Lambda P^TP.
\end{align}
Considering the LHS of Eq.\ (\ref{lambda}), and recalling that $\B{S}_0\B{R}_0 = \B{R}_0^\dagger\B{S}_0^\dagger = \frac{1}{\alpha}I$ and $u^\dagger \B{R}_0^\dagger=\B{S}_0w=0$, we find
\begin{align}
P^2 &= 	\left[wu^\dagger + \B{R}_0 \B{S}_0\right]\left[wu^\dagger + \B{R}_0 \B{S}_0\right]= wu^\dagger + \frac{1}{\alpha}\B{R}_0 \B{S}_0,
\end{align}
so that
\begin{align}
P^2 -P &=  wu^\dagger + \frac{1}{\alpha}\B{R}_0 \B{S}_0-wu^\dagger - \B{R}_0 \B{S}_0 = \frac{1-\alpha}{\alpha}\B{R}_0 \B{S}_0.
\end{align}
As for the RHS of Eq.\ (\ref{lambda}), we first note that
\begin{align}
P P^T &= 	\left[ wu^\dagger + \B{R}_0 \B{S}_0\right]\left[ uw^\dagger + \B{S}_0^\dagger \B{R}_0^\dagger\right]\\
&= w u^\dagger u w^\dagger + wu^\dagger \B{S}_0^\dagger \B{R}_0^\dagger+\B{R}_0 \B{S}_0uw^\dagger + \B{R}_0 \B{S}_0\B{S}_0^\dagger \B{R}_0^\dagger\\
P^TP &= 	\left[ uw^\dagger + \B{S}_0^\dagger \B{R}_0^\dagger\right]\left[ wu^\dagger + \B{R}_0 \B{S}_0\right]\\
&= u w^\dagger w u^\dagger + uw^\dagger \B{R}_0 \B{S}_0+\B{S}_0^\dagger \B{R}_0^\dagger wu^\dagger + \B{S}_0^\dagger \B{R}_0^\dagger\B{R}_0 \B{S}_0.
\end{align}
Let us assume that $\Lambda = \B{S}_0^\dagger \Lambda^\prime \B{R}_0^\dagger$. We will justify this after the fact, since we will show that it leads to the unique solution to the optimization problem. With this ansatz, the terms in $PP^T$ which end in $w^\dagger$ and the terms in $P^T P$ which begin with $u$ all vanish, leaving
\begin{align}
	&=\frac{1}{2}\Big[ wu^\dagger \B{S}_0^\dagger \B{R}_0^\dagger+ \B{R}_0 \B{S}_0\B{S}_0^\dagger \B{R}_0^\dagger\Big]\B{S}_0^\dagger \Lambda^\prime \B{R}_0^\dagger\Big[\B{S}_0^\dagger \B{R}_0^\dagger wu^\dagger+ \B{S}_0^\dagger \B{R}_0^\dagger\B{R}_0 \B{S}_0\Big]\\
	&=\frac{1}{2\alpha^2}\Big[ wu^\dagger \B{S}_0^\dagger + \B{R}_0 \B{S}_0\B{S}_0^\dagger \Big] \Lambda^\prime \Big[\B{R}_0^\dagger wu^\dagger+ \B{R}_0^\dagger\B{R}_0 \B{S}_0\Big].
\end{align}
If we further assume that the measurement is unbiased, so that $w=\frac{1}{n}u$, then this becomes simply $\frac{1}{2\alpha^2}\Big[ \B{R}_0 \B{S}_0\B{S}_0^\dagger \Big] \Lambda^\prime \Big[\B{R}_0^\dagger\B{R}_0 \B{S}_0\Big]$.
Recall that informational completeness implies that $\B{S}_0\B{S}_0^\dagger$ and $\B{R}_0^\dagger \B{R}_0$ are invertible. So let $\Lambda^\prime = \beta(\B{S}_0\B{S}_0^\dagger)^{-1} (\B{R}_0^\dagger\B{R}_0)^{-1}$. As for the constant $\beta$, we require
\begin{align}
 \frac{1}{2\alpha^2} \Big[\B{R}_0 \B{S}_0 \B{S}_0^\dagger \Big]\beta (\B{S}_0\B{S}_0^\dagger)^{-1} (\B{R}_0^\dagger\B{R}_0)^{-1}\Big[ \B{R}_0^\dagger\B{R}_0 \B{S}_0\Big]=\frac{\beta}{2\alpha^2}\B{R}_0 \B{S}_0 \overset{!}{=}\frac{1-\alpha}{\alpha}\B{R}_0 \B{S}_0.
\end{align}
Thus $\beta = 2\alpha(1-\alpha)$, and
\begin{align} 
\Lambda &= 2\alpha(1-\alpha)\B{S}_0^\dagger (\B{S}_0\B{S}_0^\dagger)^{-1}(\B{R}_0^\dagger\B{R}_0)^{-1}\B{R}_0^\dagger.
\end{align}
Putting it all together, the unique Born matrix which minimizes the LTP deformation with respect to the Frobenius norm for an unbiased depolarizing reference measurement is
\begin{small}
\begin{align}
\Phi &= I - \frac{1}{2}	P^T\Lambda P^T\\
&= I - \frac{1}{2}\left[ \frac{1}{n}uu^\dagger + \B{S}_0^\dagger \B{R}_0^\dagger\right]\left[2\alpha(1-\alpha)\B{S}_0^\dagger (\B{S}_0\B{S}_0^\dagger)^{-1}(\B{R}_0^\dagger\B{R}_0)^{-1}\B{R}_0^\dagger\right]\left[ \frac{1}{n}uu^\dagger + \B{S}_0^\dagger \B{R}_0^\dagger\right]\\
&= I+\alpha(\alpha-1)\B{S}_0^\dagger \B{R}_0^\dagger\Big[ \B{S}_0^\dagger (\B{S}_0\B{S}_0^\dagger)^{-1}(\B{R}_0^\dagger\B{R}_0)^{-1}\B{R}_0^\dagger\Big] \B{S}_0^\dagger \B{R}_0^\dagger\\
&= I + \frac{\alpha-1}{\alpha}\B{S}_0^\dagger (\B{S}_0\B{S}_0^\dagger)^{-1}(\B{R}_0^\dagger\B{R}_0)^{-1}\B{R}_0^\dagger.
\end{align}
\end{small}
Finally, we show this this gives us the unique solution. The Frobenius norm is strictly convex: letting $f(\Phi) = \lVert I - \Phi \rVert^2$, for $\Phi_1 \neq \Phi_2$ and $0 < t < 1$, we have $f(t \Phi_1 + (1-t)\Phi_2) < tf(\Phi_1) + (1-t)f(\Phi_2)$. Moreover, the feasible set is $\{\Phi: P\Phi P = P\}$. This is affine since it is the set of solutions of a linear equation: indeed, if $\Phi_1, \Phi_2$ are in the set, one may easily check that so is $t\Phi_1 + (1-t)\Phi_2$. Suppose that $\Phi_1$ and $\Phi_2$ were distinct minimizers. But then
\begin{align}
f\left(\frac{1}{2}(\Phi_1 + \Phi_2\right) < \frac{1}{2}f(\Phi_1) + \frac{1}{2}f(\Phi_2),
\end{align}
which contradicts minimality, and so $\Phi_1=\Phi_2$. Thus the solution to the optimization problem is unique.
\end{proof}

But how does this Born matrix act on probability vectors? Recalling that $P(R|\rho)=\B{R}|\rho)=w + \B{R}_0 \rho_0 = \frac{1}{n} u + \B{R}_0 \rho_0$ in the unbiased case, we have
\begin{align}
\Phi P(R|\rho) &= \Big[ I + \frac{\alpha-1}{\alpha}\B{S}_0^\dagger (\B{S}_0\B{S}_0^\dagger)^{-1}(\B{R}_0^\dagger\B{R}_0)^{-1}\B{R}_0^\dagger\Big]\Big[	\frac{1}{n}u + \B{R}_0 \rho_0\Big]\\
&= \frac{1}{n} u + \B{R}_0 \rho_0 + \frac{\alpha-1}{\alpha}\B{S}_0^\dagger (\B{S}_0\B{S}_0^\dagger)^{-1}(\B{R}_0^\dagger\B{R}_0)^{-1}\B{R}_0^\dagger\B{R}_0 \rho_0\\
&= \frac{1}{n} u + \Big(\B{R}_0 + \frac{\alpha-1}{\alpha}\B{S}_0^\dagger (\B{S}_0\B{S}_0^\dagger)^{-1}\Big)\rho_0.
\end{align}
In other words, this Born matrix does \emph{not} act as an Protourgleichung! That said, there is a special case. We can always choose $\B{S}_0 = \frac{1}{\alpha}(\B{R}_0)^+=\frac{1}{\alpha}(\B{R}_0^\dagger\B{R}_0)^{-1}\B{R}_0^\dagger$---that is, $\B{S}_0$ can be taken proportional to the pseudoinverse of $\B{R}_0$, which is a left inverse since $\B{R}_0$ has linearly independent columns. Symmetrically, we have $\B{R}_0=\frac{1}{\alpha}\B{S}_0^\dagger(\B{S}_0\B{S}_0^\dagger)^{-1}$, and so 
\begin{align}
\Phi &= I + \alpha(\alpha-1)\B{R}_0 \B{S}_0.
\end{align}
and
\begin{align}
\Phi P(R|\rho) &= \frac{1}{n} u + \Big(\B{R}_0 + (\alpha-1)\B{R}_0 \Big)\rho_0= \frac{1}{n} u + \alpha\B{R}_0\rho_0.
\end{align}
In this special case, when the reference states derive from the pseudoinverse, the $\Phi$ which minimizes $\lVert I - \Phi \rVert_F$ acts as a Protourgleichung. In fact, even if the measurement is biased and the states were not obtained from the pseudoinverse, we can  nevertheless always use a Born matrix of the form $\Phi =  I + \alpha(\alpha-1)\B{R}_0 \B{S}_0$. Indeed,
\begin{align}
P\Phi P &= \left[ wu^\dagger + \B{R}_0 \B{S}_0\right] \left[I + \alpha(\alpha-1)\B{R}_0 \B{S}_0 \right] \left[ wu^\dagger + \B{R}_0 \B{S}_0\right] \\
&=wu^\dagger +\frac{1}{\alpha}\B{R}_0 \B{S}_0+\alpha(\alpha-1)\left(\frac{1}{\alpha^2}\right)\B{R}_0 \B{S}_0\\
&= wu^\dagger +\B{R}_0 \B{S}_0 = P.
\end{align}
Indeed, notice that  $\Phi =  I + \alpha(\alpha-1)\B{R}_0 \B{S}_0$ is  precisely the Born matrix we derived in Eq. (\ref{turns-out-min}), by finding a $\Phi$ supported entirely on $\col(P)$, and then adding in the projector onto its complement.

\subsection{A choice of measures}

Recall that for a MIC, we have
\begin{align}
 P(R|R)&=\frac{1}{\alpha}I + \left(1-\frac{1}{\alpha}\right)wu^\dagger.
\end{align}
It is not hard to confirm that the unique $\Phi$ matrix is
\begin{align}
\Phi &= \alpha I + (1-\alpha)wu^\dagger,	
\end{align}
which clearly acts as a Protourgleichung. For overcomplete reference measurements, however, we must make a choice of $\Phi$. We've seen many examples,
\begin{align}
\Phi_1 = \alpha I + (1-\alpha)w u^\dagger,
\end{align}
which is algebraically simplest,
\begin{align}
\Phi_2 = w u^\dagger + \alpha^2 \B{R}_0 \B{S}_0,
\end{align}
which is supported entirely on $\col(P)$,
\begin{align}
\Phi_3 = I + \alpha(\alpha-1)\B{R}_0\B{S}_0
\end{align}
which achieves smaller LTP deformation by adding to the former the projector onto the complement of $\col(P)$, and which in certain circumstances (unbiasedness, and the reference states derive from the pseudoinverse) coincides with 
\begin{align}
\Phi_{\min} = I + \frac{\alpha-1}{\alpha}\B{S}_0^\dagger (\B{S}_0\B{S}_0^\dagger)^{-1}(\B{R}_0^\dagger\B{R}_0)^{-1}\B{R}_0^\dagger,
\end{align}
which we have shown achieves the smallest LTP deformation with respect to the Frobenius norm. The first three act as a Protourgleichung on $\col(P)$, and yet achieve different values for the LTP deformation. The final one, the one which explicitly minimizes $\lVert I - \Phi \rVert_F$, need not. 
These examples show that the two concepts (minimal deformation with respect to a matrix norm, and the form of the Protourgleichung itself) are in fact distinct. We could amend the measure, e.g., to consider the distance not from the identity per se, but from the identity on the relevant subspace, that is, $\col(P)$: then e.g., the first three $\Phi$ matrices would be scored the same. But if we can achieve algebraic simplicity by adding support on the complement, why shouldn't we? We conclude simply that care must be taken to choose the measure proper to the circumstance at hand.

Happily, however, the exercise has allowed us to identify a distinguished set of reference states for any unbiased informationally complete reference measurement, namely those given by
\begin{align}
\B{S} =\left(\begin{array}{ccc}  
1 & \dots & 1 \\
\hline
& \\
& \frac{1}{\alpha}(\B{R}_0^\dagger\B{R}_0)^{-1}\B{R}_0^\dagger \\
& &\\
\end{array}\right)
\end{align}
for which the minimal $\Phi$ (with respect to the 2-norm) does in fact act as a Protourgleichung. In fact, if the measurement is weighted morphophoric, then this just gives us a parallel update reference measurement.

\section{Conclusion}

In this chapter, we have established that the Protourgleichung 
\begin{align}
P(E|\rho) = \sum_{j=1}^{n} P(E|R_j)\Big[ \alpha P(R_j|\rho) + (1-\alpha)P(R_j|\mu)\Big]
\end{align}
is ubiquitous. In any no-restriction GPT (which allows all mathematically possible states and effects), and for any reference measurement, one can always find a set of reference states which yield an expression of this form. The key requirement is that the channel associated with the reference measurement must be depolarizing---and this requirement can always be satisfied. Considering the case where the reference states can be chosen proportional to the effects (modulo a depolarization parameter), we related our construction to so-called ``morphophoric measurements,'' and ultimately to complex projective 2-designs, from which we recovered the standard Urgleichung. In light of these results, we reconsidered a matrix norm based measure of LTP deformation, showing by example that it fails to adequately capture the essence of the Protourgleichung, even as it still holds interesting lessons about the nature of reference measurements to teach us.

The Protourgleichung is the simplest nonclassical normative constraint between probability assignments $P(E|\rho)$, $P(E|R)$, and $P(R|\rho)$: it is universally available whether one works in quantum theory, classical probability theory, or any of the more exotic members of the GPT community. Nevertheless, the Urgleichung itself is a child of quantum theory over $\mathbb{C}$, its particular parameter ($\alpha=d+1$) and the cardinality of its outcomes ($d^2$) hinting at quantum theory's special structure, and whispering of the possibility of \emph{pure} SIC-POVMs. Indeed, these latter arguably represent the simplest nonclassical generalization of the trivial classical reference measurement, which simply reads off an objective property of a system without disturbing it.

Consider that, in the terms we've developed here, the trivial classical reference measurement has states and effects which are simply the $r$-dimensional basis vectors. It is a parallel update measurement, with reference states proportional to reference effects. The corresponding channel is simply the identity channel, which is trivially depolarizing, and the Protourgleichung collapses to the LTP. The measurement is trivially morphophoric, simply mapping the probability simplex exactly to itself. It is unbiased. It is a MIC. It is equiangular. Its reference states are extremal, corresponding to the vertices of the probability simplex itself. The reference effects are extremal too. Each reference state can be perfectly distinguished by some effect\footnote{That is, there is some effect which yields probability 1 on that state.}, and each reference effect can be perfectly distinguished by some state. Moreover, the reference states and reference effects perfectly distinguish each other. Finally, such a measurement exists in any dimension, where the maximal number of pairwise perfectly distinguishable states scales with the dimension.

Compare this to the properties of a SIC reference measurement in quantum theory over $\mathbb{C}$, which is a parallel update measurement, whose associated channel is depolarizing with parameter $\frac{1}{\alpha} = \frac{1}{d+1}$. It is morphophoric, unbiased, a MIC with $d^2$ elements. It is equiangular. Its reference states and reference effects are extremal, corresponding to the vertices of a simplex. Each reference state can be perfectly distinguished by some effect, and each reference effect can be perfectly distinguished by some state. And finally, such a measurement (apparently) exists in any dimension, where the maximal number of pairwise perfectly distinguishable states scales with the dimension\footnote{The point of emphasizing that the maximal number of pairwise perfectly distinguishable states scales with the dimension is to exclude, for example, so-called \emph{spin-factor} theories, whose state spaces are $n$-dimensional balls, and for which SICs exists trivially. In such theories, regardless of the dimension, the maximal number of pairwise perfectly distinguishable states is always 2. Such a theory is an example of a \emph{Euclidean Jordan Algebra} \cite{staceyQuantumTheorySymmetry2019, wilceRoyalRoadQuantum2018} of which quantum theory over $\mathbb{C}$ is also an example, as well as quantum theory over $\mathbb{R}, \mathbb{H}$, and $d=3$ quantum theory over $\mathbb{O}$. For quantum theory over $\mathbb{R}$, pure SICs are only known to exist in $d=2,3,7,23$, and are known \emph{not} to exist in most dimensions \cite{fuchsSICQuestionHistory2017}. For quantum theory over $\mathbb{H}$, pure SICs are only known to exist in $d=2,3$, and there is numerical evidence suggesting that they don't exist already in $d=4$ \cite{cohnOptimalSimplicesCodes2016}. There is a pure SIC in octonionic $d=3$, but this is the only dimension in which such a theory can be constructed. We shall hear more about Euclidean Jordan algebras in subsequent chapters.}.

The only differences, then, are the values of the depolarizing parameter, the fact that the state space is mapped to a subset of the probability simplex, and the fact that the reference states and reference effects don't perfectly distinguish each other. Thus pure SIC reference measurements are arguably the closest analogue to the trivial classical reference measurement, and this is true not just in one theory, but in a whole infinite sequence of them, for each Hilbert space dimension $d$. From these considerations, if one were trying to generalize classical probability theory as minimally as possible, incorporating the idea that nature forgoes hidden variables\footnote{Or more positively, as Blake Stacey would put it, respecting nature's \emph{vitality} \cite{staceyQuantumTheorySymmetry2019}.}, one could imagine nothing simpler than positing the existence of pure SIC reference measurements, with their attendant, the Urgleichung---and indeed, this has been the basis of QBist reconstructions of quantum mechanics hitherto.

\section{Appendix: Real Vector Space QM}
\label{realqm}
A close cousin of standard quantum theory is quantum mechanics over $\mathbb{R}$. Here there is a subtlety if one works with vectorized operators. For quantum mechanics over $\mathbb{C}$, if we take $P(R|R) = \B{RS}$ where the rows of $\B{R}$ are vectorized effects $(R_i|$ and the columns of $\B{S}$ are vectorized states $|R_i)$, then $\B{RS}$ will be a full rank factorization: this is because the real space of Hermitian matrices is $d^2$, but also the vectorized operators are $d^2$ dimensional. On the other hand, if one works over $\mathbb{R}$, the state space is the space of real symmetric matrices which has dimension $d(d+1)/2$ even as the vectorized operators remain $d^2$ dimensional. In this latter case, $\B{RS}$ built from vectorized operators will \emph{not} be a full rank factorization. One could instead work with the states and effects expanded in an operator basis of $d(d+1)/2$ elements; otherwise, one will have to modify the Born identity to $\B{S}\Phi \B{R} = \Pi_{\text{sym}^2}$ where $\Pi_{\text{sym}^2}$ is the projector onto symmetric subspace on two factors, i.e.\ onto the space of vectorized symmetric matrices. This issue is quite general if one doesn't work with full rank factorizations.

Thus for a depolarizing channel in real vector space quantum mechanics, we have (in terms of vectorized operators)
\begin{align}
	\sum_{i=1}^n |S_i)(R_i|&=	 \frac{1}{\alpha}\Pi_{\text{sym}^2} + \left(1-\frac{1}{\alpha}\right)\frac{1}{d}|I_d)(I_d|.
\end{align}
This theory is also self dual, so  taking the trace of both sides, and assuming our states are pure, we find
\begin{align}
	d =	 \frac{1}{\alpha}\frac{d(d+1)}{2} + 1-\frac{1}{\alpha} \Longrightarrow \alpha = \frac{d+2}{2},
\end{align}
and
\begin{align}
	\sum_{i=1}^n |S_i)(R_i|&=	 \frac{2}{d+2}\Pi_{\text{sym}^2} + \frac{1}{d+2}|I_d)(I_d|\\
	&= \frac{1}{d+2}\Big(I_{d^2} +\text{SWAP} +|I_d)(I_d|\Big).
\end{align}
Again dividing by $1/d$ so the trace of both sides is 1, and taking the partial transpose, which actually leaves both sides of the equation invariant, we obtain
\begin{align}
	 \sum_{i=1}^n w_i S_i \otimes S_i =\frac{1}{d(d+2)}\Big(I_{d^2} +\text{SWAP} +|I_d)(I_d|\Big),
\end{align}
which according to \cite{harrowChurchSymmetricSubspace2013} coincides precisely with $\int |\psi\rangle\langle\psi|^{\otimes 2} d\psi$ where the integral is taken over pure states in the real vector space theory. Thus  weighted morphophoric references measurements for quantum theory over $\mathbb{R}$ coincide with real projective 2-designs.

On the subject of real vector space quantum mechanics, \cite{Weiss2021} conducted a numerical search for MIC reference measurements minimizing $\lVert I - \Phi \rVert$ in $d=4$. It is known that in this dimension one cannot construct a SIC-POVM out of pure effects: the maximum of 6 equiangular lines is not enough to furnish the 10 elements required for a MIC. In fact, in $d=4$, the minimum number of elements in a pure biased 2-design is 11, and the minimum number of elements in a pure unbiased 2-design is 12 \cite{hughesSphericalTtdesignsSmall2021}. These efforts were unable to yield minimal reference measurements whose Born matrix acts as a Protourgleichung, a fact which is now contextualized by the present work. Indeed, such reference measurements \emph{can} be found, although they will not minimize $\lVert I - \Phi \rVert$. It is worth noting that subsequent numerical searches suggest that in this case, if the reference effects are pure, then the depolarizing reference states must be mixed. It is also worth noting that unlike in \cite{debrotaSymmetricInformationallyComplete2020}, the minimal reference measurement does depend on the choice of matrix norm, which is likely to be true in the general overcomplete case.

%% file: sections/classicality.tex
\UMBchapter{(No) Return to classicality}
\label{ch:classicality}

\section{Introduction}

Having supplemented standard probability with a set of coherence conditions rich enough to encompass the brand of nonclassicality exemplified by quantum mechanics, we may then ask: when is a gambler justified in using vanilla probability theory after all? When is the gambler justified in acting \emph{as if} objects have definite properties before they are measured and which are revealed by measurement, and which do not depend on the context in which they are measured? To put it more precisely, when is the gambler justified in supposing that there is a reference measurement $\{\lambda_i\}$ satisfying $\Phi_\lambda = I$ so that regardless of whether the reference measurement is  actually performed, the gambler ought to appeal to the law of total probability $P(E|\rho) = P(E|\lambda)P(\lambda|\rho)$ for guidance. To answer this, we assimilate Spekkens's notion of \emph{noncontextual ontological models} to our general philosophy. Following \cite{shahandehUnifiedLinearAlgebraic2025}, we show how an equirank nonnegative matrix factorization of the probability table characterizing a scenario corresponds to so-called simplex embedding \cite{selbyLinearProgramTesting2024}, which furnishes a noncontextual ontological model of the scenario. In particular, we show that if the probability table characterizing the scenario captures a reference measurement $\{R_i\}$, then the existence of a noncontextual ontological model can be understood as the existence of a classical reference measurement $\{\lambda_i\}$ for which $\{R_i\}$ itself is a reference. Thus we can understand noncontextual ontological models in terms more suitable to the QBist: indeed, this provides a motivation for the idea of a noncontextual ontological model to begin with. We also show how simplex embeddings implicate a variant of the Born matrix whose nonnegativity implies classicality, and discuss the general conditions under which $\Phi$ may be taken to be a stochastic matrix. Finally, we show how violations of Bell inequalities and noncontextuality inequalities more generally can be interpreted as the cost of abandoning the fundamental nonclassical coherence condition.

\section{Noncontextuality}

We have developed the notion of a \emph{state} as an equivalence class of consequences conditional upon any of which the gambler would assign the same probabilities to the reference measurement. In short, for us the concept of a state is fully grounded in a gambler's state of expectation about reference outcomes regardless of the peculiarities of prior circumstances. If the gambler would assign probabilities $P(R|A)$ given circumstance $A$ and probabilities $P(R|B)$ given circumstance $B$, and $P(R|A)=P(R|B)$, we would identify the circumstances $A$ and $B$, saying: they amount to the same state. Calling $A$ or $B$ the \emph{context}, we see that our definition of a state is precisely \emph{noncontextual}. The state is what for the gambler is invariant across contexts. And of course, by coherence, assigning the same state of expectation to the reference measurement conditional on either circumstances means that the gambler ought to assign the same probabilities to \emph{any} measurement---and the same could be said of the effects.

We also showed how we may form a GPT-style representation using a rank decomposition, and we will now see how this preserves noncontextuality. To this end, let us consider an $m \times n$ table of conditional probabilities where the columns correspond to outcomes of preparatory measurements and the rows correspond to outcomes of subsequent measurements,
\begin{align*}
	P(E|\pi) &= \begin{pmatrix} 
 	P(E_1|\pi_1) & \cdots & P(E_1|\pi_n)\\
 	\vdots & \ddots &\vdots \\
 	P(E_m|\pi_1) & \cdots & P(E_m|\pi_n)
 	\end{pmatrix}.
\end{align*}
This matrix $P(E|\pi)$ will have some rank $r$, and we may form a rank decomposition $P(E|\pi) = \B{EP}$ for $\B{E}\in \mathbb{R}^{m \times r}_r, \boldpi\in \mathbb{R}^{r\times n}_r$. Now suppose two rows of $P(E|\pi)$ are identical: $P(E_i|\pi) = P(E_j|\pi)$. Then in the rank decomposition $(E_i|\boldpi = (E_j|\boldpi$, or $( (E_i| - (E_j| )\boldpi=0$. But $\boldpi\in \mathbb{R}^{r\times n}_r$: it has $r$ linearly independent rows. Thus if a linear combination of them yields the zero vector, each component in the linear combination must be identically 0, and so $(E_i|=(E_j|$. Similarly, if we assign identical probabilities to two outcomes $E_i$ and $E_j$ given any $\pi$, then we assign the same row vector $(E_i|=(E_j|$. In this way, the representation of context becomes simply the label $i$ or $j$ which remembers which consequence in the equivalence class actually occurred. More generally, suppose that a row is linearly dependent on a subset of other rows, $P(E_i|\pi) = \sum_{j\in J} \alpha_j P(E_j|\pi)$ for $J$ some subset of rows. Then $(E_i|\boldpi = \sum_{j\in J} \alpha_j (E_j|\boldpi$ where the same argument implies $(E_i| = \sum_{j\in J} \alpha_j (E_j|$. In this way, a rank factorization preserves linear equivalences in probability assignments, guaranteeing the new representation is noncontextual. In particular, if e.g., two rows of $P(E_i|\pi)$ and $P(E_j|\pi)$ are identical, we ought to identify the two outcomes even though they may appear in two different measurements:  in a contextual representation, however, the vector representatives $(E_i|$ and $(E_j|$ might depend on which measurement the outcome appears in.

As we will see, the feature that guarantees that $P(E|\pi)=\B{EP}$ is a noncontextual representation is that $\rank(P(E|\pi)) = \rank(\B{E})=\rank(\boldpi)$. Otherwise, the argument from linear independence would not hold: we might have $P(E_i|\pi) = P(E_j|\pi)$, but it would not necessarily follow that $(E_i| = (E_j|$. Similarly, we might have $P(E|\pi_i) = P(E|\pi_j)$, but it would not necessarily follow that $|\pi_i) = |\pi_j)$. We would not want to call $|\pi_i)$ or $|\pi_j)$ themselves states, since according to the gambler's own lights, assigning $|\pi_i)$ or $|\pi_j)$ would lead precisely to the same probabilities for all measurement outcomes. The gambler would be completely indifferent to the choice of representation $|\pi_i)$ or $|\pi_j)$ as far as their mesh of beliefs is concerned: the only difference between them is in the irrelevant context, that it was in fact outcome $\pi_i$ as opposed to $\pi_j$ which occurred. In contrast, a rank decomposition yields one vector representative for equivalence class.

Thus if we forget the labels, in the representation furnished by a rank decomposition, each effect is assigned an $(E|$ and each state a $|\pi)$. In the language of Spekkens \cite{schmidStructureTheoremGeneralizednoncontextual2024}, the table $P(E|\pi)$ gives an \emph{operational theory}, assigning a probability to each pairing of $E_i$ and $\pi_j$. Quotienting by ``operational equivalence'' yields a set of states and effects which \emph{separate} each other: two separate states must yield different probabilities on some effect, and two separate effects must yield different probabilities on some state. This mutual separation of states and effects is a prerequisite for passing to a GPT, or generalized probabilistic theory. (One must also complete the theory by taking conic combinations of states and effects, and adding in opposite effects $(\neg E| = (1| - (E|$ for each effect, if they are not in the table.) In the language of Spekkens, a GPT is then a noncontextual model of an operational theory\footnote{For us, of course, we begin from states as distributions $P(R|\pi)$ and effects as distributions $P(E|R)$. This is a noncontextual representation even though the dimension $n$ of the probability vectors might be larger than the rank of $P(R|R)$: as we have seen the coherence conditions $P(R|\pi) = P(R|R)\Phi P(R|\rho)$ and $P(E|R) = P(E|R)\Phi P(R|R)$ ensure that any $P(R|\pi)$ and $P(E|R)$ live in the column and row spaces of $P(R|R)$.}.

If instead we use an arbitrary decomposition $P(E|\pi)= \B{EP}$, however, even if we ``forget the labels,'' we may end up with vector representatives $(E_i|$ and $|\pi_i)$ that remember their context even though that context makes no ``operational difference'' in terms of the gambler's own probability assignments. Indeed, even if $P(E|\pi)$ has no repeated rows or columns, if we use an arbitrary decomposition, because of the linear dependence, there will be many different choices of vector representatives that make no difference at all, thus introducing context dependence where there was none before. So which decompositions guarantee a noncontextual representation? We have the following lemma:

\begin{lemma}
   \label{separating-decomposition}
   Let  $M=AB$ and let $\{(A_i|\}$ denote the rows of $A$ and $\{|B_i)\}$ denote the columns of $B$. Then $\{(A_i|\}$ separates points in $\linspan(\{|B_i\})$ and $\{|B_i)\}$ separates points in $\linspan(\{(A_i|\})$ iff  $\rank(M) = \rank(A) = \rank(B)$.
\end{lemma}
\begin{proof}
We first note that $\linspan(\{(A_i|\}) = \row(A)$ and $\linspan(\{|B_i\}) = \col(B)$. Moreover, $\dim \row(A) = \rank(A)$ and $\dim \col(B) = \rank(B)$. Now to say that $\{(A_i|\}$ separates points in $\linspan(\{|B_i)\}$ is to say that for 
\begin{align}
\forall |x), |y) \in \col(B), A|x) = A|y) \Longleftrightarrow |x) = |y).
\end{align}
Thus $A$ is 1-to-1 on $\col(B)$. 1-to-1 linear maps must preserve the dimensionality of the subspace on which they act. Thus $\dim A \col(B) = \dim \col(B)=\rank(B)$. But since $M=AB$, $\col(M) = A \col(B)$ and so $\rank(M) = \dim A \col(B)$. We conclude that $\rank(B) = \rank(M)$. Similarly, $\row(M)=\row(A) B$ so $\rank(M) = \dim \row(A) B$. To say that $\{|B_i)\}$ separates points in $\linspan(\{(A_i|\})$ is to say that 
\begin{align}
\forall (x|, (y| \in \row(A): (x|B = (y|B \Longleftrightarrow (x| = (y|.
\end{align}
So $\dim \row(A) B = \dim \row(A)=\rank(A)$ and since $\row(M) = \row(A) B$,  $\rank(M) = \dim \row(A) B$ and $\rank(A)=\rank(M)$.

Conversely, suppose $\rank(M) = \rank(A) = \rank(B)$. We have $\rank(M) = \dim A \col(B)=\rank(B)=\dim \col(B)$ so that $A$ is 1-to-1 on $\col(B)$, and so the rows of $A$ separate vectors in $\col(B)$. Similarly, $\rank(M) = \dim \row(A) B=\rank(A)=\dim \row(A)$ so that $B$ is 1-to-1 on $\row(A)$, and the columns of $B$ separate vectors in $\row(A)$.
\end{proof}

Thus to preserve noncontextuality we ought to confine ourselves to decompositions $P(E|\pi) = \B{EP}$ where $\rank(P(E|\pi))=\rank(\B{E})=\rank(\boldpi)$. Indeed, there would not seem to be any good reason for exploring contextual representations, unless of course they had some other perhaps desirable structure. With this in mind, we now turn to the possibility of constructing a classical model of $P(E|\pi)$. This would require us to be able to write for each state $\pi$ and effect $E$,
\begin{align}
P(E|\pi) = \sum_i P(E|\lambda_i) P(\lambda_i|\pi),
\end{align}
that is, we factorize $P(E|\pi)=P(E|\lambda)P(\lambda|\pi)$. In the language of Spekkens, we thereby construct an \emph{ontological model}: we introduce a set of ``hidden variables'' which take values $\{\lambda_i\}$, we map effects to response functions $P(E|\lambda)$, and we map states to distributions $P(\lambda|\rho)$. In our terms, this is equivalent to assuming there exists a reference measurement with outcomes $\{\lambda_i\}$ whose performance the gambler is indifferent to so that $\Phi_\lambda = I$. It is this indifference that licenses the gambler to act as if the system is characterized by some underlying properties which the reference measurement $\{\lambda_i\}$ reads off. Those without a commitment to expressing everything in terms of coherence between the beliefs of a gambler betting on the consequences of their own actions, would simply want to say: if I can give an ontological model, then the system must really ``have'' such properties.

But in fact, constructing such ontological models is trivial. Indeed, there is an ontological model for every such table $P(E|\pi)$, namely, $P(E|\pi)=P(E|\pi) I_n$. Here we introduce as many reference outcomes as $\pi$'s. Each $\pi_j$ is associated with a state of complete certainty about the outcome of the reference measurement: a column vector of all 0's and a 1 in the $j$th place. We take the response function $P(E_i|\lambda)$ to simply be the $i$th row of $P(E|\pi)$ so that indeed $P(E_i|\lambda) P(\lambda|\pi_j) = P(E_i|\pi_j)$ as desired. In fact, in the language of Spekkens, we have in this model measurement noncontextuality but preparation contextuality. 

Now it may be that $P(E|\pi)$ is invertible. Then this trivial decompositions will be equirank, and by Lemma \ref{separating-decomposition} the construction will be noncontextual. In the language of Spekkens, we would have a \emph{noncontextual ontological model}. But in general the trivial decomposition will yield a contextual representation. For example, suppose the third column of $P(E|\pi)$, that is, $P(E|\pi_3)$ were a convex combination of the first two: $P(E|\pi_3) = pP(E|\pi_1) + (1-p)P(E|\pi_2)$. This is a linear dependency, and so $P(E|\pi)$ cannot have full rank. In the decomposition $P(E|\pi)=P(E|\lambda)P(\lambda |\pi) = P(E|\pi)I_n$, we map $P(\lambda|\pi_1)= (1,0,\dots)^\dagger$, $P(\lambda|\pi_2) = (0,1,0,\dots)^\dagger$, and $P(\lambda|\pi_3) = (0,0,1,\dots)^\dagger$ so that each $P(\lambda|\pi_j)$ picks out the $j$th column $P(E_i|\pi_j)$. But since $P(E|\pi_3) = pP(E|\pi_1) + (1-p)P(E|\pi_2)$, it follows that $P(\lambda|\pi_3)=(0,0,1,0,\dots)^\dagger$ is indistinguishable from $pP(\lambda|\pi_1)+(1-p)P(\lambda|\pi_2) = (p, 1-p, 0, \dots)^\dagger$. In other words, there are two completely different states of uncertainty about the reference outcomes $\{\lambda_i\}$ which by the gambler's own lights are consistent with assigning identical probabilities to all measurements. The difference between the two we call \emph{context}. Now this particular example could be avoided by identifying which columns of $P(E|\pi)$ are extremal (in the sense of convex independence) and which rows are extremal (in the sense of conic independence) and restricting attention to those: but the underlying problem is quite general.

This trivial example shows that a \emph{nonnegative matrix factorization} (NMF) is always possible, that is, a decomposition $M=AB$ where $M,A,B$ are elementwise nonnegative. But it is generally desirable to have more compact such factorizations.
\begin{definition}
 Let $M\in \mathbb{R}^{m\times n}_{\ge 0}$ be a nonnegative matrix.
 The \emph{nonnegative rank}  $\rank_+(M)=r_+$ is the smallest inner dimension such that $M$ can be factored as $M=AB$ where $A \in \mathbb{R}^{m\times r_+}$ and $B \in \mathbb{R}^{r_+ \times n}$. It satisfies $\rank(M) \leq \rank_+(M) \leq \min(m, n)$. 
\end{definition}
\noindent In particular, we would like to find an NMF $P(E|\pi)=P(E|\lambda)P(\lambda|\pi)$ where the columns $P(\lambda|\pi_i)$ are proper probability distributions summing to 1. The nonnegative matrix rank puts a lower bound on the number of the outcomes of the classical reference measurement.
\begin{lemma}
Let $M$ be column stochastic and $M=AB$ a nonnegative matrix factorization. After deleting any zero columns of $A$, there is an equivalent factorization $M=A^\prime B^\prime$ with $A^\prime$ and $B^\prime$ column stochastic.
\end{lemma}
\begin{proof}
Let $a_i$ denote the columns of $A$ and let $\overline{a}_i$ denote $i$th column sum. Now if $\overline{a}_i=0$, delete that column as well as the $i$th row of $B$: since $A$ is nonnegative, the whole column must be 0, and since $M = a_1 b_1^\dagger + \dots $, neither will contribute to $M$. Now let $K=\text{diag}(\overline{a}_1, \dots)$: then $A^\prime = A K^{-1}$ will be column stochastic. $B^\prime = K B$ will ensure $M = A^\prime B^\prime$. Meanwhile, let $u=(1,\dots,1)^\dagger$. Clearly, $u^\dagger B^\prime = u^\dagger K B$. But $u^\dagger K = (\overline{a}_1, \dots) = u^\dagger A$, so $u^\dagger B^\prime = u^\dagger A B = u^\dagger M = u^\dagger$ since $M$ is assumed column stochastic.
\end{proof}
\begin{corollary}
We may always construct an NMF of $P(E|\pi)= P(E|\lambda)P(\lambda|\pi)$ so that $P(\lambda|\pi)$ is column stochastic.
\end{corollary}
\begin{proof}
 We may always organize the rows of $P(E|\pi)$ in terms of measurements, i.e., $E_1^{(1)}, E_2^{(1)}, \dots, E_1^{(2)}, E_2^{(2)}, \dots$ where the superscript refers to which measurement the outcome is a part of. If a measurement has a full set of outcomes, then $\forall j: \sum_i P(E_i^{(k)}|\pi_j) = 1$: that is, the column sums of the rectangular block $P(E^{(k)}|\pi)$ are all 1. If a measurement doesn't have a full set of outcomes, we can always add a row namely $P(1|\pi) - \sum_i P(E_i^{(k)}|\pi)$ which completes the measurement: we may always consider the outcome that none of the other outcomes occurs. Suppose there are $h$ measurements. In any column, there are $h$ blocks which sum to 1: the column sums of $P(E|\pi)$ are therefore $h$, so that $h^{-1}P(E|\pi)$ is column stochastic and so by the previous lemma, $h^{-1}P(E|\pi) = P^\prime(E|\lambda)P(\lambda |\pi)$ are all column stochastic. Thus $P(E|\pi) = [hP(E^\prime|\lambda)]P(\lambda |\pi)$ is a decomposition of the original matrix, and $P(\lambda|\pi)$ is column stochastic, so its columns can be interpreted as distributions of the outcomes of the reference measurement $\{\lambda_i\}$.
\end{proof}

Invoking Lemma \ref{separating-decomposition}, we may now characterize when a noncontextual ontological model exists of a probability table $P(E|\pi)$.
\begin{corollary}[Theorem 1 of \cite{shahandehUnifiedLinearAlgebraic2025}]
   \label{equirank}
$P(E|\pi)$ admits a noncontextual ontological model iff there exists a nonnegative matrix factorization $P(E|\pi) = P(E|\lambda)P(\lambda|\pi)$ satisfying $\rank(P(E|\pi)) = \rank(P(E|\lambda)) = \rank(P(\lambda|\pi))$.
\end{corollary}

Finally, it may be that $P(E|\pi)$ contains within it probabilities $P(R|R)$ characterizing a reference measurement $\{R_i\}$. We shall see that the existence of a noncontextual ontological model amounts to the assumption that a classical reference measurement $\{\lambda_i\}$ exists for which $\{R_i\}$ is a reference.

\subsection{An example}

An example might be helpful here. Let us consider a probability table with three measurements each with three outcomes, and four preparatory outcomes. 
\begin{align}
P(E|\pi)
&=
\left(\begin{array}{c|cccc}
& \pi_1 & \pi_2 & \pi_3 & \pi_4\\
\hline
A_{AB} & 1/2 & 1/2 & 1/2 & 1/2\\
B_{AB} & 1/2 & 0 & 1/2 & 0\\
N_{AB} & 0 & 1/2 & 0 & 1/2\\
\hline
B_{BC} & 1/2 & 0 & 1/2 & 0\\
C_{BC} & 1/2 & 1/2 & 0 & 0\\
N_{BC} & 0 & 1/2 & 1/2 & 1\\
\hline
C_{AC} & 1/2 & 1/2 & 0 & 0\\
A_{AC} & 1/2 & 1/2 & 1/2 & 1/2\\
N_{AC} & 0 & 0 & 1/2 & 1/2
\end{array}\right).
\end{align}
We will label the three measurements $AB$, $BC$ and $AC$. Indeed, notice that $P(A_{AB}|\pi) = P(A_{AC}|\pi)$: thus we are justified in identifying an effect $A$ whose equivalence class is $\{A_{AB}, A_{AC}\}$ since we assign the same probabilities regardless of which $\pi$ occurs. Similarly, we may identify effects $B$ and $C$. The measurement labels then serve to remind us which pair of effects appear together in a single measurement, along with a third effect to ensure probabilities sum to 1. This table has the following contextual factorization with a classical reference measurement of four outcomes,
\begin{align}
P(E|\pi) &= P(E|\lambda) P(\lambda|\pi) \\
&=\left(\begin{array}{c|cccc}
& \lambda_1 & \lambda_2 & \lambda_3 & \lambda_4\\
\hline
A_{AB} & 1 & 0 & 0 & 1\\
B_{AB} & 0 & 1 & 0 & 0\\
N_{AB} & 0 & 0 & 1 & 0\\
\hline
B_{BC} & 0 & 1 & 0 & 0\\
C_{BC} & 1 & 0 & 0 & 0\\
N_{BC} & 0 & 0 & 1 & 1\\
\hline
C_{AC} & 1 & 0 & 0 & 0\\
A_{AC} & 0 & 1 & 1 & 0\\
N_{AC} & 0 & 0 & 0 & 1
\end{array}\right)\left(\begin{array}{c|cccc}
& \pi_1 & \pi_2 & \pi_3 & \pi_4\\
\hline
\lambda_1 & 1/2 & 1/2 & 0 & 0\\
\lambda_2 & 1/2 & 0 & 1/2 & 0\\
\lambda_3 & 0 & 1/2 & 0 & 1/2\\
\lambda_4 & 0 & 0 & 1/2 & 1/2
\end{array}\right).
\end{align}
Indeed, notice for example that 
\begin{align}
P(A_{AB}|\lambda) = \begin{pmatrix} 1 & 0 & 0 & 1 \end{pmatrix},
\end{align}
but
\begin{align}
P(A_{AC}|\lambda) = \begin{pmatrix} 0 & 1 & 1 & 0 \end{pmatrix}.
\end{align}
The response function assigned to the effect $A$ depends on context: whether $A$ appears in the measurement $AB$ or the measurement $AC$. By construction $P(A_{AB}|\lambda)P(\lambda|\pi) = P(A_{AC}|\lambda)P(\lambda|\pi)$: this is necessary in order to reproduce the table. But if the gambler were able to perform the classical reference measurement and obtain outcome $\lambda_1$,  notice that $P(A_{AB}|\lambda_1) = 1$ but $P(A_{AC}|\lambda_1) = 0$: they would be certain that $A$ would occur in the measurement $AB$ but certain that $A$ would \emph{not} occur in the measurement $AC$. Thus the gambler would not regard ``$A$'' as a property of the system independent of the choice of measurement: whether ``$A$'' depends on the context in which it is measured.

Moreover, $\rank(P(E|\pi)) = 3, \rank(P(E|\lambda))=4, \rank(P(\lambda|\pi))=3$. Thus indeed the equirank condition is violated. Denote the four columns of $P(E|\pi)$ as $p_1, p_2, p_3, p_4$, and note that $p_1-p_2-p_3+p_4=0$. We can therefore write an arbitrary vector in the column space of $P(E|\pi)$ as e.g., $x=ap_1 + b p_2 + c p_3$. Forming the explicit expression for $x$ and demanding that it be elementwise nonnegative leads to several inequalities which $a, b, c$ must satisfy, not all of which are independent: they can be reduced down to four, namely $b\ge 0, c \ge0, a+b \ge 0, a+c \ge 0$. In three dimensions, extreme rays are where two independent inequalities are both saturated \cite{schrijver1986theory}. Solving for these explicitly and discarding solutions where $(a,b,c)=(0,0,0)$, we find four extreme rays: $(1,0,0), (0,1,0), (0,0,1), (-1,1,1)$. But assigning these values to $a,b,c$ yields just $p_1, p_2, p_3,$ and $p_4$ again. Suppose that $P(E|\lambda)$ only had three columns. Since $P(E|\pi) = P(E|\lambda)P(\lambda|\pi)$, these three columns would have to generate a cone with four extreme rays, which is contradictory. Thus $\rank_+(P(E|\pi))\ge 4$, and since we have an example with inner dimension 4, it follows that $\rank_+(P(E|\pi))=4$: we cannot find an NMF with fewer hidden variables, and so any classical model must be contextual.

\subsection{Simplex embedding}

We now show that we can understand the construction of a noncontextual classical model in terms of a linear embedding of states and effects into the simplex and its dual hypercube in such a way that all probabilities are preserved. 

\begin{lemma}
   \label{embedding-lemma}
Let $M=AB$ be a rank decomposition so that $M\in \mathbb{F}^{m \times n}_r$, $A \in \mathbb{F}^{m\times r}_r$, and $B \in \mathbb{F}^{r \times n}_r$. At the same time, let $M=CD$ be an equirank decomposition with arbitrary inner dimension: $C\in \mathbb{F}^{m\times k}$, and $D \in \mathbb{F}^{k\times n}$ with $r=\rank(M)=\rank(C)=\rank(D)$. We may express this decomposition equivalently as
\begin{align}
C = A T_A && D = T_B B
\end{align}
for $T_A, T_B$ satisfying $T_A T_B = I_r$. Moreover, $T_B T_A$ projects column vectors into $\col(D)$ and row vectors into $\row(C)$. 
\end{lemma}
\begin{proof}
Since $M=CD$, every column of $M$ is a linear combination of the columns of $C$. Hence $\col(M) \subseteq \col(C)$. But since $\rank(M)=\rank(C)$, the column spaces have the same dimension, and so in fact $\col(M)=\col(C)$. Meanwhile, because $M=AB$ is a rank decomposition, $\col(M) = \col(A)$ and thus $\col(A)=\col(C)$. Since $A$ and $C$ share the same column space, and the columns of $A$ form a basis for $\col(A)$, we can write $C= A T_A$ for some operator $T_A$ which specifies the (unique) expansion coefficients. Similarly, $\row(M) \subseteq \row(D)$. But since $\rank(M)=\rank(D)$, we have in fact $\row(M) = \row(D)$. Meanwhile, because $M=AB$ is a rank decomposition, $\row(M)=\row(B)$ and thus $\row(B)=\row(D)$ and we can write $D = T_B B$ for some operator $T_B$. Finally, since $M=AB$ is a rank decomposition, $A$ has a left inverse and $B$ has a right inverse. Thus explicitly $T_A = A^L C$ and $T_B = D B^R$ and so $T_A T_B = A^L CD B^R = A^L M B^R = A^L AB B^R = I$. Conversely, suppose $C=A T_A$ and $D = T_B B$, then
\begin{align}
\rank(C) = \rank(A T_A)\leq \rank(A) = \rank(M) = \rank(CD) \leq \rank(C),
\end{align}
and similarly for $\rank(D)$, so that $M=CD$ is an equirank decomposition.

Finally, we note $T_B T_A=(T_B T_A)^2$ is a projector. Clearly, $\col(T_B T_A) \subseteq \col(T_B)$. Conversely, if $y \in \col(T_B)$, then $y=T_B z$ for some $z$. But $(T_B T_A)y = T_B T_A T_B z = T_B z = y$ so every vector in $\col(T_B)$ is fixed by $T_B T_A$. Thus $\col(T_B T_A) = \col(T_B)$. Meanwhile since $D= T_B B$ and $B$ has full row rank, its columns span all of $\mathbb{F}^r$. So $\col(D) = T_B \col(B) = T_B(\mathbb{F}^r) = \col(T_B)$. Thus $\col(T_B T_A)=\col(D)$ and so $T_B T_A$ projects onto $\col(D)$. At the same time, $\row(T_B T_A) \subseteq \row(T_A)$. Conversely, if $y \in \row(T_A)$, then $y = z T_A$. But $y (T_B T_A) = z T_A T_B T_A = z T_A = y$ so every vector in $\row(T_A)$ is fixed by $T_B T_A$. Thus $\row(T_B T_A) = \row(T_A)$. Meanwhile since $C = A T_A$ and $A$ has full column rank, its rows span all of $\mathbb{F}^r$. So $\row(C) = \row(A)  T_A = (\mathbb{F}^r)T_A = \row(T_A)$. Thus $\row(T_B T_A) = \row(C)$ and so $T_B T_A$ projects onto $\row(C)$. 

\end{proof}

\begin{corollary}
   \label{simplex_embedding_matrices}
   Let $P(E|\pi) = P(E|\lambda)P(\lambda|\pi)$ be an equirank stochastic NMF and let $P(E|\pi) = \B{E}\boldpi$ be a rank decomposition. Then $P(E|\lambda) = \B{E}T_E$ and $P(\lambda|\pi) = T_\Pi \boldpi$ where $T_E = \B{E}^L P(E|\lambda)$ and $T_\Pi = P(\lambda|\pi)\boldpi^R$ satisfy $T_E T_\Pi = I_r$. Moreover, $\Pi_\lambda= T_\Pi T_E$ projects column vectors into $\col(P(\lambda|\pi))$ and row vectors into $\row(P(E|\lambda))$.
\end{corollary}

\begin{remark}
This formulation in terms of embedding maps $T_E$ and $T_\Pi$ is more appropriate if one wishes to consider whether one can embed an entire GPT (and not just a probability table $P(E|\pi)$) into the simplex and its dual. 
\end{remark}

\begin{remark}
We note that the more general idea of embedding one GPT into another GPT has been treated systematically in \cite{mullerTestingQuantumTheory2023, schmidShadowsSubsystemsGeneralized2025} although finding such embeddings, if they exist, will not in general be as computationally tractable as finding a simplex embedding. In particular, one may ask when a GPT can be embedded into quantum theory itself \cite{mullerTestingQuantumTheory2023} so that the state and effect spaces can be mapped into $d \times d$ Hermitian matrices, where $d = \sqrt{r}$. One then may test quantitatively whether $P(E|\pi)$, encompassing one's best judgement of the behavior of a system, is compatible with the quantum formalism.
\end{remark}

\subsection{QBist noncontextuality}

While so far we have hewn closely to the standard account of noncontextual ontological models, we now show that if among the scenarios captured by the table $P(E|\pi)$ there is a reference measurement $\{R_i\}$, then we can understand noncontextuality in terms more amenable to the QBist mode of thought. For a QBist, the ultimate classical reference measurement $\{\lambda_i\}$ is one for which $\Phi_\lambda=I$: the fundamental nonclassical coherence relation collapses to the law of total probability, signifying the gambler's indifference to whether the classical reference measurement is performed or not. Let $P(E|\pi) = P(E|\lambda)P(\lambda|\pi)$ be a decomposition of $P(E|\pi)$. We will show that this representation is noncontextual if and only if $\{R_i\}$ forms a reference for $\{\lambda_i\}$. This makes sense conceptually: since the the states $P(E|R)$ and effects $P(R|\lambda)$ are by definition noncontextual, if $\{R_i\}$ forms a reference for $\{\lambda_i\}$ itself, then $P(E|\lambda)$ and $P(\lambda|\pi)$ must inherit this noncontextuality. For QBists, then, this gives a motivation for the idea of a noncontextual ontological model to begin with.

To see this, suppose that the gambler relies upon a reference measurement with outcomes $\{R_i\}$, but that they further suppose there is an ultimate classical reference measurement $\{\lambda_i\}$ (with $c$ outcomes, taken to be a preparatory measurement) whose actual performance the gambler is indifferent to. Specifically, we suppose that $P(R|R)$ is a submatrix of $P(E|\pi)$, the probability table that characterizes the scenarios the gambler is invested in. Let $P(E|\pi) = P(E|\lambda)P(\lambda|\pi)$ be an equirank decomposition $P(E|\pi)$ into response functions and probability vectors. As we have seen, there then exist simplex embedding matrices such that $P(E|\lambda) = \B{E}T_E$ and $P(\lambda|\pi) = T_\Pi \boldpi$, where $P(E|\pi) = \B{E}\boldpi$ is a rank decomposition. We then have $P(R|R)= \B{RS}$, where $\B{R}$ is constructed from the relevant rows of $\B{E}$ and $\B{S}$ is constructed from the relevant colums of $\boldpi$. Since $\{R_i\}$ forms a reference for the scenarios summarized by $P(E|\pi)$, we have $\rank( P(R|R)) = \rank(P(E|\pi))$, so that $P(R|R)=\B{RS}$ is a rank decomposition and $\B{S}\Phi \B{R} = I$. Thus in particular,
\begin{align}
P(R|\lambda) = \B{R} T_E && P(\lambda|R) = T_\Pi \B{S},
\end{align}
so that
\begin{align}
P(E|\lambda) =  \B{E}T_E = \B{ES}\Phi \B{R} T_E = P(E|R)\Phi P(R|\lambda)
\end{align}
and
\begin{align}
P(\lambda|\pi) = T_{\Pi} \boldpi  = T_\Pi \B{S} \Phi \B{R}\boldpi = P(\lambda|R)\Phi P(R|\pi).
\end{align}
We conclude that if $P(E|\pi)=P(E|\lambda)P(\lambda|\pi)$ is an equirank decomposition, furnishing a noncontextual ontological model, then
\begin{align}
 P(E|\lambda) = P(E|R)\Phi P(R|\lambda)  && P(\lambda|\pi) = P(\lambda|R)\Phi P(R|\pi),
\end{align}
which shows that $\{R_i\}$ is a reference for $\{\lambda_i\}$: we can treat its states and effects like any other in the fundamental nonclassical coherence relation. 

There is, however, a catch. Notice that
\begin{align}
``P(\lambda|\lambda)'' = P(\lambda|R)\Phi P(R|\lambda) = T_\pi \B{R} \Phi \B{S} T_E = T_\pi T_E = \Pi_\lambda,
\end{align}
where $\Pi_\lambda$ projects column vectors onto $\col(P(\lambda|\pi))$ and row vectors onto $\row(P(E|\lambda))$. But if $\{\lambda_i\}$ is an ultimate classical reference measurement, we would expect that $P(\lambda|\lambda) = I_c$, where $c$ is the number of outcomes of the classical reference measurement! In a special case, this is possible. Suppose $ c=r=\rank(P(E|\pi))$. Then $P(E|\pi) = P(E|\lambda)P(\lambda|R)$ would be not just an equirank decomposition, but a rank decomposition. In this case, $T_\Pi$ and $T_E$ would be $r\times r$ matrices and since $T_E T_\Pi = I_r$, we must have $T_E = T_\Pi^{-1}$ so that $\Pi_\lambda = T_\Pi T_E = I_r$, as desired. In other words, if $P(E|\pi) = P(E|\lambda)P(\lambda|\pi)$ is in fact a rank decomposition, $P(\lambda|\lambda) = I$, as expected. Otherwise if $c > r$, although $``P(\lambda|\lambda)'' = \Pi_\lambda$, the matrix nevertheless behaves exactly as one would want,
\begin{align}
P(E|\lambda)P(\lambda|\lambda)P(\lambda|\pi) = P(E|\pi),
\end{align}
since $\Pi_\lambda$ projects column vectors into $\col(P(\lambda|\pi))$ and row vectors into $\row(P(E|\lambda))$. Thus in the relevant sense, $\Pi_\lambda$ acts just like the identity. In fact, it couldn't be otherwise. Since $P(R|\lambda)\Pi_\lambda = P(R|\lambda)$ and $\Pi_\lambda P(\lambda|R) = P(\lambda|R)$, we have
\begin{align}
 P(R|\lambda) =  P(R|\lambda)P(\lambda|R)\Phi P(R|\lambda) = P(R|R)\Phi P(R|\lambda)\\
 P(\lambda|R) = P(\lambda|R)\Phi P(R|\lambda)P(\lambda|R) = P(\lambda|R)\Phi P(R|R),
\end{align}
so that $P(R|\lambda) \in \col(P(R|R))$ and $P(\lambda|R) \in \row(P(R|R))$. If $\rank(P(R|R)) = r < c$, there is no way that $P(\lambda|R)\Phi P(R|\lambda) = I_c$ since this would imply that $c \leq \rank(P(\lambda|R)) \leq \rank(P(R|R)) = r$, a contradiction. If we like, however, we could extend $P(\lambda|\lambda)$ to the full identity matrix: clearly it will make no difference since $\Pi_\lambda$ already acts on the identity on our response functions and probability vectors. Moreover, we may adopt $\Phi_\lambda = I$, since
\begin{align}
``P(\lambda|\lambda)''\Phi_\lambda ``P(\lambda|\lambda)'' = \Pi_{\lambda}^2 = ``P(\lambda|\lambda)''.
\end{align}

With this technicality aside, we now show the converse. Let $P(E|\pi)=P(E|\lambda)P(\lambda|\pi)$. If $P(R|R)$ is a submatrix of $P(E|\pi)$ which characterizes a reference measurement, and $\{R_i\}$ forms a reference for $\{\lambda_i\}$, in other words, $P(\lambda|R)\Phi P(R|\lambda) = \Pi_\lambda$, then $P(E|\pi) = P(E|\lambda)P(\lambda|\pi)$ must be an equirank decomposition, constituting a noncontextual ontological model. 

\begin{theorem}
Let $P(E|\pi) = P(E|\lambda)P(\lambda |E)$ and assume that a conditional probability matrix $P(R|R)$ characterizing a reference measurement is a submatrix of $P(E|\pi)$. Thus in particular, $P(R|R)= P(R|\lambda)P(\lambda|R)$. Let $\Phi$ be a Born matrix for $P(R|R)$. Then $P(\lambda|R)\Phi P(R|\lambda) = \Pi_\lambda$ where $\Pi_\lambda$ projects column vectors onto $\col(P(\lambda|\pi))$ and row vectors onto $\row(P(E|\lambda))$ iff $P(E|\pi) = P(E|\lambda)P(\lambda |E)$  is a noncontextual ontological model.
\end{theorem}
\begin{proof}
On the one hand, we have shown that the existence of a noncontextual ontological model is equivalent to $P(E|\pi) = P(E|\lambda)P(\lambda |E)$ being an equirank decomposition. Then $P(E|\lambda) = E T_\B{E}, P(\lambda|\pi) = T_\Pi \boldpi$, and $T_\Pi T_E = \Pi_\lambda$ for simplex embedding matrices $T_E$ and $T_\Pi$ (Corollary \ref{simplex_embedding_matrices}). Here $P(E|\pi) = \B{E}\boldpi$ is a rank decomposition. Since $\{R_i\}$ is a reference measurement, $\rank(P(R|R)) = \rank(P(E|\pi))$. Thus $P(R|R)= \B{RS}$ is a rank decomposition, where $\B{R}$ is constructed from the corresponding rows of $\B{E}$ and $\B{S}$ is constructed from the corresponding columns of $\boldpi$. Since $\Phi$ is a $\{1\}$-inverse of $P(R|R)$, we have $\B{S}\Phi \B{R} = I$ (Lemma \ref{born_id_lemma}). Thus
\begin{align}
\Pi_\lambda = T_\Pi T_E = T_\Pi \B{S} \Phi \B{R} T_E = P(\lambda|R)\Phi P(R|\lambda).
\end{align}
Conversely, let $P(\lambda|R)\Phi P(R|\lambda) =\Pi_{\lambda}$. Then since $\{P(\lambda|R)\} \subseteq \{P(\lambda|\pi)\}$, 
\begin{align}
P(\lambda|R) &= \Pi_\lambda P(\lambda|R)= P(\lambda|R)\Phi P(R|\lambda) P(\lambda|R)= P(\lambda|R)\Phi P(R|R). 
\end{align} 
Thus $\rank(P(\lambda|R)) \leq \rank(P(R|R))$. But from $P(R|R) = P(R|\lambda)P(\lambda|R)$, we have $\rank(P(R|R)) \leq \rank(P(\lambda|R))$. We conclude that $\rank(P(\lambda|R)) = \rank(P(R|R))=\rank(P(E|\pi))$. Similarly, since $\{P(R|\lambda) \}\subseteq \{P(E|\lambda)\}$,
\begin{align}
P(R|\lambda) &= P(R|\lambda)\Pi_\lambda =P(R|\lambda)  P(\lambda|R)\Phi P(R|\lambda) = P(R|R)\Phi P(R|\lambda).
\end{align}
Thus $\rank(P(R|\lambda)) \leq \rank(P(R|R))$. But from $P(R|R) = P(R|\lambda)P(\lambda|R)$, we have $\rank(P(R|R)) \leq \rank(P(\lambda|R))$. We conclude that $\rank(P(R|\lambda))=\rank(P(R|R))=\rank(P(E|\pi))$ and so the decomposition $P(E|\pi) = P(E|\lambda)P(\lambda|\pi)$ is equirank, corresponding to a noncontextual ontological model.
\end{proof}

\section{Computational interlude}

\subsection{Alternating projections}

We now discuss some of the practicalities of computing such matrix factorizations. We have already seen that one may calculate a rank decomposition using the singular value decomposition,
\begin{align}
P(E|\pi) &= U\Sigma V^{\dagger} = \left[\begin{matrix} U_1 & U_2 \end{matrix}\right] \left[ \begin{matrix} \Sigma_r & 0 \\ 0 & 0 \end{matrix} \right]\left[ \begin{matrix} V_1^{\dagger} \\ V_2^{\dagger} \end{matrix}\right] =  U_1(\Sigma_r V_1^{\dagger}) = \B{ES},
\end{align}
where $\Sigma_r$ is a diagonal matrix of the $r$ nonzero singular values. But in practice, even as one refines one's probabilities, one will generically find $P(E|\pi)$ is in fact full rank, perhaps with a few large singular values, and many more smaller singular values. One could set these small singular values to 0 (and which singular values to zero out is a matter of judgement), and from the SVD decomposition obtain a low rank approximation to $P(E|\pi)$. But then it is possible that the low rank approximation will contain numbers that lie outside [0,1]. To deal with this, the literature \cite{markovsky2011low} suggests using an alternating series of optimization. Fixing a choice of $r$, one initializes $\B{E}$ at random, and then solves numerically the following convex program (with e.g. python's \texttt{cvxpy} \cite{diamondCVXPYPythonEmbeddedModeling2016}),
\begin{align}
\min_{\boldpi} ||P(E|\pi)-\B{EP}||^2 \text{ such that } \forall i,j: 0 \le [\B{EP}]_{ij}\le 1,
\end{align}
which gives the closest elementwise nonnegative approximation to $P(E|\pi)$ holding $\B{E}$ fixed and varying $\boldpi$. Then holding $\boldpi$ fixed, one solves
\begin{align}
\min_{\B{E}} ||P(E|\pi)-\B{EP}||^2 \text{ such that } \forall i,j: 0 \le [\B{EP}]_{ij}\le 1,
\end{align}
finding the closest approximation over choices of $\B{E}$. One alternates between these two programs until one finds a reasonable enough rank $r$ approximation of $P(E|\pi)$ which is elementwise nonnegative. 

Each individual step is a convex quadratic program and can be efficiently computed, large dimensionality notwithstanding; the optimal product may be unique even when the factor itself is not. The same method may be extended to find a nonnegative matrix factorization if one further demands that the elements of $\B{E}$ and $\boldpi$ lie in $[0,1]$. 

\subsection{A linear program for simplex embedding}

\cite{selbyLinearProgramTesting2024} offers an alternative, principled way of calculating a simplex embedding, if it exists: here we give it our own spin, showing that it amounts to the nonnegativity of a variant of the Born matrix. Let $P(E|\pi) = \B{EP}$ be a rank decomposition yielding effects $\{(E_i|\}$ and states $\{ |\pi_i) \}$. On the one hand, we can consider the cone generated by the states
\begin{align}
\text{cone}(\boldpi) = \Big\{ |\rho) \ \Big| \ |\rho) = \sum_i \alpha_i |\pi_i), \alpha_i \ge 0 \Big\};
\end{align}
on the other hand, we can consider its dual cone, the set of linear functionals nonnegative on the states,
\begin{align}
\text{cone}^*(\boldpi) = \Big\{ (\eta| \ \Big| (\eta|\boldpi \ge_e 0 \Big\}.
\end{align}
Let $\boldpi^*$ be the matrix whose $n_\pi^*$ rows correspond to representatives of the dual cone's extreme rays. Then $\text{cone}^*(\boldpi) = \text{cone}(\boldpi^*)$. Clearly, then, we can write any row vector nonnegative on every column of $\boldpi$ as some conic combination of the rows of $\boldpi^*$, i.e. $x^\dagger \boldpi^*$ for $x\ge_e 0$.

For a geometric intuition: consider the convex hull of several points. Such a convex set may be described in two alternative ways \cite{ziegler2012lectures}. The first is called the \emph{vertex representation}: it specifies the set by giving its extremal points, those points which cannot be written as convex combinations of any other points in the set. The second is called the \emph{halfspace representation}: here each facet of the convex set corresponds to an inequality $a_i \cdot x + b_i \leq 0$ defining a halfspace, and the convex set is expressed as the intersection of all these halfspaces. For example, suppose we have the following 4 states given in Bloch representation so that $(1|= (1,0,0)$,
\begin{align}
|S_1) = \begin{pmatrix} 1 \\ 1 \\ 1 \end{pmatrix} &&
|S_2)= \begin{pmatrix} 1 \\ 1 \\ -1 \end{pmatrix} &&
|S_3)= \begin{pmatrix} 1 \\ -1 \\ -1 \end{pmatrix} &&
|S_4) = \begin{pmatrix} 1 \\ -1 \\ 1 \end{pmatrix}:
\end{align}
the traceless part of the states form the vertices of a square. These are clearly the extreme points of the square: any point in the interior of the square can be written as a convex combination of these four states. We can consider more generally the cone they generate, whose base is cut by $(1|=(1,0,0)$. The four states then lie on extremal rays of the cone which terminate at $|0)$. At the same time, we could describe the same square as an intersection of halfspaces: all the points $(x,y)$ satisfying $-1 \leq x \leq 1$ and $-1 \leq y \leq 1$. Putting these in standard form, 
\begin{align}
y-1 \leq 0 && -y -1 \leq 0 && z -1 \leq 0 && -z -1 \leq 0.
\end{align}
The square is the intersection of all points satisfying these four inequalities. Introducing an auxiliary dimension, we can write these affine inequalities as linear inequalities, indeed, as a single matrix inequality,
\begin{align}
\begin{pmatrix}-1 & 1 & 0 \\ -1 & -1 & 0 \\ -1 & 0 & 1 \\ -1 & 0 & -1 \end{pmatrix} \begin{pmatrix}1 \\ x \\ y \end{pmatrix} \leq 0,
\end{align}
which, incidentally, also allows one to appreciate the role of the ``traceful'' part of the state in a new light. Multiplying by -1 gives
\begin{align}
\begin{pmatrix}1 & -1 & 0 \\ 1 & 1 & 0 \\ 1 & 0 & -1 \\ 1 & 0 & 1 \end{pmatrix} \begin{pmatrix}1 \\ x \\ y \end{pmatrix} \ge 0,
\end{align}
which shows that any conic combination of the rows of the left hand matrix will be nonnegative on the square. In fact, these rows represent extremal rays of the dual cone $\cone^*(\boldpi)$. Geometrically, they correspond also to a square, albeit rotated 45 degrees: by duality, extremal points have been exchanged with facets. In GPT-language, this dual square correspond to all of the logically possible effects in the theory.

Returning to the main thread, we can go in the opposite direction, considering the cone generated by the effects $\B{E}$ in the rank decomposition $P(E|\pi) = \B{EP}$,
\begin{align}
\text{cone}(\B{E}) = \Big\{ (\eta| \ \Big| \ (\eta| = \sum_i \beta_i (E_i|, \beta_i \ge 0 \Big\},
\end{align}
as well as its dual cone
\begin{align}
\text{cone}^*(\B{E}) = \Big\{ |\sigma) \ \Big| \B{E}|\sigma) \ge_e 0 \Big\},
\end{align}
for which $\text{cone}^*(\B{E}) = \text{cone}(\B{E}^*)$, the $n_E^*$ columns of $\B{E}^*$ being representatives of the extremal rays. Then we can write any column vector nonnegative on every row of $\B{E}$ as some conic combination of the columns of $\B{E}^*$, i.e., $\B{E}^* y$ for $y \ge_e 0$. Suppose now that we can find a matrix $\Phi^*$ of shape $n_E^* \times n_\pi^*$ which is elementwise nonnegative and which satisfies
\begin{align}
\B{E}^* \Phi^* \boldpi^* = I.
\end{align}
Then since by construction $\B{E}\B{E}^* \ge_e 0$ and $\boldpi^*\boldpi \ge_e 0$, we have
\begin{align}
(\B{E}\B{E}^*) \Phi^* (\boldpi^* \boldpi) = P(E|\pi),
\end{align}
where each parenthesized matrix is nonnegative. Consolidating this into the product of two matrices, we therefore obtain a nonnegative matrix factorization, and after possibly renormalizing it, a decomposition $P(E|\pi) = P(E|\lambda)P(\lambda|\pi)$. Indeed, letting e.g., $T_E = \B{E}^*\Phi^*$ and $T_\Pi = \boldpi^*$, we have therefore constructed by Lemma \ref{embedding-lemma} a simplex embedding, and thus a noncontextual ontological model.

Given $\B{E}^*$ and $\boldpi^*$, which may be computed using standard software packages (e.g., \texttt{cdd} \cite{PycddlibPythonWrapper}), it is a simple linear program to find an elementwise nonnegative $\Phi^*$ satisfying $\B{E}^* \Phi^* \boldpi^* = I$ if one exists. If it does not exist, following \cite{selbyLinearProgramTesting2024}, we may relax the problem and seek a $\Phi^*$ satisfying the following program:
\begin{align*}
&\text{minimize } p \\
&\ \ \ \text{over  } p, \Phi^*\\
&\text{subject to } \B{E}^*\Phi^* \boldpi^* = (1-p)I + p|\mu)(1|\\
& \ \ \ \Phi^* \ge_e 0, 0 \leq p \leq 1,
\end{align*}
where $|\mu)(1|$ is the completely depolarizing channel. Given sufficient noise, any $P(E|\pi)$ has a noncontextual ontological model: depolarizing noise will eventually shrink the (traceless part of the) state and effect spaces sufficiently that they can fit simultaneously into the simplex and the dual hypercube of response functions. The minimum amount of depolarizing noise that makes this possible is a measure of the nonclassicality of the scenario.

Finally, we note that $\B{E}^* \Phi^* \boldpi^* = I$ is a kind of cousin of the Born identity $\boldpi\Phi \B{E} = I$ where instead of actual states $\boldpi$, we have representatives of the extremal rays of the cone dual to the effects $\B{E}^*$ and instead of actual effects $\B{E}$, we have representatives of the extremal rays of the cone dual to the states $\boldpi^*$. In the special case that $\B{E}^*=\boldpi$ and $\boldpi^*=\B{E}$, then $\Phi^* = \Phi$ is a nonnegative $\{1\}$-inverse $P(E|\pi)$, if it exists. In the next section, we will pursue this idea further.

\section{Stochastic $\Phi$}

We showed in Lemma  \ref{taken_quasistochastic} that $\Phi$ may always be taken to be quasistochastic. When might it be possible to go further and take $\Phi$ to be stochastic? Now if $P(R|R)$ is invertible, the answer is very simple: the only stochastic matrices with stochastic inverses are permutation matrices \cite{dingTeachingTipWhen2013}. To handle the more general case, we adapt an argument given in \cite{tamGeometricTreatmentGeneralized1981}.

\begin{theorem}
   \label{stochastic-phi-theorem}
Let $P\equiv P(X|Y)$ be a column stochastic matrix. Then there exists a column stochastic $\{1\}$-inverse $\Phi$, satisfying $P\Phi P = P$ iff
\begin{enumerate}
\item $P(X|Y)$ factors into the product of two stochastic matrices $P(X|Y)=P(X|\lambda)P(\lambda|Y)$.
\item The columns of $P(X|\lambda)$ have pairwise disjoint support: $\forall i\neq j: \operatorname{supp}P(X|\lambda_i)\cap \operatorname{supp}P(X|\lambda_j)=\emptyset$.
\item Every column of $P(X|\lambda)$ is also a column of $P(X|Y)$.
\end{enumerate}
\end{theorem}
\begin{proof}
   First, we assume a column stochastic $\Phi$ exists satisfying $P\Phi P = P$. Recall that $\Pi = P \Phi$ is a projector onto the column space of $P$. The $j$th column of $\Pi$ is $\pi_j = P \Phi e_j = P \phi_j$ where $\phi_j$ is the $j$th column of $\Phi$. We assumed $\Phi$ is column stochastic so $\pi_j$ is a convex combination of columns of $P$. Now denote by $P(X|\lambda_i)$ the extreme columns of $P(X|Y)$, that is, the distinct extreme points of the convex hull of the columns of $P(X|Y)$. Now in general, $\Pi P(X|\lambda_i ) = P(X|\lambda_i)$ and so 
\begin{align}
P(X|\lambda_i) = \sum_j P(X_j|\lambda_i) \pi_j = \sum_j P(X_j|\lambda_i) [P \phi_j],
\end{align}
 that is, $P(X|\lambda_i)$ must be a convex combination of the $\pi_j$ which are themselves convex combinations of columns of $P$. But we assumed that $P(X|\lambda_i)$ is extremal. Thus if $P(X_j|\lambda_i)$ is nonzero, $\pi_j = P(X|\lambda_i)$ since $P(X|\lambda_i)$ can't be expressed as a convex combination of any other columns. Now suppose $P(X|\lambda_i)$ and $P(X|\lambda_k)$ share support at index $j$. Then by the above argument, $\pi_j = P(X|\lambda_i)=P(X|\lambda_k)$: they must be equal. We conclude that if two extreme columns are distinct, they must have disjoint supports, which gives us $(2)$. We already constructed $P(X|\lambda)$ from the extreme columns of $P(X|Y)$, giving us $(3)$. And since every column of $P(X|Y)$ is a convex combination of its extreme columns, there must be some column stochastic matrix $P(\lambda|Y)$ such that $P(X|Y) = P(X|\lambda)P(\lambda|Y)$, which gives us $(1)$.

Conversely, assume $(1)$, $(2)$, and $(3)$. By $(1)$, $P(X|Y)= P(X|\lambda)P(\lambda|Y)$. By $(3)$, for each $\lambda_i$ there is some column index $f(i)$ such that $P(X|\lambda_i)=P(X|Y_{f(i)})$. Let $P(Y|\lambda)$ satisfy $P(Y_i|\lambda_j):=\delta_{i,f(j)}$ so that $P(X|\lambda)= P(X|Y)P(Y|\lambda)$. Meanwhile, define $P(\lambda|X)$ by first setting
\begin{align}
P(\lambda_i|X_j)=
\begin{cases}
1, & j \in \operatorname{supp}P(X|\lambda_i),\\
0, & \text{otherwise},
\end{cases}
\end{align}
and then, if there is some index $j$ which doesn't lie in the support of any column $P(X|\lambda_i)$, put a single $1$ anywhere in that column. In the end, because the supports are disjoint, each column of $P(\lambda|X)$ will have exactly one nonzero entry, and $P(\lambda|X)P(X|\lambda) = I$. To see this, first assume $i\neq k$. Then 
\begin{align}
\sum_j P(\lambda_i|X_j) P(X_j |\lambda_k) = 0,
\end{align}
since by construction $P(\lambda_i|X)$ and $P(X|\lambda_k)$ have disjoint supports for $i\neq k$. If $i=k$, 
\begin{align}
\sum_j P(\lambda_i|X_j) P(X_j |\lambda_i) = \sum_{j \in \operatorname{supp} P(X|\lambda_i)} P(X_j|\lambda_i) = 1,
\end{align}
since $P(X|\lambda)$ is column stochastic, and we are summing just the nonzero entries in a column. Finally, let $\Phi = P(Y|\lambda)P(\lambda|X)$. Clearly, $\Phi$ is column stochastic and
\begin{align}
P \Phi P &= P(X|Y)\big\{P(Y|\lambda)P(\lambda|X)\big\} P(X|Y)\\
&= \big\{ P(X|Y)P(Y|\lambda)\big\} P(\lambda|X)\big\{P(X|\lambda)P(\lambda|Y)\big\}\\
&= P(X|\lambda) P(\lambda|Y) = P.
\end{align}
\end{proof}
\begin{corollary}
   In the above construction, $P(X|Y) = P(X|\lambda)P(\lambda|Y)$ is a rank decomposition.
\end{corollary}
\begin{proof}
    By $(2)$ the columns of $P(X|\lambda)$ have pairwise disjoint supports, and so they must be linearly independent. Thus $P(X|\lambda) \in \mathbb{R}^{n\times r}_r$ and $\rank(P) = \rank\big(P(X|\lambda)P(\lambda|Y)\big) \leq \rank\big(P(X|\lambda)\big) = r$. But by $(3)$ every column of $P(X|\lambda)$ is also a column of $P$, and so $\rank\big(P(X|\lambda)\big) \leq \rank(P)$, from which we conclude $\rank(P)=r$. Finally, since $P(\lambda|Y) \in \mathbb{R}^{r \times m}$, $\rank\big( P(\lambda|Y)\big) \leq \min(r, n) \leq r$. But $\rank(P) \leq \rank\big(P(\lambda|Y)\big) \leq r = \rank(P)$ so that $\rank\big(P(\lambda|Y)\big)=r$ as well.
\end{proof}

\begin{remark}
Notice that just as $P(X|Y) = P(X|\lambda)P(\lambda |Y)$, $\Phi = P(Y|\lambda)P(\lambda|X)=P(Y|X)$, which we could call a stochastic retrodiction map. Moreover, the key step is that $P(\lambda|X)P(X|\lambda)= I$. In other words, the events $\{X_i\}$ allow one to completely retrodict the $\lambda_i$'s. 
\end{remark}

\begin{corollary}
In the above construction, $P(\lambda|X)P(X|\lambda) =I$: so $P(X|\lambda)$ has a stochastic left inverse. Similarly, $P(Y|\lambda)$ is a stochastic right inverse of $P(\lambda|Y)$.
\end{corollary}
\begin{proof}
We defined $P(Y_i|\lambda_j) = \delta_{i, f(j)}$ so that $P(X|\lambda) = P(X|Y)P(Y|\lambda)$. But that means $P(X|\lambda) = P(X|\lambda)P(\lambda|Y)P(Y|\lambda)$. Acting from the left with $P(\lambda|X)$, we have $I=P(\lambda|Y)P(Y|\lambda)$ as desired.
\end{proof}

\begin{corollary}
Let $P(R|R)$ be the conditional probability matrix that characterizes the reference measurement. If $P(R|R)$ has a stochastic Born matrix, then the corresponding GPT is classical.
\end{corollary}
\begin{proof}
From Theorem \ref{stochastic-phi-theorem} and its corollaries, we know that $P(R|R)$ has a stochastic rank decomposition $P(R|R) = P(R|\lambda)P(\lambda|R)$. Then $P(\lambda|\pi) = P(\lambda|R)\Phi P(R|\pi)$ is a probability distribution and $P(E|\lambda) = P(E|R)\Phi P(R|\lambda)$ is a response function which satisfy
\begin{align}
P(E|\lambda)P(\lambda|\pi) &=  P(E|R)\Phi P(R|\lambda) P(\lambda|R)\Phi P(R|\pi) \\
&= P(E|R)\Phi P(R|R)\Phi P(R|\pi) \\
&= P(E|R)\Phi P(R|\pi) \\
&= P(E|\pi).
\end{align}

Now suppose that the effects $\{E_i\}$ and the states $\{\pi_i\}$ are informationally complete for the GPT. Then $\rank(P(E|R)) = \rank(P(R|\pi)) = \rank(P(R|R)) = r$. Now from $P(E|\lambda) = P(E|R)\Phi P(R|\lambda)$, we have $\rank(P(E|\lambda)) \leq \rank(P(E|R))$. But from $P(E|R) = P(E|\lambda)P(\lambda|R)$, we have $\rank(P(E|R))\leq \rank(P(E|\lambda))$. Thus $\rank(P(E|\lambda)) = r$. Similarly, comparing $P(\lambda|\pi) = P(\lambda|R)\Phi P(R|\pi)$ and $P(R|\pi) = P(R|\lambda)P(\lambda|\pi)$, we have $\rank(P(\lambda|\pi)) = r$. Since $P(R|R) = P(R|\lambda)P(\lambda|R)$ is a rank decomposition, there must be $r$ $\lambda$'s. Thus $P(E|\lambda)$ has full column rank and $P(\lambda|\pi)$ has full row rank; from $P(E|\pi) = P(E|\lambda)P(\lambda|\pi)$ we may conclude that $\rank(P(E|\pi)) = r$. The factorization is therefore equirank. In particular, this will hold if we consider \emph{all} the states and effects of the GPT: the GPT is therefore entirely classical.
\end{proof}

\subsection{An example} 

For concreteness, we give an example of a situation when a stochastic $\Phi$ matrix is possible. Let 
\begin{align}
P(X|Y)=P(X|\lambda)P(\lambda|Y) = 
\begin{pmatrix}
\frac23 & 0\\
\frac13 & 0\\
0 & \frac14\\
0 & \frac34
\end{pmatrix}\begin{pmatrix}
1 & 0 & \frac12 & \frac34\\
0 & 1 & \frac12 & \frac14
\end{pmatrix}=
\begin{pmatrix}
\frac23 & 0 & \frac13 & \frac12\\
\frac13 & 0 & \frac16 & \frac14\\
0 & \frac14 & \frac18 & \frac18\\
0 & \frac34 & \frac38 & \frac38 
\end{pmatrix}.
\end{align}
Then
\begin{align}
\Phi = P(Y|\lambda)P(\lambda|X) = 
\begin{pmatrix}
1 & 0\\
0 & 1\\
0 & 0\\
0 & 0
\end{pmatrix}\begin{pmatrix}
1 & 1 & 0 & 0\\
0 & 0 & 1 & 1
\end{pmatrix}=
 \begin{pmatrix}
1 & 1 & 0 & 0\\
0 & 0 & 1 & 1\\
0 & 0 & 0 & 0\\
0 & 0 & 0 & 0
\end{pmatrix}.
\end{align}
Slightly more generally, suppose that 
\begin{align}
P(X|\lambda) = \begin{pmatrix} p & 0 \\ (1-p) & 0 \\ 0 & q \\ 0 & 1-q \end{pmatrix} && P(\lambda) = \begin{pmatrix} a \\ b \end{pmatrix} && P(\lambda|X) = \begin{pmatrix}
1 & 1 & 0 & 0\\
0 & 0 & 1 & 1
\end{pmatrix},
\end{align}
which yields 
\begin{align}
P(X) = P(X|\lambda)P(\lambda) = \begin{pmatrix} a p \\ a (1-p) \\ bq \\ b(1-q) \end{pmatrix}.
\end{align}
Clearly, $P(\lambda) = P(\lambda|X)P(X)$. Thinking of $P(\lambda)$ as a state and $P(X|\lambda)$ as a set of effects, we see that we can tell a very classical story. First $\lambda_1$ or $\lambda_2$ is chosen with some probability. Conditional on $\lambda_1$, a coin is flipped with bias $p$, and if it lands heads, the measurement yields $X_1$ and if it lands tails, the measurement yields $X_2$. Similarly, conditional on $\lambda_2$, a coin is flipped with bias $q$, and this gives either $X_3$ or $X_4$. If you like, first the ultimate classical reference measurement is performed, reading off the property of the system (which $\lambda$ is the case), and then conditional on that value, one of several outcomes associated to that property is chosen at random. Meanwhile, the $n$ distributions over the $r$ hidden variables that give rise to $P$, that is, $P(\lambda|Y_i)$, are constrained: all the states of complete certainty about which $\lambda$ is the case must be among them. 

\section{Negativity in quantum theory}

For completeness we reproduce here the nice proof of \cite{ferrieFrameRepresentationsQuantum2008} that negativity is necessary in any frame representation of quantum mechanics, and so no noncontextual ontological model is possible for full quantum theory. See also \cite{ferrieFramedHilbertSpace2009, ferrieNecessityNegativityQuantum2010, ferrieQuasiprobabilityRepresentationsQuantum2011} for a general discussion of the relationship between quantum theory, frames and their duals, negativity, and contextuality.

\begin{theorem}[Theorem 2 of \cite{ferrieFrameRepresentationsQuantum2008}]
There does not exist a dual frame of positive semidefinite operators for a frame of positive semidefinite operators for $d>1$.
\end{theorem}
\begin{proof}
   Let $\{E_i\}$ be a PSD operator frame, and $\{\tilde{E}_i\}$ its dual. Consider the map 
   \begin{align}
   M(\rho) = \sum_i \tilde{E}_i \tr(E_i\rho) = \rho.
   \end{align}
Forming its Choi state,
\begin{align}
C_M &= \frac{1}{d}\sum_{ij} M(|i\rangle\langle j|) \otimes |i\rangle\langle j|\\
&= \frac{1}{d}\sum_{ijk} \tilde{E}_k \tr(E_k |i\rangle\langle j|) \otimes |i\rangle\langle j|\\
&=\frac{1}{d}\sum_k \tilde{E}_k \otimes \sum_{ij} \langle j|E_k|i\rangle |i\rangle\langle j|\\
&=\frac{1}{d} \sum_k \tilde{E}_k \otimes E_k^T,
\end{align}
we see that it is a sum of tensor products of positive semidefinite operators: in other words, $C_M$ is a separable state. But $M$ is supposed to be the identity map, whose Choi state is maximally entangled. Thus the dual frame of a positive semidefinite operator frame cannot itself be composed entirely of positive semidefinite operators.
\end{proof}

\begin{corollary}
   Since the dual frame contains operators with negative eigenvalues, $W(E|\rho) = \B{E}|\rho)$ will in general take negative values. Thus in quantum theory, there must be negativity in any frame representation, and so no noncontextual ontological model is possible \cite{schmidStructureTheoremGeneralizednoncontextual2024}.
\end{corollary}

\section{QBist Bell Inequalities}

As we noted in the previous chapter, the authors of \cite{debrotaSymmetricInformationallyComplete2020} explored $\lVert I-\Phi \rVert$ with respect to any unitarily invariant norm as a measure of nonclassicality. The intuition is that classically it is always possible to pick a reference measurement such that $\Phi=I$: but this is precisely \emph{forbidden} by quantum theory.  This is suggestive of the idea that quantum systems do not have underlying properties which are read off by  measurements, even reference measurements, but that, as QBism holds, the outcomes of measurement are better regarded as the possible consequences for an individual agent gambling upon a nature which is being continually cocreated. But for a single system, this is a heuristic argument and not at all a proof: after all, the outcomes of any single system quantum experiment can be spoofed by a laptop computer which is generally regarded to operate on definite properties by perfectly definite rules. From this point of view, there is nothing to exclude the hypothesis that a single quantum system has some definite, yet unknown properties which are however disturbed by measurement. The situation is different, however, when we consider multipartite scenarios. There the violation of Bell's inequalities force us into a quandary: either quantum systems have underlying properties that condition the results of measurement, but these properties may be disturbed \emph{without regard for locality};  or else the outcomes of measurements are created in the act of measurement for the individual observer. (We exclude here the evasion of the quandary by the hypothesis of a multiverse.)
For QBists, the point is that when one formulates a Bell inequality, one in fact appeals to the law of total probability in defiance of the fact that performing a reference measurement makes a difference. We now show that violations of Bell inequalities, and noncontextuality inequalities more generally, may be understood as quantifying the cost of abandoning the fundamental nonclassical coherence condition. 

\subsection{Local realism?}

The usual way of introducing Bell inequalities is to consider a multipartite scenario, where $n$ distant parties share parts of a system, and each party agrees to choose from a menu of potential measurements on the part local to them. In the simplest case, suppose Alice has a choice of measurements $\{A_i\}$ each of which has outcomes $\{a_i\}$, and Bob has a choice of measurements $\{B_i\}$ each of which has outcomes $\{b_i\}$. We can then consider the joint probability distribution $P(A_k=a_i, B_l=b_j)$. In particular, following Bell, we may consider the subset of all joint probability distributions which can be expressed
\begin{align}
   \label{hidden-variables-Bell}
P(A_k=a_i, B_l=b_j) = \sum_m P(A_k=a_i|\lambda_m)P(B_l=b_j| \lambda_m)P(\lambda_m),	
\end{align}
where $P(\lambda_m)$ is a distribution over some underlying ``hidden variables'' $\{\lambda_m\}$, while $P(A_k=a_i|\lambda_m)$ and $ P(B_l=b_j| \lambda_m)$ are response functions which depend only upon Alice (or Bob's, respectively) local choice of measurement---and the hidden variables. The point is that, for Bell, any correlations in Alice and Bob's measurement outcomes ought to be explained by correlations between the hidden variables which their systems are supposed to have agreed upon before they were separated. The set of all such ``classical'' joint probability distributions for a given multipartite scenario forms a so-called \emph{Bell polytope} $\mathcal{L}$ \cite{wernerAllmultipartiteBellcorrelationInequalities2001, pironioAllClauserHorne2014}. 

For us, we take the point of view of a gambler who expects to act upon Alice and upon Bob, and who believes them when they promise that they will try to communicate the results of their measurements. But for the gambler, there is no measurement until they themselves act upon Alice and upon Bob \cite{fuchsQBismPolishingPoints2025}. Introducing reference measurements $\{R^{A}_i\}$ and $\{R^{B}_i\}$ for each of them, the gambler ought in general to appeal to the coherence condition,
\begin{align}
   \label{bell-born}
P(A_k = a_i, B_l = b_j) = \Big\{P(A_k=a_i|R^A) \otimes P(B_l=b_j|R^B) \Big\}\Big\{\Phi^A \otimes \Phi^B\Big\}P(R^A, R^B|\rho).
\end{align}
If the gambler is indifferent to whether the reference measurements are performed or not, then this reduces to
\begin{align}
   \label{visible-variables-Bell}
P(A_k = a_i, B_l = b_j) &= \Big\{P(A_k=a_i|R^A) \otimes P(B_l=b_j|R^B) \Big\}P(R^A, R^B|\rho)\\
&= \sum_{ru}P(A_k=a_i|R^A_r)  P(B_l=b_j|R^B_u) P(R_r^A, R_u^B|\rho),
\end{align}
which indeed has Bell form, where the outcomes of the reference measurements play the role of the so-called hidden variables\footnote{This is perfectly sensible if we analogize to classical physics. Classically, the ultimate reference device is simply reading off the positions and momenta of all the particles: these are the ``hidden variables'' upon which any other observable supervenes---and in principle they need not be hidden at all! And this is as it should be: what is the point of hypothesizing some underlying properties which could never in principle be measured? But to consider something measurable according to e.g., quantum theory is quite restrictive: distributions $P(R_r^A, R^B_s|\rho)$ are not as general as $P(\lambda_m)$. The former are constrained e.g., to correspond to quantum states, which rules out large swathes of the probability simplex, whereas the latter are otherwise unconstrained. Thus joint probability distributions in the form of Eq.~\ref{visible-variables-Bell} are a subset of those defined by Eq.~\ref{hidden-variables-Bell}. Luckily, this is all we will need.}. In this way, the assumption of ``Bell locality'' amounts for us to the assumption that the gambler can introduce a reference measurement for Alice and for Bob, and that they are indifferent to whether these reference measurements are performed or not. The key point is, as always, that the gambler may very well \emph{not} be indifferent to whether the reference measurements are performed. Turning it around, and as we will see, we can bound $\lVert I - \Phi \rVert$ in terms of the magnitude of a Bell inequality violation.

\subsection{CHSH}

For sake of intuition, consider for example the CHSH scenario. This is a bipartite scenario in which the gambler can ask Alice and Bob for the outcomes of two possible measurements they made. The outcomes are valued in $\{\pm 1\}$. Supposing that $P(A_k=a_i, B_l=b_j)$ takes Bell form, one may derive the famous inequality
\begin{align}
-2 \leq \langle A_1 B_1\rangle + \langle A_1 B_2\rangle + \langle A_2B_1\rangle - \langle A_2 B_2\rangle  \leq 2.
\end{align}
Let us consider $A_1, A_2$ first. Suppose they take opposite values. In that case, $(-1) + (+1) = (+1) + (-1) = 0$ so that $A_1 + A_2 = 0$, and since either we have $(+1) - (-1)=+2$ or $(-1) - (+1)=-2$, we have $A_1 - A_2 = \pm 2$. On the other hand, suppose they take the same values. In that case, $(+1) - (+1) = (-1) - (-1)=0$ so that $A_1 - A_2 = 0$, and since either we have $(+1) + (+1) = +2$, or $(-1) + (-1) = -2$, we conclude $A_1 + A_2 = \pm 2$. Let us now consider the quantity
\begin{align}
C &= (A_1 + A_2)B_1 + (A_1-A_2)B_2.
\end{align}
If $A_1, A_2$ take opposite values, then $A_1+A_2=0$ and $A_1-A_2=\pm 2$ so that $C = 0 + (\pm 2)(\pm1)=\pm 2$. On the other hand, if $A_1, A_2$ take the same values, then $A_1-A_2=0$ and $A_1+A_2=\pm 2$ so that $C = (\pm 2)(\pm 1) + 0=\pm 2$. Either way, $C=\pm 2$. Assuming a joint probability distribution over the outcomes of the four measurements, we may consider the expectation value of $C$:
\begin{align}
\langle C\rangle &= \langle 	 (A_1 + A_2)B_1 + (A_1-A_2)B_2\rangle\\
&=\langle A_1B_1\rangle + \langle A_2 B_1\rangle + \langle A_1 B_2\rangle -\langle A_2 B_2\rangle.
\end{align}
Since $C=\pm2$, we must have $-2\leq \langle C\rangle \leq 2$, and so
\begin{align}
-2	\leq \langle A_1B_1\rangle + \langle A_2 B_1\rangle + \langle A_1 B_2\rangle -\langle A_2 B_2\rangle \leq 2,
\end{align}
as desired. We derived the inequality solely by assuming that $A_1, A_2, B_1, B_2$ all take preexisting values in $\pm 1$, and it holds \emph{whatever} joint probability distribution we assign to the outcomes of the four measurements.

Let us take another point of view, however. Since $\langle A_k B_l\rangle = \sum_{ij} a_i b_j P(A_k=a_i, B_l=b_j)$,  letting
\begin{align}
s=\begin{pmatrix} 1 & 1 \\ 1 & -1 \end{pmatrix}	&& c_{ijkl}= a_i b_j s_{kl},
\end{align}
and denoting by $\B{p}$ and  $\B{c}$ the tensors $P(A_k=a_i, B_l=b_j)$ and $c_{ijkl}$ flattened into vectors, we may rewrite the CHSH inequality as
\begin{align}
-2 \leq \B{c}^\dagger \B{p} \leq 2.
\end{align}
Here $\B{p}$ is often called a \emph{behavior}. This is quite general: a Bell inequality always takes the form
\begin{align}
\B{c}^\dagger \B{p}\leq \beta,	
\end{align}
where $\B{c}$ represents a particular observable and $\beta$ is a classical bound derived by supposing that $\B{p}$ takes Bell form. Geometrically speaking, each Bell inequality $(\B{c}, \beta)$ defines a halfspace: their collective intersection is the Bell polytope $\mathcal{L}$, the set of all behaviors which can be written in Bell form. In fact, the CHSH inequality corresponds to a facet of this polytope.

Of course, quantum mechanics generally violates Bell inequalities. In the CHSH scenario, let  $A_1=Z, A_2=X$, and $B_1=(Z+X)/\sqrt{2}, B_2=(Z-X)/\sqrt{2}$, where $Z,X$ are Pauli observables and let the initial state be $\rho=|\psi\rangle\langle \psi|$ where $|\psi\rangle=(|00\rangle+|11\rangle)/\sqrt{2}$. Then by the Born rule
\begin{align}
	P(A_k=a_i, B_l=b_j)=\tr\Big((\Pi_{a_i|A_k}\otimes \Pi_{b_j|B_l})\rho\Big), 
\end{align}
where $\Pi_{a_i|A_k}$ is the rank-1 projector onto the $a_i$ outcome of observable $A_k$. Famously, this choice of state and observables leads to $|\B{c}^\dagger \B{p}|=2\sqrt{2}$, the maximal violation compatible with quantum mechanics.

\subsection{Bounding $\lVert I - \Phi \rVert$}

In order to show how a Bell inequality violation can bound $\lVert I - \Phi \rVert$, we first rewrite Eq. \ref{bell-born} as
\begin{align}
\label{composite_born}
	&P(A_k=a_i, B_l=b_j)\nonumber\\
   &= \sum_{rs,uv} P(A_k=a_i|R^A_r)\Phi^A_{rs}P(B_l=b_j| R_u^B)\Phi^B_{uv}P(R_s^A, R_v^B|\rho)\\
	&=\sum_{rs,uv} \Big[P(A_k=a_i|R^A_r)P(B_l=b_j|R_u^B)P(R^A_s, R^B_v|\rho) \Big]\Big[\Phi^A_{rs}\Phi^B_{uv}\Big]\\
	&=\sum_{rs,uv}  T_{ijkl,rsuv}\phi_{rsuv},
\end{align}
so that denoting by $\B{p}$ the tensor $P(A_k=a_i, B_l=b_j)$ flattened into a vector, reshaping the tensor $T$ into a rectangular matrix $\B{T}$, and finally flattening $\Phi^A \otimes \Phi^B$ into a vector $\phi=|\Phi^A \otimes \Phi^B)$, we can achieve a yet more compact expression $\B{p} = \B{T}\phi$. In other words, the reference probabilities which specify the Bell scenario are packaged into the matrix $\B{T}$, separate from the choice of Born matrices. Consequently, we can write any Bell inequality as
\begin{align}
\B{c}^\dagger \B{T} \phi \leq \beta.
\end{align}
Now let $\phi_Q=|\Phi^A \otimes \Phi^B)$ and $\phi_C=|I\otimes I)$, where the soft brackets denote the vectorization of the matrices. On the one hand, $\B{p}_Q=\B{T}\phi_Q$ gives us our nonclassical distribution, while $\B{p}_C=\B{T}\phi_C$ lies in the Bell polytope $\mathcal{L}$. We now recall that the operator norm associated with a vector norm $\lVert \cdot \rVert$ is 
\begin{align}
\lVert \B{A} \rVert_{\text{op}} = \sup_{\lVert \B{v}\rVert=1}	\lVert \B{Av} \rVert,
\end{align}
so that immediately we have $\lVert \B{Av}\rVert \leq \lVert \B{A}\rVert_\text{op}\lVert \B{v} \rVert$. Putting this together, we have
\begin{align}
\lVert \B{p}_Q	-\B{p}_C\rVert=\lVert \B{T} \phi_Q	-\B{T} \phi_C\rVert &= \lVert \B{T}(\phi_Q-\phi_C)\rVert \leq\lVert \B{T}\rVert_\text{op}\lVert \phi_Q-\phi_C\rVert,
\end{align}
or
\begin{align}
	\lVert \phi_Q-\phi_C \rVert	\ge \frac{\lVert \B{p}_Q-\B{p}_C\rVert}{\lVert \B{T}\rVert_\text{op}}.
\end{align}
But $\lVert \B{p}_Q-\B{p}_C\rVert \ge \text{dist}(\B{p}_Q, \mathcal{L})$ since $\B{p}_C$ lies in $\mathcal{L}$. We then have the following theorem.

\begin{theorem}
Let $\B{p}$ be a behavior which violates  $\B{c}^\dagger \B{p} \leq \beta$. Then $\B{c}^\dagger \B{p}=\beta + \Delta$ for $\Delta \ge 0$, and 
\begin{align}
\label{polytope_inequality}
\text{dist}(\B{p}, \mathcal{L})\ge \frac{\Delta}{\lVert \B{c}\rVert_*}	,
\end{align}
where $\lVert \cdot \rVert_*$ is the dual norm\footnote{In particular, for a $p$-norm, $\lVert \B{x}\rVert_p = \left(\sum_i x_i^p\right)^{1/p}$, the dual norm is the H\"older dual $\lVert \B{x} \rVert_q$ satisfying $1/p + 1/q=1$ \cite{boydConvexOptimization2004}. Notice that the Euclidean norm is self-dual in this sense.} of $\lVert \cdot \rVert$.
\end{theorem}
\begin{proof}
We  first reprove a standard result. Let $H=\{\B{x}\  |\ \langle \B{y},\B{x}\rangle =\beta \}$ be a hyperplane. The distance of a point $\B{x}_0$ to $H$ (with respect to a choice of norm $\lVert \cdot \rVert$) is
\begin{align}
	\text{dist}(\B{x}_0, H)=\frac{|\Delta|}{\lVert \B{y}\rVert_*},
\end{align}
where $\lVert \cdot \rVert_*$ is the dual norm and $\Delta = \langle \B{y}, \B{x}_0\rangle-\beta$. To see this, we observe that for a norm $\lVert \cdot \rVert$, its \emph{dual norm} \cite{boydConvexOptimization2004} is defined as
\begin{align}
\lVert \B{x} \rVert_* = \sup_{\lVert \B{y}\rVert \leq 1} \langle \B{x}, \B{y}\rangle.
\end{align}
Immediately, we have the tight inequality
\begin{align}
\langle \B{x},\B{y}\rangle =\lVert \B{y}\rVert \left\langle \B{x}, \frac{\B{y}}{\lVert \B{y}\rVert}\right\rangle \leq \lVert \B{y}\rVert \lVert \B{x}\rVert_*.
\end{align}
Let us now  consider the distance from a point $\B{x}_0\in \mathbb{R}^n$ to a hyperplane $H=\{\B{x} \ |\ \langle \B{y},\B{x}\rangle =\beta \}$. In other words, we seek the point in $H$ closest to $\B{x}_0$,
\begin{align}
 \text{dist}(\B{x}_0, H)=\min_{\B{x}\in H}\lVert \B{x}_0 - \B{x}\rVert.
\end{align}
 By definition, for $\B{x}\in H$, $\langle\B{y},\B{x}\rangle=\beta$; by the same token, for $\B{x}_0\notin H$, $\langle \B{y},\B{x}_0\rangle=\beta + \Delta$. Subtracting these two expressions yields 
\begin{align}
\langle \B{y},\B{d}\rangle =\Delta && \B{d}=\B{x}_0-\B{x},
\end{align}
so that we can reformulate our problem as $\min_{\langle \B{y},\B{d}\rangle =\Delta} \lVert \B{d} \rVert$. But $|\Delta|=|\langle \B{y},\B{d}\rangle| \leq \lVert \B{y}\rVert_* \lVert \B{d}\rVert$, or
\begin{align}
\lVert \B{d}\rVert 	\ge \frac{|\Delta|}{\lVert \B{y}\rVert_*}.
\end{align}
Since the inequality is tight, we can achieve the minimum, and thus
\begin{align}
 \text{dist}(\B{x}_0, H)=\frac{|\Delta|}{\lVert \B{y}\rVert_*}.
 \end{align}
Finally, since every such inequality defines a halfspace $\{\B{x}|\B{c}^\dagger \B{x}\leq \beta\}$ bounded by the corresponding hyperplane $\{\B{x}|\B{c}^\dagger \B{x}=\beta\}$, and since the polytope $\mathcal{L}$ lies entirely in the halfspace, the distance to the hyperplane gives a lower bound on the distance to the whole polytope. In other
\begin{align}
\text{dist}(\B{p}, \mathcal{L})\ge \text{dist}(\B{p}, H)=\frac{\Delta}{\lVert \B{c}\rVert_*}.
\end{align}
\end{proof}
\noindent With this result in hand, we conclude that
\begin{align}
	\lVert \phi_Q-\phi_C \rVert	\ge \frac{\lVert \B{p}_Q-\B{p}_C\rVert}{\lVert \B{T}\rVert_{\text{op}}} \ge \frac{\text{dist}(\B{p}_Q, \mathcal{L})}{\lVert \B{T}\rVert_{\text{op}}} \ge \frac{\Delta}{\lVert \B{T}\rVert_{\text{op}}\lVert \B{c}\rVert^*} = \frac{\B{c}^\dagger \B{p} - \beta }{\lVert \B{T}\rVert_{\text{op}}\lVert \B{c}\rVert_*},
\end{align}
where $\Delta$ is the Bell inequality violation. Specializing to the Euclidean norm $\lVert \cdot \rVert_2$, we have that the associated operator norm is $\lVert \B{T} \rVert_\text{op}= \sigma_{\max}(\B{T})$, the largest singular value \cite{hornMatrixAnalysis1985}; that the Euclidean norm is self-dual; and finally, that $\lVert \text{vec}(\B{A})\rVert_2= \lVert \B{A}\rVert_F$, the Frobenius norm $\sqrt{\tr(\B{A}^\dagger\B{A})}$. Letting $\Phi = \Phi^A \otimes \Phi^B$, we arrive at the following theorem:

\begin{theorem}
   Consider a two party Bell scenario. Introduce a reference measurement for each party and consider the joint reference measurement with Born matrix $\Phi = \Phi_A \otimes \Phi_B$. Let $\B{c}$ be the Bell observable, $\Delta$ be the Bell inequality violation, and $\sigma_{\max}$ denote the largest singular value. Finally, let $\B{T}$ be the matrix with entries
   \begin{align}
   T_{ijkl, rsuv}=P(A_k=a_i|R^A_r)P(B_l=b_j|R_u^B)P(R^A_s, R^B_v|\rho),
   \end{align}
where $P(A_k=a_i|R^A_r)$ and $P(B_l=b_j|R_u^B)$ are the response functions characterizing Alice and Bob's effects and $P(R^A_s, R^B_v|\rho)$ is the probability distribution characterizing the initial state. Then
\begin{align}
\lVert I - \Phi\rVert_F \ge 	\frac{\Delta}{\sigma_{\max}(\B{T})\lVert \B{c}\rVert_2}.
\end{align}
The generalization to $m$ parties is straightforward.
\end{theorem}

Returning to the CHSH example, assigning a qubit SIC reference measurement $\{R_i\}$ to Alice and Bob each, we can characterize the quantum state of their two qubits with the joint probability distribution
\begin{align}
P(R_i, R_j|\rho)=\tr\Big((R_i\otimes R_j)\rho\Big),	
\end{align}
and characterize their measurements with
\begin{align}
P(A_k=a_i|R_u)=\tr(\Pi_{a_i|A_k}\sigma_u) && P(B_l=b_j|R_v)=\tr(\Pi_{b_j|B_l}\sigma_v),
\end{align}
where  $\sigma_u = R_u/\tr(R_u)$ are reference states. Maximal violation of the CHSH inequality gives $\Delta = 2\sqrt{2}-2$. Meanwhile, for two qubit SIC reference devices, $\sigma_{\max}(\B{T})=\sqrt{4/3}$, yielding $\lVert I - \Phi\rVert_F \ge 0.1794$. Now in fact, $\lVert I - \Phi\rVert_F \approx 24.4949$, so the bound is not particularly informative as to its exact value. The point is rather that \emph{it cannot be 0}. Indeed, we may take 
\begin{align}
	\mathcal{B}=\frac{\Delta}{\sigma_{\max}(\B{T})\lVert \B{c}\rVert_2}
\end{align}
itself to be a measure of the nonclassicality of the scenario as captured by the reference measurement. This result opens up a number of interesting further directions for research. In particular, one may ask: optimizing over number of parties, reference measurements, initial states, choices of measurements, and Bell inequalities, which scenario gives the largest bound? Is it possible to find a bound which is equal to the actual value of $\lVert  I-\Phi \rVert$? Fixing a reference device (e.g., a SIC), which Bell scenario gives the largest bound? Conversely, fixing a Bell scenario, which reference measurement gives the largest bound? 

\subsection{Noncontextuality inequalities}

The wonderful review article \cite{brunnerBellNonlocality2014} gives an explicit algorithm for constructing a Bell inequality which witnesses the nonclassicality of a behavior $\B{p}$ or else affirms its classicality. The first observation is that any uncertainty in the response functions (e.g., $P(A_j=a_i|\lambda_k)$) can be shunted instead into the hidden variables $\lambda$: thus it suffices to consider models where the local response functions take values in $\{0,1\}$, that is to say, it suffices to consider deterministic local hidden variable models. In such a model, the hidden variables simply assign outcomes to measurements in a particular run. For example,  let there be two parties, Alice and Bob, and let  $\lambda = (\lambda_{A_2}, \dots, \lambda_{A_m}; \lambda_{B_2}, \dots, \lambda_{B_n})$ be an assignment of outcomes to each of $m$ measurements for Alice and each of $n$ measurements for Bob in a given run. To each possible $\lambda$ corresponds a deterministic behavior
\begin{align}
P_\lambda (A_k=a_i, B_l=b_j) = \delta_{a_i, \lambda_{A_k}}  \delta_{b_j, \lambda_{B_l}}.
\end{align}
A behavior $\B{p}$ then has Bell form iff it can be written as a convex combination of these deterministic behaviors,
\begin{align}
\B{p} = \sum_\lambda \B{p}_\lambda p(\lambda)  && p(\lambda)\ge0, \ \  \sum_\lambda p(\lambda)=1	.
\end{align}
Given a behavior $\B{p}$, consider the following linear (and so efficiently solvable) program 
\begin{align}
\max_{\B{c}, \beta}	\ S=\B{c}^\dagger \B{p} - \beta \ \text{ such that }\  \forall \lambda: \B{c}^T\B{p}_\lambda - \beta \leq 0 \text{ and } \B{c}^T\B{p}-\beta \leq 1.
\end{align}
Now if $\B{p}$ has Bell form, from $\forall \lambda: \B{c}^\dagger\B{p}_\lambda - \beta \leq 0$, we have, summing over all $\lambda$'s,
\begin{align}
\sum_\lambda p(\lambda) \B{c}^\dagger \B{p}_\lambda - \beta \leq 0	\Longrightarrow \B{c}^\dagger \B{p}\leq \beta. 
\end{align}
Thus $S \leq 0$. If $\B{p}$ is not of Bell form, then by the second constraint $\B{c}^\dagger\B{p}-\beta = S \leq 1$. Since $\B{p} \notin \mathcal{L}$, a separating hyperplane can be rescaled so that the optimum achieves $S=1$. The point is that any deterministic behavior satisfies $\B{c}^\dagger \B{p}_\lambda  \leq \beta $, and thus so does \emph{any} behavior in Bell form. At the same time $\B{c}^\dagger\B{p}\leq \beta +1$: the classical bound is violated by $\B{p}$ itself. Thus the Bell functional $\B{c}$ witnesses the nonclassicality of $\B{p}$. We note that the Bell polytope $\mathcal{L}$ is the convex hull of the deterministic behaviors, which are its extreme points: switching from the vertex representation to the halfspace representation of the polytope, we see that $\mathcal{L}$ is characterized by a finite set of Bell inequalities. If $\B{c}^\dagger\B{p}\leq \beta $ is true for any point $\B{p}$ in the polytope, then the set $\{\B{p}\in \mathcal{L}: \B{c}^\dagger\B{p}=\beta\}$ is a \emph{face}. Faces of dimension $\dim\mathcal{L}-1$ are called \emph{facets}, and the corresponding Bell inequalities are called \emph{tight}: any other Bell inequality can be written as a non-negative combination of facet inequalities.

In fact, exactly the same technique works for the more general class of \emph{noncontextuality inequalities}. 
The simplest example of a contextuality inequality is provided by the Specker triangle \cite{kunjwalMinimalStatedependentProof2014}. Suppose we have three measurements $A, B, C$ each of which has two outcomes to which we assign valuations in $\{-1, +1\}$. Each pair of measurements can be made jointly, but not all three together. Suppose that these measurements reveal some preexisting properties of the system: $a, b, c \in \{-1, +1\}$, that is, the $A$ measurement reads off whether $a$ is $\pm 1$, and so forth. Clearly, $(ab)(ac)(bc)=a^2 b^2 c^2 = 1$. In particular, this means that the products of each pair cannot all be $-1$: we cannot have $ab = ac = bc = -1$. Either none of them or two of them can be $-1$, and so at most two of the outcomes of the $A, B, C$ measurements must disagree. Consequently,
\begin{align}
\begin{cases} 1, & \text{ if } A \neq B \\ 0, & \text{ if } A = B \end{cases}\Bigg\} + \begin{cases} 1, & \text{ if } A \neq C \\ 0, & \text{ if } A = C \end{cases}\Bigg\} + \begin{cases} 1, & \text{ if } B \neq C \\ 0, & \text{ if } B = C \end{cases}\Bigg\} \leq 2.
\end{align}
Taking expectations, this implies
\begin{align}
P(A \neq B) + P(A \neq C) + P (B \neq C) \leq 2,
\end{align}
and using the fact that $P(A\neq B) = P(A=+1, B =-1) + P(A=-1, B=+1)$ as well as the fact that outcomes are valued in $\{-1, +1\}$ so that $\langle AB \rangle = 1 - 2P(A\neq B)$, we may rewrite the inequality as
\begin{align}
\langle AB\rangle + \langle BC\rangle + \langle CA \rangle \ge -1.
\end{align}
Finally, letting
\begin{align}
\B{p} = \begin{pmatrix}P(A=+1, B=+1) \\ P(A=+1, B=-1) \\ P(A = -1, B=+1) \\ P(A=-1, B=-1) \\ P(A=+1, C=+1) \\ P(A=+1, C=-1) \\ P(A = -1, C=+1) \\ P(A=-1, C=-1) \\ P(B=+1, C=+1) \\ P(B=+1, C=-1) \\ P(B = -1, C=+1) \\ P(B=-1, C=-1)\end{pmatrix}
\end{align}
and
\begin{align}
\B{c} = \begin{pmatrix}
    1 & -1 & -1 & +1 & +1 & -1 & -1 & +1 & +1 & -1 & -1 & +1  
   \end{pmatrix},
\end{align}
we see that we can write the inequality as
\begin{align}
\B{c} \cdot \B{p} \ge -1.
\end{align}
More generally, we may define response functions
\begin{align}
P(A=a_0, B=b_0|a,b,c) &= \delta_{a_0, a}\delta_{b_0, b}\\
P(A=a_0, C=c_0|a,b,c) &= \delta_{a_0, a}\delta_{c_0, c}\\
P(B=b_0, C=c_0|a,b,c) &= \delta_{b_0, b}\delta_{c_0, c},
\end{align}
which formalize the idea that the measurements $A,B,C$ simply read off the values of the properties $a,b,c$, so that given a distribution over their values $P(a,b,c)$, we can write 
\begin{align}
P(A=a_0, B=b_0) &= \sum_{abc} P(A=a_0, B=b_0|a,b,c) P(a,b,c)\\
P(A=a_0, C=c_0) &= \sum_{abc} P(A=a_0, C=c_0|a,b,c) P(a,b,c)\\
P(B=b_0, C=c_0) &= \sum_{abc} P(B=b_0, C=c_0|a,b,c) P(a,b,c).
\end{align}
Gathering up these three distributions into a behavior $\B{p}$, we may then consider the geometry of the allowed behaviors implied by the law of total probability. As before, they form a polytope $\mathcal{L}$ of allowed classical behaviors, and the inequality $\B{c} \cdot \B{p} \ge -1$ corresponds to a facet of this polytope. We could have considered nondeterministic response functions, and more general types of hidden variables, but as long as the pairwise distributions may be obtained by marginalizing over a distribution $P(a,b,c)$, the resulting behavior will lie in the abovementioned polytope whose vertices are precisely states of certainty about which values $a,b,c$ take. Converting to the halfspace representation yields all the noncontextuality inequalities, of the form $\B{c} \cdot \B{p} \leq \beta$.

As we discussed earlier, if we introduce a reference measurement for the system, to say that $A, B, C$ are pairwise jointly measurable is to say that there exist three measurements $\{J^{AB}\}, \{J^{AC}\}, \{J^{BC}\}$ such that
\begin{align}
P(A,B|\rho) &= P(A,B|J^{AB})P(J^{AB}|R)\Phi P(R|\rho) \\
P(A,C|\rho) &= P(A,C|J^{AC})P(J^{AC}|R)\Phi P(R|\rho) \\
P(B,C|\rho) &= P(B,C|J^{BC})P(J^{BC}|R)\Phi P(R|\rho)
\end{align}
satisfying
\begin{align}
P(A|\rho) &= \sum_i P(A, B_i|\rho) = \sum_i P(A, C_i|\rho) \\
P(B|\rho) &= \sum_i P(A_i, B|\rho) = \sum_i P(B, C_i|\rho) \\
P(C|\rho) &= \sum_i P(A_i, C|\rho) = \sum_i P(B_i, C|\rho).
\end{align}
Gathering up $P(A,B|\rho),P(A,C|\rho),P(B,C|\rho)$ into a behavior $\B{p}_Q$, we can compare it to the behavior $\B{p}_C$ obtained by setting $\Phi = I$, and relate the violation of any noncontextuality inequality for the scenario to $\lVert I - \Phi \rVert$ by the same argument we gave for Bell inequalities.

\section{Conclusion}

In this chapter, we have given a QBist-friendly account of classicality, adapting work on so-called noncontextual ontological models to the QBist framework. We have shown how such models correspond to nonnegative matrix factorizations, or equivalently simplex embeddings, relating such constructions to the existence (or not) of a nonnegative cousin of the Born matrix, emphasizing along the way the QBist idea that classicality involves indifference to whether or not a reference measurement is performed. Finally, we have shown that for QBists, violations of Bell inequalities and the like may be understood as quantifying the cost of the failure to adopt the fundamental nonclassical coherence condition. 

%% file: sections/characterizing.tex
\UMBchapter{Characterizing quantum state space with a single quantum measurement}
\label{ch:characterizing}

\section{Introduction}

In 1927, Niels Bohr introduced the notion of \emph{complementarity}, that not all aspects of a physical system may be simultaneously definite, as the distinctive feature of the new quantum mechanics \cite{Bohr1928-vp, baggott2011quantum}. For example, the Heisenberg uncertainty principle, $\sigma_x \sigma_p\ge \frac{1}{2}\hbar$, tells us that if we experiment upon an ensemble of identically prepared particles, then a small variance in the measured position of the particles implies a large variance in their momentum, and vice versa. In particular, regardless of the choice of ensemble, the variance of the two quantities cannot be made arbitrarily small together while remaining consistent with quantum theory. The uncertainty principle may be generalized e.g.,\ to arbitrary pairs of observables, and even to collections of observables \cite{hou2016uncertaintyrelationsmultiobservables}. If Bohr was right that complementarity is the defining feature of quantum theory, then it ought to be possible to characterize ``quantum states'' as nothing other than probability assignments which satisfy appropriate uncertainty relations for all possible observables.

From this point of view, what is essential is not the traditional Hilbert space formalism, but instead the constraints quantum theory urges on probability assignments: indeed, the former can be seen as a convenient mathematical technique for imposing those very constraints. As we have discussed, this is the central contention of QBism \cite{fuchs2019qbismquantumtheoryheros}, a subjective Bayesian interpretation of quantum mechanics, which holds that quantum theory should not be viewed as a description of physical reality but rather as a set of consistency constraints on probability assignments motivated by nature's lack of hidden variables. In particular, QBists side with Schr\"odinger in viewing the quantum state as nothing more than a ``catalogue of expectations'' \cite{cat}. Indeed, while Bohr and Heisenberg focused on relations between observables like position and momentum, contemporary quantum information theory contemplates a more general class of measurements, so-called \emph{informationally complete} (IC) measurements \cite{cavesUnknownQuantumStates2002}. Remarkably, assigning appropriate probabilities to the outcomes of a single IC measurement is equivalent to assigning a quantum state. Thus one may take Schr\"odinger's  ``catalogue of expectations'' a step further, and identify quantum states with probability distributions directly.

 The caveat is that while all quantum states correspond to probability distributions, not all probability distributions correspond to quantum states \cite{appleby2011propertiesqbiststatespaces}. On the Hilbert space side, such invalid distributions correspond to self adjoint matrices which are not positive semidefinite, and thus cannot be regarded as density matrices. A picture thereby emerges of quantum state space as a privileged \emph{subset} of the probability simplex, corresponding to just those probability distributions which map back to valid states. The shape of the subset depends on the choice of informationally complete measurement used as a ``reference,'' and can be derived by calculating certain quantities associated with the Hilbert space representation of the measurement. From a foundational perspective, however, one might wonder whether it is possible to characterize the shape of such a privileged subset without reference to Hilbert space as such. Related questions have recently been asked in the context of GPT tomography \cite{mazurekExperimentallyBoundingDeviations2021, graboweckyExperimentallyBoundingDeviations2022}, which provides a theory agnostic approach to reconstructing state and effect spaces from the experimental data collected from a wide range of measurements. Certainly, a state is invalid if it implies a negative probability for some effect. From a QBist point of view, the Born rule is just an example of a general nonclassical coherence constraint, expressed entirely in terms of probabilities with respect to a reference measurement. One may thus rule out a probability distribution if the Born rule, formulated in terms of probabilities, yields a negative number for some measurement outcome. In this way, the shape of state space would emerge out of the demand for probabilistic consistency.
  
 Nevertheless, just as a single IC reference measurement may characterize a quantum system entirely in terms of reference probabilities, one might wonder whether a single specially chosen reference measurement could perform ``state space tomography'' by appealing only to probabilities assigned to that single measurement. Probabilistic consistency with this one measurement would imply consistency with every possible measurement, and provide an important proof of principle for the QBist approach to quantum mechanics. In this chapter, we demonstrate that this can in fact be done. Taking one's reference measurement to be a \emph{complex projective 3-design} allows the set of valid probability distributions---that is, the shape of quantum state space---to be elegantly characterized by an uncertainty principle\footnote{We note that whereas the usual uncertainty principle relates probability distributions with respect to different measurements, our generalization is formulated with respect to a single measurement.}, precisely in the spirit of Bohr. Crucially, the only building blocks needed to formulate this principle are reference measurement probabilities.

  In particular, we will show that probability assignments to the outcomes of the reference measurement cannot be too sharp in a prescribed way: they must satisfy a lower bound on the variance with respect to any of a natural class of observables. This is so even as those same probability assignments may imply a sharp distribution on the outcomes of some alternative measurement via the Born rule. This is a dramatic reversal of the situation classically, where any alternative measurement may be regarded as a coarse graining of a reference measurement e.g., of the positions and momenta of a set of particles. In the classical case, on the one hand, certainty is achievable for the reference; on the other hand, one cannot achieve more certainty about alternative measurements than about the reference. Thus our result underscores in a novel and perspicacious way the degree to which quantum mechanics resists hidden variable interpretations. Moreover, the notion of defining the set of valid probability distributions in terms of an uncertainty principle is a promising approach to constructing generalizations of quantum theory in the spirit of  \emph{generalized probabilistic theories} \cite{mullerProbabilisticTheoriesReconstructions2021, barnum2013postclassicalprobabilitytheory} or in the QBist literature, the \emph{qplex} research program \cite{Appleby_2017}. Independently motivating the features of a 3-design representation, and demonstrating how a Hilbert space representation arises from them (rather than the other way around), would represent a significant advance in the ongoing quest to derive the quantum formalism from satisfying quantum information theoretic principles, and at the same time likely lead to a novel characterization of 3-designs themselves. Indeed, it is to this subject we turn in the next and final chapter. Moreover, we note that 3-designs are of great contemporary interest due to their special role in the theory of classical shadow estimation, where employing a 3-design allows one for example to achieve constant sample complexity in fidelity estimation independent of system size \cite{Huang_2020,mao2024magicquditshadowestimation,Kliesch_2021, chen2024nonstabilizernessenhancesthriftyshadow}. We hope the present work places these developments in a broader context.
  
In the spirit of completeness, we begin by reviewing the probability-first formalism for quantum mechanics furnished by an \emph{informationally complete reference measurement}, and then lay out the basic features of complex projective $3$-designs. These mathematical preliminaries aside, we derive scalar constraints on probability assignments corresponding to pure states: pure state probability assignments turn out to live in the intersection of 2-norm and 3-norm spheres of specified radii restricted to a natural subspace. Operationally, these constraints can be interpreted as bounds on the agreement between several copies of the reference measurement; conceptually, they can be understood as entropic uncertainty principles \cite{PhysRevLett.60.1103}. We then establish a vector constraint on pure probability assignments, which draws our attention to a particular 3-index tensor built out of the probabilities which characterize the reference measurement itself. To handle the case of mixed states, we report the key insight that for a 3-design measurement, one can relate the variance of an observable as measured directly to the variance of the observable as measured by the reference. This is what allows quantum state space to be characterized in its entirety by a single uncertainty principle, which lower bounds the variance of any of a natural class of observables. In closing, we observe that the reason 3-designs play such a privileged role is that for just these measurements, the structure coefficients for the Jordan algebra of observables \cite{farautAnalysisSymmetricCones1994} can be expressed solely in terms of the probabilities which characterize the reference measurement.  The essence of this observation was already made in \cite{obst2024wignerstheoremstabilizerstates, Heinrich_2019}, but its significance for probabilistic representations of quantum mechanics was left unexplored. In fact, we show that just as a 2-design allows the Born rule to be expressed as a gentle modification of the law of total probability, 3-designs allow the quantum Jordan product on observables to be expressed as a gentle modification of the classical Jordan product, the elementwise product of valuations on reference outcomes.

\section{The reference measurement formalism} 

In quantum mechanics, the most general form of a measurement with a finite number of outcomes consists of a set $\{E_i\}_{i=1}^n$ of positive semidefinite matrices called \emph{effects}, acting on a Hilbert space $\mathcal{H}_d$ satisfying $\sum_{i} E_i = I$. If a set of effects $\{R_i\}$ span the $d^2$-dimensional operator space, we call the measurement \emph{informationally complete}:  the probabilities $P(R_i|\rho)=\tr(R_i\rho)$ fully characterize the density matrix $\rho$ representing a quantum state, which must itself be positive semidefinite with $\tr(\rho)=1$. We suppose that upon obtaining an outcome $R_i$, an agent would assign a corresponding state $\sigma_i$ to a subsequent reference measurement\footnote{In the language of the previous chapters, we take the reference measurement to be a preparatory measurement.}. We suppose that $\{\sigma_i\}$ also span the $d^2$-dimensional operator space, and in fact we will always assume that for our reference measurement $\sigma_i = \tr(R_i)^{-1} R_i$,  that $\forall i: \tr(R_i)^{-1}=n/d$, and that each $\sigma_i$ is a pure state. In other words, we specialize to the case of unbiased rank one measurements $R_i = (d/n)|\psi_i\rangle\langle \psi_i|$.

The minimal number of effects in an IC measurement is $d^2$: moreover, at best an IC-measurement may furnish a linearly independent, but not orthonormal basis \cite{e16031484, cuffaro2024quantumstatesmaximalmagic, debrota2020varietiesminimaltomographicallycomplete}, and more generally an informationally \emph{over}complete set. We must therefore take up the matter of its dual representation---with a probabilistic  twist. Let $|\sigma_i) =\text{vec}(\sigma_i)= (I \otimes \sigma_i)\sum_j |j,j\rangle$ be the vectorization of a reference state $\sigma_i$, and similarly let $(R_i|=\sum_i\langle i,i|(I \otimes R_i)$ be the vectorization of a reference effect. In general,  $(A|B)=\tr(A^\dagger B)$, and so arranging $(R_i|$ into the rows of a matrix $\B{R}$, and $|\sigma_i)$ into the columns of a matrix $\textbf{S}$, we can write the conditional probability matrix with elements $P(R_i|R_j)$, for the probability of a subsequent reference outcome given an initial reference outcome, $P\equiv \B{RS}$. By informational completeness, these probabilities fully characterize the reference measurement itself, and they will play a fundamental role in the sequel.

We call a \emph{Born matrix} any matrix $\Phi$ which satisfies $P\Phi P = P$, the defining equation of a $\{1\}$-inverse of $P$ \cite{GeneralizedInverses2003a}. As we have seen, it follows from informational completeness that $P\Phi P = P \Longleftrightarrow \B{S}\Phi \B{R}=I$ \cite{weiss2024depolarizingreferencedevicesgeneralized}, which provides a resolution of the identity, and thus a dual representation $|\rho) = \B{S}\Phi \B{R}|\rho)$ or,
\begin{align}
\rho=\sum_{ij}\Phi_{ij}P(R_j|\rho)\sigma_i.
\end{align}
If the measurement operators (and states) are linearly independent, then $P$ will be invertible, and thus $\Phi = P^{-1}$. Otherwise, there will be a variety of choices for the Born matrix\footnote{Recall that the $\{1\}$-inverses of a matrix $P=U\Sigma V^\dagger$ may all be calculated from its singular value decomposition via 
\begin{align}
	\Phi = V \begin{pmatrix}\sigma^{-1} & A \\ B & C \end{pmatrix}U^\dagger\nonumber,
\end{align}
where $A, B, C$ are completely arbitrary matrices, and $\sigma$ is the diagonal matrix of nonzero singular values. A typical example is the Moore-Penrose pseudoinverse, for which $A=B=C=0$ \cite{GeneralizedInverses2003a}.}, and different assignments of probabilities will lead to the same ascription of a density matrix. 

Consider now some alternative measurement $\{E_i\}_{i=1}^m$. We can write the Born rule probability $P(E_i|\rho)=\tr(E_i\rho)$ in terms of reference probabilities,
\begin{align}
P(E_i|\rho) &= \tr(E_i\rho)=(E_i|\B{S}\Phi \B{R}|\rho)\\
&=\sum_{jk}P(E_i|R_j)\Phi_{jk}P(R_k|\rho)\nonumber,
\end{align}
and appreciate that the Born rule has become a simple \emph{deformation} \cite{ferrieQuasiprobabilityRepresentationsQuantum2011} of the law of total probability $P(E_i) = \sum_j P(E_i|R_j)P(R_j)$.

In particular, taking $\B{E}=\B{R}$, we have that $P(R_i|\rho)=\sum_{jk}P(R_i|R_j)\Phi_{jk}P(R_k|\rho)$. This is a fundamental consistency criterion in an overcomplete probability representation. As we saw in Chapter \ref{ch:born_identity}, it follows from the defining equation of a $\{1\}$-inverse that $P\Phi$ is a projector onto $\col(P)$: recall the column space of $P$ (or equivalently, its range) is the span of its columns. Now $P=\B{RS}$ is in fact a full rank factorization of $P$, and hence the columns of $\B{R}$ form a basis for $\text{col}(P)$ \cite{piziakFullRankFactorization1999}. Thus any vector with components $x_i =\tr(R_iX)$, that is, $x=\B{R}|X)$, must lie in $\text{col}(P)$. Conversely, if $x\in \text{col}(P)$, there must exist a vector $\tilde{x}$ such that $x_i= \sum_j P(R_i|R_j)\tilde{x}_j=\tr\left(R_i \sum_j \tilde{x}_j\sigma_j\right)=\tr(R_i \tilde{X})$ for some operator $\tilde{X}$. By the same token, just as distributions $P(R|\rho)$ must live in $\col(P)$, response functions $P(E|R)$ ought to live in $\row(P)$. Moreover without loss of generality, valuations on reference outcomes may be taken to be vectors in $\col(P)$. Finally, taking $\sigma_i = \tr(R_i)^{-1} R_i =(n/d)R_i$ means that $P=P^\dagger$ and so $\col(P)=\row(P)$. 

\section{Making designs}

It follows from the representation theory of the unitary and symmetric groups \cite{harrowChurchSymmetricSubspace2013, bacon2005quantumschurtransformi} that the \emph{$t$th moment of quantum state space}, that is, the $t$th tensor power of a pure state averaged over all pure states is 
\begin{align}
	\label{integral_pure_states}
\int |\psi\rangle\langle \psi|^{\otimes t} d\psi = \binom{d+t-1}{t}^{-1}\Pi_{\text{sym}^t},
\end{align}
where $d\psi$ denotes the Haar measure on pure states, and $\Pi_{\text{sym}^t}$ is the projector onto the permutation symmetric subspace on $t$ tensor factors \cite{waldronIntroductionFiniteTight2018}. Note $\tr(\Pi_{\text{sym}^t})=\binom{d+t-1}{t}$ is just the dimension of that subspace. As $\Pi_{\text{sym}^t}$ can be expressed as a sum over all permutation operators, we have in fact
\begin{align}
	\int |\psi\rangle\langle \psi|^{\otimes t} d\psi &=\binom{d+t-1}{t}^{-1}\frac{1}{t!}\sum_{\pi\in S_t}T_\pi
\end{align}
where $T_{\pi} = \sum_{a_1, \dots, a_t}|a_{\pi^{-1}(1)} ,\dots, a_{\pi^{-1}(t)}\rangle\langle a_1, \dots, a_t|$.

A \emph{quantum  state $t$-design}, also called a \emph{complex projective t-design} \cite{waldronIntroductionFiniteTight2018}, is an ensemble of pure states $\{p_i, |\psi_i\rangle\}_{i=1}^n$ which satisfy
\begin{align}
	\label{sum_haar_integral}
\sum_{i=1}^n p_i|\psi_i\rangle\langle \psi_i|^{\otimes t} =\int |\psi\rangle\langle \psi|^{\otimes t} d\psi,
\end{align}
that is, the average over the design ensemble mimics the average over all pure states up to the $t$-th moment. We will call a design \emph{unbiased} or \emph{unweighted} if $\forall i: p_i=\frac{1}{n}$. The number of elements in a $t$-design satisfies 
\begin{align}
n \ge \binom{d-1+\lfloor t/2 \rfloor}{\lfloor t/2 \rfloor }	\binom{d-1 +\lceil t/2 \rceil}{\lceil t/2 \rceil} \nonumber.
\end{align}
We note that a $t$-design of any order always exists for sufficiently large $n$ \cite{SEYMOUR1984213}, and a $t$-design is also a $(t-1)$-design. For $t=1$, we have $\sum_i p_i|\psi_i\rangle\langle \psi_i| =\frac{1}{d}I$ which shows that a $1$-design furnishes a set of rank-1 projectors which, rescaled, sum to the identity: thus a 1-design is a quantum measurement. For a 2-design 
\begin{align}
\label{swappy}
\sum_i p_i |\psi_i\rangle\langle \psi_i|^{\otimes 2}=\frac{1}{d(d+1)}(I \otimes I + T_{21}),
\end{align}
where $T_{21}$ is the swap operator: two typical examples are symmetric informationally complete (SIC) states and states corresponding to a complete set of mutually unbiased bases (MUBs) \cite{bengtsson2017discretestructuresfinitehilbert}. In fact, it  follows from Eq.\ \ref{swappy} that for any unbiased 2-design reference device, we can take $\Phi=(d+1)I - \frac{d}{n}J$, where $J$ is the Hadamard identity, the matrix of all 1's (Appendix \ref{app3})\footnote{Moreover, viewing $\B{R}$ as a frame analysis operator and $\BT{S}=\B{S}\Phi$ as an frame synthesis operator, we see that the dual analysis operator $\BT{S}^\dagger$ has the same range as $\B{R}$ itself. Thus by Lemma \ref{dual_analysis_range}, the columns of $\BT{S}$ furnish canonical dual elements}.

The Born rule then takes the profoundly elegant form
\begin{align}
P(E|\rho) &=\sum_i P(E|R_i)\left[(d+1)P(R_i|\rho)-\frac{d}{n}\right],
\end{align}
 which was given an independent motivation in the case that $n=d^2$ in \cite{debrotaSymmetricInformationallyComplete2020}, and is related to the fact that 2-designs are optimal for linear quantum state tomography \cite{scottTightInformationallyComplete2006, Zhu_2011}. The terrain of 3-designs is an area of active investigation: several examples were presented in \cite{slomczynskiMorphophoricPOVMsGeneralised2020}, and more general constructions were given in \cite{Gross_2021, zhu2024momentsquditcliffordorbits}. In particular, 3-designs have been studied for their use in the theory of classical shadows \cite{Huang_2020,mao2024magicquditshadowestimation,Kliesch_2021} since they lead to well-controlled variance in the estimation of expectation values. In $d=2$, the complete set of MUBs forms a $3$-design as does the union of the tetrahedral SIC and the antipodal SIC formed by Bloch sphere inversion. Notably, the set of $n$-qubit stabilizer states \cite{Zhu_2017} also forms an unbiased 3-design. When $t=3$, $n\ge \frac{1}{2}d^2(d+1)$, a bound which is not tight \cite{waldronIntroductionFiniteTight2018}.

 Finally, we note that the \emph{order-$t$ frame potential} $\mathcal{F}_t$ provides a variational characterization of $t$-designs \cite{waldronIntroductionFiniteTight2018, renesSymmetricInformationallyComplete2004}.
\begin{theorem}
\begin{align}
\mathcal{F}_t[\{p_i,|\psi_i\rangle\}]&=	\sum_{ij}p_ip_j|\langle \psi_i|\psi_j\rangle|^{2t}\ge \binom{d+t-1}{t}^{-1}\nonumber,
\end{align}
with equality if and only if $\{p_i,|\psi_i\rangle\}$ forms a complex-projective t-design.
\end{theorem}
\begin{proof}
 Theorem \ref{frame_pot_lower_bound}, we showed that the frame potential $\tr(F^2)\ge \tr(F)/d$ achieves its lower bound only for a tight frame. Here $F$ is the frame operator. From Eq. (\ref{integral_pure_states}) and Eq. (\ref{sum_haar_integral}), we have
\begin{align}
F_t = \sum_i \sqrt{p_i}|\psi_i\rangle^{\otimes t} \sqrt{p_i}\langle \psi_i|^{\otimes t} = \binom{d+t-1}{t}^{-1}\Pi_{\text{sym}^t}.
\end{align}
This implies that $\{\sqrt{p_i}|\psi_i\rangle^{\otimes t}\}$ forms a tight frame for the symmetric subspace on which $\Pi_{\text{sym}^t}$ is the identity operator. Applying Theorem \ref{frame_pot_lower_bound} to $\tr(F_t^2)$ gives the result.
\end{proof}

\section{The shape of quantum state space}

\subsection{Bounding agreement}
\label{Agreement-probabilities}

We begin by providing a characterization of pure state probability distributions in terms of a set of entropic uncertainty principles. They can be motivated by considering the following scenario. Suppose an agent performs $t$ preparatory measurements $\rho_1, \dots, \rho_t$, and afterwards performs a reference measurement on each of the $t$ systems. What is the probability that all $t$ reference measurements give the same outcome? In other words, we are interested in the agreement probability
\begin{align}
P(\text{agree}|\rho_1, \dots, \rho_t)=\sum_{i=1}^n \prod_{j=1}^t P(R_i|\rho_j)= \tr\left(\sum_{i=1}^n R_i^{\otimes t} \otimes_{j=1}^t \rho_j\right)\nonumber.
\end{align}
Assuming the reference measurement is unbiased and that the reference states are proportional to effects $R_i = \frac{d}{n}\sigma_i$, we have $P(\text{agree}|\rho_1, \dots, \rho_t)= \frac{d^t}{n^{t-1}}\tr\left(\frac{1}{n}\sum_{i=1}^n \sigma_i^{\otimes t} \otimes_{j=1}^t \rho_j\right)$. If we further assume that the reference measurement forms a $t$-design,  $\frac{1}{n}\sum_i\sigma_i^{\otimes t} = \binom{d+t-1}{t}^{-1}\Pi_{\text{sym}^t}$, and so the agreement probability can be written
\begin{align}
P(\text{agree}|\rho_1, \dots, \rho_t) &=\frac{d^t}{n^{t-1}}\binom{d+t-1}{t}^{-1}\frac{1}{t!}\sum_{\pi\in S_t}\tr(T_\pi\otimes_{j=1}^t \rho_j)\nonumber.
\end{align}
To evaluate expressions like this, it suffices to consider traces with cyclic permutations. For $t=2$, the swap operator $T_{21}=\sum_{ab}|b,a\rangle\langle a, b|$ yields
\begin{align}
\tr\left(	(X\otimes Y)\sum_{ab}|b,a\rangle\langle a, b|\right)=\sum_{ab}\langle a|X|b\rangle \langle b|Y|a\rangle=\tr(XY)\nonumber,
\end{align}
while similarly, for $t=3$, a cyclic permutation of three elements delivers
\begin{align}
\tr\left(	(X \otimes Y \otimes Z)\sum_{abc}|b,c,a\rangle\langle a, b,c|\right)=\sum_{abc} \langle a|X|b\rangle\langle b|Y|c\rangle\langle c|Z|a\rangle=\tr(XYZ)\nonumber.
\end{align}
In light of this, we have for the order-2 agreement probability,
\begin{align}
P(\text{agree}|\rho_1, \rho_2)
&=\frac{1}{d+1}\left(\frac{d}{n}\right)\Big[\tr(\rho_1)\tr(\rho_2)+\tr(\rho_1\rho_2)\Big].
\end{align}
Now $\tr(\rho_1\rho_2)\leq \sqrt{\tr(\rho_1^2)\tr(\rho_2^2)}$, while the purity satisfies $\tr(\rho^2)\leq 1$. Thus when $\rho_1=\rho_2=\rho$ pure, we saturate the upper bound of this quantity. For the lower bound, we note $\tr(\rho_1\rho_2)\ge0$ with equality if and only if $\rho_1, \rho_2$ are orthogonal, which leads to bounds
\begin{align}
\left(\frac{d}{n}\right)\frac{1}{d+1}\leq\sum_i P(R_i|\rho_1)P(R_i|\rho_2)\leq \left(\frac{d}{n}\right)\frac{2}{d+1}.
\end{align}
Similarly, for $t=3$, we find
\begin{small}
\begin{align}
P(\text{agree}|\rho_1, \rho_2, \rho_3) &= \frac{1}{(d+1)(d+2)}\left(\frac{d}{n}\right)^2\Big[\tr(\rho_1)\tr(\rho_2)\tr(\rho_3) +\tr(\rho_1)\tr(\rho_2\rho_3)\nonumber\\
&+\tr(\rho_2)\tr(\rho_1\rho_3)+\tr(\rho_3)\tr(\rho_1\rho_2)+\tr(\rho_1\rho_2\rho_3)+\tr(\rho_1\rho_3\rho_2)\Big]\nonumber,
\end{align}
\end{small}
\!\!which is maximized when $\rho_1=\rho_2=\rho_3=\rho$ pure, so that $\tr(\rho^2)=\tr(\rho^3)=1$, and minimized when $\rho_1, \rho_2, \rho_3$ are mutually orthogonal (when the dimension permits), delivering bounds
\begin{align}
\left(\frac{d}{n}\right)^2\frac{1}{(d+1)(d+2)} &\leq \sum_i P(R_i|\rho_1)P(R_i|\rho_2)P(R_i|\rho_3) \leq \left(\frac{d}{n}\right)^2\frac{6}{(d+1)(d+2)}.
\end{align}
Since in both cases, the upper bounds are saturated by identical pure states, we conclude that pure state probability assignments with respect to a 3-design lie in the nonnegative orthant, in the intersection of three kinds of spheres. From $\sum_i P(R_i|\rho)$, they live on a 1-norm sphere of radius 1; from $\sum_i P(R_i|\rho)^2$, they live on a 2-norm sphere of radius $\sqrt{\left(\frac{d}{n}\right)\frac{2}{d+1}}$; and from $\sum_i P(R_i|\rho)^3$, they live on a 3-norm sphere of radius $\sqrt[3]{\left(\frac{d}{n}\right)^2\frac{6}{(d+1)(d+2)}}$. Finally, since we derived all probabilities from the trace of a reference effect on a state, our probability distributions live in $\text{col}(P)$.

We could continue on, contemplating the $t$-fold agreement probability for an unbiased $t$-design reference device for any $t$. Since a pure state satisfies $\forall t: \tr(\rho^t)=1$, we would find that pure probability vectors live on $t$-norm spheres with fixed radii determined by the agreement probability
\begin{align}
P(\text{agree}|\rho^{\otimes t}) &= \sum_i P(R_i|\rho)^t =\frac{d^t}{n^{t-1}}\binom{d+t-1}{t}^{-1}\frac{1}{t!}\sum_{\pi\in S_t}\tr(T_\pi\rho^{\otimes t})= \frac{d^t}{n^{t-1}}\binom{d+t-1}{t}^{-1},
\end{align}
where the last follows from the fact that since $\forall t: \tr(\rho^t)=1$, all of the $t!$ terms in the sum will be 1.

The following lemma \cite{Jones_2005, fuchs2015strugglesblockuniverse}, however, assures us that a $3$-design is all we need.
\begin{lemma}
	\label{rank1-lemma}
	A Hermitian operator $A$ is a rank-1 projector if and only if $\tr(A^2) = \tr(A^3)=1$. 
\end{lemma}
\begin{proof}
Let $\{\lambda_i\}$ be the eigenvalues of $A$. $\tr(A^2) = \tr(A^3)=1$  means that $\sum_i \lambda_i^2 = \sum_i \lambda_i^3 = 1$.
On the one hand, $\sum_i \lambda_i^2 = 1$ implies that $\forall i: -1 \leq \lambda_i \leq 1$. On the other hand, $\sum_i \lambda_i^3 \leq \sum_i \lambda_i^2$ with equality if and only if $\forall i: \lambda_i \in \{0, 1\}$. But since the whole sum must be 1, we must have exactly one $\lambda_i=1$ and the rest  0. Thus $A$ is a rank-1 projector, or equivalently a pure state $\rho$.
\end{proof}
\noindent In light of this, as long as $t\ge 3$, we can fully characterize the pure states of quantum theory with respect to an unbiased $t$-design reference device by $\forall i: P(R_i|\rho)\ge0$ and 
\begin{align}
\sum_i P(R_i|\rho)&= 1	\\
\sum_i P(R_i|\rho)^2&= \left(\frac{d}{n}\right)\frac{2}{d+1}	\\
\sum_i P(R_i|\rho)^3 &= \left(\frac{d}{n}\right)^2\frac{6}{(d+1)(d+2)},
\end{align}
along with $P(R_i|\rho)=\sum_{jk}P(R_i|R_j)\Phi_{jk}P(R_k|\rho)$, that is, the probability distribution lives in $\text{col}(P)$\footnote{In fact, we can show more specifically that a probability distribution corresponds to a pure state if and only if its component that lies in $\text{col}(P)$ satisfies the quadratic and cubic constraints. To see this, let $\rho=\sum_{ij}\Phi_{ij}P(R_j|\rho)\sigma_i$. Then $\tr(\rho^2)=\frac{n}{d}\sum_{ij}\Phi_{ij}P(R_j|\rho)\sum_{kl}P(R_i|R_k)\Phi_{kl}P(R_l|\rho)$. Decomposing $P(R|\rho)=x+y$, for $x\in \text{col}(P)$ and $y \in \text{col}(P)^\perp$,  and using the form of $\Phi$ for an unbiased 2-design, we find $\tr(\rho^2)=(d+1)\left(\frac{n}{d}\right)\sum_ix_i^2 - 1$. Thus if and only if $\sum_i x_i^2=\left(\frac{d}{n}\right)\frac{2}{d+1}$ does $\tr(\rho^2)=1$. The result is analogous for $\tr(\rho^3)=1$, using the expression for $\Re \big[\tr(R_i \sigma_j \sigma_k)\big]$  derived in section \ref{idempotents}.}.

 Thus in a sense we have already achieved our goal since quantum state space is the convex hull of all pure state probability assignments. The bounds on agreement probabilities, from which the constraints on pure states were derived, already suggest a kind of uncertainty principle: after all, classically, there is nothing in principle preventing such an agreement probability from being 1 when $\rho_1=\dots=\rho_t$, or 0 for perfectly distinguishable states. In fact, the upper bounds may be understood as implying a set of \emph{entropic} uncertainty principles. Defining the order-$t$ R\'enyi entropy of a probability-distribution $\{P(R_i|\rho)\}$ as
\begin{align}
H_t\Big(\{P(R_i|\rho)\}\Big) &= \frac{1}{1-t}\log\left(\sum_i P(R_i|\rho)^t\right)	
\end{align}
for $0< t <\infty$, it is clear from the above discussion that pure state probability distributions achieve the lower bound on the R\'enyi entropies of order $t=2,3$ over all states, and this, along with the restriction to $\text{col}(P)$, is enough to characterize them completely.

\subsection{The contour of idempotents}
\label{idempotents}

As an alternative approach, we can derive a single equation picking out pure state probability assignments by appealing to the fact that, for a normalized state, $\rho=\rho^2$ if and only if $\rho=\rho^\dagger$ is pure, that is, a rank one projector. Substituting the resolution of the identity $\rho=\sum_{ij}\Phi_{ij}P(R_j|\rho)\sigma_i$ into $P(R_i|\rho^2)=\tr(R_i\rho^2)$, we find that
\begin{align}
\label{rho_rhosq}
P(R_i|\rho)  = 	\sum_{lm} P(R_l|\rho)P(R_m|\rho)\sum_{jk}\Phi_{jl}\Phi_{km}\Re\big[\tr(R_i\sigma_j\sigma_k)\big]\nonumber.
\end{align}
We need only consider the real part since  $\tr(R_i \sigma_k\sigma_j)=\tr\big( R_i (\sigma_j \sigma_k)^\dagger\big)=\tr\big( R_i (\sigma_j^* \sigma_k^*)^T\big)=\tr\big( R_i^T \sigma_j^* \sigma_k^*\big)=\tr\big( R_i^* \sigma_j^* \sigma_k^*\big)=\tr(R_i\sigma_j \sigma_k)^*$ as $R_i, \sigma_j, \sigma_k$ are all positive semidefinite and so every term in Eq.\ \ref{rho_rhosq} is added to its complex conjugate.

 Let $\mathcal{M}_t= \int |\psi\rangle\langle\psi|^{\otimes t}d\psi$ be the $t$th moment of quantum state space. This is itself a valid state, and so we can consider its probability distribution $P(R_i, R_j, R_k, \dots|\mathcal{M}_t)$ with respect to $t$ copies of the reference measurement. If we assume an unbiased set of effects, by the same argument as in Section \ref{Agreement-probabilities}, it follows that
\begin{align}
&P(R_i|\mathcal{M}_1) = \frac{1}{n}	\\
&P(R_i, R_j|\mathcal{M}_2) =\frac{1}{d+1}\left(\frac{1}{n}\right)\left[\frac{d}{n} + P(R_i|R_j)\right]\label{m2}\\
\label{any_triples}&P(R_i, R_j, R_k|\mathcal{M}_3)=\frac{1}{(d+1)(d+2)}\left(\frac{d}{n^2}\right)\times \\
&\Bigg[ \frac{d}{n}+  P(R_j|R_k)  +P(R_i|R_j) +P(R_i|R_k) + 2\Re \big[\tr(R_i \sigma_j \sigma_k)\big]\Bigg].\nonumber	
\end{align}
\!\!If we further assume that the reference states form a 3-design then $\mathcal{M}_3=\frac{1}{n}\sum_i \sigma_i^{\otimes 3}$ and so $P(R_i, R_j, R_k|\mathcal{M}_3)= \frac{1}{n}\sum_{m}P(R_i|R_m)P(R_j|R_m)P(R_k|R_m)$. Equating these two expressions allows us to calculate $\Re \big[\tr(R_i \sigma_j \sigma_k)\big]$ directly from the conditional probability matrix $P(R_i|R_j)$ which characterizes the reference measurement itself,
\begin{align}
\Re \big[\tr(R_i \sigma_j \sigma_k)\big] &=	\frac{1}{2}\Bigg[(d+1)(d+2)\left(\frac{n}{d}\right)\sum_mP(R_i|R_m)P(R_j|R_m)P(R_k|R_m)\nonumber\\
&-  P(R_j|R_k)-P(R_i|R_j)-P(R_i|R_k)-\frac{d}{n}\Bigg].\nonumber
\end{align}
Then Eq.\ \ref{rho_rhosq}, which expresses $\rho=\rho^2$ in terms of probability-assignments,  simplifies to
\begin{align}
\label{quadratic}
P(R_i|\rho) =\frac{1}{2}\Bigg[\frac{1}{2}(d+1)(d+2)\left(\frac{n}{d}\right)\sum_m P(R_i|R_m)P(R_m|\rho)^2-\frac{d}{n}\Bigg]\nonumber,
\end{align}
which depends only upon $P(R_i|R_j)$ and $P(R_i|\rho)$. A probability distribution satisfying Eq.\ \ref{quadratic} is clearly in $\text{col}(P)$, and it is straightforward to check that Eq.\ \ref{quadratic} implies the scalar constraints proved in the previous section.

Finally, we note that Eq.\! \ref{any_triples} implies a state assignment $\mathcal{M}_3$ allows one to extract $\Re[\tr(E_i\sigma_j\sigma_k)]$ (where $\sigma_i = \tr(E_i)^{-1}E_i$) from the joint probability distribution $P(E_i, E_j, E_k|\mathcal{M}_3)$ for \emph{any} measurement $\{E_i\}$. One may compare this method to the procedure described in \cite{oszmaniec2021measuringrelationalinformationquantum}, which exploits the backaction on a control qubit after a controlled cyclic permutation to extract the real part of a trace of an $n$-product of states from an expectation value. In particular, having estimated $\Re[\tr(E_i\sigma_j\sigma_k)]$ for any informationally complete measurement, we may characterize pure state probability distributions with respect to that measurement according to Eq.\! \ref{rho_rhosq}. What makes a 3-design distinctive is that $\Re[\tr(R_i\sigma_j\sigma_k)]$ can be calculated from $P(R_i|R_j)$ alone, and so pure state probability distributions can be characterized by appealing to probabilities assigned to one single measurement.

\subsection{A variance-based uncertainty principle}

We now give a condition for the validity of \emph{any} probability assignment, pure or mixed. For a 3-design measurement, the variance of an observable as measured by a standard von Neumann measurement can be directly related to the variance of the same observable as estimated by the reference measurement. From this consideration, we can characterize the validity of any probability distribution in terms of a lower bound on the variance of all observables of a natural class, that is, those that live in $\col(P)$.

We begin by noting that a Hermitian matrix $\rho$ is positive semidefinite if and only if its second moment with respect to all Hermitian observables $X$ is nonnegative:
\begin{align}
\forall X: \tr(X^2\rho)\ge0 \Longleftrightarrow \rho \ge0.	
\end{align}
This follows from the fact that $X^2 \ge 0$ and the fact that the cone of positive semidefinite matrices is self-dual \cite{farautAnalysisSymmetricCones1994}. More simply, one can observe that $\rho\ge 0$ is equivalent to $\forall \psi: \langle\psi|\rho|\psi\rangle=\tr(|\psi\rangle\langle \psi|\rho)\ge0$, and any $X^2$ can be decomposed into a sum of rank-1 projectors weighted by nonnegative numbers via the spectral decomposition. 

Substituting  $X=\sum x_i R_i$ and $\rho=\sum_{ij}\Phi_{ij}P(R_j|\rho)\sigma_i$, we find that
\begin{align}
\label{vn_var}
\forall X: \tr(X^2\rho) =\left(\frac{d}{n}\right)\sum_{ijkl} x_ix_j\Re[\tr(R_i\sigma_j\sigma_k)]\Phi_{kl}P(R_l|\rho) \ge 0.
\end{align}
If we assume that $x \in \text{col}(P)$, and exploit the expression for $\Re[\tr(R_i\sigma_j\sigma_k)]$ in terms of $P(R_i|R_j)$, we may simplify the expression for $\tr(X^2\rho)$.

\begin{lemma}
   For an unbiased complex-projective 3-design reference measurement, 
\begin{align}
\label{simplified_var}
\tr(X^2\rho) =\frac{1}{2}\left(\frac{d+2}{d+1}\right)\Bigg[\langle x^2\rangle_\rho-\frac{d}{d+2}\Big(\langle x^2\rangle_\mu-2\langle x\rangle_\mu\langle x\rangle_\rho\Big)\Bigg],
\end{align}
where e.g. $\langle x^2\rangle_\rho = \sum_i x_i^2 P(R_i|\rho)$ and $\forall i: P(R_i|\mu)=\frac{1}{n}$ are the probabilities for the maximally mixed state.
\end{lemma}
\begin{proof}
Let $x$ be a valuation on reference outcomes, that is, an assignment of real numerical values to the outcomes of the reference measurement. This is equivalent to the assignment of a self-adjoint operator $X$ since $\langle x\rangle_\rho=\sum_i x_i P(R_i|\rho)=\tr\left(\sum_i x_i R_i\rho\right)=\tr(X\rho)=\langle X\rangle_\rho^{(\text{vn})}$, although in an overcomplete representation different choices of $x$ will yield the same operator $X$. In other words, any real valuation $x$ determines a self-adjoint operator $X=\sum_i x_i R_i$, though the representation is not unique. The variance with respect to a standard von Neumann measurement of $X$ is $\text{Var}[X]^{\text{(vn)}}_\rho = \tr(X^2\rho)-\tr(X\rho)^2$, where $\tr(X^2\rho)= \left(\frac{d}{n}\right)\sum_{ijkl} x_ix_j\Re[\tr(R_i\sigma_j\sigma_k)]\Phi_{kl}P(R_l|\rho)$. Substituting in the expression for $\Re[\tr(R_i\sigma_j\sigma_k)]$ yields
	\begin{align}
	\label{ugly_variance}
\tr(X^2\rho)&=\frac{1}{2}\Bigg[(d+1)(d+2)\sum_{k} \left(\sum_i P(R_k|R_i)x_i\right)^2P(R_k|\rho)\nonumber\\
&-\left(\frac{d}{n}\right)\sum_{ij}x_iP(R_i|R_j)x_j-2d\langle x\rangle_\mu\langle x\rangle_\rho-d^2\langle x\rangle_\mu^2\Bigg].
\end{align}	
We assume that $x \in \text{col}(P)$ so that $x_i=\tr(R_i \tilde{X})$ for some Hermitian $\tilde{X}$. By the 2-design property then,
\begin{align}
\sum_jP(R_i|R_j)x_j&=
	\sum_j	P(R_i|R_j)\tr(R_j\tilde{X}) =d\tr\left(\frac{1}{n}\sum_j \sigma_j^{\otimes 2} (R_i \otimes \tilde{X}) \right)\\
	&=\frac{1}{d+1}\tr\big((I + T_{21})(R_i \otimes \tilde{X})\big)=\frac{1}{d+1}\left(\frac{d}{n}\tr(\tilde{X})+\tr(R_i\tilde{X})\right),
\end{align}
where $\tr(\tilde{X})=\sum_i \tr(R_i\tilde{X})=\sum_i x_i=n\langle x\rangle_\mu$, so that 
\begin{align}
	\sum_j P(R_i|R_j)x_j&=\frac{d\langle x\rangle_\mu +x_i}{d+1}\\
	\left(\sum_j P(R_i|R_j)x_j \right)^2&=\frac{d^2\langle x\rangle_\mu^2+2d\langle x\rangle_\mu x_i+x_i^2}{(d+1)^2}\\
	\sum_{ij}x_iP(R_i|R_j)x_j&=\frac{n\big(d\langle x\rangle_\mu^2+\langle x^2\rangle_\mu\big)}{d+1}.
\end{align}
 Substituting these expressions into Eq.\ \ref{ugly_variance} yields Eq.\ \ref{simplified_var}.
\end{proof}

As we have seen, the restriction that $x \in \text{col}(P)$ is natural since already any Hermitian observable $X$ can be expressed with respect to the reference device in this form. Remarkably, this expression relates the second moment of $X$ with respect to a standard von Neumann measurement, whose outcomes are eigenvalues of $X$, to the second moment of $X$ \emph{with respect to the reference measurement}, where now the $x_i$'s are interpreted as numerical values assigned to reference measurement outcomes. Only for a 3-design is such a simple relationship possible, and this is what will allow us to characterize valid probability distributions in terms of a lower bound on the variance with respect to the reference measurement.

The restriction  $x \in\text{col}(P)$ (for real $x$) is equivalent to the assumption that $x_i=\tr(R_i\tilde{X})$ for some Hermitian $\tilde{X}$. By the 2-design property, if $X=\sum_i x_i R_i$ for $x_i=\tr(R_i\tilde{X})$, then $\tilde{X}=\left(\frac{n}{d}\right)\Big[(d+1)X - \tr(X)I\Big]$. Consequently, if the RHS of Eq.\ \ref{simplified_var} is nonnegative for some real $x \in \text{col}(P)$, then $\tr(X^2\rho)\ge0$ for $X=\frac{1}{d+1}\left(\frac{d}{n}\right)\Big[\tilde{X}+\tr(\tilde{X})I\Big]$; and if the RHS is nonnegative for \emph{all} real $x \in \text{col}(P)$, $\tr(X^2\rho)\ge0$ for all Hermitian $X$. We conclude that probability assignments $P(R_i|\rho)$ are valid if and only if
\begin{align}
\label{the_uncertainty_principle}
\forall x\in \text{col}(P): \text{Var}[x]_\rho\ge \frac{d}{d+2}\Big(\langle x^2\rangle_\mu -2\langle x\rangle_\mu\langle x\rangle_\rho\Big)-\langle x\rangle_\rho^2,
\end{align}
where  $\text{Var}[x]_\rho=\sum_i x_i^2 P(R_i|\rho) - \left(\sum_i x_i P(R_i|\rho)\right)^2$. In this way, the shape of quantum state space can be understood in terms of a variance based uncertainty principle: valid probability assignments on reference outcomes cannot be too sharp lest they violate a lower bound on the variance for any observable in $\text{col}(P)$.  We note that a related inequality was derived recently in \cite{chen2024nonstabilizernessenhancesthriftyshadow} in the context of bounding the variance of expectation values in shadow estimation.

\section{The Jordan product}

We can shed further light on the special role that 3-designs play in encoding the shape of quantum state space by considering their relationship to the \emph{Jordan algebra} of observables. To see this, let us return to the inequality in Eq.\! \ref{vn_var}, which we may reshape  into $\forall x: \sum_i x_i [\mathcal{L}_\rho]_{ij} x_j \ge0$, where
\begin{align} 
	\label{L}
	[\mathcal{L}_\rho]_{ij}=\sum_{kl} \Re[\tr(R_i\sigma_j\sigma_k)]\Phi_{kl}P(R_l|\rho).
\end{align}
	 From this we conclude that probability assignments $P(R_i|\rho)$ are valid if and only if $\mathcal{L}_\rho$ is positive semidefinite. In particular, for an unbiased 3-design reference device, Eq.\! \ref{L} simplifies to
\begin{align}
	[\mathcal{L}_\rho]_{ij}&=\frac{1}{2}\Bigg[(d+1)(d+2)\left(\frac{n}{d}\right)\sum_mP(R_m|R_i)P(R_m|R_j)P(R_m|\rho)\nonumber\\
	&- P(R_i|R_j) -P(R_i|\rho)-P(R_j|\rho)-\frac{d}{n}\Bigg],\nonumber
\end{align}
which has the virtue of depending only upon reference measurement probabilities, and which provides a straightforward way to check whether the variance bound is satisfied.

But there is another interpretation of the operator $\mathcal{L}_\rho$. Recall that under the Jordan product $A\odot B=\frac{1}{2}(AB+BA)$, Hermitian matrices over $\mathbb{C}$ form a \emph{Euclidean Jordan algebra} \cite{farautAnalysisSymmetricCones1994,staceyQuantumTheorySymmetry2019, wilceRoyalRoadQuantum2018}. A Jordan algebra is a nonassociative algebra which satisfies commutativity and the Jordan identity,
\begin{align}
A \odot B &= B \odot A \nonumber\\
A^2 \odot (B \odot A) &= A \odot (B \odot A^2) \nonumber.
\end{align}
If we define $L_A$ to be the linear operator which takes the Jordan product with $A$, that is, $L_A(B)=A\odot B$, the Jordan identity is equivalent to $\big[L_A, L_{A^2}\big]=0$. A \emph{Euclidean} Jordan algebra enjoys the additional property that there exists an inner product on the underlying vector space $\mathcal{V}$ such that $\forall A, B, C \in \mathcal{V}: \langle L_A (B), C\rangle=\langle B, L_A(C)\rangle$.

 By introducing an informationally complete reference measurement, we identify quantum states with probability distributions. We may then ask: how can we represent the Jordan product in terms of probabilities? Treating states $\rho$ and $\tau$ as observables of the Jordan algebra, if $|\rho \odot \tau)=L_\rho|\tau)$,  using the resolution of the identity $\B{S}\Phi \B{R}=I$, we have
\begin{align}
\B{R}L_\rho|\tau)=\B{R}L_\rho\B{S}\Phi \B{R}|\tau)=\mathcal{L}_\rho\Phi P(R|\tau),
\end{align}
where $\mathcal{L}_\rho=\B{R}L_\rho\B{S}$, whose matrix elements are
\begin{align}
\label{Lelements}
[\mathcal{L}_\rho]_{ij} &= \tr(R_i L_\rho(\sigma_j))=\frac{1}{2}\big(\tr(R_i\rho \sigma_j) + \tr(R_i\sigma_j\rho)\big)\nonumber\\
&=	\sum_{kl} \Re[\tr(R_i\sigma_j\sigma_k)]\Phi_{kl}P(R_l|\rho).
\end{align}
Indeed, this is precisely the matrix we developed earlier, whose positive semidefiniteness diagnoses the validity of probability assignments $P(R_i|\rho)$. 

Significantly, Eq.\! \ref{Lelements} reveals that the three-index tensor $ \Re[\tr(R_i\sigma_j\sigma_k)]$ encodes the structure coefficients for the Jordan product on $d\times d$ Hermitian matrices over $\mathbb{C}$, and thus fully defines it by its action on the reference states and effects. Since pure states are idempotents of the Jordan algebra  with trace 1, the structure coefficients implicitly determine the geometry of the state space. At the same time, we have shown that the components of this tensor can be extracted from the joint probability distribution $P(R_i,R_j,R_k|\mathcal{M}_3)$.  Finally, taking our reference measurement to be an unbiased 3-design means that  $P(R_i,R_j,R_k|\mathcal{M}_3)=\frac{1}{n}\sum_m P(R_i|R_m)P(R_j|R_m)P(R_k|R_m)$ so that $P(R_i|R_j)$ \emph{alone} is sufficient to characterize the Jordan product, and through this algebraic structure, the entire state space. 

\section{A gentle modification}

As we have seen for a 2-design, the Born rule appears as a remarkably gentle modification of the classical law of total probability,
\begin{align}
P(E|\rho) &= \sum_i P(E|R_i)\Big\{(d+1)P(R_i|\rho) - \frac{d}{n}\Big\}.
\end{align}
We will now see that our results imply that for a 3-design, the quantum Jordan product appears as a similarly gentle modification of the classical Jordan product. In the previous section, we worked out the formula for the quantum Jordan product on probability vectors: expressing it instead directly in terms of valuations on the reference measurement makes the correspondence with the classical rule most manifest.

But first let us remind ourselves what the classical Jordan product is in this case. As we discussed in Chapter \ref{ch:probabilities}, classically, real valued random variables, which are valuations $\Omega \rightarrow \mathbb{R}$ on a sample space $\Omega$, form an algebra. Given valuations $x(\omega)$ and $y(\omega)$, we may form their product $(x \circ y)(\omega) = x(\omega)y(\omega)$. Treating $x$ and $y$ as vectors, this amounts to the elementwise product $x \circ y$. Quantum mechanically, the role of the classical sample space is played by the outcomes of the reference measurement, and if $x$ and $y$ are two valuations on reference outcomes, we may form their classical product $x \circ y$: in fact, since $\circ$ is commutative and  associative, this is nothing other than the Jordan product on $\mathbb{R} \oplus \cdots \oplus \mathbb{R} \simeq \mathbb{R}^n$. As we have seen, valuations on the reference measurement are equivalent to the assignment of a quantum mechanical observable. But of course, quantum mechanically there is another way of multiplying observables: the Jordan product $X \odot Y = (XY + YX)/2$. Expressing this latter directly in terms of valuations yields the classical formula with a gentle quantum correction.
\begin{theorem}
	\label{jordan-gentle}
For $x, y\in \col(P)$, the quantum Jordan product $X \odot Y = (XY+YX)/2$ for $X=\sum_i x_i R_i$ and $Y=\sum_i y_i R_i$ may be expressed with respect to an unbiased 3-design reference measurement as
\begin{align}
x \odot y = \gamma(x \circ y) + (1-\gamma)\Big( \overline{x}y + \overline{y}x - (x \cdot y)u\Big)/n,
\end{align}
where $x \circ y$ is the Hadamard or elementwise vector product, $\overline{x}$ denotes the sum of the vector $x$, and $\gamma = \frac{1}{2}\frac{d+2}{d+1}$. 
\end{theorem}
\begin{proof}
See appendix \ref{app3}.
\end{proof}
\begin{remark}
 This formula is valid for $x, y \in \col(P)$: $x \odot y$ may not be in $\col(P)$, and hence ought to be projected into it.  Even if $x \odot y \notin \col(P)$, however, since $P(R|\rho) \in \col(P)$, we have $\langle X \odot Y \rangle = \sum_i (x\odot y)_i P(R_i|\rho)$, that is, the formula properly reproduces the expectation value since contracting with the probability vector will kill any components in $\col(P)^\perp$. 
\end{remark}
\begin{remark}
Continuing to count, a $4$-design will make the symmetric product of three operators look as close as possible to $x \circ y \circ z$, and so on. But for characterizing quantum mechanics, counting to 3 is enough.
\end{remark}
\begin{remark}
In appendix \ref{associator}, we show how unitary maps may be expressed in terms of this formula.
\end{remark}

\section{Conclusion}

We have thus shown that the shape of quantum state space can be understood in terms of an uncertainty principle which constrains the probabilities one ought to assign to the outcomes of a 3-design reference measurement. Compatibility with this uncertainty principle can be diagnosed through the positive semidefiniteness of a particular operator $\mathcal{L}_\rho$ constructed from reference probabilities. We have also provided a set of scalar constraints that pick out pure state probability distributions, which can alternatively be summarized by a single vector constraint. Conceptually, these constraints can be understood as entropic uncertainty principles, and operationally they relate to the agreement probability on multiple copies of the reference measurement.

 Crucially, each term that appears in our equations is grounded in a probability assignment, and even better, these probabilities refer to the behavior of a single reference measurement. The possibility of achieving this rests on the delicate interplay between unitary symmetry and the Jordan algebra of observables.  The algebraic structure of quantum theory implies that the 3rd moment of quantum state space determines them all, and so does a reference measurement furnished by a 3-design. This further vindicates the centrality of 3-designs already suggested by their optimality in classical shadow estimation tasks.

 Our result holds particular significance for the QBist research program in the foundations of quantum mechanics. As we have explained, QBism argues that quantum theory is best understood not as a description of physical reality, but rather as a set of normative guidelines for gambling on the consequences of one's actions in a world undergoing ceaseless creation \cite{fuchs2019qbismquantumtheoryheros, fuchsQbismWhereNext2023, debrota2024quantumdynamicshappenspaper}. Consequently, QBist ``reconstructions'' of quantum mechanics proceed \cite{PhysRevA.104.022207, Appleby_2017} by motivating the constraints on probability assignments implied by quantum theory in the same spirit in which de Finetti derived the usual rules of probability theory by contemplating what constraints a gambler ought to place on their different bets in order to prevent a sure loss.

The simplicity of our result is very much in the spirit of \cite{staceyQuantumTheorySymmetry2019}, which suggests that the constraints implied by quantum theory are in some sense the ``most symmetrical'' compatible with what the author calls ``vitality,'' i.e.\ the nonexistence of a hidden variable model. Indeed, our result shows that these constraints may be understood as a fundamental expression of complementarity. That said, our derivation presumes a prior knowledge of traditional quantum theory: the significance of our work here is that it exposes the structure that must be aimed for in any future reconstructive effort, and it is to the latter that we turn in Chapter \ref{ch:reconstruction}.

Here we observe that previous efforts at QBist reconstruction  \cite{Appleby_2017} took as their starting place the generalization of the image of quantum state space within the probability simplex induced by a symmetric informationally complete (SIC) reference measurement. SIC measurements have a host of virtues: the corresponding states form a simplex in quantum state space whose vertices are pure states; the conjecture of their existence in any Hilbert space dimension has led to a fruitful and unexpected interplay between physics and algebraic number theory \cite{Appleby_2017}. SIC states, however, only form 2-designs, and thus the Jordan structure coefficients cannot be extracted directly from probability assignments assigned to a single reference device. Breaking this barrier is the central innovation of the present work. 
 
More specifically, the SIC based reconstructive effort began from the observation that the 2-design condition implies  the inner product between any two probability vectors must lie between certain upper and lower bounds. Inspired by this, \cite{Appleby_2017} defined a \emph{qplex} to be set of probability vectors which mutually satisfy these bounds, to which no more elements can be added without inconsistency, and which contains a simplex of pure states (corresponding to a SIC). The goal of the program was to motivate these bounds on independent grounds, situate quantum theory in the vaster landscape of qplexes, and provide a principle by which quantum theory could be identified within this landscape. This approach is consonant with developments in quantum foundations over the last 25 years where the study of so-called \emph{generalized probabilistic theories} \cite{mullerProbabilisticTheoriesReconstructions2021, barnum2013postclassicalprobabilitytheory} has played a central role in providing new perspectives on quantum theory.

 The authors of \cite{Appleby_2017} demonstrated that any qplex whose symmetry group is a stochastic subgroup of the orthogonal group isomorphic to the unitary group must correspond to quantum theory, and vice versa: moreover, the existence of such a subgroup is equivalent to the existence of a particular SIC. At the same time, the authors left open the possibility of there being a simpler principle which could pick out quantum theory among the qplexes. The present work shows that demanding the fundamental reference measurement to have the properties of a 3-design, rather than a  2-design, means that a single finite set of probability distributions is sufficient to characterize the theory, which is much more tractable, analytically and computationally, as well as more conceptually satisfying.
 
 Just as a qplex generalizes the representation of quantum mechanics according to a SIC, it is natural to consider analogous generalizations of a 3-design representation, what we might call \emph{3-qplexes}. For instance, one could explore the landscape of all state spaces defined by an uncertainty principle as in Eq.\! \ref{the_uncertainty_principle}, without at first restricting $P(R_i|R_j)$, the probabilities that characterize the reference measurement, to correspond to an actual quantum 3-design. Which properties of quantum theory are preserved in such generalized theories, and which fall by the wayside? Just as identifying quantum theory among the qplexes led to an alternative characterization of SICs themselves, picking out quantum theory among the 3-qplexes would lead to an alternative characterization of 3-designs. A new approach to identifying and constructing 3-designs, in particular, of self-testing them, would have significant practical application in quantum computing and beyond---it is to this subject that we now turn.
 
\section{Appendix: The first three moments}
\label{app3}

For reference, we first prove in one place some basic properties of the first three moments of an unbiased quantum 3-design. In general, for an unbiased $t$-design $\{\sigma_i\}$, we have
\begin{align*}
 \mathcal{M}_t = \frac{1}{n} \sum_i \sigma_i^{\otimes t} = \int |\psi\rangle\langle \psi|^{\otimes t} d\psi = \binom{d+t-1}{t}^{-1}\Pi_{\text{sym}^t} = \binom{d+t-1}{t}^{-1} \frac{1}{t!} \sum_{\pi \in S_t} T_\pi.
\end{align*}
We will again and again use the generalized swap trick. Let $T_{\pi}$ be the unitary operator which performs a cyclic permutation of $m$ subsystems. Then
\begin{align}
\tr\big(T_\pi (A_1 \otimes \dots \otimes A_m)\big) = \tr(A_1 \dots A_m).
\end{align}
We now examine the implications for unbiased $t$-designs that follow from the structure of the moment operators $\mathcal{M}_1, \mathcal{M}_2,$ and $\mathcal{M}_3$. 
\begin{enumerate}
\item $\mathcal{M}_1 = \frac{1}{d} I $. For an unbiased $1$-design, therefore, $\frac{1}{n}\sum_i \sigma_i = \frac{1}{d}I$. Letting $R_i = \frac{d}{n}\sigma_i$ yields $\sum_i R_i = I$, from we conclude that $\{R_i\}$ may be viewed as POVM elements corresponding to a measurement.
\item $\mathcal{M}_2 = \frac{1}{d(d+1)}(I + T_{21})$. This yields 
\begin{align}
\tr(\mathcal{M}_2(X \otimes I))= \frac{1}{n}\sum_i \tr(\sigma_i X)\sigma_i = \frac{1}{d(d+1)}\Big(\tr(X) I + X\Big),
\end{align}
which we may rearrange into a resolution of the identity,
\begin{align}
X &= (d+1)\sum_i \tr(R_i X)\sigma_i - \tr(X) I\\
&= \sum_i \tr(R_i X) \Big[ (d+1)\sigma_i - I \Big] \\
&= \sum_i \tr(R_i X) \tilde{R_i},
\end{align}
where $\{\tilde{R}_i\}$ are elements of the frame dual to the frame provided by the POVM. Alternatively, 
\begin{align}
X &= \sum_i \left[(d+1)\tr(R_i X) - \frac{d}{n}\tr(X)\right]\sigma_i = \sum_{ij} \Phi_{ij}\tr(R_j X) \sigma_i,
\end{align}
where $\Phi = (d+1)I - \frac{d}{n}J$: here $J$ is the matrix of all 1's. In this way, the form of the Born matrix may be derived from the second moment operator $\mathcal{M}_2$. Indeed,
\begin{align}
P(E_i|\rho) &= \tr(E_i \rho) = \tr\left( E_i \sum_{jk} \Phi_{jk}P(R_k|\rho)\sigma_j \right) \\
&= \sum_{jk} P(E_i|R_j) \Phi_{jk}P(R_k|\rho)\\
&=\sum_j P(E_i|R_j)\left[(d+1)P(R_j|\rho) - \frac{d}{n}\right],
\end{align}
reproduces the Born rule. At the same time, 
\begin{align}
\tr(\mathcal{M}_2(X \otimes Y)) = \frac{1}{n}\sum_i \tr(\sigma_i X)\tr(\sigma_i Y) = \frac{1}{d(d+1)}\Big(\tr(X) \tr(Y) + \tr(XY)\Big).
\end{align}
Let $x_i = \tr(\sigma_i \tilde{X})$. Noticing that $\sum_i x_i = (n/d)\tr(\tilde{X})$, the action of $P(R_i|R_j)=\tr(R_i \sigma_j)$ on any such vector $x$ can be expressed
\begin{align}
\sum_j P(R_i|R_j)x_j &= \frac{d}{n}\sum_j \tr(\sigma_i \sigma_j)\tr(\sigma_j \tilde{X}) =  \frac{1}{(d+1)}\Big(\tr(\tilde{X}) + \tr(\sigma_i \tilde{X})\Big) \\
&= \frac{1}{(d+1)}\Big((d/n)\sum_j x_j + x_i\Big).
\end{align}
This holds for any vector $x \in \col(P(R|R))$ since if $x \in \col(P(R|R))$, then $x_i = \sum_j P(R_i|R_j)y_j = \tr\big(\sigma_i (d/n)\sum_j y_j \sigma_j\big) = \tr(\sigma_i \tilde{X})$.  
\item $\mathcal{M}_3 = \frac{1}{d(d+1)(d+2)}(I + T_{132} + T_{321} + T_{213} + T_{312} + T_{231})$. On the one hand,
\begin{small}
\begin{align}
& \tr_1(\mathcal{M}_3(X \otimes I \otimes I))=\frac{1}{n}\sum_i \tr(\sigma_i X) \sigma_i \otimes \sigma_i\nonumber \\
&=\frac{1}{d(d+1)(d+2)}\Bigg(\tr(X) I \otimes I + \tr(X) T_{21} + I \otimes X + X \otimes I + (I \otimes X)T_{21} + (X \otimes I)T_{21}\Bigg)\\
&=  \frac{2}{d(d+1)(d+2)}\Big(\big(\tr(X)I\otimes I + I \otimes X + X \otimes I\big)\Pi_{\text{sym}^2}\Big).
\end{align}
\end{small}
On the other hand,
\begin{align}
&\tr_{12}(\mathcal{M}_3(X \otimes Y \otimes I)) = \frac{1}{n}\sum_i \tr(\sigma_i X) \tr(\sigma_i Y)\sigma_i\nonumber\\
&= \frac{1}{d(d+1)(d+2)} \Big((\tr(X)\tr(Y) + \tr(XY))I + \tr(X)Y + \tr(Y) X + 2 X \odot Y\Big),
\end{align}
where $X \odot Y = (XY+YX)/2$ is the Jordan product on $d\times d$ Hermitian matrices over $\mathbb{C}$.
\end{enumerate}
\noindent We now turn to the proof of Theorem \ref{jordan-gentle}.
\begin{proof}
Let $X = \sum_i x_i R_i$ and $Y=\sum_i y_i R_i$ for $x,y\in \col(P)$. We will rewrite 
\begin{align}
	\label{to_rewrite}
&\frac{1}{n}\sum_i \tr(\sigma_i X) \tr(\sigma_i Y) \sigma_i \nonumber\\
&=\frac{1}{d(d+1)(d+2)} \Big((\tr(X)\tr(Y) + \tr(XY))I + \tr(X)Y + \tr(Y) X + 2 X \odot Y\Big)
\end{align}
entirely in terms of $x$ and $y$.  Let $u = (1,\dots,1)^\dagger$ so that 
\begin{align}
\tr\left(\sigma_i X \right) &= \sum_j\tr\left(\sigma_i R_j \right)x_j =\frac{1}{d+1}\Big(\frac{d}{n}u^\dagger x + x_i\Big),
\end{align}
and
\begin{align}
\tr(XY) &= \sum_{ij}x_i \tr(R_i R_j) y_j =\frac{d}{n} \sum_{i}x_i \left[\frac{1}{d+1}\left(\frac{d}{n}u^\dagger y + y_i\right)\right]\\
&=\frac{1}{d+1}\frac{d}{n}\Bigg( \frac{d}{n} (u^\dagger x)(u^\dagger y) + x^\dagger y\Bigg).
\end{align}
Substituting these identities into Eq. (\ref{to_rewrite}) and expressing each term as linear combination of the POVM elements $\{R_i\}$ yields
\begin{align}
&\sum_i \frac{1}{d+1}\Big(\frac{d}{n}u^\dagger x + x_i\Big) \frac{1}{d+1}\Big(\frac{d}{n}u^\dagger y + y_i \Big) R_i  \nonumber \\
&=\frac{1}{(d+1)(d+2)} \Bigg[\left(\frac{d^2}{n^2}(u^\dagger x)(u^\dagger y)+ \frac{1}{d+1}\frac{d}{n}\Bigg( \frac{d}{n} (u^\dagger x)(u^\dagger y) + x^\dagger y\Bigg)\right)\sum_i R_i\nonumber\\
& + \frac{d}{n}(u^\dagger x) \sum_i y_i R_i + \frac{d}{n}(u^\dagger y) \sum_i x_i R_i + 2 \sum_i (x \odot y)_i R_i\Bigg].
\end{align}
Matching coefficients of $R_i$ yields
\begin{align}
&\frac{1}{(d+1)^2}\left(\frac{d^2}{n^2}(u^\dagger x)(u^\dagger y) + \frac{d}{n} (u^\dagger x) y_i + \frac{d}{n} (u^\dagger y) x_i + x_i y_i\right) \nonumber \\
&= \frac{1}{(d+1)(d+2)}\Bigg(\frac{d^2}{n^2}(u^\dagger x)(u^\dagger y)+  \frac{d^2}{n^2}\frac{1}{d+1} (u^\dagger x)(u^\dagger y) \nonumber \\
&+ \frac{1}{d+1} \frac{d}{n}x^\dagger y + \frac{d}{n}(u^\dagger x) y_i + \frac{d}{n}(u^\dagger y) x_i + 2(x \odot y)_i\Bigg),
\end{align}
which simplifies to
\begin{align}
x \odot y 
&= \frac{1}{2}\Bigg(
\frac{d+2}{d+1}\,(x \circ y) 
+ \frac{1}{d+1}\frac{d}{n}
\Big( (u^\dagger x) y 
+ (u^\dagger y) x 
- (x^\dagger y) u\Big)
\Bigg)\\
&=  \gamma(x \circ y) + (1-\gamma)\Big( \overline{x}y + \overline{y}x - (x \cdot y)u\Big)/n,
\end{align}
where $x \circ y$ is the Hadamard or elementwise vector product and $\gamma = \frac{1}{2}\frac{d+2}{d+1}$. Since a 3-design furnishes an overcomplete representation, however, this expression for the quantum Jordan product is not unique. Letting $\Pi=P\Phi$ be the projector onto $\col(P)$, then $\Pi(x \odot y)$ gives the canonical representation of the product on $\col(P)$.
\end{proof}

\section{Appendix: The associator and the von Neumann equation}
\label{associator}

Unitary state updates in quantum mechanics are usually expressed in terms of the commutator $[X,Y] = XY - YX$ which measures how noncommutative the product of two Jordan elements is. Given an operator $O$ and a Hamiltonian $H$, one may express the time derivative of $O$ in terms of the von Neumann equation,
\begin{align}
\frac{d}{dt} O(t) &= -i[H, O(t)], 
\end{align}
so that $O(t) = U(t) O U(t)^\dagger$ where $U= e^{-i H t}$ is a unitary operator, and we note that in the Heisenberg picture, we ought to flip the sign so that $O(t)=U(t)^\dagger O U(t)$.  In this appendix, using the results of \cite{townsendJordanFormulationQuantum2016, akhiezerCommutatorMapReal2015, NCategoryCafe,farautAnalysisSymmetricCones1994, McCrimmon2003-kp}, we show how unitary state updates may be alternatively expressed in entirely Jordan algebraic terms using the \emph{associator},
\begin{align}
(X, Y, Z) &= (X\odot Y)\odot Z - X \odot (Y \odot Z),
\end{align}
which measures how nonassociative is the product of three Jordan elements.

First, let us give a little context. Suppose we are working with an arbitrary Euclidean Jordan algebra, and we consider a one parameter family of automorphisms $G_t$. We can then consider a continuous reversible ``evolution'' of an element $x$, $x(t) = G_t x$. By definition, $G_0=u$, where $u$ is the identity element of the EJA; $G_{s+t} = G_s G_t$; and $G_t(x \odot y) = G_t x\odot G_t y$. Now for small $t$, $G_t = I + t D + O(t^2)$ for some linear map $D$. Suppose $G_t$ acts on the Jordan product $x \odot y$. On the one hand,
\begin{align}
G_t(x \odot y) &= x \odot y + t D(x \odot y) + O(t^2).
\end{align}
On the other hand,
\begin{align}
G_tx  \odot G_t y &= (x + t Dx) \odot (y + t Dy) + O(t^2)\\
 &= x \odot y + t(Dx \odot y + x \odot Dy) + O(t^2).
\end{align}
Equating coefficients of $t$ gives
\begin{align}
D(x \odot y) = Dx \odot y + x \odot Dy,
\end{align}
that is, $D$ satisfies the \emph{Leibniz rule}, and so is called a \emph{derivation}. It represents an infinitesimal change that preserves a product, in this case, the Jordan product. Indeed, if we consider the differential equation
\begin{align}
\frac{d}{dt}x(t) = Dx(t),
\end{align}
since $D$ is a derivation, we have
\begin{align}
\frac{d}{dt}(x \odot y)(t) = \frac{d}{dt}x(t) \odot y + x \odot \frac{d}{dt}y(t).
\end{align}
Conversely, every derivation integrates to a one-parameter family of automorphisms. The differential equation $\frac{d}{dt}x(t) = Dx(t)$ has the solution $x(t) = e^{tD}x$. We must show that this preserves the Jordan product. On the one hand, let $f(t) = e^{tD}(x \odot y)$. Then $\frac{d}{dt}f(t) = Df(t)$. On the other hand, let $g(t) = e^{tD}x \odot e^{tD}y$. Differentiating gives $\frac{d}{dt}g(t) = De^{tD}x \odot e^{tD}y + e^{tD}x \odot D e^{tD}y = D g(t)$ by the Leibniz rule. Since $f$ and $g$ satisfy the same linear ODE ($\frac{d}{dt}z(t) = Dz(t)$) with the same initial conditions $f(0)=g(0)= x\odot y$, they must be equal. Thus $e^{tD}(x \odot y) = e^{tD}x \odot e^{tD}y$, as desired.

Now let $L_a x = a \odot x$ be the linear operator that performs the Jordan product. It is a remarkable theorem of Jacobson and Koecker that any derivation of an EJA $D$ lies in $ \linspan\{[L_a, L_b]\}$ \cite{McCrimmon2003-kp,farautAnalysisSymmetricCones1994,jacobson1968structure}: in other words, the derivations of an EJA are precisely given by linear combinations of commutators of Jordan product operators. But 
\begin{align}
[L_a, L_b]c = a \odot (b \odot c) - b \odot (a \odot c) = (a,b,c),
\end{align}
which is precisely the associator we defined above.

For associative Jordan algebras, for example, $\mathbb{R}^n$ equipped with the elementwise product $\circ$, corresponding to classical probability theory, the associator always vanishes. While one can formulate continuous ``evolution'' in classical theory using the formalism of continuous Markov chains and their rate matrices, this evolution does not originate intrinsically from the theory itself. In contrast, the geometric meaning of derivations is particularly transparent for the so-called spin-factor Jordan algebras which are defined on $
V_n = \mathbb{R} \oplus \mathbb{R}^n$ and whose Jordan product is
\begin{align}
(x_1,x_0)\odot(y_1,y_0)
=
(x_1y_1+x_0\cdot y_0,\;
x_1 y_0+ y_1 x_0).
\end{align}
For simplicity, let us pick elements $a=(0,a_0)$ and $b=(0, b_0)$. Then
\begin{align}
L_a(x_1,x_0)
=
(a_0 \cdot x_0,x_1 a_0),
\qquad
L_b(x_1,x_0)
=
(b_0\cdot x_0,x_1 b_0).
\end{align}
A straightforward calculation gives
\begin{align}
[L_a,L_b](x_1,x_0)
=
\left(
0,\;
(b_0\cdot x_0)u-(a_0\cdot x_0)b_0
\right).
\end{align}
Notice that the scalar component is unchanged. As for the vector part, suppose that $n=3$, and recall that in three dimensions $(b_0\cdot x_0)a_0-(a_0\cdot x_0)b_0
=
(a_0\times b_0)\times x_0$, 
so that if $\frac{d}{dt}(x_1, x_0)(t) = [L_a, L_b](x_1, x_0)$, we have
\begin{align}
\frac{d}{dt}x(t)
=
\omega\times x(t),
\end{align}
where $\omega=a_0\times b_0$.
This is precisely the equation for a rigid-body rotation around the axis $\omega$.

The further remarkable fact about quantum mechanics over $\mathbb{C}$ is that all these equations can be rewritten not in terms of an associator with two observables, but in terms of commutators with a \emph{single} observable. In fact, the $n=3$ spin-factor case is just such an example as it coincides with a qubit. Quantum mechanics over $\mathbb{C}$ thus generalizes in a surprising way the fact that in three dimensions, any vector representing an axis of rotation can be expressed as the cross product of two vectors. We first observe that the associator can be rewritten as a nested commutator. Letting $X \odot Y = (XY + YX)/2$ be the Jordan product on Hermitian matrices, we have
\begin{align}
(X, O, Y) &= \frac{1}{2}(XO + OX) \odot Y - X \odot \frac{1}{2}(OY + YO)\\
&= \frac{1}{4}\Big(XOY + YXO + OXY + YOX - XOY - OYX - XYO - YOX\Big)\\
&= \frac{1}{4}\Big(YXO + OXY - OYX - XYO\Big)\\
&= \frac{1}{4}\Big(O[X,Y] - [X,Y]O \Big)\\
&= - \frac{1}{4}[[X,Y], O].
\end{align}
This holds for any EJA built out of Hermitian matrices. For quantum mechanics over $\mathbb{C}$ in particular, however, we have the von Neumann equation $\frac{d}{dt} O(t) = -i[H, O(t)]$. Writing $H = \tr(H) I/d + H_0$ for $\tr(H_0)=0$, we see that the von Neumann equation cares only about the traceless part of the Hamiltonian,
\begin{align}
\frac{d}{dt} O(t) &=  -i [\tr(H) I/d + H_0, O]= -i[H_0, O].
\end{align}
What is not trivial is that over $\mathbb{C}$, \emph{any} traceless Hermitian matrix $H_0$ can be written as $i$ times the commutator between two Hermitian matrices $X,Y$: $H_0 = i[X, Y]$ \cite{akhiezerCommutatorMapReal2015}. Thus let $H_0 = -\frac{1}{4}i[X,Y]$ so that
\begin{align}
\frac{d}{dt}O(t)&=-i\left[-\frac{1}{4}i[X,Y], O(t)\right]=-\frac{1}{4}[[X,Y], O(t)]= (X, O(t), Y).
\end{align}
As desired, we have reexpressed the von Neumann equation in terms of the associator instead of the commutator, which in fact is the form that generalizes to any EJA.

What to make of this in the light of 3-designs? We have for $x,y \in \col(P)$,
\begin{align}
x\odot y = \gamma(x \circ y) + (1-\gamma)\Big( \overline{x}y + \overline{y}x - (x \cdot y)u\Big)/n.
\end{align}
 Let $\Pi=P \Phi $ be the projector onto $\col(P)$. Making sure to project onto $\col(P)$ after an application of $\odot$,  we find after some algebra, 
\begin{align}
&(x,y,z)\\
 &= \frac{1}{4(d+1)^2}\Bigg\{(d+2)^2 \Big( \Pi(x \circ y) \circ z - x \circ \Pi (y \circ z)\Big) \nonumber \\
 &+ \left(\frac{d}{n}\right)^2\Big( \big( n (y \cdot z) -  \overline{y}\overline{z}\big) x - \big(n (x \cdot y) - \overline{x}\overline{y}\big)z + \big(\overline{z} (x \cdot y) - \overline{x} (y \cdot z)\big)u\Big)\Bigg\}.\nonumber
\end{align}
Meanwhile, let $L_x y = x \odot y$ so that
\begin{align}
L_x = \frac{1}{2}\frac{1}{d+1}\Big( (d+2)\diag{x} + \frac{d}{n}\big( \overline{x} I + x u^\dagger - u x^\dagger \big)\Big).
\end{align}
Similarly, we can consider the linear operator $A_{x,y}$ such that $A_{x,y}o = (x,o,y)$,
\begin{align}
A_{x,y}
&= \frac{1}{4(d+1)^2}\Bigg\{
(d+2)^2\big(\diag{y}\Pi \diag{x} - \diag{x}\Pi \diag{y}\big)\\
&+ \left(\frac{d}{n}\right)^2
\Big(
n\,(x y^{\dagger} - y x^{\dagger})
+ \overline{x}\,(y u^{\dagger} - u y^{\dagger})
+ \overline{y}\,(u x^{\dagger} - x u^{\dagger})
\Big)
\Bigg\}.\nonumber
\end{align}
For compactness, writing $x \wedge y = xy^{\dagger} - yx^{\dagger}$, we have
\begin{align}
\Pi A_{x,y}
=
\frac{1}{4(d+1)^2}\left\{
(d+2)^2 [\Pi \diag{y}, \Pi \diag{x}]
+\left(\frac{d}{n}\right)^2\Big(
n\,x\wedge y+\overline{x}\,y\wedge u+\overline{y}\,u\wedge x
\Big)
\right\},
\end{align}
so that
\begin{align}
O(t) = U(t) O U(t)^\dagger = e^{-iHt} O e^{iHt} = \sum_i o(t)_i R_i,
\end{align}
where $o(t) = \exp(\Pi A_{x,y} t)o$, $O = \sum_i o_i R_i$ for $o\in \col(P)$, and  $H = -\frac{1}{4}i[X,Y]$ for  $X= \sum_i x_i R_i$ and $Y = \sum_i y_i R_i$. We may connect this to the general reference measurement representation of a unitary map. Let $P(R_i|R_j^{(U)}) = \tr(R_i U \sigma_j U^\dagger)$. Then
\begin{align}
\sum_{jk}P(R_i|R_j^{(U)})\Phi_{jk} P(R_k|\rho)
&= \sum_j \tr\left( R_i U \sum_{jk} \Phi_{jk} \tr(R_k\rho)\sigma_j U^\dagger\right)\\
&=\tr(R_i U \rho U^\dagger) \\
&= P(R_i|\rho^{(U)}).
\end{align}
At the same time, from the above considerations,
\begin{align}
P(R|R^{(U)})\Phi = \Pi \exp(\Pi A_{x,y} t)\Pi ,
\end{align}
where we sandwich the operator with $\Pi$'s since the expression $P(R_i|R_j^{(U)}) = \tr(R_i U \sigma_j U^\dagger)$ does not assume that $P(R_i|R_j^{(U)})\Phi$ acts solely on vectors in $\col(P)$ although it produces only vectors in $\col(P)$. In this way, unitary maps can be represented entirely in Jordan algebraic terms.

%% file: sections/reconstruction.tex
\UMBchapter{Reconstruction}
\label{ch:reconstruction}

\section{Introduction}

In the last chapter, we began within quantum mechanics, assumed our reference measurement was constructed from a complex projective 3-design, and then presented an image of quantum state space within the probability simplex, showing that it can be defined in terms of an uncertainty principle. In particular, we showed that the geometry of quantum mechanics as a whole is encoded in the single matrix $P(R|R)$ which characterizes the reference measurement itself. The keystone in the argument was the connection between the third tensor moment of quantum state space and the Jordan algebra of observables. In this chapter, we go in the opposite direction. We begin from almost nothing: the gambler equipped with the nonclassical coherence criterion $P(E|\rho) = P(E|R)\Phi P(R|\rho)$ which we have seen can be derived on very general grounds. We then proceed to place increasingly strong constraints on the matrix $P(R|R)$ until we arrive back at quantum mechanics, and can conclude that the reference measurement in fact forms a 3-design. The conceptual innovation is that just as we may consider the gentlest possible modification of the law of total probability, the Protourgleichung, in the same way we may consider the gentlest possible modification of the classical rule for multiplying valuations on the reference outcomes. We make a series of assumptions which then allow us to identify this rule with the quantum Jordan product. On the one hand, this reconstruction of quantum mechanics allows us to see precisely how it is, from the inside as it were, that 3-designs make the quantum Jordan product look as close as possible to the classical Jordan product. In the QBist spirit, it is grounded entirely in constraints on the reference measurement. On the other hand, our result may be interpreted in practical terms as a way of \emph{self-testing} \cite{supicSelftestingQuantumSystems2020} 3-design measurements. If $P(R|R)$ satisfies our constraints, we show it must have a Hilbert space representation, and in fact correspond to an unbiased complex-projective $3$-design: other methods, e.g., calculating the frame potential, do not in themselves guarantee that such a representation exists. Moreover, once our conditions are met, if $P(R|R)$ minimizes the $t$'th order frame potential, the same framework certifies $t$-designs for $t \ge 3$.

\section{The Urgleichung}

We begin as always with the gambler trying to make better decisions. We assume they have identified a reference measurement $\{R_i\}_{i=1}^n$ for a domain they are interested in so that they adopt the nonclassical coherence condition
\begin{align}
P(E|\rho) = P(E|R)\Phi P(R|\rho),
\end{align}
where $\Phi$ is a chosen $\{1\}$-inverse of $P(R|R)$, the conditional probability matrix which characterizes the reference measurement. Guided by simplicity, we assume that $P\equiv P(R|R)$ is in fact symmetric $(P=P^T)$, and so bistochastic. Thus $\col(P)=\row(P)$, and so probability vectors $P(R|\rho)$, response functions $P(E|R)$, and valuations all live in the same subspace: $\col(P)$. We denote the projector onto this subspace $\Pi = P\Phi$. Finally, we also assume that $P$ is constant along its diagonal: $\forall i: P(R_i|R_i) = \text{const}$. 

Next, we assume that $\Phi=\alpha I + \beta J$, where $J$ is the matrix of all 1's, is a Born matrix for $P$: thus the nonclassical coherence condition takes Protourgleichung form. If we require that $\Phi$ is quasistochastic, having negative entries, but with columns summing to 1, so that it preserves the normalization of probability vectors, then we must have $\alpha + n \beta = 1$, so that $\beta = (1-\alpha)/n$. 

\begin{lemma}
Suppose $P\Phi P = P$ for $\Phi = \alpha I + \beta J$ where $\beta = (1-\alpha)/n$. Then 
\begin{align}
\forall y \in \col(P) : P y = \frac{1}{\alpha}y + \left(1-\frac{1}{\alpha}\right)\overline{y}u/n,
\end{align}
for $u=(1,\dots, 1)^\dagger$.
\end{lemma}
\begin{proof}
We have $P \Phi P = \alpha P^2 + \beta J = P$ since $PJP=J$. Moreover, since $JP x = Jx$, it follows that if we let $y=Px \in \col(P)$, then $\alpha Py + \beta Jy = y$. Thus $Py = \frac{1}{\alpha}(y - \beta u u^\dagger y)$, from which the result follows.
\end{proof}

\begin{lemma}
   \label{eigs}
Suppose $P$ is stochastic, symmetric, and $P\Phi P = P$ for $\Phi = \alpha I + \beta J$ where $\beta = (1-\alpha)/n$. Then the eigenvalues of $P$ are $\in \{1, 1/\alpha, 0\}$.
\end{lemma}
\begin{proof}
Since $P$ is bistochastic, $Pu = u$, showing that 1 is an eigenvalue. Suppose that $v\neq 0$ is an eigenvector with eigenvalue $\lambda$ satisfying $u^\dagger v = 0$. Applying $v$ to $\alpha P^2 + \beta J = P$ gives $\alpha \lambda^2 v = \lambda v$ so that $\lambda(\alpha \lambda -1) = 0$. We conclude that $\lambda =0$ or $\lambda=1/\alpha$. 
\end{proof}

\begin{corollary}
   \label{diag_entries}
   If we further assume that $P$ is constant along its diagonal, then $\forall i: P(R_i|R_i) = \big(1+(r-1)/\alpha\big)/n$.
\end{corollary}
\begin{proof}
$\tr(P) = n P(R_i|R_i)=\sum_i\lambda_i=1+(r-1)/\alpha$, where $r=\rank(P)$, from which the result follows.
\end{proof}

\begin{lemma}
$\Pi = P\Phi$ is an orthogonal projector onto $\col(P)$.
\end{lemma}
\begin{proof}
Since $P\Phi P = P$, we have $\Pi^2 = P \Phi P \Phi = P\Phi = \Pi$, so $\Pi$ is a projector. Then since $P=P^\dagger$ (so that $P$ is bistochastic) and $\Phi = \alpha I + (1-\alpha)J/n$, we have $P\Phi = \Phi P$. Moreover, $\Pi^\dagger = (P\Phi)^\dagger = \Phi^\dagger P^\dagger  = \Phi P = P \Phi$. Finally, since $\Pi = P\Phi$, $\range(\Pi) \subseteq \col(P)$. Let $y\in \col(P)$ so that $y=Px$. Then $\Pi y = P \Phi Px = Px = y$, so that $\Pi$ fixes $\col(P)$. 
\end{proof}

\section{A nonclassical product}

As we discussed in Chapter \ref{ch:probabilities}, a real-valued random variable is a map from a sample space $\Omega\rightarrow \mathbb{R}$. Viewing the elements of the sample space as the outcomes of the finest grained classical reference measurement, it becomes clear that more generally we ought to define a real valued random variable as a map from reference outcomes to the reals: $\{R_i\}\rightarrow \mathbb{R}$. Indeed, if we have a valuation $\mathfrak{x}$ on any measurement $\{X_i\}$, then $x =\mathfrak{x} P(X|R)\Phi$ is an equivalent valuation on $\{R_i\}$, where $\mathfrak{x}$ and $x$ are understood as row vectors\footnote{In general we will write valuations $x$ as row vectors. In order to simplify notation, however, we may sometimes treat $x$ as a column vector when convenient.}. By equivalent, we mean that $\sum_i \mathfrak{x}_i P(X_i|\rho) = \sum_i x_i P(R_i|\rho)$ so that the gambler ought to assign the same price to both. 

Moreover, such random variables form an algebra: multiplication of random variables is achieved by the elementwise or Hadamard product $x \circ y$. Clearly, the product should be valued at $\sum_i  x_i y_i P(R_i|\rho)$.  What we would like, although we do not assume it yet\footnote{We introduce it as a guiding hypothesis, and later, by restricting the form of $P(R|R)$, will we justify it.}, is that there exists a measurement $\{M^{(x \circ y)}_i\}$ and a valuation $\mathfrak{m}^{(x \circ y)}$ such that $x \circ y = \mathfrak{m}^{(x \circ y)} P(M^{(x \circ y)}|R)\Phi$ which we would value at the same price,
\begin{align}
\sum_i \mathfrak{m}^{(x \circ y)}_i P(M^{(x \circ y)}|\rho) = \sum_i  x_i y_i P(R_i|\rho).
\end{align}
Since $x$ and $y$ are valuations on $\{R_i\}$, whatever this measurement $\{M^{(x \circ y)}_i\}$ is, should it exist, it will depend on the particular choice of reference measurement. But we could imagine introducing an alternative product on valuations, $x \odot y$, such that there exists a measurement $\{M^{(x \odot y)}_i\}$ and a valuation $\mathfrak{m}^{(x \odot y)}$ such that $x \odot y = \mathfrak{m}^{(x \odot y)} P(M^{(x \odot y)}|R)\Phi$, so that
\begin{align}
\sum_i \mathfrak{m}^{(x \odot y)}_i P(M^{(x \odot y)}|\rho) = \sum_i [x \odot y]_i P(R_i|\rho),
\end{align}
but where the measurement $\{M^{(x \odot y)}_i\}$ \emph{would not depend in any way on the reference measurement} used in the definition of the product $\odot$. 

For motivation, recall that we ought to use $P(E|R, \rho) = P(E|R)P(R|\rho)$ in the case that we perform an intermediate reference measurement: in this way, $P(E|R, \rho)$ carries a reference measurement dependence. In contrast, we ought to use $P(E|\rho) = P(E|R)\Phi P(R|\rho)$ in the case that the reference measurement remains hypothetical: here $P(E|\rho)$ does \emph{not} depend on the choice of reference measurement, even as it can be expressed in terms of reference probabilities. At the same time, treating $P(E|R)\Phi$ as a valuation $v$ on reference outcomes, we can interpret $P(E|\rho)$ as the price at which we value the random variable $v$ on a reference measurement we actually perform.

Analogously, the random variable $x \circ y$ can be understood in terms of a reference measurement the gambler actually performs. At the same time, by hypothesis, it is equivalent to a valuation on a measurement $\{M^{(x \circ y)}_i\}$, but the nature of the measurement depends on the choice of reference. In contrast, while $x \odot y$ can be understood as a valuation on the reference which is equivalent to a valuation on a measurement $\{M^{(x \odot y)}_i\}$, we would like the latter to not depend at all on the choice of reference. The reference measurement should remain in this sense ``hypothetical,'' in that it merely provides a convenient way of expressing the observable, which is independent of it. In seeking such a product, we will be guided by this analogy: just as the Protourgleichung 
\begin{align}
P(E|\rho) = \sum_i P(E|R_i)\big\{ \alpha P(R_i|\rho) + (1-\alpha)P(R_i|\mu)\big\},
\end{align}
brings the fundamental nonclassical coherence relation as close as possible to the classical law of total probability, we will seek a nonclassical product on random variables $\odot$ which looks as close as possible to the classical product $\circ$. In other words, we have already assumed that $\Phi = \alpha I + (1-\alpha)J/n$, that is, the nonclassical coherence relation is a minimal deformation of the classical rule, the law of total probability. We now similarly attempt to deform the classical rule for the multiplication of random variables in as gentle a way as possible. 

Notice that the Hadamard product is: \emph{commutative}, that is, $x \circ y = y \circ x$; \emph{permutation equivariant}, that is, for any permutation matrix $P_\pi$, it satisfies $(P_\pi x) \circ (P_\pi y) = P_\pi(x \circ y)$; and finally, it enjoys $u=(1,\dots,1)^\dagger$ as a \emph{multiplicative identity}: $u \circ a = a$. We will now explore the space of products which enjoy these same properties, with a further demand: just as $P(E|R)\Phi P(R|\rho) = \alpha P(E|R)P(R|\rho) + (1-\alpha)P(E|R)P(R|\mu)$ where $P(R_i|\mu)=1/n$, gives a one-parameter mixture with the classically expected probabilities, we want our product $\odot$ to be a one-parameter mixture with the classically expected product $\circ$ in the simplest possible way.

\begin{theorem}
Let $\odot: \mathbb{R}^n \times \mathbb{R}^n \rightarrow \mathbb{R}^n$ be a bilinear map which
\begin{enumerate}
\item is commutative: $x \odot y = y \odot x$,
\item is permutation equivariant: $\forall \pi\in S_n: P_\pi(x \odot y) = (P_\pi x) \odot (P_\pi y)$,
\item is unital with unit $u$: $\forall x \in \mathbb{R}^n: u \odot x = x$, where $u=(1,\dots, 1)^\dagger$,
\item is a one-parameter mixture with the classical product: $x \odot y = \gamma (x \circ y) + (1-\gamma) z$ where $z\neq 0$ contains as few terms as possible.
\end{enumerate}
Then any such product $\odot$ must take the form
\begin{align}
x \odot y &= \gamma (x \circ y) + (1-\gamma)(\overline{y} x + \overline{x} y - (x \cdot y)u)/n,
\end{align}
where $\circ$ is the entrywise or Hadamard product, $\overline{x}=\sum_i x_i$, $\cdot$ is the dot product, and $\gamma$ is an arbitrary parameter.
\end{theorem}
\begin{proof}
Since $\odot$ is bilinear, there must be a rank-3 tensor $T_{ijk}$ such that
\begin{align}
(x \odot y)_i = \sum_{jk} T_{ijk}x_j y_k.
\end{align}
Commutativity is equivalent to invariance under swapping of the last two indices,
\begin{align}
T_{ijk} = T_{ikj}.
\end{align}
Permutation equivariance means that for any permutation $\pi \in S_n$, 
\begin{align}
   T_{\pi(i)jk} = T_{i\pi^{-1}(j)\pi^{-1}(k)} \Longleftrightarrow T_{ijk} = T_{\pi(i)\pi(j)\pi(k)},
\end{align}
that is, $T_{ijk}$ must be invariant under simultaneous permutations of its indices. Such permutations can shuffle the indices arbitrarily, but can't change the patterns of equality between the indices. There are five possibilities,
\begin{align}
i = j = k, && i = j \neq k, && i = k \neq j, && j = k \neq i, && i, j, k \text{distinct}
\end{align}
and so the most general form for a permutation covariant tensor is 
\begin{align}
T_{ijk} = c_1 \delta_{ij}\delta_{ik} + c_2 \delta_{ij} + c_3 \delta_{ik} + c_4 \delta_{jk} + c_5,
\end{align}
which yields
\begin{align}
[x \odot y]_i=\sum_{jk} T_{ijk}x_j y_k = c_1 x_i y_i + c_2 \overline{y}x_i + c_3 \overline{x} y_i + c_4 x \cdot y + c_5 \overline{x}\overline{y},
\end{align}
or
\begin{align}
x \odot y = c_1 x \circ y + c_2 \overline{y}x + c_3 \overline{x} y + c_4 (x \cdot y)u + c_5 \overline{x}\overline{y}u.
\end{align}
From commutativity, $T_{ijk}=T_{ikj}$, so that $c_2 = c_3$. Demanding that $u$ is a multiplicative identity gives
\begin{align}
u \odot y &= c_1 u \circ y + c_{23} (\overline{y}u + \overline{u} y ) + c_4 (u \cdot y)u + c_5 \overline{u}\overline{y}u\\
&= c_1 y + c_{23}(\overline{y} u + n y) + c_4 \overline{y} u + n c_5 \overline{y} u\\
&= (c_1 + n c_{23}) y + (c_{23}+c_4 + n c_5 )\overline{y} u=y.
\end{align}
We conclude
\begin{align}
c_1 + n c_{23} = 1 && c_{23} + c_4 + n c_5 = 0,
\end{align}
and 
\begin{align}
x \odot y &= (1- n c_{23})x \circ y  + c_{23}(\overline{y}x + \overline{x} y) + c_4 (x \cdot y) u - \frac{1}{n}(c_{23} + c_4)\overline{x}\overline{y} u\\
&= (1-n c_{23})x \circ y + c_{23}(\overline{y} x + \overline{x} y - \overline{x}\overline{y}u/n) + c_4( x \cdot y - \overline{x}\overline{y}/n)u.
\end{align}
The final demand, that $x \odot y$ is a one-parameter mixture with $x \circ y$ forces $c_{23}\propto c_4$. If we take $c_4 = - c_{23}$, then $ \overline{x}\overline{y} u/n$ term drops out, leading to the simplest expression with the fewest terms,
\begin{align}
x \odot y &= (1- n c_{23}) x \circ y + c_{23}(\overline{y} x + \overline{x}y - (x \cdot y)u)\\
&= \gamma (x \circ y) + (1-\gamma)(\overline{y} x + \overline{x} y - (x \cdot y)u)/n.
\end{align}
\end{proof}

\begin{remark}
We defined $x \odot y$ on $\mathbb{R}^n \times \mathbb{R}^n \rightarrow \mathbb{R}^n$, but this isn't quite right for our purposes. On the one hand, we always assume that $x,y\in \col(P)$; on the other hand, even if $x, y\in \col(P)$, $x \odot y$ need not be in $\col(P)$. Thus for consistency, we ought to project the result back into $\col(P)$: therefore let $x \odotP y = \Pi(x \odot y)$ which takes $\col(P)\times \col(P) \rightarrow \col(P)$. Through this column space restriction, the product $\bar{\odot}$ then depends on $P$. 
\end{remark}

We may now fix the value of $\gamma$ in terms of $\alpha$. Notice that if we insert the projector $P\Phi$ into the nonclassical coherence relation,
\begin{align}
\sum_i \mathfrak{x}_i P(X_i|\rho)= \mathfrak{x} P(X|R)\Phi P(R|\rho) = \Big\{\mathfrak{x} P(X|R)\Phi\Big\} P \Big\{ \Phi P(R|\rho) \Big\} = x^\dagger P y = \langle x, y\rangle_P,
\end{align}
where $x^\dagger = \mathfrak{x} P(X|R)\Phi$ and $y=\Phi P(R|\rho)$, we can rewrite the relation as an inner product on vectors with metric $P$. By Lemma \ref{eigs}, $P$ is positive semidefinite: and on $\col(P)$ it is positive definite. Thus $\langle \cdot, \cdot\rangle_P$ is a proper inner product on $\col(P)$. Clearly, $x$ is a valuation: what about $y$? From
\begin{align}
\sum_j y_j P(R_j|R_i) = \sum_{jk} \Phi_{jk}P(R_k|\rho) P(R_j|R_i) = P(R_i|\rho),
\end{align}
we see that we can interpret $y$ as reference valuation such that $\langle y\rangle_{R_i} = P(R_i|\rho)$: the expectation of $y$ on the reference measurement, conditional on the preparatory outcome $R_i$, is equivalent to the probability for the $R_i$'th reference outcome conditional on $\rho$. Finally, notice that $\langle (P(X_i|R)\Phi)^\dagger, \Phi P(R|\rho)\rangle_P = P(X_i|\rho)$ reproduces the fundamental nonclassical coherence rule itself.

Now we would like our product $\odot$ to be compatible with this inner product. 
\begin{lemma} 
Let $x \odot y$ be a column vector and $u=(1,\dots, 1)^\dagger$. Demanding
   \begin{align}
u^\dagger (x \odot y) = \langle x, y\rangle_P.
\end{align}
fixes  $\gamma = \frac{1}{2}(1+1/\alpha)$.
\end{lemma}
\begin{proof}
On the one hand,
\begin{align}
u^\dagger (x \odot y) &= \gamma (x \cdot y) + (1-\gamma)(2\overline{x}\overline{y}/n - (x \cdot y))\\
&= (2\gamma -1)(x \cdot y) + 2(1-\gamma)\overline{x}\overline{y}/n.
\end{align}
On the other hand,
\begin{align}
\langle x, y\rangle_P = x^\dagger P y = \frac{1}{\alpha}(x \cdot y) + \left(1-\frac{1}{\alpha}\right) \overline{x}\overline{y}/n.
\end{align}
Equating terms implies that $2\gamma -1 = 1/\alpha$ from which the result follows.
\end{proof}

\begin{corollary}
   Fixing $\gamma = \frac{1}{2}(1+1/\alpha)$ implies that 
   \begin{align}
   \forall x, y, z \in \col(P):\langle x, y \odot z \rangle_P = \langle x \odot y, z\rangle_P,
   \end{align}
   so that $\odot$ is self-adjoint with respect to this inner product.
\end{corollary}
\begin{proof}
Since for $x, y \in \col(P)$,
\begin{align}
\langle x, y\rangle_P = x^\dagger P y = \frac{1}{\alpha}(x \cdot y) + \left(1-\frac{1}{\alpha}\right) \overline{x}\overline{y}/n,
\end{align}
we have
\begin{align}
\langle x,y\odot z\rangle_P
&=
\frac{1}{\alpha}\,x\cdot (y\odot z)
+
\left(1-\frac{1}{\alpha}\right)
\overline{x}\,\overline{y\odot z}/n,
\\
\langle x\odot y,z\rangle_P
&=
\frac{1}{\alpha}\,(x\odot y)\cdot z
+
\left(1-\frac{1}{\alpha}\right)
\overline{x\odot y}\,\overline{z}/n.
\end{align}
Now,
\begin{align}
x\cdot (y\odot z)
&=
\gamma \sum_i x_i y_i z_i
+
(1-\gamma)
\left(
\overline{z}(x\cdot y)
+
\overline{y}(x\cdot z)
-
\overline{x}(y\cdot z)
\right)/n,
\\
(x\odot y)\cdot z
&=
\gamma \sum_i x_i y_i z_i
+
(1-\gamma)
\left(
\overline{y}(x\cdot z)
+
\overline{x}(y\cdot z)
-
\overline{z}(x\cdot y)
\right)/n.
\end{align}
Thus
\begin{align}
x\cdot (y\odot z)-(x\odot y)\cdot z
&=
2(1-\gamma)
\left(
\overline{z}(x\cdot y)
-
\overline{x}(y\cdot z)
\right)/n \\
&=
\left(1-\frac{1}{\alpha}\right)
\left(
\overline{z}(x\cdot y)
-
\overline{x}(y\cdot z)
\right)/n.
\end{align}
Taking component sums gives
\begin{align}
\overline{y\odot z}
=
\frac{1}{\alpha}(y\cdot z)
+
\left(1-\frac{1}{\alpha}\right)
\overline{y}\,\overline{z}/n,
&&
\overline{x\odot y}
=
\frac{1}{\alpha}(x\cdot y)
+
\left(1-\frac{1}{\alpha}\right)
\overline{x}\,\overline{y}/n,
\end{align}
so that
\begin{align}
\overline{x}\,\overline{y\odot z}
-
\overline{x\odot y}\,\overline{z}
&=
\frac{1}{\alpha}
\left(
\overline{x}(y\cdot z)
-
\overline{z}(x\cdot y)
\right).
\end{align}
Putting this all together, we find
\begin{align}
&\langle x,y\odot z\rangle_P-\langle x\odot y,z\rangle_P
\nonumber \\
&=
\frac{1}{\alpha}
\left(1-\frac{1}{\alpha}\right)
\left(
\overline{z}(x\cdot y)
-
\overline{x}(y\cdot z)
\right)/n
+
\left(1-\frac{1}{\alpha}\right)
\frac{1}{\alpha}
\left(
\overline{x}(y\cdot z)
-
\overline{z}(x\cdot y)
\right)/n
\\
&=0,
\end{align}
as desired.
\end{proof}

\begin{remark}
The projected product $x \odotP y$ is also self-adjoint with respect to $\langle \cdot, \cdot\rangle_P$. Since $P\Pi = P P \Phi = P \Phi P = P$ and $\Pi P = P\Phi P = P$,
\begin{align}
\langle x, y \odotP z\rangle_P &= x^\dagger P \Pi (y \odot z) =x^\dagger P (y \odot z) = \langle x, y \odot z\rangle_P\\
\langle x \odotP y, z \rangle_P &= (x \odot y)^\dagger \Pi^\dagger P z = \langle x \odot y, z\rangle_P.
\end{align}

\end{remark}

\begin{lemma}
The algebra on $\col(P)$ defined by $\odotP$ along with vector addition and scalar multiplication is \emph{formally real}: if a sum of squares vanishes, each term must individually vanish.
\end{lemma}
\begin{proof}
Let $\{v_i\}$ be a set of valuations. Suppose that $\sum_i v_i^{\odot 2} = 0$. Then
\begin{align}
0 &= \left\langle \sum_i v_i^{\odot 2}, u\right\rangle_P = \sum_i \langle v_i \odot v_i, u\rangle_P = \sum_i \langle v_i, v_i \odot u\rangle_P = \sum_i \langle v_i, v_i\rangle_P = \sum_i \lVert v_i \rVert_P^2,
\end{align}
which implies that $\forall i: v_i=0$.
\end{proof}

\section{Moment matching}

Suppose we have a measurement $\{X_i\}$ with a valuation $\mathfrak{x}$. As we have seen, $x=\mathfrak{x} P(X|R)\Phi$ is an equivalent valuation on the reference measurement in the sense that
\begin{align}
\sum_i \mathfrak{x}_i P(X_i|\rho) = \sum_i x_i P(R_i|\rho).
\end{align}
Similarly, higher moments of $\mathfrak{x}$ can be expressed as valuations on the reference measurement. Letting $x^{(m)} = \mathfrak{x}^{\circ m} P(X|R)\Phi$ where  $\mathfrak{x}^{\circ m}$ denotes the $m$-th Hadamard power of the valuation vector, we have
\begin{align}
\sum_i \mathfrak{x}_i^m P(X_i|\rho) = \sum_i x^{(m)}_i P(R_i|\rho).
\end{align}
At the same time, we supposed that the reference valuation $x \odotP y$ is equivalent to a valuation $\mathfrak{m}^{(x \odotP y)}$ on a measurement $\{M_i^{(x \odotP y)}\}$ independent of the choice of reference. In particular, we can consider $x^{\odotP m}$ for some power $m$: this too should correspond to a valuation on a measurement independent of the choice of reference. What meaning can we give to $\sum_i [x^{\odotP m}]_i P(R_i|\rho)$, the $m$-th moment of $x$ with respect to $\odotP$? By assumption, there must exist a measurement $\{M_i^{(x^{\odotP m})}\}$, independent of the choice of reference, on which there exists an equivalent valuation. The most minimal supposition would be that this measurement is $\{X_i\}$ itself and further that $x^{\odotP m}$ is equivalent to $\mathfrak{x}^m$: this would certainly be independent of the choice of reference. But that would be too restrictive: after all, there may be multiple measurements and valuations which imply the same reference valuation $x = \mathfrak{x^\prime}P(X^\prime|R)\Phi$. We thus merely assume that there exists some class of measurements for which 
\begin{align}
\sum_i \mathfrak{x}_i^m P(X_i|\rho) = \sum_i x^{\odotP m}_i P(R_i|\rho),
\end{align}
that is, $x^{(m)} = \mathfrak{x}^{\circ m} P(X|R)\Phi = x^{\odotP m}$. For this class of measurements, then, $\langle \mathfrak{x}^m\rangle = \langle x^{\odotP m}\rangle$: we can calculate the higher moments of $\mathfrak{x}$ entirely in terms of the reference valuation $x$. This is in the spirit of $\odotP$ being a minimal modification of $\circ$: whereas the classical powers of a reference valuation $x^{\circ m}$ carry a reference measurement dependence, we assume that the nonclassical powers $x^{\odotP m}$ remove this dependence, and the simplest assumption is that they reflect the higher moments of the original valuation $\mathfrak{x}^m$.

This assumption, perhaps innocuous, will turn out to have profound consequences. The reason is that, as we have defined it, $x \odotP y$ is commutative, but it need not be \emph{associative}: it may be that $(x \odotP y) \odotP z \neq x \odotP (y \odotP z)$. In order for $x^{\odotP m}= x \odotP x \odotP x \odotP \dots $ to have an unambiguous meaning, we must therefore require that $\odotP$ powers of $x$ associate with each other. Then even if the algebra so defined is in general nonassociative, $x \odotP x \odotP x \odotP \dots $ will have an univocal meaning, independent of how it is parenthesized. An algebra with such a product is called \emph{power associative}. But how can we implement the constraint that $\odotP$ is power associative? Concretely, since $\odotP$ is defined in terms of $P$, what restrictions must be placed upon the structure of $P$ to guarantee power associativity? There is a very elegant answer to this question: we may appeal to a remarkable theorem\footnote{One of many.} of von Neumann, Jordan, and Wigner \cite{jordan1933verallgemeinerungsmöglichkeiten, Jordan1934-cm} to show that our product $\odotP$ is power associative if and only if it satisfies an identity known as the \emph{Jordan identity}.

\begin{theorem}[von Neumann, Jordan, Wigner \cite{Jordan1934-cm}]
Let $V$ be a finite dimensional real vector space equipped with a commutative bilinear product $\odotP$, not assumed associative. Further assume that $V$ is formally real with respect to this product. Then defining powers recursively by $x^{\odotP 1} = x$ and $x^{\odotP a} = x \odotP x^{\odotP (a-1)}$, the following are equivalent:
\begin{gather}
\forall x\in V, a,b \in \mathbb{N}_{\ge 1}: x^{\odotP a} \odotP x^{\odotP b} = x^{\odotP (a+b)} \label{pow_asc}\\
\Longleftrightarrow \nonumber \\
\forall x, y\in V: ((x \odotP x) \odotP y) \odotP x = (x \odotP x) \odotP (y\odotP x).\label{jord_ide}
\end{gather}
\end{theorem}
We have already shown that $x \odotP y = \Pi(x \odot y)$ defines a commutative bilinear product on $\col(P)$, a finite dimensional real vector space. And moreover, because $\odotP$ is self-adjoint with respect to the inner product $\langle \cdot, \cdot\rangle_P$ on $\col(P)$, it defines a formally real algebra. The theorem then tells us that if this algebra is power associative (Eq. (\ref{pow_asc})), then $\odotP$ must satisfy the Jordan identity (Eq. (\ref{jord_ide})), and conversely, if $\odotP$ satisfies the Jordan identity, the algebra must be power associative. Therefore we can ensure that $x^{\odotP m}$ has a univocal meaning by imposing the Jordan identity on $\odotP$. Since $\odotP$ is fixed up to the choice of $P$, this ultimately means putting a constraint on $P$ itself.

\subsection{Polarization}

We would like to massage the Jordan identity into a form which makes it easy to check whether a given $P$ induces a product $\odotP$ which satisfies it. In particular, it would be convenient if one could introduce an orthonormal basis $\{e_i\}$ for $\col(P)$ and check the identity on just those basis vectors. This is the right idea: but in order to formulate it correctly, we will have to \emph{polarize} the Jordan identity.

Let $L_x y = x \odotP y$ so that $L_x$ is the linear operator which performs $\odotP$ with $x$. Explicitly,
\begin{align}
L_x &= \Pi \Big\{\gamma \diag{x} + (1-\gamma)(x u^\dagger + \overline{x}I - ux^\dagger)/n\Big\},
\end{align}
where $D_x$ is the diagonal matrix with $x$ along its diagonal. Rearranging the Jordan identity using commutativity, we may reexpress it in terms of $L_x$ and $L_{x^{\odotP 2}}$,
\begin{align}
x \odotP ((x \odotP x) \odotP y) = (x \odotP x) \odotP (x\odotP y) \Longleftrightarrow L_x L_{x^{\odotP 2}} y = L_{x^{\odotP 2}}L_x y.
\end{align}
Since this must hold for all $x,y \in \col(P)$, we see that the Jordan identity is equivalent to 
\begin{align}
   \forall x\in \col(P): \mathcal{J}(x) = [L_x, L_{x^{\odotP 2}}]\Pi = 0,
\end{align}
an expression which is cubic in $x$. By the bilinearity of $\odotP$, we have for any scalar $\lambda$, $L_{\lambda x}=\lambda L_x$ and $L_{(\lambda x)^{\odotP 2}}=\lambda^2  L_{x^{\odotP 2}}$ so that $[L_{\lambda x}, L_{(\lambda x)^{\odotP 2}}]\Pi = \lambda^3 [L_x, L_{x^{\odotP 2}}]\Pi$. Thus  $\mathcal{J}(x)=0$ is a homogeneous cubic identity in $x$. A real homogeneous polynomial identity of degree $d$ is equivalent to its full $d$-linear polarization: since $d=3$, it suffices to consider the third order polarization \cite{McCrimmon2003-kp}. Indeed, checking only $\forall i: [L_{e_i}, L_{e_i^{\odotP 2}}]\Pi=0$ on an orthonormal basis $\{e_i\}$ for $\col(P)$ is not enough. To check the Jordan identity in terms of basis vectors $\{e_i\}$, we must polarize, or linearize, the Jordan identity so that it becomes a \emph{trilinear identity} \cite{McCrimmon2003-kp}: then it suffices to check it on all triples of basis vectors. The standard procedure for doing so is the following. 

Let $a,b,c\in \col(P)$ be arbitrary, and let $x = ta + sb + rc$. Then $L_x = tL_a + s L_b + r L_c$, and because $\odotP$ is commutative and bilinear, we have
\begin{align}
L_{x^{\odotP 2}} = t^2 L_{a^{\odotP 2}} + s^2 L_{b^{\odotP 2}} + r^2 L_{c^{\odotP 2}} + 2ts L_{a \odotP b} + 2srL_{b \odotP c} + 2rt L_{c \odotP a},
\end{align}
and so
\begin{align}
   \label{polar}
\mathcal{J}(x)=[L_x, L_{x^{\odotP 2}}]\Pi &= 
\Big\{t^3[L_a,L_{a^{\odotP 2}}]
+s^3[L_b,L_{b^{\odotP 2}}]
+r^3[L_c,L_{c^{\odotP 2}}]
\nonumber\\
&+t^2s\Big(
2[L_a,L_{a\odotP b}]
+[L_b,L_{a^{\odotP 2}}]
\Big)
\nonumber\\
&+ts^2\Big(
[L_a,L_{b^{\odotP 2}}]
+2[L_b,L_{a\odotP b}]
\Big)
\nonumber\\
&+t^2r\Big(
2[L_a,L_{a\odotP c}]
+[L_c,L_{a^{\odotP 2}}]
\Big)
\nonumber\\
&+tr^2\Big(
[L_a,L_{c^{\odotP 2}}]
+2[L_c,L_{a\odotP c}]
\Big)
\nonumber\\
&+s^2r\Big(
2[L_b,L_{b\odotP c}]
+[L_c,L_{b^{\odotP 2}}]
\Big)
\nonumber\\
&+sr^2\Big(
[L_b,L_{c^{\odotP 2}}]
+2[L_c,L_{b\odotP c}]
\Big)
\nonumber\\
&+2tsr\Big(
[L_a,L_{b\odotP c}]
+[L_b,L_{a\odotP c}]
+[L_c,L_{a\odotP b}]
\Big)\Big\}\Pi.
\end{align}
The full third order polarization is obtained from the $tsr$ coefficient. Indeed, define the symmetric trilinear map
\begin{align}
   \label{tripolar}
T(a,b,c) = \frac{1}{3}\Big\{[L_a,L_{b\odotP c}] +[L_b,L_{c\odotP a}]+[L_c,L_{a\odotP b}]\Big\}\Pi.
\end{align}
Then $T(x,x,x) = [L_x, L_{x^{\odotP 2}}]\Pi=\mathcal{J}(x)$, showing that $T$ is the full third-order polarization of the homogeneous cubic map $\mathcal{J}(x)$. We now show that $\forall a, b, c \in \col(P): T(a,b,c)=0$ is equivalent to the original Jordan identity. On the one hand, suppose the Jordan identity holds. Then $\forall x \in \col(P): [L_x, L_{x^{\odotP 2}}]\Pi=0$. Since the expression in Eq. (\ref{polar}) is a polynomial in $t,s,r$, and it must vanish for any choice of $t,s,r$, every coefficient in fact must be identically zero. In particular, this means that the $tsr$ coefficient vanishes. Thus $T(a,b,c)=0$. Since $a,b,c$ are arbitrary, the Jordan identity therefore implies $\forall a,b,c \in \col(P): T(a,b,c)=0$. Conversely, suppose $\forall a,b,c \in \col(P): T(a,b,c)=0$. Then in particular, $\forall x \in \col(P): T(x,x,x)= [L_x, L_{x^{\odotP 2}}]\Pi=0$, so the Jordan identity holds. We conclude that $\forall x \in \col(P): \mathcal{J}(x)=0$ is equivalent to $\forall a,b,c \in \col(P): T(a,b,c)=0$. The benefit of the latter trilinear identity is that if $\{e_i\}$ is an orthonormal basis and $a=\sum_i a_i e_i, b= \sum_i b_i e_i, c=\sum_i c_i e_i$, we have
\begin{align}
T(a,b,c) = \sum_{ijk}a_i b_j c_k T(e_i, e_j, e_k).
\end{align}
Since $T$ is trilinear, $\forall a,b,c \in \col(P): T(a,b,c)=0$ if and only if 
\begin{align}
\forall i, j, k: T(e_i, e_j, e_k) = 0.
\end{align}
Instead of using an orthonormal basis $\{e_i\}$ for $\col(P)$, we could simply use the columns of $P$ itself (or just $r$ linearly independent columns of $P$). We may then express the Jordan identity as a relation, while complicated, entirely among the matrix elements of $P(R|R)$.

\begin{theorem}
Let $P\equiv P(R|R)$, $\Phi = \alpha I + (1-\alpha)J/n$, and let
 \begin{align}
L_x &= P \Phi \Big\{\gamma \diag{x} + (1-\gamma)(x u^\dagger + \overline{x}I - ux^\dagger)/n\Big\},
\end{align}
where $\gamma = (1+1/\alpha)/2$ so that $L_x y = x \odotP y$. Finally let
\begin{align}
T(a,b,c) = \frac{1}{3}\Big\{[L_a,L_{b\odotP c}] +[L_b,L_{c\odotP a}]+[L_c,L_{a\odotP b}]\Big\}P \Phi.
\end{align}
Then $\odotP$ satisfies the Jordan identity iff
\begin{align}
\forall i, j, k: T(P(R|R_i), P(R|R_j), P(R|R_k)) = 0.
\end{align}
\end{theorem}

\subsection{Euclidean Jordan algebras}

We arrived at the Jordan identity by first minimally modifying the classical product $x \circ y$ on reference valuations to the nonclassical product $x \odotP y$. Inspired by the relationship between the law of total probability and the fundamental nonclassical coherence rule, we conceived a hope: that both $x \circ y$ and $x \odotP y$ would correspond to equivalent valuations on \emph{some measurements}, and that while the former would depend on the choice of reference measurement, the latter would not. We then hoped that for some class of measurements, 
\begin{align}
   \label{moment-matching}
\sum_i \mathfrak{x}_i^m P(X_i|\rho) = \sum_i x^{\odotP m}_i P(R_i|\rho),
\end{align}
that is, that $x^{\odotP m} = \mathfrak{x}^{\circ m} P(X|R)\Phi$. In assuming this, we were guided by simplicity: the reference-independent measurement with an equivalent valuation to $x^{\odotP m}$ ought to be $\{X_i\}$ itself, the equivalent valuation being $\mathfrak{x}^{\circ m}$. For this to have any hope of being realized, however, $\odotP$ would have to be power associative: otherwise $x^{\odotP m}$ would have no unambiguous meaning. But since the bilinear product $\odotP$ defines a real finite dimensional formally real commutative algebra, assuming power associativity is equivalent to assuming that $\odotP$ satisfies the Jordan identity. Finally, we were able to express this identity entirely in terms of constraints on the matrix $P(R|R)$ which characterized the reference measurement itself. Thus $\odotP$ defines a formally real or \emph{Euclidean} Jordan algebra on $\col(P)$. 

As we will see, the supposition that reference valuations form a Euclidean Jordan algebra places remarkably profound constraints on the state and effect spaces compatible with the reference measurement. Indeed, it means that our hopes are already realized: not only will there exist measurements with valuations equivalent to $x \circ y$ and $x \odotP y$, the latter being reference measurement independent, but also there will exist measurements whose moments satisfy Eq. (\ref{moment-matching}). In fact, those measurements will turn out to be implied entirely by the assignment of the reference valuation $x$, and their effects will correspond to a set of \emph{mutually orthogonal idempotents} of the Jordan algebra. States and effects will turn out to correspond to squares $x^{\odotP 2}$ of reference valuations, and in fact Euclidean Jordan algebras are self-dual, so that the state cone $\mathscr{S}$ will be equal to the effect cone $\mathscr{E}$. Finally, we avail ourselves of the seminal result of Jordan, von Neumann, and Wigner: the classification of the Euclidean Jordan algebras  \cite{Jordan1934-cm,farautAnalysisSymmetricCones1994, McCrimmon2003-kp,schafer1995introduction}. Remarkably, all finite dimensional Euclidean Jordan algebras are isomorphic to direct sums of the so-called simple Euclidean Jordan algebras. These fundamental building blocks come in just a few shapes and sizes:
\begin{itemize}
\item $\text{Herm}_d(\mathbb{R})$, the algebra of $d \times d$ self-adjoint real matrices with product $a \star b = \frac{1}{2}(ab + ba).$
\item $\text{Herm}_d(\mathbb{C})$, the algebra of $d\times d$ self-adjoint complex matrices with product $a \star b = \frac{1}{2}(ab + ba).$
\item $\text{Herm}_d(\mathbb{H})$, the algebra of $d\times d$ self-adjoint quaternionic matrices with product $a \star b = \frac{1}{2}(ab + ba).$
\item $\text{Herm}_3(\mathbb{O})$, the algebra of $3 \times 3$ self-adjoint octonionic matrices with product $a \star b = \frac{1}{2}(ab + ba).$
\item The spin factors $\mathbb{R} \oplus \mathbb{R}^n$ with product $(t, x) \star (t^\prime, x^\prime) = (t t^\prime + \langle x, x^\prime\rangle, t x^\prime + t^\prime x)$.
\end{itemize}
We are thus well on our way along the royal road \cite{wilceRoyalRoadQuantum2018} to quantum mechanics. We now turn to unpacking the consequences of Jordan identity; then we will try to narrow our focus to $\text{Herm}_d(\mathbb{C})$ specifically. But already we have learned something crucial to the QBist project: Euclidean Jordan algebras may be straightforwardly motivated by considering what is the most natural nonclassical product one can define on reference valuations. By centering the interplay between the reference measurement and the measurements for which they form a reference, we hope to have given new meaning to the formal algebra of observables defined by Jordan et al.

\section{Lifting the product}

Before we move on, however, it will be useful in the sequel to lift our product $\odotP$, which is defined on valuations, to a product on probability vectors and response functions. Recall that earlier we defined the inner product $\langle \cdot, \cdot\rangle_P$ so that 
\begin{align}
\sum_i \mathfrak{x}_i P(X_i|\rho)= \mathfrak{x} P(X|R)\Phi P(R|\rho) = \Big\{\mathfrak{x} P(X|R)\Phi\Big\} P \Big\{ \Phi P(R|\rho) \Big\} = x^\dagger P y = \langle x, y\rangle_P,
\end{align}
where $x^\dagger = \mathfrak{x} P(X|R)\Phi$ and $y=\Phi P(R|\rho)$, and demanded that $u^\dagger(x \odotP y) = \langle x, y\rangle_P$. Moreover, $\langle (P(X_i|R)\Phi)^\dagger, \Phi P(R|\rho)\rangle_P = P(X_i|\rho)$ reproduces the fundamental nonclassical coherence relation. At the same time, however, we could define an inner product on $\col(P)$, $\langle \cdot, \cdot\rangle_\Phi$ such that
\begin{align}
\sum_i \mathfrak{x}_i P(X_i|\rho) = \Big\{\mathfrak{x} P(X|R)\Big\} \Phi \Big\{P(R|\rho)\Big\} = g^\dagger \Phi h = \langle g, h\rangle_\Phi,
\end{align}
which reproduces $P(X_i|\rho) = \langle P(X_i|R)^\dagger, P(R|\rho)\rangle_\Phi$. One way of putting it is that from
\begin{align}
P(X_i|\rho) = \langle (P(X_i|R)\Phi)^\dagger, \Phi P(R|\rho)\rangle_P,
\end{align}
$\langle \cdot, \cdot\rangle_P$ defines an inner product on \emph{dual coordinates} while from
\begin{align}
P(X_i|\rho) = \langle P(X_i|R)^\dagger, P(R|\rho)\rangle_\Phi,
\end{align}
$\langle \cdot, \cdot\rangle_\Phi$ defines an inner product on \emph{regular coordinates}. Thus implicitly we defined $\odotP$ on dual coordinates: we would thus like to lift it to a product $\hat{\odot}$ on regular coordinates, and thus on response functions and probability vectors directly.

Now if we have dual coordinates $x = \Phi h$, then acting with $P$ gives us regular coordinates $h = Px$ since $h = P\Phi h$ for $h \in \col(P)$. Similarly, if $y^\dagger = g^\dagger \Phi$, then $g^\dagger = y^\dagger P$. We are thus led to consider
\begin{align}
g \hat{\odot} h = P(\Phi g \odotP \Phi h) = P\Pi(\Phi g \odot \Phi h) = P(\Phi g \odot \Phi h),
\end{align}
where since $P \Pi = P P \Phi = P\Phi P = P$ it is inessential whether we use the original product $\odot$ or the projected product $\odotP$. Just as $\odot$ is compatible with $\langle \cdot, \cdot\rangle_P$, $\hat{\odot}$ is compatible with $\langle \cdot, \cdot\rangle_\Phi$. On the one hand, 
\begin{align}
u^\dagger(g \hat{\odot} h) &= u^\dagger P(\Phi g \odot \Phi h)= u^\dagger(\Phi g \odot \Phi h) \\
&= \langle \Phi g, \Phi h\rangle_P = g^\dagger \Phi P \Phi h = g^\dagger \Phi h = \langle g, h\rangle_\Phi,
\end{align}
and on the other hand, 
\begin{align}
\langle f, g \hat{\odot} h\rangle_\Phi &= f^\dagger \Phi P(\Phi g \odot \Phi h) = \langle \Phi f, \Phi g \odot \Phi h\rangle_P\\
&= \langle \Phi f \odot \Phi g, \Phi h\rangle_P = (\Phi f \odot \Phi g)^\dagger P \Phi h = \langle f \hat{\odot} g, h\rangle_\Phi,
\end{align}
so that $\hat{\odot}$ is self-adjoint with respect to $\langle \cdot, \cdot\rangle_\Phi$. Working out the explicit form for $\hat{\odot}$, we find the following.
\begin{lemma}
   Lifting $\odot$ from a product on dual coordinates to a product $\hat{\odot}$ on regular coordinates gives
   \begin{align}
g \hat{\odot} h &= \frac12\left\{
\alpha(\alpha+1)P(g \circ h)
-
(\alpha-1)
\left(
\overline{h} g+\overline{g} h  + \alpha (g \cdot h) u
\right)/n.
\right\}
\end{align}
Moreover, let $\hat{L}_g$ be the linear operator such that $\hat{L}_g h = g \hat{\odot} h$. It may be expressed 
\begin{align}
\hat{L}_g &=  P L_{\Phi g}\Phi
=
\frac12\left\{
\alpha(\alpha+1)P\diag{g}
-
(\alpha-1)
\left(
gu^\dagger+\overline g\,I + \alpha ug^\dagger
\right)/n
\right\}.
\end{align}
\end{lemma}

\begin{proof}
We had $L_x y = x \odotP y$ where
\begin{align}
L_x &= \Pi \Big\{\gamma \diag{x} + (1-\gamma)(x u^\dagger + \overline{x}I - ux^\dagger)/n\Big\}.
\end{align}
Let $x = \Phi g= \alpha g+(1-\alpha)\overline g\,u/n$ so that $\diag{x}=\diag{\Phi g} = \alpha\diag{g}+(1-\alpha)\overline g\,I/n$. Moreover, notice that $\Phi g u^\dagger-u g^\dagger \Phi = \alpha(gu^\dagger-ug^\dagger)$. Using $\gamma=(1+1/\alpha)/2$, we arrive at
\begin{align}
   \label{LPhig}
L_{\Phi g}
=
\frac12\Pi \left\{
(\alpha+1)\diag{g}
+
(\alpha-1)
\left(
gu^\dagger-ug^\dagger-\overline g\,I
\right)/n
\right\},
\end{align}
so that
\begin{align}
\hat{L}_g = P L_{\Phi g}\Phi
=
\frac12\left\{
(\alpha+1)P\diag{g}\Phi
+
(\alpha-1)P
\left(
gu^\dagger-ug^\dagger-\overline g\,I
\right)\Phi/n
\right\}.
\end{align}
Now $P\diag{g}\Phi = \alpha P\diag{g} + (1-\alpha)Pg\,u^\dagger/n$ and $Pg=(1/\alpha)g+(1-1/\alpha)\overline g\,u/n$ which yields
\begin{align}
P\diag{g}\Phi =
\alpha P\diag{g} -
(\alpha-1)(1/\alpha)gu^\dagger/n - (\alpha-1)^2(1/\alpha)\overline g\,J/n^2.
\end{align}
Meanwhile
\begin{align}
P
\left(
gu^\dagger-ug^\dagger-\overline g\,I
\right)\Phi
&=
Pg\,u^\dagger
-
u g^\dagger\Phi
-
\overline g\,P\Phi
\\
&=
(1/\alpha)gu^\dagger
-
\alpha ug^\dagger
-
\overline g\,P\Phi
+
(\alpha^2-1)(1/\alpha)\overline g\,J/n.
\end{align}
We conclude that
\begin{align}
\hat{L}_g = P L_{\Phi g}\Phi
=
\frac12\left\{
\alpha(\alpha+1)P\diag{g}
-
(\alpha-1)
\left(
gu^\dagger+\alpha ug^\dagger+\overline g\,P\Phi
\right)/n
\right\}.
\end{align}
By assumption $\hat{L}_g$ acts on vectors in $\col(P)$ so we can substitute $P\Phi \rightarrow I$, from which the result follows.
\end{proof}

Finally, we may formulate the Jordan identity in terms of $\hat{\odot}$ instead of $\odotP$. First let $\mathcal{L}_g = \hat{L}_g P$ so that $\mathcal{L}_g \Phi h = \hat{L}_g P \Phi h = \hat{L}_g h = g \hat{\odot} h$ for $h \in \col(P)$. We have
\begin{align}
\mathcal{L}_g &= 
\frac12\left\{
\alpha(\alpha+1)P\diag{g}P
-
(\alpha-1)
\left(
gu^\dagger P+\alpha ug^\dagger P +\overline g\,P
\right)/n
\right\}\\
&=
\frac12\left\{
\alpha(\alpha+1)P\diag{g}P
-
(\alpha-1)
\left(
  gu^\dagger+ug^\dagger + \overline g\,P + (\alpha-1)\overline{g}J /n
\right)/n
\right\}.
\end{align}
Notice that unlike $L_g$, $\mathcal{L}_g=\mathcal{L}_g^\dagger$ so that the latter is self-adjoint in the usual sense. Moreover, we note in passing that since $P^2 = \frac{1}{\alpha}P + (1-1/\alpha)J/n$, we have
   \begin{align}
\mathcal{L}_u  
&=
\frac{1}{2}\left\{\alpha(\alpha+1)P^2-(\alpha-1)\left(uu^\dagger+uu^\dagger+nP+(\alpha-1)J/n
\right)/n\right\}\\
&=\frac{1}{2}\left\{\alpha(\alpha+1)\left[\frac{1}{\alpha}P + (1-1/\alpha)J/n\right]
- (\alpha-1)\left(nP + (\alpha+1)J \right)/n\right\}\\
&= \frac{1}{2}\left\{2 P + 2(\alpha-1)J/n - 2(\alpha-1)J/n \right\} = P,
   \end{align}
so that $\mathcal{L}_u \Phi = \Pi$, which indeed acts as the identity on $\col(P)$. Finally, consider in particular $\mathcal{L}_{P(R|R_k)}$. Treated as a three index tensor, this object fully defines $\hat{\odot}$. Explicitly,
   \begin{align}
  [\mathcal{L}_{P}]_{ijk} &= 
   \frac{1}{2}\Bigg\{\alpha(\alpha+1)\sum_m P(R_m|R_i)P(R_m|R_j)P(R_m|R_k)
\\&-(\alpha-1)
\Big(
 P(R_i|R_k) + P(R_j|R_k) + P(R_i|R_j) 
+
(\alpha-1)/n
\Big)/n\Bigg\}.\nonumber
   \end{align}
Then $[\mathcal{L}_g]_{ij} =  \sum_{kl}[\mathcal{L}_{P}]_{ijk} \Phi_{kl} g_l$ so that
\begin{align}
[g \hat{\odot} h]_i = \sum_{jklm} [\mathcal{L}_{P(R|R)}]_{ijk} \Phi_{kl} \Phi_{jm} g_l  h_m,
\end{align}
and
\begin{align}
[P(R|R_j) \hat{\odot} P(R|R_k)]_i = \sum_{ab} [\mathcal{L}_P]_{ajk}\Phi_{ab}P(R_b|R_i),
\end{align}
or equivalently, $P(R|R_j) \hat{\odot} P(R|R_k) = \sum_{i} C_{ijk}P(R|R_i)$. Here $C_{ijk} = \sum_{a} [\mathcal{L}_P]_{ajk}\Phi_{ai}$ encodes the structure coefficients for $\hat{\odot}$, and we may formulate the Jordan identity in terms of them.
\begin{lemma}
Let $C_{ijk} = \sum_{a} [\mathcal{L}_P]_{ajk}\Phi_{ai}$ be the structure coefficients for $\hat{\odot}$. Then $\hat{\odot}$ satisfies the Jordan identity iff
\begin{align} &\forall t,i,j,k,\ell: \sum_{rs} C_{rij} \left( C_{sk\ell}C_{trs} - C_{sr\ell}C_{tks} \right) \nonumber\\ &\qquad+ \sum_{rs} C_{rjk} \left( C_{si\ell}C_{trs} - C_{sr\ell}C_{tis} \right) \nonumber\\ &\qquad+ \sum_{rs} C_{rki} \left( C_{sj\ell}C_{trs} - C_{sr\ell}C_{tjs} \right) =0.\end{align}
\end{lemma}
\begin{proof}
Analogously to our earlier construction, using the third order polarization, $\hat{\odot}$ satisfies the Jordan identity iff
\begin{align}
\forall a,b, c \in \col(P): \frac{1}{3}\Big\{[\mathcal{L}_a \Phi ,\mathcal{L}_{b\hat{\odot} c}\Phi ] +[\mathcal{L}_b\Phi ,\mathcal{L}_{c\hat{\odot} a}\Phi ]+[\mathcal{L}_c \Phi,\mathcal{L}_{a\hat{\odot} b}\Phi ]\Big\} = 0,
\end{align}
where we note that unlike in Eq. (\ref{tripolar}) there is no need to insert a trailing projector $\Pi$ since $\mathcal{L}_g \Phi \Pi = (\hat{L}_g P) \Phi P \Phi = \hat{L}_g P \Phi =\mathcal{L}_g \Phi$. Using trilinearity, we can formulate the Jordan identity entirely in terms of $P(R|R)$ as
\begin{align}
\forall i,j,k: &[\mathcal{L}_{P(R|R_i)\hat{\odot}P(R|R_j)}\Phi,
\mathcal{L}_{P(R|R_k)}\Phi]
\nonumber\\
&\qquad+
[\mathcal{L}_{P(R|R_j)\hat{\odot}P(R|R_k)}\Phi,
\mathcal{L}_{P(R|R_i)}\Phi]
\nonumber\\
&\qquad+
[\mathcal{L}_{P(R|R_k)\hat{\odot}P(R|R_i)}\Phi,
\mathcal{L}_{P(R|R_j)}\Phi]
=0.
\end{align}
Since $P(R|R_j) \hat{\odot} P(R|R_k) = \sum_{i} C_{ijk}P(R|R_i)$, we have $
\mathcal{L}_{P(R|R_j)\hat{\odot}P(R|R_k)} = \sum_{i} C_{ijk} \mathcal{L}_{P(R|R_i)}$ and so
\begin{align}
\forall i,j,k: & \sum_r \Big\{C_{rij} [\mathcal{L}_{P(R|R_r)}\Phi,
\mathcal{L}_{P(R|R_k)}\Phi]
\nonumber\\
&\qquad+
C_{rjk} [\mathcal{L}_{P(R|R_r)}\Phi,
\mathcal{L}_{P(R|R_i)}\Phi]
\nonumber\\
&\qquad+
 C_{rik}[\mathcal{L}_{P(R|R_r)}\Phi,
\mathcal{L}_{P(R|R_j)}\Phi]\Big\}
=0.
\end{align}
To further resolve this expression, we will apply the LHS to some $P(R|R_\ell)$. First,
\begin{align}
\mathcal{L}_{P(R|R_k)}\Phi P(R|R_\ell)
=
P(R|R_k)\hat{\odot}P(R|R_\ell)
=
\sum_s C_{sk\ell}P(R|R_s),
\end{align}
so that
\begin{align}
\mathcal{L}_{P(R|R_r)}\Phi
\mathcal{L}_{P(R|R_k)}\Phi P(R|R_\ell)
&=
\sum_{st}
C_{sk\ell}C_{trs}P(R|R_t),
\\
\mathcal{L}_{P(R|R_k)}\Phi
\mathcal{L}_{P(R|R_r)}\Phi P(R|R_\ell)
&=
\sum_{st}
C_{sr\ell}C_{tks}P(R|R_t),
\end{align}
which leads to
\begin{align}
[\mathcal{L}_{P(R|R_r)}\Phi,\mathcal{L}_{P(R|R_k)}\Phi]P(R|R_\ell)
=
\sum_{st}
\left(
C_{sk\ell}C_{trs}
-
C_{sr\ell}C_{tks}
\right)
P(R|R_t).
\end{align}
Rewriting all the commutators in this way, we arrive at
\begin{align}
&\forall i,j, k, \ell: \sum_t
\Bigg[
\sum_{rs}
C_{rij}
\left(
C_{sk\ell}C_{trs}
-
C_{sr\ell}C_{tks}
\right)
\nonumber\\
&\qquad\qquad+
\sum_{rs}
C_{rjk}
\left(
C_{si\ell}C_{trs}
-
C_{sr\ell}C_{tis}
\right)
\nonumber\\
&\qquad\qquad+
\sum_{rs}
C_{rki}
\left(
C_{sj\ell}C_{trs}
-
C_{sr\ell}C_{tjs}
\right)
\Bigg]
P(R|R_t)
=0.
\end{align}
The expression in the brackets is a vector $v$ indexed by $t$. Therefore we can reexpress the constraint simply as $Pv=0$. But from the form of $C_{ijk} = \sum_{a} [\mathcal{L}_P]_{ajk}\Phi_{ai}$, and in particular the expression for $[\mathcal{L}_P]_{ijk}$, we see that $v\in \col(P)$ automatically. If $Pv=0$ and $v \notin \nullspace(P)$, we must have $v=0$, from which the result follows.
\end{proof}

\section{Self-duality}

Now that we know that our nonclassical product defines a Euclidean Jordan algebra, we may appeal to many powerful results. For instance, the famed Koecher-Vinberg theorem \cite{kriegMinnesotaNotesJordan1999,farautAnalysisSymmetricCones1994, barnum2013localtomographyjordanstructure} provides an alternative, geometric characterization of a Euclidean Jordan algebra: the squares of elements of an EJA form a \emph{symmetric cone}, essentially a convex cone which is \emph{homogeneous} and \emph{self-dual}, and conversely any symmetric cone arises from the squares of the elements of some Euclidean Jordan algebra. Homogeneity tells us that there is an invertible transformation  which takes any element in the interior of the cone to any other element while preserving the ordering of elements: in other words, the automorphism group of the cone acts transitively\footnote{In quantum theory, this corresponds to the fact that we can map any state $\rho$ to any state $\sigma$ by a suitable Kraus operator: $K=\sigma^{1/2}\rho^{-1/2}$, where $\rho, \sigma$ are full-rank.}. Meanwhile, let $C$ be the cone of squares of elements of an EJA. Then self-duality means that if one considers $C^*$, the dual cone of $C$, that is, the space of linear functionals nonnegative on the cone, then $C=C^*$. For us, this means that we can interpret squares of elements of our EJA precisely as states or equivalently, effects. Indeed, if
\begin{align}
P(E|R) = x \hat{\odot} x && P(R|\rho) = y \hat{\odot} y,
\end{align}
then
\begin{align}
P(E|\rho) = P(E|R)\Phi P(R|\rho) = \langle P(E|R), P(R|\rho)\rangle_\Phi = \langle x^{\hat{\odot}2}, y^{\hat{\odot}2}\rangle_\Phi \ge 0,
\end{align}
where the nonnegativity follows from self-duality. Indeed, the same principle guarantees that e.g., $P(R|\rho)$ is indeed a probability vector since by the same token,
\begin{align}
P(R_i|\rho) = P(R_i|R)\Phi P(R|\rho) = \langle P(R_i|R), P(R|\rho)\rangle_\Phi =\langle r_i^{\hat{\odot} 2}, y^{\hat{\odot} 2}\rangle_\Phi \ge 0.
\end{align}
Similarly, in terms of $\odotP$, if
\begin{align}
\Phi P(E|R)^\dagger = x \odotP x && \Phi P(R|\rho) = y \odotP y,
\end{align}
then
\begin{align}
P(E|\rho) = \big\{P(E|R)\Phi\big\} P \big\{\Phi P(R|\rho) \big\}= \langle \Phi P(E|R)^\dagger, \Phi P(R|\rho)\rangle_P = \langle x^{\odotP 2}, y^{\odotP 2}\rangle_P \ge 0.
\end{align}

\subsection{The uncertainty principle}

Letting $\mathfrak{x}$ be a valuation on a measurement $\{X_i\}$ and $x=\mathfrak{x}P(X|R)\Phi$ be the equivalent valuation on the reference measurement, we supposed that
\begin{align}
\sum_i \mathfrak{x}_i^2 P(X_i|\rho) &= \sum_i (x \odot x)_i P(R_i|\rho)
\end{align}
for some class of measurements. More concisely,
\begin{align}
\langle \mathfrak{x}^2\rangle_\rho &= \langle x^{\odot 2}\rangle_\rho = \gamma \langle x^2\rangle_\rho + (1-\gamma)\Big( 2 \langle x\rangle_\mu \langle x\rangle_\rho - \langle x^2\rangle_\mu\Big),
\end{align}
where e.g., $\langle x^2\rangle_\rho = \sum_i x_i^2 P(R_i|\rho)$ and $\forall i: P(R_i|\mu) =1/n$\footnote{Indeed, a Euclidean Jordan algebra picks out a distinguished state $P(R|\mu)$ which is proportional to the unit of the algebra.}. We will soon show how the structure of a Euclidean Jordan algebra itself will pick out exactly for which class of measurements this relation holds. But at the moment, let us take it for granted, and notice that because the second moment of a random variable is always nonnegative $\langle \mathfrak{x}^2\rangle_\rho \ge 0$, we immediately have a lower bound on the second moment of $x$ with respect to the reference,
\begin{align}
\langle x^2\rangle_\rho  &\ge \frac{\alpha-1}{\alpha +1}\Big( \langle x^2\rangle_\mu - 2 \langle x\rangle_\mu \langle x\rangle_\rho \Big).
\end{align}
Since this should hold for any $x \in \col(P)$, we therefore have a restriction on  probability assignments $P(R|\rho)$ on reference outcomes. In fact, this restriction follows entirely from the fact that the cone of squares of the EJA is self-dual without separately assuming $\langle \mathfrak{x}^2\rangle_\rho = \langle x^{\odot 2}\rangle_\rho$, and in fact is a necessary and sufficient condition for $P(R|\rho)$ to be a valid state. Indeed, for $P(R|\rho) \in \col(P)$,
\begin{align}
(x \odot x)^\dagger P(R|\rho) = (x \odot x)^\dagger P \Phi P(R|\rho) = \langle x^{\odot 2}, \Phi P(R|\rho)\rangle_P.
\end{align}
By self-duality, $\langle x^{\odot 2}, \Phi P(R|\rho)\rangle_P \ge 0$ iff $\Phi P(R|\rho) = y^{\odotP 2}$ for some $y$, that is, $P(R|\rho)$ is a valid state. (Notice we've used $\odotP$, the projected product, since by assumption $P(R|\rho) \in \col(P)$). Thus without even specifying anything about the nature of the measurement $\{X_i\}$, we have the constraint $\langle x^{\odot 2}\rangle_\rho \ge 0$. We can therefore fully characterize the state space via
\begin{align}
   \label{var_low_bound_recon}
P(R|\rho) \in \Delta_n \cap \col(P) \text{ valid} \Longleftrightarrow \forall x \in \col(P): \langle x^2\rangle_\rho \ge \frac{\alpha-1}{\alpha +1}\Big( \langle x^2\rangle_\mu - 2 \langle x\rangle_\mu \langle x\rangle_\rho \Big),
\end{align}
Thinking of the lower bound as a bound on the variance of any observable, we see the shape of the state space is determined entirely by a kind of \emph{uncertainty principle}: to respect this uncertainty principle, probability distributions on reference outcomes cannot be too sharp in a prescribed way.

But how can we check straightforwardly whether a given distribution $p\equiv P(R|\rho)$ is in fact valid? Notice that we can rewrite the lower bound in matrix form as
\begin{align}
\forall x\in \col(P): \sum_i (x \odot x)_i  p_i &= x^\dagger \Big\{ \gamma \diag{p} + (1-\gamma)\big( up^\dagger + p u^\dagger - \overline{p}I \big)/n \Big\}x\\
&= x^\dagger V_p x \ge 0,
\end{align}
so that the validity of $p\in \Delta_n \cap \col(P)$ amounts to the positive semidefiniteness of $V_p=V_p^\dagger$ on $\col(P)$. Thus if we let $\mathcal{L}_p = \Pi V_p \Pi$, where $\Pi$ is the projector onto $\col(P)$, then the validity of $p$ corresponds to the positive semidefiniteness of $\mathcal{L}_p$ in the full sense, so that
\begin{align}
p \in \Delta_n \cap \col(P) \text{ valid} \Longleftrightarrow \mathcal{L}_p \ge 0.
\end{align}
In this way, the positive semidefiniteness of $\mathcal{L}_p$ allows one to diagnose whether a given distribution $p$ is consistent with the reference measurement. 

The astute reader may have noticed that we are using the same symbol $\mathcal{L}$ as in the previous section where we defined $\mathcal{L}_x$ to act as $\mathcal{L}_x \Phi y= x \hat{\odot} y$. This is no accident! Indeed, let
\begin{align}
V_x = \gamma \diag{x} + (1-\gamma)\big( ux^\dagger + x u^\dagger - \overline{x}I \big)/n,
\end{align}
which coincides with the previous definition when $\overline{x}=1$ as it does for probability vectors. Now on the one hand, we earlier defined
\begin{align}
\mathcal{L}_x &= \hat{L}_x P = P L_{\Phi x} \Phi P\\
&=\frac{1}{2} P \left\{
(\alpha+1)\diag{x}
+
(\alpha-1)
\left(
xu^\dagger-ux^\dagger-\overline x\,I
\right)/n
\right\}\Phi P,
\end{align}
where we have used the expression we already worked out for $L_{\Phi x}$ in Eq. \ref{LPhig}. On the other hand, we now propose that
\begin{align}
\mathcal{L}_x = \Pi V_x \Pi &= P \Phi \Big\{\gamma D_x + (1-\gamma)(u x^\dagger + x u^\dagger - \overline{x} I)/n\Big\} P \Phi:
\end{align}
indeed, one can easily verify that the two expressions are the same by working out $\Phi V_x$. Thus $\mathcal{L}_x$ has two interpretations. On the one hand, $\mathcal{L}_x \Phi y = x \hat{\odot} y$ in fact performs the Jordan product itself. On the other hand, the positive semidefiniteness of $\mathcal{L}_{P(R|\rho)}$ diagnoses the validity of a distribution $P(R|\rho)$ with respect to the reference measurement. This applies not just to states but also to effects. In fact, from
\begin{align}
P(E|R)^\dagger y^{\odot 2} = P(E|R)^\dagger \Phi P y^{\odot 2} = \langle \Phi P(E|R), y^{\odot 2}\rangle_P \ge 0 \text{ iff } \Phi P(E|R) = x \odotP x,
\end{align}
we see that exactly the same argument applies. Indeed, by self-duality, the positive semidefiniteness of $\mathcal{L}_{P(E|R)}$ diagnoses the validity of $P(E|R)$. 

\subsection{The self-duality constant}

Indeed, by self-duality, states can be rescaled into effects and vice versa. Defining the \emph{self-duality constant} allows us to introduce a rescaled product $\check{\odot}$, which will prove particularly useful when we discuss idempotents of the Jordan algebra. To begin, let $P(R|\rho)$ be a state. Self-duality implies there must be a corresponding effect $P(\rho|R)$. By Bayes's theorem,
\begin{align}
P(\rho|R_i) &= \frac{P(R_i|\rho)P(\rho)}{P(R_i)} = \kappa P(R_i|\rho).
\end{align}
In particular, let us apply this to the reference states and effects. For clarity, we will write $P(R_i|R_j) \equiv P(R_i|\sigma_j)$ where $\sigma_j$ denotes a reference state. Considering when $i=j$, we have
\begin{align}
P(\sigma_i|R_i) = \kappa P(R_i|\sigma_i),
\end{align}
where we will call $\kappa$ the \emph{self-duality constant}. Now we supposed that $P(R_i|\sigma_i)$ is a constant, and in Corollary \ref{diag_entries}, we fixed its value to $P(R_i|\sigma_i) = (1 + (r-1)/\alpha)/n$. Let us further suppose that in fact $P(\sigma_i|R_i) = 1$: conditional on obtaining outcome $\{R_i\}$ on the reference measurement, if we then perform a measurement among whose outcomes is $\sigma_i$, we will be certain of getting that outcome. In fact, requiring that each reference state be certain for its corresponding effect fixes the value of $\kappa$,
\begin{align}
\kappa = \frac{n}{1 + (r-1)/\alpha},
\end{align}
where $r=\rank(P)$. By linearity then we have in general, that
\begin{align}
P(\rho|R) &= \kappa P(R|\rho)^\dagger.
\end{align}
In fact, this places a constraint on all states $P(R|\rho)$ such that $P(\rho|\rho)=1$.
\begin{align}
P(\rho|\rho) &= P(\rho|R)\Phi P(R|\rho)\\
&=\kappa \Big\{ \alpha P(R|\rho)^\dagger P(R|\rho) + (1-\alpha)/n\Big\} = 1,
\end{align}
from which it follows that
\begin{align}
\sum_i P(R_i|\rho)^2 &= \frac{\alpha^2 + r -1}{\alpha^2 n} .
\end{align}
At the same time,
\begin{align}
 P(\rho|\rho) &= \langle P(\rho|R)^\dagger, P(R|\rho)\rangle_\Phi \\
 &= u^\dagger(P(\rho|R)^\dagger \hat{\odot} P(R|\rho))\\
 &= u^\dagger(\kappa P(R|\rho) \hat{\odot} P(R|\rho)).
\end{align}
We are thus motivated to introduce a rescaled product $\check{\odot} = \kappa \hat{\odot}$ and a rescaled inner product $\langle \cdot, \cdot\rangle_{\kappa \Phi}$ such that
\begin{align}
P(\rho|\rho) &= u^\dagger (P(R|\rho) \check{\odot} P(R|\rho))\\
&= \langle P(R|\rho), P(R|\rho)\rangle_{\kappa \Phi } = \lVert P(R|\rho)\rVert_{\kappa \Phi}^2,
\end{align}
which will turn out to be the right tool to use when discussing idempotents of the Jordan product.

\section{Idempotents}

In a Euclidean Jordan algebra, idempotents $x = x\odotP x$ of the Jordan product play a crucial role. Many key results from standard linear algebra may be ported to the Euclidean Jordan algebraic setting---in particular, the spectral theorem. We say two idempotents are orthogonal iff $x \odotP y = 0$. A \emph{primitive idempotent} \cite{farautAnalysisSymmetricCones1994,McCrimmon2003-kp, schafer1995introduction} is a nonzero element $e$ of the algebra such that $e \odotP e = e$ and which cannot be expressed as a sum of two (or more) nonzero orthogonal idempotents. That is, an idempotent is primitive iff there does \emph{not} exist any $x,y$ such that $e = x + y$ and $x \odotP y = 0$. Euclidean Jordan algebras enjoy a \emph{spectral theorem}: any element $x$ of the algebra may be decomposed as a real linear combination of mutually orthogonal primitive idempotents,
\begin{align}
x = \sum_i \lambda_i e_i && e_i \odotP e_j = \delta_{ij}e_i && \sum_i e_i = u
\end{align}
where we may call the $\lambda_i$'s eigenvalues, and $u$ is the unit of the algebra, such that $\forall x: u \odotP x = x$. Not only that, but every element can be expressed as a linear combination of no more than $d$ mutually orthogonal primitive idempotents, where $d$ is called the \emph{rank} of the EJA. We may define the \emph{trace} of an element $\tr(x)$, and indeed, $\tr(x) = \sum_i \lambda_i$ is the sum of the eigenvalues\footnote{For us $\tr(x) = u^\dagger x$, and so e.g., $P(\rho|\rho) = \tr(P(R|\rho) \check{\odot} P(R|\rho))$.}. In particular, the only allowed eigenvalues for idempotents are 0 and 1, and so $\tr(e)=1$ for any primitive idempotent. It follows also that any square $x \odotP x$ must have all nonnegative eigenvalues, and conversely any element with nonnegative eigenvalues can be expressed as a square. The squares form the \emph{nonnegative cone}, $C=\{x\odot x : x \in V \}$ and this cone is self-dual with respect to the trace inner product, so that we have as well that $C= \{y \in V: \forall x \in C : \tr(x \odotP y)\ge 0\}$. Clearly all idempotents are in the nonnegative cone: in particular, the primitive idempotents generate its extremal rays. Finally, we have observed a consequence of these considerations is that in a Euclidean Jordan algebraic theory, probabilities are given by the trace inner product of two nonnegative elements, one representing a state and the other an effect.

We briefly sketch how an element $x$ of a Euclidean Jordan algebra can be spectrally decomposed \cite{farautAnalysisSymmetricCones1994,schafer1995introduction, McCrimmon2003-kp}. Defining $x^{\odotP 0} = u$, first form the powers of $x$: $u, x, x^{\odotP 2}, x^{\odotP 3}, \dots$. Since the algebra is finite dimensional, these powers will eventually become linearly dependent. Concretely, one may arrange the powers as the columns of a matrix, adding column after column, until the matrix becomes singular. When this happens, we can write the final column as a linear combination of the earlier columns: let $M$ be the matrix with the linearly independent powers of $x$ and let $y$ be the first linearly dependent power. Then $y = Mz$ for some $z$, and we can invert the equation by taking for example the Moore-Penrose pseudoinverse of $M$: $z = M^+ y$. This first relation of linear dependence
\begin{align}
x^{\odotP m} + a_{m-1}x^{\odotP (m-1)} + \dots + a_1x + a_0 u = 0
\end{align} 
gives the minimal polynomial of $x$,
\begin{align}
p_x(t) &= t^{\odotP m} + a_{m-1}t^{\odotP (m-1)} + \dots + a_1t + a_0 = \prod_{i} (t-\lambda_i),
\end{align}
where $\{\lambda_i\}$ are the distinct eigenvalues of $x$.
\begin{lemma}
In a Euclidean Jordan algebra, the roots $\{\lambda_i\}$ will be real \cite{farautAnalysisSymmetricCones1994,McCrimmon2003-kp,schafer1995introduction}.
\end{lemma}
\begin{proof}
Suppose $p_x(t)$ had a complex root $z=a+ib$. Since $p_x(t)$ has all real coefficients, then $z^* = a-ib$ must also be a root. Let $f(t) = (t -z)(t-z^*) = (t-a)^2 + b^2$. Since both $f(t)$ and $p_x(t)$ vanish at $z$ and $z^*$, $f(t)$ divides $p_x(t)$ and we may write $p_x(t) = f(t) r(t)$ for some polynomial $r(t)$. Now
\begin{align}
r(x) \odotP p_x(x) &= r(x) \odotP f(x) \odotP r(x) \\
&=r(x) \odotP \Big((x-au)^{\odotP 2} + b^2 u \Big)\odot r(x) \\
&= \big((x-au)r(x)\big)^{\odotP 2} + b^2 r(x)^{\odotP 2} = 0,
\end{align}
since $p_x(x)=0$ and we are working in the associative subalgebra of powers of $x$. But this is a vanishing sum of squares and so by formal reality, $(x-a u)\odotP r(x) = 0$ and $br(x)=0$. Suppose $r(x)=0$. Since $p_x(t) = f(t) r(t)$ where $f(t)$ has degree 2, the degree of $r(t)$ must be less than that of $p_x(t)$. But if $r(x)=0$, then this is a polynomial that vanishes on $x$ and which has smaller degree that the minimal polynomial---in other words, it must be the minimal polynomial itself! So $r(x)\neq 0$. Thus in fact, $b=0$: the root was real to begin with.
\end{proof}

Now for each eigenvalue, using Lagrange interpolation, we may calculate the idempotents in the spectral decomposition in accordance with the following theorem \cite{farautAnalysisSymmetricCones1994,McCrimmon2003-kp,schafer1995introduction}.
\begin{theorem}
Let $x$ be an element of the Jordan algebra, and let $\{\lambda_j\}$ be its distinct eigenvalues. If we let
\begin{align}
e_i = \overline{\bigodot}_{i \neq j} \frac{x-\lambda_j u}{\lambda_i -\lambda_j},
\end{align}
then  $x = \sum_i \lambda_i e_i$ for $e_i \odotP e_j = \delta_{ij}e_i$ and $\sum_i e_i = u$.
\end{theorem}
\begin{proof}
Suppose that $x=\sum_{k=1}^s \lambda_k e_k$ for $e_k \odotP e_l = \delta_{kl} e_k$ and $\sum_k e_k = u$, for distinct eigenvalues $\lambda_k$. Let $q_i(t) = \prod_{j \neq i} \frac{t-\lambda_j}{\lambda_i-\lambda_j}$. Clearly, $q_i(\lambda_k) = \delta_{ik}$. By power associativity, we can consider
\begin{align}
q_i(x) = \overline{\bigodot}_{i \neq j} \frac{x-\lambda_j u}{\lambda_i -\lambda_j}.
\end{align}
Since $x = \sum_k \lambda_k e_k$ and $u =\sum_k e_k$, we have $x-\lambda_j u = \sum_k \lambda_k e_k - \lambda_k \sum_k e_k = \sum_k (\lambda_k - \lambda_j)e_k$ so that
\begin{align}
q_i(x) &= \overline{\bigodot}_{i \neq j} \sum_k \frac{\lambda_k -\lambda_j}{\lambda_i - \lambda_j}e_k = \sum_k \left(\prod_{j\neq i} \frac{\lambda_k -\lambda_j}{\lambda_i - \lambda_j}\right) e_k = \sum_k \delta_{ik} e_k = e_i,
\end{align}
where we have used the fact that if $a = \sum_k a_k e_k$ and $b= \sum_l b_l e_l$, then $a \odotP b = \sum_k a_k b_k e_k$.
Conversely, suppose
\begin{align}
e_i = \overline{\bigodot}_{i \neq j} \frac{x-\lambda_j u}{\lambda_i -\lambda_j}.
\end{align}
Now $\sum_i q_i(\lambda_k) = \sum_i \delta_{ik} = 1$ so that $\sum_i q_i(t) -1 = 0$ at the $s$ eigenvalues $\lambda_1, \dots, \lambda_s$. But since each $q_i$ has degree $s-1$, the degree of this polynomial must be $\leq s-1$: at the same time, it vanishes at $s$ points. Therefore it must be identically zero, and we conclude that $\sum_i q_i(t) = 1$, and therefore $\sum_i q_i(x) = u$. Similarly, consider $\sum_i \lambda_i q_i(\lambda_k) = \sum_i \lambda_i \delta_{ik} = \lambda_k$. We conclude that $\sum_i \lambda_i q_i(t) -t$ vanishes at the $s$ eigenvalues, but since its degree cannot be more than $s-1$, it must vanish identically. We conclude $\sum_i \lambda_i q_i(t) = t$, or $\sum_i \lambda_i q_i(x) = \sum_i \lambda_i e_i = x$.

Finally, suppose that $i\neq j$, and consider that $q_i(\lambda_k)q_j(\lambda_k) = (q_i q_j)(\lambda_k) = \delta_{ik}\delta_{jk} = 0$. Thus $(q_i q_j)(t)$ vanishes at the $s$ eigenvalues. The minimal polynomial of $x$ is $p_x(t) = \prod_k (t-\lambda_k)$ which also vanishes at the $s$ distinct eigenvalues. Thus $p_x(t)$ divides $(q_i q_j)(t)$, that is, we can write $(q_i q_j)(t) = p_x(t)h(t)$ for some polynomial $h(t)$. We conclude
\begin{align}
e_i \odotP e_j = q_i(x) \odotP q_j(x)  = (q_i q_j)(x) = p_x(x) \odot h(x) = 0,
\end{align}
since $p_x(x) = 0$. Now suppose $i = j$. Observe that $q_i(\lambda_k)^2 - q_i(\lambda_k) = (q_i^2 - q_i)(\lambda_k) = \delta_{ik}^2 - \delta_{ik} = 0$ so that the polynomial vanishes at the $s$ eigenvalues. Thus the minimal polynomial  $p_x(t)$ divides it, and so $q_i(t)^2 - q_i(t) = p_x(t)g(t)$ for some polynomial $g(t)$. We conclude
\begin{align}
e_i \odot e_i - e_i = q_i(x) \odot q_i(x) - q_i(x) = (q_i^2 - q_i)(x) = p_x(x) \odot g(x) = 0,
\end{align}
which completes the proof.
\end{proof}
Finally, we note that the idempotents in the spectral decomposition will not be primitive if the eigenvalues are degenerate. To resolve this, let $e$ be an idempotent, and let $V_1(e) = \{y : e \odotP y = y\}$: this is the eigenspace with eigenvalue 1 of the idempotent. If $e$ is primitive, then $\dim V_1(e) =1$. Letting $L_e y = e \odotP y$, it then suffices to check whether the $+1$ eigenspace of $L_e$ is one dimensional. If not, then notice that $V_1(e)$ defines its own Euclidean Jordan algebra with unit $e$. Pick a random vector $y \in V_1(e)$, and consider its powers $e, y, y^{\odotP 2}, \dots$: by the same procedure as above, one may construct its spectral decomposition. One recursively repeats this recipe until one has (not uniquely) $d$ mutually orthogonal primitive idempotents, where $d$ is the rank of the EJA.

\subsection{From valuations to measurements}

From the beginning, we had tentatively supposed that if we have two measurements $\{X_i\}$ and $\{Y_i\}$ with valuations $\mathfrak{x}$ and $\mathfrak{y}$, and we assign equivalent valuations to the reference measurement $x=\mathfrak{x}P(X|R)\Phi$ and $y=\mathfrak{y}P(Y|R)\Phi$, then there ought to be a third measurement $\{M^{(x \odotP y)}_i\}$ and a valuation $\mathfrak{m}^{x \odotP y}$ such that $\mathfrak{m}^{x \odotP y} P(M^{(x \odotP y)}|R)\Phi = x \odotP y$. We promised that this is guaranteed by the fact that reference valuations form a Euclidean Jordan algebra. We will now see that the spectral decomposition implies something even stronger is true.

Take \emph{any} valuation $x$ with respect to the reference measurement, and decompose it into orthogonal idempotents $x = \sum_i \mathfrak{x}_i e_i$ with respect to $\odotP$. Since the $e_i$'s are their own squares, they correspond to effects: we take the state and effect spaces to be subsets of the self-dual cone of squares of the EJA. We are working with $\odotP$, that is, in dual coordinates, where $P(E|\rho) = \langle P(E|R)\Phi, \Phi P(R|\rho)\rangle_P$. Thus we must have $e_i = P(X_i|R)\Phi$ for some $P(X_i|R)$. Switching to regular coordinates, where $P(E|\rho) = \langle P(E|R), P(R|\rho)\rangle_\Phi$, we have $P(X_i|R) = e_i P$. Indeed, $\sum_i P(X_i|R) = \sum_i e_i P = u^\dagger P = u^\dagger=P(1|R)$, so that these effects indeed form a measurement $\{X_i\}$ and the eigenvalues $\mathfrak{x}_i$ represent a valuation on its outcomes. Indeed, $\{X_i\}$ is not just any measurement. Because its effects are a set of mutually orthogonal idempotents,
\begin{align}
P(X_i|X_j) = P(X_i|R)\Phi P(R|X_j) = u^\dagger(P(R|X_i) \check{\odot} P(R|X_j)) = \delta_{ij},
\end{align}
so that if one assigns the state $P(R|X_j)$, then one will expect outcome $X_j$ with certainty upon a measurement of $\{X_i\}$: the states and effects perfectly distinguish each other.

Moreover, because in the decomposition $x = \sum_i \mathfrak{x}_i P(X_i|R)\Phi$, the valuations $\mathfrak{x}_i$ are eigenvalues, we have $x^{\odotP m} = \sum_i \mathfrak{x}^m_i P(X_i|R)\Phi$, and so our moment matching condition is assured,
\begin{align}
\sum_i \mathfrak{x}_i^m P(X_i|\rho) = \sum_i [x^{\odotP m}]_iP(R_i|\rho).
\end{align}
Thus the class of measurements for which $\langle \mathfrak{x}^m\rangle_\rho = \langle x^{\odotP m}\rangle_\rho$ are precisely those whose effects are mutually orthogonal idempotents, that is, those measurements which perfectly distinguish their corresponding states.

Finally, let us compare $x \circ y$ and $x \odotP y$. In either case, we may spectrally decompose the valuation to find an equivalent idempotent measurement. What is the difference between them? If one chose a different reference measurement, one could always express the product $x \circ y$ according to the original reference measurement with respect to the new one. But this would entirely depend on the relationship between the original reference and the new one. In contrast, for \emph{any} reference measurement, the product $\odotP$ is picked out as the one which defines the state space itself. Indeed, if one chose a different reference measurement, the expression for $x \odotP y$ would change, but the product's relationship to the very consistency rules which ought to govern the gambler's probability assignments would not. In this sense, the valuation and measurement equivalent to $x \circ y$ depends on the choice of reference, while $x \odotP y$ is independent of that choice. Indeed, more generally, since any reference valuation $x$ implies, through $\odotP$, an equivalent ``spectral measurement,'' valuations have an autonomous existence, as it were, regardless of the choice of reference.

\subsection{Characterizing idempotents}

Let us now turn to an explicit characterization of the idempotents with respect to our product $\check{\odot}$. 

\begin{lemma}
If $x \check{\odot} x = x$, then
\begin{align}
x = \frac{1}{2}\left\{\frac{\alpha(\alpha+1)}{ (\kappa^{-1} + (\alpha-1)\overline{x}/n)}P(x \circ x) -(\alpha-1)\overline{x}u/n\right\}.
\end{align}
\end{lemma}
\begin{proof}
From 
\begin{align}
g \check{\odot} h &= \frac{\kappa}{2}\left[
\alpha(\alpha+1)P(g \circ h)
-
(\alpha-1)
\left(
\overline{h} g+\overline{g} h  + \alpha (g \cdot h) u
\right)/n
\right],
\end{align}
we have for $x = x \check{\odot} x$,
\begin{align}
x &= \frac{\kappa}{2}\left[
\alpha(\alpha+1)P(x \circ x)
-
(\alpha-1)
\left(
2\overline{x} x + \alpha (x \cdot x) u
\right)/n
\right],
\end{align}
Taking the sum of both sides of the equation gives
\begin{align}
\overline{x} &= \frac{\kappa}{2}\left[
\alpha(\alpha+1)(x \cdot x)
-
(\alpha-1)
\left(
2\overline{x}^2/n + \alpha (x \cdot x)
\right)
\right]\\
&=\kappa\left[
\alpha(x \cdot x)
-
(\alpha-1)
\overline{x}^2/n 
\right],
\end{align}
so that
\begin{align}
x \cdot x = \frac{1}{\kappa\alpha}\overline{x} + \left(1-\frac{1}{\alpha}\right) \overline{x}^2/n.
\end{align}
Substituting this in, we find
\begin{align}
x &= \frac{\kappa}{2}\left[
\alpha(\alpha+1)P(x \circ x)
-
(\alpha-1)
\left(
2\overline{x} x + \alpha \left\{\frac{1}{\kappa\alpha}\overline{x} + \left(1-\frac{1}{\alpha}\right) \overline{x}^2/n\right\} u
\right)/n
\right],
\end{align}
so that collecting terms, we have
\begin{align}
(1+ \kappa(\alpha-1)\overline{x}/n)x &= \frac{1}{2\kappa}\left[
\alpha(\alpha+1)P(x \circ x)
-
\kappa^{-1} (\alpha-1)
 \overline{x} \left(1 + \kappa\left(\alpha-1\right) \overline{x}/n  \right)u/n
\right],
\end{align}
or
\begin{align}
x = \frac{1}{2}\left\{\frac{\alpha(\alpha+1)}{ (\kappa^{-1} + (\alpha-1)\overline{x}/n)}P(x \circ x) -(\alpha-1)\overline{x}u/n\right\}
\end{align}
\end{proof}
\begin{remark}
Another way of interpreting this formula is that just as for $x\in \col(P)$, the action of $P$ on $x$ is very simple, simply mixing $x$ and $u$, the action of $P$ on $x\circ x$ is also just as simple for idempotents $x = x \check{\odot} x$: $x$ is simply mixed with $u$.
\end{remark}

\begin{corollary}
For $x \check{\odot} x = x$, 
\begin{align}
\sum_i x_i^2 = \frac{1}{\kappa\alpha}\overline{x} + \left(1-\frac{1}{\alpha}\right) \overline{x}^2/n.
\end{align}
\end{corollary}

\begin{corollary}
   \label{trips}
   For $x \check{\odot}x = x$, 
   \begin{align}
   \sum_i x_i^3
&=
\frac{\overline{x}\left(
\kappa^{-1}+(\alpha-1)\overline{x}/n
\right)
\left(
2\kappa^{-1}+(\alpha-1)\overline{x}/n
\right)}{\alpha(\alpha+1)}.
\end{align}
\end{corollary}
\begin{proof}
Substituting our expression for an idempotent $x$ into $x \cdot x$, we obtain
\begin{align}
x\cdot x
&=
\frac12\left\{
\frac{\alpha(\alpha+1)}
{\kappa^{-1}+(\alpha-1)\overline{x}/n}
x^\dagger P(x\circ x)
-
(\alpha-1)\overline{x}^2/n
\right\}.
\end{align}
First,
\begin{align}
x^\dagger P(x\circ x)
&=
\frac{1}{\alpha}\sum_i x_i^3
+
\left(1-\frac{1}{\alpha}\right)\overline{x}(x\cdot x)/n .
\end{align}
Then, solving for \(\sum_i x_i^3\) yields
\begin{align}
\sum_i x_i^3
&=
\frac{\kappa^{-1}+(\alpha-1)\overline{x}/n}{\alpha+1}
\left[
2(x\cdot x)+(\alpha-1)\overline{x}^2/n
\right]
-
(\alpha-1)\overline{x}(x\cdot x)/n .
\end{align}
Using
\begin{align}
x\cdot x
=
\frac{1}{\kappa\alpha}\overline{x}
+
\left(1-\frac{1}{\alpha}\right)\overline{x}^2/n
=
\frac{\overline{x}}{\alpha}
\left(
\kappa^{-1}+(\alpha-1)\overline{x}/n
\right),
\end{align}
and simplifying leads to the result.
\end{proof}

In particular, a primitive idempotent satisfies $\tr(x) = u^\dagger x = \overline{x}=1$, and since every state is represented by a probability distribution, primitive idempotents may be identified with distributions $P(R|\rho)$.

\begin{corollary}
$P(R|\rho)$ is a primitive idempotent of $\hat{\odot}$ iff
\begin{align}
   \label{best-prim}
P(R_i|\rho) = \frac{1}{2}\left\{\frac{\alpha(\alpha+1)}{ (\kappa^{-1} + (\alpha-1)/n)}\sum_{j}P(R_i|R_j)P(R_j|\rho)^2 -(\alpha-1)P(R_i|\mu)\right\},
\end{align}
where $\kappa = \frac{n}{1 + (r-1)/\alpha}$.
\end{corollary}

Moreover, the proof of Lemma \ref{rank1-lemma} from the previous chapter is still valid in the broader Euclidean Jordan algebraic setting due to the existence of the spectral decomposition. Thus an element of an EJA is a primitive idempotent iff $\tr(x^{\check{\odot} 2})=\tr(x^{\check{\odot}3})=1$. But this means
\begin{align}
\tr\Big( P(R|\rho)^{\check{\odot}2}\Big) = u^\dagger ( P(R|\rho) \check{\odot} P(R|\rho)) = P(\rho|R)\Phi P(R|\rho) = P(\rho|\rho) = 1,
\end{align}
and we already saw that this implies
\begin{align}
\sum_i P(R_i|\rho)^2 = \frac{1}{\kappa\alpha} + \left(1-\frac{1}{\alpha}\right)/n.
\end{align}

\begin{remark}
In fact, because we took $P(\sigma_i|\sigma_i) = \tr\Big(P(R|\sigma_i)^{\check{\odot} 2}\Big) = 1$ for reference states, and the distributions $P(R|\sigma_i)$ are by construction valid states, we have that the reference states must be primitive idempotents already. $ \tr\Big(P(R|\sigma_i)^{\check{\odot}2}\Big)= \sum_i \lambda_i^2 = 1$ is only possible given the constraint that $\tr(P(R|\sigma_i)) = \sum_i \lambda_i = 1$ and $\lambda_i \ge 0$, if just a single $\lambda_i$ is 1 and the rest 0.
\end{remark}

\noindent Similarly, 
\begin{align}
\tr\Big( P(R|\rho)^{\check{\odot}3}\Big) &= u^\dagger \Big\{  P(R|\rho) \check{\odot}\big(P(R|\rho)  \check{\odot} P(R|\rho)\big) \Big\} \\ 
&= P(\rho|R)\Phi \big(P(R|\rho)  \check{\odot} P(R|\rho)\big) = P(\rho|\rho) = 1,
\end{align}
so that substituting Eq. (\ref{best-prim}) for $P(R|\rho)  \check{\odot} P(R|\rho)$, we find as in Corollary \ref{trips},
 \begin{align}
   \sum_i P(R_i|\rho)^3
&=
\frac{\left(
\kappa^{-1}+(\alpha-1)/n
\right)
\left(
2\kappa^{-1}+(\alpha-1)/n
\right)}{\alpha(\alpha+1)}.
\end{align}
\begin{corollary}
   \label{prim_conds}
   $P(R|\rho)$ is a primitive idempotent of $\hat{\odot}$ iff $P(R|\rho) \in \Delta_n \cap  \col(P)$ and
   \begin{align}
   \sum_i P(R_i|\rho)^2 &= \frac{\kappa^{-1} + (\alpha-1)/n}{\alpha}\\
    \sum_i P(R_i|\rho)^3
&=
\frac{\left(
\kappa^{-1}+(\alpha-1)/n
\right)
\left(
2\kappa^{-1}+(\alpha-1)/n
\right)}{\alpha(\alpha+1)}.
   \end{align}
\end{corollary}
\noindent Along with the normalization $\sum_i P(R_i|\rho) = 1$ condition, we see that the primitive idempotents, or \emph{pure states}, live in the intersection of three kinds of spheres: a 1-norm, 2-norm, and a 3-norm sphere of specified radii, intersected with $\col(P)$ and lying in the nonnegative orthant. Any valid state can be written as a convex combination of such pure states.

\subsection{Orthogonality}

In order to check whether primitive idempotents are mutually orthogonal, the following lemma will help.

\begin{lemma}
Let $P(\rho|R)\Phi P(R|\tau) = \tr\big( P(R|\rho) \check{\odot} P(R|\tau)\big) = 0$. Then
\begin{align}
   \label{orth_cond}
P(R|\rho) \cdot P(R|\tau) = \frac{\alpha-1}{\alpha n}.
\end{align}
Moreover if $P(R|\rho) \check{\odot} P(R|\tau)=0$,
\begin{align}
\sum_j P(R_i|R_j)P(R_j|\rho)P(R_j|\tau) &= \frac{\alpha-1}{\alpha(\alpha+1)}\Big(P(R_i|\rho) + P(R_i|\tau) + (\alpha-1)P(R_i|\mu)\Big)/n.
\end{align}
\end{lemma}
\begin{proof}
   Let $p, q$ be the probability distributions. Our product is
\begin{align}
p \check{\odot} q&= \frac{\kappa}{2}\left[ \alpha(\alpha+1)P(p \circ q) - (\alpha-1) \left(  p+q+ \alpha (p \cdot q) u \right)/n \right].
\end{align}
The condition that $p \check{\odot} q = 0$ is just that
\begin{align}
P(p \circ q) &= \frac{\alpha-1}{\alpha(\alpha+1)}\Big(  p+q+ \alpha (p \cdot q) u \Big)/n.
\end{align}
Summing over both sides gives 
\begin{align}
p \cdot q = \frac{\alpha-1}{\alpha(\alpha+1)}\Big( 2/n + \alpha (p \cdot q) \Big) \Longrightarrow p \cdot q = \frac{\alpha-1}{\alpha n}.
\end{align}
Thus
\begin{align}
P(p \circ q) &= \frac{\alpha-1}{\alpha(\alpha+1)}\Big(  p+q+ (\alpha-1) u/n \Big)/n.
\end{align}
\end{proof}
\begin{remark}
   In fact, for a Jordan algebra, if $P(R|\rho)$ and $P(R|\tau)$ are assumed to be idempotents, and $\tr(P(R|\rho) \check{\odot} P(R|\tau)) = 0$, then $P(R|\rho) \check{\odot} P(R|\tau)=0$ \cite{McCrimmon2003-kp}. Thus the condition that $P(R|\rho)$ and $P(R|\tau)$ are mutually orthogonal primitive idempotents reduces down entirely to scalar constraints, along with the subspace restriction, and the nonnegativity of the entries of the probability vectors.
\end{remark}
\begin{remark}
   Just as $P(p \circ p)$ acts simply on an idempotent $p$, so too it acts simply on $P(p \circ q)$ for $p \check{\odot} q = 0$.
\end{remark}
\begin{corollary}
   For any two reference distributions, since the state space is the convex hull of the primitive idempotents,
   \begin{align}
 \frac{\alpha-1}{\alpha n} \leq  P(R|\rho) \cdot P(R|\tau) \leq \frac{\kappa^{-1} + (\alpha-1)/n}{\alpha}
   \end{align}
\end{corollary}
\begin{remark}
In a more general setting, these are the pairwise constraints defining a qplex \cite{Appleby_2017}.
\end{remark}

\subsection{Testing the rank of an EJA}
\label{testing_rank}

The \emph{rank} of an EJA is the maximum number of mutually orthogonal primitive idempotents. One could determine this, for example, by searching directly for sets of distributions satisfying $P(\rho_i|\rho_j) = P(\rho_i|R)\Phi P(R|\rho_j) = \delta_{ij}$ subject to the primitive idempotent constraints we just derived. More simply, one may rely upon the spectral decomposition. As we have noted, every element of an EJA has a spectral decomposition $x = \sum_i \lambda_i e_i$ where $e_i$ are mutually orthogonal idempotents and $\lambda_i$ are the distinct eigenvalues of $x$ \cite{farautAnalysisSymmetricCones1994}. For a generic element of the EJA, the eigenvalues will all be distinct, and so there will be $d$ of them, where $d$ is the rank of the EJA \cite{orlitzkyRankComputationEuclidean2022}. We can thus determine the rank of the EJA with a simple recipe: pick many elements of the algebra at random, take their spectral decompositions, and determine the maximum number of distinct eigenvalues. This will almost certainly be the rank of the Euclidean Jordan algebra.

\section{A simple EJA}
\label{simple_eja}
As we have said, Jordan--von Neumann--Wigner theorem of 1934 \cite{Jordan1934-cm, McCrimmon2003-kp} classifies the Euclidean Jordan algebras, showing they must be direct sums of the \emph{simple Euclidean Jordan algebras}, which we now define. A simple EJA is one whose only ideals are $\{0\}$ and the entire vector space $V$ on which the algebra is defined, in our case $\col(P)$. In this case, an \emph{ideal} is a subspace $\mathcal{I} \subseteq V$ such that $y\in \mathcal{I}$ implies $\forall x\in V: x \odotP y \in \mathcal{I}$. In other words, if $y$ is in the ideal, then under the Jordan product, it traps any other vector in that subspace with it. For us, then, an ideal satisfies e.g., $\forall x \in \col(P): \mathcal{L}_x \mathcal{I} \subseteq \mathcal{I}$. A subspace $\mathcal{I}$ is left invariant by a matrix $M=M^\dagger$ iff the orthogonal projector onto that subspace commutes \cite{hornMatrixAnalysis1985}: $[ M, \Pi_{\mathcal{I}}]=0$. Thus we require
\begin{align}
\nexists \Pi_{\mathcal{I}} \text{ such that } \forall x \in \col(P): [\mathcal{L}_x , \Pi_{\mathcal{I}}] = 0 \text{ unless } \Pi_{\mathcal{I}}=0 \text{ or } \Pi_{\mathcal{I}}=\Pi. 
\end{align}
where $\Pi$ is the projector onto $\col(P)$ and where $\mathcal{L}_x = \mathcal{L}_x^\dagger$. We presume as well that $\Pi_{\mathcal{I}}\Pi = \Pi \Pi_{\mathcal{I}} = \Pi_{\mathcal{I}}$ so that $\Pi_{\mathcal{I}}$ projects inside $\col(P)$. Moreover, since $\mathcal{L}_x$ was constructed by sandwiching $V_x$ with the projector $\Pi$, the commutator is already restricted to $\col(P)$ (otherwise, we might need to project again). By linearity, it suffices to check the above condition for each reference state $P(R|R_i)$. Practically speaking, what one ought to do is introduce a basis for $\col(P)$. Let $B$ be the matrix whose columns form an orthonormal basis $\{b_i\}$ for $\col(P)$. We may then parameterize the space of symmetric matrices $Z=Z^\dagger$, and solve for
\begin{align}
\forall i: [B^\dagger  \mathcal{L}_{b_i} B, Z] = 0.
\end{align}
If such a $Z$ exists, then its eigenspaces will be preserved by $B^\dagger  \mathcal{L}_{b_i} B$: thus one can construct the projector onto those eigenspaces to get the projector onto the ideal. Simplicity of the EJA is equivalent to the only solution being $Z=\lambda I$, a multiple of the identity. 

Another way of thinking about simplicity is in terms of the \emph{center} of the EJA. Let $V = \col(P)$ and let $L_x$ be the linear operator which performs the Jordan product with $x$. The center of a Euclidean Jordan algebra may be defined as $Z=\{z \in V: [L_z, L_x] = 0 \ \forall x \in V\}$ \cite{McCrimmon2003-kp}. In other words, it is just those elements whose Jordan product operator commutes with all other Jordan product operators. If the EJA is composite $V=V_1 \oplus V_2 \oplus \dots \oplus V_k$, the center is spanned by the units of each summand $u_1, u_2, \dots, u_k$. Thus $\dim Z = k$. The EJA is simple iff $Z$ is spanned by just a single unit $u$ \cite{farautAnalysisSymmetricCones1994}. To test this, let $\{b_i\}_{i=1}^r$ be a basis for $\col(P)$ and $B$ be the matrix with the basis elements as its columns. Let $\tilde{L}_i = B^\dagger L_{b_i} B$ be the Jordan product operator for the $b_i$th element represented as an $r\times r$ matrix. We can write an arbitrary element $z = \sum_l z_l b_l$ so that $\tilde{L}_z = \sum_l z_l \tilde{L}_l$. Then $z$ will be central iff
\begin{align}
\forall i: [\tilde{L}_z, \tilde{L}_i] &= \sum_l z_l [\tilde{L}_l, \tilde{L}_i]  = 0.
\end{align}
Alternatively, let $Z_{ijkl} = [\tilde{L}_l, \tilde{L}_i]_{jk}$ so that we require $\forall i,j,k: \sum_l Z_{ijkl} z_l = 0$. Reshaping this into an $r^3 \times r$ matrix, this becomes $Zz = 0$. The EJA is simple, then, if $\dim \nullspace(Z) = 1$: there is just one element $Z=u$ whose Jordan product operator commutes with all others.

While the above considerations yield a conceptually straightforward way of calculating whether the EJA is simple, when the rank of $P$ is large, the tensor $Z_{ijkl}$, which has $r^4$ entries, will become intractable to work with. We therefore provide an algorithm which will certify simplicity with overwhelming probability. We begin from the observation that if an EJA is composite, a primitive idempotent must lie entirely in one of the factors \cite{farautAnalysisSymmetricCones1994}. 

\begin{lemma}
   \label{no_composite_primitives}
Let $V$ be the vector space on which a composite EJA is defined, and let $\odot$ denote the Jordan product. $V$ may therefore be decomposed as $V=V_1 \oplus V_2 \oplus \dots$. Primitive idempotents must live entirely in one of the summands.
\end{lemma}
\begin{proof}
Without loss of generality, suppose $V= V_1 \oplus V_2$. Suppose $e$ were a primitive idempotent supported in both factors. Thus $e=e_1 \oplus e_2$ satisfies $e \odot e = e$. Since $(x_1 \oplus x_2)\odot (y_1 \oplus y_2)=(x_1 \odot y_1) \oplus (x_2 \odot y_2)$, we have $e_1 \odot e_1 = e_1 $ and $e_2 \odot e_2 = e_2$. In other words, $e_1$ and $e_2$ must themselves be idempotent. Thus $e=(e_1 \oplus 0) + (0 \oplus e_2)$ is a sum of idempotents. Since the summands are supported on different sectors, they are mutually orthogonal. Thus $e$ can be expressed as a sum of mutually orthogonal idempotents,  which contradicts primitivity.
\end{proof}

Suppose we pick a primitive idempotent $e$ at random. By the above lemma, it must lie entirely in one sector of the EJA. Now select random elements $\{x_i\}$ of the algebra and take the Jordan product of each with $e$, obtaining new elements $\{x_i^\prime\}$. Then repeat the process, selecting random elements of the algebra, and taking the Jordan product with each of the elements $\{x_i^\prime\}$ to obtain $\{x_i^{\prime\prime}\}$, hitting those with random elements of the algebra, for some number of repetitions. Finally, consider the linear span of the elements so obtained: this is an approximation of the ideal of which $e$ is a part. Indeed, if $e$ were supported on a single sector of the EJA, then the linear span of the elements obtained throughout this process should remain in that same sector. If the dimension of the linear span is in fact the dimension of $V$, the whole vector space, then the EJA must be simple.

To obtain a primitive idempotent $e$, one may find some element which satisfies the constraints we derived above: or one may obtain $e$ by spectrally decompositing a random element of the Jordan algebra. Generically, such an element will have $d$ distinct eigenvalues: thus its spectral decomposition will yield $d$ mutually orthogonal primitive idempotents $\{e_i\}$. Once one has selected a starting primitive idempotent $e$, instead of hitting it with random elements of the algebra, one may more systematically hit $e$ with elements of a basis $\{b_i\}$ for $V$. Then introduce a basis $\{b_i^\prime\}$ for the linear span of $\{e \odot b_i\}$. Once this is done, hit each element of $\{b_i^\prime\}$ with elements of the original basis $\{b_i\}$, and consider the linear span of $\{b_i \odot b_j^{\prime}\}$. Again, introduce a basis $\{b_i^{\prime\prime}\}$ for this linear span, and then consider $\{b_i \odot b_i^{\prime\prime}\}$, and so on. Eventually the dimension of these subspaces will stabilize, and if it in fact is $\dim V$, then the EJA is simple. 

Thus given $P(R|R)$ it is a matter of computation to determine whether the Jordan algebra it encodes is simple. And for simplicity's sake, we now suppose that it is. After all, if it were not, we could build it up from simpler components, considering a reference measurement for each part: the union of the reference measurement for each will be a reference measurement for the whole. (Later we will consider another type of motivation.)

\section{Full circle}

The 1934 paper of Jordan, Wigner, and von Neumann \cite{Jordan1934-cm} not only shows that Euclidean Jordan algebras must be composites of simple EJAs, but also gives a direct way of characterizing these fundamental building blocks. Up to some special isomorphisms between the different classes of EJAs\footnote{For example, the $n=3$ spin factor is isomorphic to $\text{Herm}_2(\mathbb{C})$.}, a simple EJA is uniquely defined by two numbers. The first is its \emph{rank}, that is, the maximum number of mutually orthogonal primitive idempotents, and the second is its \emph{dimension}, the smallest vector space on which the algebra may be defined. 

\begin{center}
\begin{tabular}{c|c|c}
\textbf{Euclidean Jordan algebra} & \textbf{rank} & \textbf{dimension} \\
\hline
$\text{Herm}_d(\mathbb{R})$ & $d$ & $d(d+1)/2$ \\
$\text{Herm}_d(\mathbb{C})$ & $d$ & $d^2$ \\
$\text{Herm}_d(\mathbb{H})$ & $d$ & $d(2d-1)$ \\
$\text{Herm}_3(\mathbb{O})$ & $3$ & $27$ \\
spin factor & $2$ & $n+1$
\end{tabular}
\end{center}
We thus make one final assumption about the conditional probability matrix $P(R|R)$ which characterizes the reference measurement, that $r=\rank(P)=d^2$ where $d$ is the maximum number of mutually orthogonal idempotents. With this final piece in place, we conclude that reference valuations have a representation as $d \times d$ Hermitian matrices over $\mathbb{C}$, and that states $P(R|\rho)$ and effects $P(E|R)$ have a representation as $d\times d$ positive semidefinite Hermitian matrices over $\mathbb{C}$. In other words, we have arrived at quantum mechanics.

We may now fix our constants. By self-duality, we had $P(\rho|R) = \kappa P(R|\rho)^\dagger$ where $\kappa = \frac{n}{1+(r-1)/\alpha}$. Thus in fact
\begin{align}
\alpha = \frac{\kappa(r-1)}{n-\kappa},
\end{align}
where we now know that $r=d^2$. We also now know that there is a representation of the reference measurement in terms of $d\times d$ positive semidefinite matrices satisfying $\sum_i R_i = I$. Denoting the corresponding reference states as $\sigma_i$, we have $\forall \rho: P(\rho|R_i)= \tr(\rho\sigma_i)=\kappa P(R_i|\rho) = \kappa \tr(R_i \rho)$, from which we conclude that $R_i = \kappa^{-1}\sigma_i$. Thus 
\begin{align}
\tr\left(\sum_i R_i\right) = \sum_i \tr(R_i) = n \kappa^{-1} = \tr(I) = d,
\end{align}
so that $\kappa = n/d$. Substituting $\kappa = n/d$ and $r=d^2$ into our expression for $\alpha$, we find
\begin{align}
\alpha = \frac{(n/d)(d^2-1)}{n - n/d} = \frac{d^2-1}{d-1}=d+1.
\end{align}
We may now express our formulas with all the constants determined. We have 
\begin{align}
g \check{\odot} h &= \frac12\left\{
(d+1)(d+2)\left(\frac{n}{d}\right)P(g \circ h)
-
\left(
\overline{h} g+\overline{g} h  + (d+1) (g \cdot h) u
\right)
\right\},
\end{align}
so that the Jordan product tensor (rescaled to implement $\check{\odot}$) is 
\begin{align}
\kappa [\mathcal{L}_{P}]_{ijk} &=  \frac{1}{2}\Bigg\{(d+1)(d+2)\left(\frac{n}{d}\right)\sum_m P(R_m|R_i)P(R_m|R_j)P(R_m|R_k)
\\&-
\Big(
P(R_i|R_j)+ P(R_i|R_k) + P(R_j|R_k)
+
d/n
\Big)\Bigg\}. \nonumber
\end{align}
The pure state vector condition reads
\begin{align}
P(R_i|\rho) = \frac{1}{2}\left\{\frac{1}{2}(d+1)(d+2)\left(\frac{n}{d}\right)\sum_{j}P(R_i|R_j)P(R_j|\rho)^2 -\frac{d}{n}\right\},
\end{align}
and finally, the pure state scalar constraints become
\begin{align}
\sum_i P(R_i|\rho)^2 &= \frac{d}{n}\frac{2}{d+1}\\
\sum_i P(R_i|\rho)^3 &=  \frac{d^2}{n^2}\frac{6}{(d+1)(d+2)},
\end{align}
all of which may be compared to the results of the last chapter. In particular, the cubic equation implies that
\begin{align}
\sum_{ij} P(R_i|R_j)^3 &= \sum_i \left[\sum_j P(R_i|R_j)^3 \right] = \frac{d^2}{n}\frac{6}{(d+1)(d+2)}.
\end{align}
Since $P(R_i|R_j) = \tr(R_i\sigma_j) = \frac{d}{n}\tr(\sigma_i \sigma_j)=\frac{d}{n}|\langle \psi_i|\psi_j\rangle|^2$, we have
\begin{align}
\mathcal{F}_3(\{\psi_i\})=\frac{1}{n^2}\sum_{ij}|\langle \psi_i|\psi_j\rangle|^6 &= \frac{6}{d(d+1)(d+2)}.
\end{align}
In other words, the pure states $\{\sigma_i\}$ attain the minimum of the 3rd order frame potential \cite{waldronIntroductionFiniteTight2018}, which we introduced in Chapter \ref{ch:characterizing}. We conclude at last that our reference measurement must correspond to an unweighted complex projective 3-design.

\section{Self-testing}

While each of the conditions we placed on $P(R|R)$ were motivated by foundational considerations, in hopes of giving a reconstruction of quantum mechanics which puts  the structure of the reference measurement itself in pride of place, we now observe that our considerations may be reinterpreted as providing a means of \emph{self-testing} complex projective 3-designs in a theory-agnostic way \cite{supicSelftestingQuantumSystems2020}. A set of quantum states forms a complex projective t-design iff it minimizes the order-$t$ frame potential: here the emphasis is on a set of \emph{quantum} states. One can calculate this quantity from $P(R_i|R_j)$; alternatively, if one is not able to implement the corresponding measurement, one can calculate it from $P(\sigma_i|\sigma_j)$ obtained e.g., by performing a swap test on the states \cite{fanizzaSwapTestOptimal2020,gitiauxSWAPTestArbitrary2022}. But how can one be sure that the states in fact have a Hilbert space representation? The same problem arises even if one can measure the frame potential directly \cite{nakataComputationalComplexityUnitary2025, mcginleyPostselectionFreeLearningMeasurementInduced2024}. Such certifications of $t$-designhood are conditional on the assumption that a Hilbert space representation is in fact possible. A theory-agnostic certification would guarantee that not only the frame potential is minimized, but also a Hilbert space representation can be constructed. In the course of this chapter, we have provided exactly such a means of doing this not just for 3-designs, but for t-designs for $t\ge 3$.

\begin{theorem}
   \label{self-testing-theorem}
Let $P\equiv P(R|R)$ be a column stochastic matrix satisfying:
\begin{itemize}
\item $r = \rank(P) = d^2$ for some $d$.
\item $P = P^T$.
\item $\forall i: P(R_i|R_i) = d/n$
\item $P \Phi P = P$ for $\Phi = \alpha I + (1-\alpha)J/n$ where $\alpha=d+1$.
\item Let $L_x = P \Phi \big\{\gamma \diag{x} + (1-\gamma)(x u^\dagger + \overline{x}I - ux^\dagger)/n\big\}$ where $\gamma = (1+1/\alpha)/2$ so that $L_x y = x \odotP y$. Let $\{b_i\}$ be an orthonormal basis for $\col(P)$. Then $\forall i, j, k: T(b_i, b_j, b_k) = 0$ where
\begin{align*}
T(a,b,c) = \frac{1}{3}\Big\{[L_a,L_{b\odotP c}] +[L_b,L_{c\odotP a}]+[L_c,L_{a\odotP b}]\Big\}P \Phi.
\end{align*}
\item The resulting EJA is simple, computed according to Section \ref{simple_eja}.
\item The rank of the resulting EJA is $d$, that is, the maximum number of distributions $\{P(R|\rho_j)\}$ in $\col(P)$ satisfying 
\begin{align*}
\forall j: \sum_i P(R_i|\rho_j)^2 &= \frac{d}{n}\frac{2}{d+1} && \text{and} && \sum_i P(R_i|\rho_j)^3 =  \frac{d^2}{n^2}\frac{6}{(d+1)(d+2)},
\end{align*}
and for $j \neq k$,
\begin{align*}
\sum_i P(R_i|\rho_j)P(R_i|\rho_k) &= \frac{d}{n} \frac{1}{d+1}
\end{align*}
is $d$, computed according to Section \ref{testing_rank}.
\end{itemize}
Then $P(R_i|R_j) = \frac{d}{n}\tr(\sigma_i \sigma_j)$ where $\{\sigma_i\}$ is a set of rank-1 projectors on a complex Hilbert space of dimension $d$ which forms an unbiased complex projective 3-design.
\end{theorem} 

\begin{corollary}
   Suppose the conditions in Theorem \ref{self-testing-theorem} are satisfied, and additionally that for $t\ge 3$
   \begin{align}
   \sum_{ij} P(R_i|R_j)^t = \frac{d^t}{n^{t-2}} \binom{d+t-1}{t}^{-1}.
   \end{align}
  Then $P(R|R)$ corresponds to a unbiased complex projective $t$-design.
\end{corollary}
\begin{proof}
The conditions in Theorem \ref{self-testing-theorem} guarantee that $P(R|R)$ has a Hilbert space representation as a complex projective 3-design. Recalling that a $t$-design is also a $(t-1)-$design, if $P(R|R)$ additionally minimizes the frame potential for $t> 3$, it must be a complex projective $t$-design.
\end{proof}

\section{Assume a SIC?}

We now show that if we make just one more assumption, we can rule out composite Jordan algebras, and moreover restrict the possible simple EJA's compatible with $P(R|R)$. This assumption is that the state space implied by $P(R|R)$ can host a \emph{symmetric informationally complete} set of states \cite{staceyQuantumTheorySymmetry2019}. If $\rank(P)=r$, then this means a set of $r$ states, which are primitive idempotents, and which form a regular simplex.
\begin{theorem}
Let $\{P(R|\rho_i)\}_{i=1}^r$ be a set of primitive idempotents such that $P(\rho_i|\rho_j)=c$ for $i\neq j$ and whose sum is proportional to $u$. Then $c = (\alpha-1)/(r + \alpha -1)$.
\end{theorem}
\begin{proof}
   We have
\begin{align}
c &= P(\rho_i|\rho_j)=P(\rho_i|R)\Phi P(R|\rho_j) = \kappa P(R|R_i)\Big[\alpha I + (1-\alpha)J/n\Big]P(R|R_j)\\
&= \kappa \big\{\alpha p_i \cdot p_j + (1-\alpha)/n\big\},
\end{align}
so that $p_i \cdot p_j = \frac{1}{\alpha}\{c/\kappa - (1-\alpha)/n\}$. Now on the one hand, using the formula for $\sum_i p_i^2$ for a primitive idempotent,
\begin{align}
\left\lVert \sum_i p_i \ \right\rVert^2 &= \sum_{ij} p_i \cdot p_j = \sum_i \lVert p_i\rVert^2 + \sum_{i\neq j}p_i \cdot p_j\\
&= r\left(\frac{\kappa^{-1} + (\alpha-1)/n}{\alpha} \right) + (r^2-r)\frac{1}{\alpha}\Big[c/\kappa - (1-\alpha)/n\Big].
\end{align}
On the other hand, since $\sum_i p_i = \beta u$, we have $\sum_i \tr(p_i) = \beta \tr(u)$ which implies $r=\beta n$ or $\beta = r/n$. Thus
\begin{align}
\left\lVert \sum_i p_i \ \right\rVert^2 &= \left\lVert (r/n)u \right\rVert^2 = (r^2/n^2)n = r^2/n.
\end{align}
Equating these two expressions and using $\kappa = n/(1+(r-1)/\alpha)$ yields the result.
\end{proof}
\begin{corollary}
   Such a set is informationally complete.
\end{corollary}
\begin{proof}
   Since $P(\rho_i|\rho_j) = \kappa p_i^\dagger \Phi p_j = \kappa\big(\alpha p_i \cdot p_j + (1-\alpha)/n)$, the $r\times r$ Gram matrix $G_{ij}=p_i \cdot p_j$ may be expressed 
\begin{align}
G = (A-B)I + BJ && A =\frac{1}{\alpha}\left(\frac{1}{\kappa} - (1-\alpha)/n\right) && B = \frac{1}{\alpha}\left(\frac{c}{\kappa} - \frac{1-\alpha}{n}\right).
\end{align}
The eigenvalues of $G$ are therefore $A-B$ with multiplicity $r-1$ and $A + (r-1)B$ with multiplicity 1. $A-B = (1-c)/(\alpha \kappa)$. Since $c\neq 1$, $A-B \neq 0$. Meanwhile, $r(A + (r-1)B) = \lVert \sum_i p_i\rVert^2 > 0$. Thus $G$ is full rank, and so the $\{p_i\}$ are linearly independent.
\end{proof}
\begin{lemma}
If such a SIC set exists, the EJA must be simple or else classical.
\end{lemma}
\begin{proof}
In Lemma \ref{no_composite_primitives}, we showed that primitive idempotents must be supported entirely in a single factor of a composite EJA. Thus if $\{P(R|\rho_i)\}$ are primitive idempotents, they must be supported on one or the other sector. But if $P(R|\rho_i)$ and $P(R|\rho_j)$ are supported on different sectors, $P(\rho_i|\rho_j) = 0$, which contradicts the assumed equiangularity of the set unless $c=0$. If $c=0$, the primitive idempotents must be mutually orthogonal, $\alpha=1$, $\Phi=I$, and $\odot=\circ$: the EJA is classical. Otherwise, the SIC must be supported entirely in one or the other summand: but this contradicts informational completeness.
\end{proof}

In light of this, just as we determined $d$, the cardinality of the set of mutually orthogonal primitive idempotents, by trying to find $d$ distributions satisfying the scalar primitive idempotent constraints as well as the orthogonality constraint, we can try to find $r$ distributions which satisfy the scalar primitive idempotent constraints as well as for $i \neq j$,
\begin{align}
P(\rho_i|\rho_j) = P(\rho_i|R)\Phi P(R|\rho_j) = \frac{\alpha-1}{r + \alpha -1}
\end{align}
If this can be done, then a SIC set exists, and the EJA must be simple. Moreover, the existence of a SIC set narrows down the possible choices of simple EJA's as explained in \cite{staceyQuantumTheorySymmetry2019}. Indeed, SICs are only known to exist for certain special ranks ($d=2,3,7,23$---although there may be more yet to be discovered) in the setting of quantum mechanics over $\mathbb{R}$ \cite{Weiss2021,waldronIntroductionFiniteTight2018}, and they do not appear to exist for quantum mechanics over $\mathbb{H}$ for $d>3$ \cite{cohnOptimalSimplicesCodes2016}. The same paper, however, shows that a SIC \emph{does} exist in the exceptional $3 \times 3$ octonionic case. Moreover, a SIC always exists in a spin-factor EJA: since the state spaces are spheres, it is trivial to inscribe a regular $(r-1)-$ simplex within them. But the rank of a spin-factor EJA---the maximum number of mutually orthogonal primitive idempotents---is always just 2. Thus if $d\neq 2,3,7$ or $23$, unless one stumbles upon an exceptional SIC over $\mathbb{R}$ or $\mathbb{H}$, the EJA must correspond to quantum mechanics over $\mathbb{C}$---or it would, if SICs existed over $\mathbb{C}$ for all $d$. There is very strong numerical evidence that this is the case \cite{fuchsSICQuestionHistory2017, MarkusGrasslComputing}, as well as a by now highly developed theory behind their exact construction. For the state of the art on the SIC existence question, see \cite{applebyConstructiveApproachZauners2025} which proves the existence of SICs over $\mathbb{C}$ in all dimensions conditional on the resolution of certain conjectures in algebraic number theory (related to Hilbert's twelfth problem) as well as the proof of a special function identity.

At the risk of specializing too much, we can go further and assume that among the reference states and effects themselves are not only a SIC set, but also $s \leq d$ mutually orthogonal primitive idempotents. Then we can rule out composite EJAs as well as incompatible simple EJAs from a simple inspection of $P(R|R)$. It remains to show that a 3-design with this structure can in fact always be constructed in quantum mechanics over $\mathbb{C}$, assuming SICs exist. 
\begin{theorem}
An unbiased 3-design can be constructed in quantum mechanics over $\mathbb{C}$ which contains a SIC as well as at least $s\leq d$ orthogonal states.
\end{theorem}
\begin{proof}
Analogous to quantum state 3-designs, a unitary 3-design is an ensemble of unitaries which mimic the behavior of Haar distributed unitaries up to the third moment \cite{royUnitaryDesignsCodes2009,grossEvenlyDistributedUnitaries2007,bannaiExplicitConstructionExact2022}. For sufficiently large cardinality, such sets always exist. Let $\mathcal{U}$ be an unweighted unitary 3-design. By definition,
\begin{align}
\frac{1}{|\mathcal{U}|}\sum_{U \in \mathcal{U}} (U \Pi U^\dagger)^{\otimes 3} = \int dU (U \Pi U^\dagger)^{\otimes 3} =\binom{d+2}{3}^{-1}\Pi_{\text{sym}^3},
\end{align}
so that $\{U \Pi U^\dagger\}$, the orbit of a state $\Pi$ under the unweighted unitary 3-design, forms an unweighted state 3-design. Let $\{\Pi_i\}$ be a set of SIC states. Then
\begin{align}
\frac{1}{d^2}\sum_{i=1}^{d^2}  \frac{1}{|\mathcal{U}|}\sum_{U \in \mathcal{U}}(U \Pi_i U^\dagger)^{\otimes 3} = \binom{d+2}{3}^{-1}\Pi_{\text{sym}^3},
\end{align}
so that we have constructed an unbiased 3-design which contains a SIC subset as the union of the orbit of each SIC state under the unitary 3-design (counting repeated unitaries with multiplicity). Now without loss of generality, we can take $I \in \mathcal{U}$: if not, we may take any element $U_1$ and consider the set $\{U_1^\dagger U\}$ which forms a 3-design which does contain $I$. Now take one of the SIC states $\Pi_1$ and $s-1$ unitaries $V_i$ such that $\Pi_1, V_1 \Pi_1 V_1^\dagger, V_2 \Pi_1 V_2^\dagger, \dots $ are $s$ mutually orthogonal states: this can always be done. Then take the union (again preserving repeated unitaries) $\mathcal{U}^\prime = \mathcal{U} \cup V_1 \mathcal{U} \cup V_2 \mathcal{U} \dots$. Such an equal weight union of unitary 3-designs is again an unweighted 3-design. If we then consider the union (with multiplicity) of the orbits of the $d^2$ SIC states under $\mathcal{U}^\prime$, we have an unbiased 3-design which contains not only a SIC but also $s$ orthogonal states.
\end{proof}

\section{Conclusion}

And so we have come full circle. Let us retrace our steps. We began with the gambler trying to make better decisions, and supposed they have identified a reference measurement for a domain they are invested in. As we saw in Chapter \ref{ch:probabilities}, on very general grounds, one may derive a nonclassical coherence condition $P(E|\rho) = P(E|R)\Phi P(R|\rho)$ where $\Phi$ is any $\{1\}$-inverse of the conditional probability matrix $P(R|R)$ which characterizes the reference measurement, taken to be a preparatory measurement, itself. We then proceeded to make increasingly severe restrictions on the structure of this matrix $P\equiv P(R|R)$. We assumed that it is symmetric, and so both states $P(R|\rho)$ and response functions $P(E|R)$, as well as valuations upon the reference outcomes, ought to be confined to $\col(P)$. We also assumed the $P$ is constant along its diagonal. We presumed that the Born matrix $\Phi$ has the form of a Protourgleichung, $\Phi = \alpha I + \beta J$, and assumed it to be quasistochastic, fixing the value of $\beta$. This in fact implies that $P$ depolarizes vectors in its column space, and that $P$ has a simple eigenstructure, which fixes the value along the diagonal of $P$ in terms of $r=\rank(P)$ and $\alpha$.

We then discussed how the gambler can transfer valuations on arbitrary measurements to equivalent valuations on the reference measurement. Reference valuations have a natural product, the elementwise or Hadamard product $\circ$, which in fact is the Jordan product on $\mathbb{R}^n \oplus \dots \oplus \mathbb{R}^n$. Just as $\Phi$ represents a minimal deformation of the classical law of total probability, we attempted to deform the classical product in as minimal a way as possible, in order to define a product on valuations which we hoped would be, in some sense, independent of the choice of reference measurement. We show that any product $\odot$ which is commutative, permutation equivariant, unital for $(1,\dots,1)$ and is a one parameter mixture with the classical product with as few terms as possible leads uniquely to a product of the form 
\begin{align}
x \odot y &= \gamma (x \circ y) + (1-\gamma)(\overline{y} x + \overline{x} y - (x \cdot y)u)/n.
\end{align}
But since valuations ought to be confined to $\col(P)$, we ultimately adopted $x\bar{\odot}y=\Pi(x \odot y)$, where $\Pi$ is the projector onto $\col(P)$, to be the proper definition of our minimal extension of the classical product on valuations. Demanding compatibility with the Protourgleichung in fact fixes the value of $\gamma$ in terms of $\alpha$, and shows that the product is self-adjoint with respect to an inner product defined in terms of $P$, and the algebra it defines is formally real. 

We then supposed that there exists a special class of measurements which satisfy a moment-matching condition. If $\mathfrak{x}$ is a valuation on $\{X_i\}$, then $\sum_i \mathfrak{x}_i^m P(X_i|\rho) = \sum_i x^{\odotP m}_i P(R_i|\rho)$, that is, that the $m$th moment of the reference valuation with respect to $\odotP$ in fact reproduces the $m$th moment of the valuation on the original measurement. For this moment-matching condition to make sense, we further demanded that $\odotP$ be power associative so that $\odotP$ powers have a univocal meaning. We thus confronted a finite dimensional real and formally real commutative power associative algebra of observables defined through the bilinear product $\odotP$: we then appealed to Jordan, von Neumann, and Wigner's classic result that the algebra we have defined is in fact a Euclidean Jordan algebra, and indeed, $\odotP$ satisfies the Jordan identity. Conversely, if the product satisfies the Jordan identity, it must be power associative, and so the Jordan identity (in its third order polarized form) gives us a new constraint on $P(R|R)$.

Next, we observed that $\odotP$ is defined on valuations, which may be interpreted as dual coordinates: we therefore lifted $\odotP$ to $\hat{\odot}$ a product on regular coordinates e.g., probability vectors and response functions, working out the form of the matrix $\mathcal{L}_x$ which satisfies $\mathcal{L}_x \Phi y = x \hat{\odot}y$, and for good measure, reexpressed the Jordan identity in terms of $\hat{\odot}$. We then appealed to the self-duality of the cone of squares of an EJA to derive the shape of state space via an uncertainty principle. We showed that in fact the positive semidefiniteness of the matrix $\mathcal{L}_{P(R|\rho)}$ equivalently diagnoses the validity of the state. Self-duality also implies that states $P(R|\rho)$ can be rescaled into effects $P(\rho|R)$, and we derived a self-duality constant $\kappa$ which does just this. The assumption that if we turn a reference state into an effect in this way, we should expect the corresponding outcome with certainty fixes $\kappa$ in terms of $r, n$, and $\alpha$, and shows that if $P(\rho|\rho)=1$, the corresponding reference distribution $P(R|\rho)$ must live on the surface of a 2-sphere with a particular radius. Finally, we rescaled $\hat{\odot}$ into $\check{\odot}$ using the self-duality constant, readying us for the discussion of the idempotents of the Jordan algebra.

Indeed, we then noted that any element of the algebra has a spectral decomposition into a linear combination of primitive idempotents, and sketched a recipe for doing so. This in fact realizes our original ambition: the spectral decomposition into primitive idempotents implies that any reference valuation corresponds to a valuation on a measurement whose effects are precisely a set of mutually orthogonal idempotents---these are the types of measurements whose higher moments may be derived simply from the nonclassical product on the reference. We then derived explicit equations which characterize idempotents, and in particular a vector constraint which is a necessary and sufficient condition for a distribution $P(R|\rho)$ to be a primitive idempotent as well as a set of equivalent scalar constraints: the former 2-norm sphere constraint, as well as a 3-norm sphere constraint. Moreover, we observed that our assumptions have already implied that the reference states are primitive idempotents. Finally, we derived explicit conditions for two states to be orthogonal with respect to $\check{\odot}$, noting that the rank of a Euclidean Jordan algebra is the maximum number of mutually orthogonal primitive idempotents. Given the matrix $P(R|R)$, one may check computationally what is the rank of the algebra it gives birth to. (And by the same token, the same algorithm could be used to search for a $P(R|R)$ which has a particular rank, subject to the other constraints.)

The Jordan-von Neumann-Wigner theorem characterizes the simple Euclidean Jordan algebras, the atomic building blocks of all EJA's, and we give a recipe for checking computationally whether our EJA is simple in terms of $P(R|R)$---or equivalently, for searching for such a $P(R|R)$. Assuming simplicity, we may then appeal to the classification given in the aforementioned theorem. Supposing that the $r=\rank(P)=d^2$ where $d$ is the maximum number of mutually orthogonal idempotents, our product $\check{\odot}$ must in fact correspond to the Jordan product on $\text{Herm}_d(\mathbb{C})$, that is, $d\times d$ Hermitian matrices over the complex numbers. The state and effect spaces may be understood in terms of the positive semidefinite cone: the states must have representations as $d\times d$ positive semidefinite matrices over $\mathbb{C}$ satisfying $\tr(\rho)=1$ and the effects must be $d \times d$ positive semidefinite matrices satisfying $0 \leq E \leq I$. In other words, we have arrived at quantum mechanics. This fixes the value of the self-duality constant $\kappa$, which in turn fixes the value of $\alpha$. Coming full circle, we show that with these values, the reference states must minimize the 3rd order frame potential, and so the reference measurement corresponds to an unweighted complex projective 3-design. We then showed that beyond its value as a reconstruction of quantum mechanics, our considerations amount to a theory agnostic scheme for certifying complex projective $t$-designs for $t\ge 3$. We closed by showing that the existence of a symmetric informationally complete set excludes composite EJA's and moreover, narrows down the possible simple EJA's dramatically.

In the previous chapter, we proceeded within quantum mechanics, assumed our reference measurement was built from an unbiased complex projective 3-design, and then showed how the shape of quantum state space ccould be unpacked from this single matrix $P(R|R)$. In this chapter we reverted back to the origin: after wiping the slate clean,  by placing increasingly poweful constraints on $P(R|R)$, we were ultimately able to narrow our focus \emph{back} to quantum mechanics, showing that our assumptions take us full circle, back to a 3-design. The key step was the introduction of the nonclassical product $\odot$, guided by the idea that the classical product on valuations should be deformed as minimally as possible. Indeed, this is what is remarkable about 3-designs: just as 2-designs make the Born rule look as close as possible to the law of total probability, 3-designs make the classical Jordan product $\circ$ on valuations look as close as possible to the quantum Jordan product. Since the entire theory can be defined in terms of the Jordan product, this in fact buys us everything. Moreover, our entire construction may be viewed as a means of \emph{self-testing} unbiased $t$-design measurements for $t\ge 3$. One may of course calculate a value which amounts to the frame potential from $P(R|R)$, but there is no guarantee that $P(R|R)$ has a Hilbert space representation. Checking the rank, the Jordan identity, and so forth (Theorem \ref{self-testing-theorem}), ensures that $P(R|R)$ does indeed have a Hilbert space representation, and so is certifiably quantum $t$-design.

Until the final step fixing simplicity and the rank of the algebra, our construction is compatible with any Euclidean Jordan algebra, giving a new perspective on both their geometry and their relation to classical probability theory. That said, our reconstruction is by intention modest: its main goal, one might say, is to help one understand precisely in what sense quantum theory can be understood as a probability theory supplemented with nonclassical coherence conditions which are gentle modifications of classical rules. Indeed, this is the general strategy of QBist thinking---that and putting the limelight on a wisely chosen reference measurement so that consistency with the reference measurement properly understood implies consistency across all measurements. Of course, simply stipulating the rank of the algebra and the rank of $P$ may appear ad hoc, or ulteriorly motivated: the real point, however, is to dramatize the fact that these are precisely stipulations about the reference measurement itself. Giving fuller motivation to those choices we leave to future work. Besides assuming a SIC, many plausible routes have been already taken to narrow down the simple EJA's to quantum mechanics over $\mathbb{C}$: for example, supposing observables are generators of time evolution \cite{Barnum2014} or assuming local tomography \cite{barnum2013localtomographyjordanstructure, barnumSelfdualityJordanStructure2023}. We take here a pragmatic point of view, if you will. Confronted with some phenomenon, the gambler makes many experiments, identifying a reference measurement which they characterize by $P(R|R)$. If this matrix satisfies the constraints we have developed in this chapter, then the gambler can conclude: they may use the quantum mechanical formalism to facilitate their calculations. 

A different kind of motivation, however, may become more apparent through further study: much is still not understood about the nature of our construction. For example, we showed how the pure states or primitive idempotents correspond to distributions $P(R|\rho)$ on $n$ outcomes satisfying
\begin{align}
   \forall i: P(R_i|\rho) &\ge 0 \\
   \sum_i P(R_i|\rho) &= 1\\
\sum_i P(R_i|\rho)^2 &= \frac{d}{n}\frac{2}{d+1}\\
\sum_i P(R_i|\rho)^3 &=  \frac{d^2}{n^2}\frac{6}{(d+1)(d+2)}\\
P(R|\rho)&\in \col(P),
\end{align}
where $\dim \col(P)=d^2$. Somehow these constraints pick out precisely the $2d-2$ dimensional manifold of pure quantum states\footnote{A state vector $|\psi\rangle$ is defined by $2d$ complex numbers, and after fixing phase and overall normalization, we end up with $2d-2$ parameters.}: Somehow the hyperplane defined by $P$ intersects the spheres in a very peculiar nontrivial way. Better understanding this in purely geometrical terms might lead to a novel characterization of Euclidean Jordan algebras themselves. 

Indeed, it may be in our interest to relax our assumptions in order to study this question. For example, let us assume merely that the state space is fully characterized by the uncertainty principle,
\begin{align}
P(R|\rho) \text{ valid} \Longleftrightarrow \forall x \in \col(P): \langle x^2\rangle_\rho \ge \frac{\alpha-1}{\alpha +1}\Big( \langle x^2\rangle_\mu - 2 \langle x\rangle_\mu \langle x\rangle_\rho \Big).
\end{align}
As we have seen, assuming that our nonclassical product is power associative gives us the Jordan identity, which justifies this characterization of the state space. But let us instead take the uncertainty principle as given. One can then explore the possible state spaces which emerge from different choices of $P(R|R)$. Compatibility with the uncertainty principle is still diagnosed by the positive semidefiniteness of $\mathcal{L}_{P(R|\rho)}$: nothing in that argument depended on the Jordan structure. Thus the state space in fact forms a so-called \emph{spectrahedron}.

A \emph{spectrahedron} is the intersection of an affine-linear space with the convex cone of positive semidefinite matrices \cite{Vinzant2014WhatIA}. An $n$-dimensional affine-linear subspace of real symmetric matrices may be parameterized by a vector $x\in \mathbb{R}^n$ and a collection of $n+1$ matrices $\{A_i\}_{i=0}^n$ as
\begin{align}
A_x = A_0 + \sum_i x_i A_i.
\end{align}
The spectrahedron itself $A \subseteq \mathbb{R}^n$ consists of those $x$ such that $A_x \ge 0$. Now let
\begin{align}
T_{ijk} &=  \frac{1}{2}\Bigg\{\alpha(\alpha+1)\sum_m P(R_m|R_i)P(R_m|R_j)\delta_{mk}
\\&-(\alpha-1)
\Big(
P(R_i|R_j)+ \delta_{ik} + \delta_{jk}
+
(\alpha-1)/n
\Big)/n\Bigg\}, \nonumber
\end{align}
so that $\mathcal{L}_x = \sum_i T_{ijk} x_k$. Letting $A_0=0$ and otherwise $A_k = T_{ijk}$, it is clear then that our state space is indeed a spectrahedron (in fact the intersection of a \emph{linear space} with the positive semidefinite cone. We ought also to intersect the spectrahedron with the probability simplex as well as $\col(P)$ so that the state space is $\mathscr{S} = A \cap \Delta_n \cap \col(P)$.

One may then study this broader class of ``variance-bounded'' state spaces, which are fully defined by their compatibility with the uncertainty principle. In particular, when are such state spaces self-dual? When does there in fact exist a class of measurements such that if $\mathfrak{x}$ is the valuation on such a measurement $\{X_i\}$ and $x = \mathfrak{x}P(X|R)\Phi$ is the equivalent reference valuation, $\langle \mathfrak{x}^2\rangle_\rho = \langle x^{\odot 2}\rangle_\rho$, giving some justification for the uncertainty principle itself? When does $x \odot y$ in fact have an equivalent valuation on a measurement which is reference independent? When can the extremal states of the theory be given a simple characterization? Given a choice of $P(R|R)$, what properties does the product $\mathcal{L}_x \Phi y = x \hat{\odot} y$ have, and how do they reflect the geometry of the state space? Conversely, if one chooses a $P(R|R)$ so that $\hat{\odot}$ has some interesting property, what geometry does this imply? Assuming power associativity delivered us an embarrassment of riches, the highly constrained geometry implied by the Jordan product. But there remain deep questions about how exactly the constraints we put on $P(R|R)$ allow a particular state space to emerge, through the interplay between the bare nonclassical product $\odot$ and its projection into $\col(P)$. By studying these variants, one may be better placed to understand the relationship between the algebra of observables and the geometry of the allowed reference distributions, in particular the restriction to $\col(P)$.

Indeed, one final, parting theorem can help dramatize the importance of the subspace restriction. Suppose we defined the product on $\mathbb{R}^n$ in its entirety. It turns out that the parameter $\gamma$ interpolates between two Jordan algebras: when $\gamma=1$, we have the Jordan algebra provided by the Hadamard product; when $\gamma=0$, we have in fact a spin-factor Jordan algebra, although with a negative definite inner product. When $n=2$, any choice of $\gamma$ leads to a Jordan algebra. But for $n>2$, only $\gamma=0$ or $\gamma=1$ give rise to Jordan algebras. But if we project the product into a subspace, as we have seen, then the Jordan identity may be satisfied for intermediate values of $\gamma$. 

\begin{theorem}
Let
\begin{equation}
x \odot_{\gamma} y
=
\gamma (x \circ y)
+
(1-\gamma)
\left(
\overline{y}x+\overline{x}y-(x\cdot y)u
\right)/n
\end{equation}
on $\mathbb{R}^n$, where $u=(1,\dots,1)$ and
$\overline{x}=\sum_i x_i$. Then $\odot_{\gamma}$ satisfies the Jordan
identity for every $\gamma$ when $n=2$, and for $n>2$ if and only if
$\gamma=0$ or $\gamma=1$.
\end{theorem}

\begin{proof}
First, one may confirm that $u$ is the identity element. If $\gamma=1$, then
$\odot_{\gamma}$ is the elementwise or Hadamard product and hence is associative, so satisfies the Jordan identity: indeed, it is the Jordan product on $\mathbb{R} \oplus \dots \oplus \mathbb{R} \simeq \mathbb{R}^n$. If $\gamma=0$, let $a=\overline{x}/n$ and $b=\overline{y}/n$ so that
\begin{equation}
x=au+x_0,
\qquad
y=bu+y_0,
\qquad
\overline{x_0}=\overline{y_0}=0.
\end{equation}
Then
\begin{equation}
x\odot_0 y
=
ay_0+bx_0 
+ \left(ab-(x_0\cdot y_0)/n\right)u,
\end{equation}
or, separating out the traceful and traceless components,
\begin{align}
(a, x_0) \odot_0 (b, y_0) = \Big(ab + B(x_0, y_0), a y_0 + bx_0\Big),
\end{align}
for $B(x_0, y_0) = -(x_0 \cdot y_0)/n$. In fact, this is the spin-factor Jordan product with a negative definite inner product.  Hence $\odot_0$ satisfies the
Jordan identity.

For $n=2$, let $h=e_1-e_2$ for computational basis vectors $\{e_i\}_{i=1}^n$. Then every element in the algebra is of the form $au+bh$,
and since
\begin{equation}
u\odot_{\gamma}x=x,
\qquad
h\odot_{\gamma}h=(2\gamma-1)u,
\end{equation}
the algebra is generated entirely by $h$. Thus it suffices to check
\begin{align}
(h^{\odot_\gamma 2}) \odot_\gamma h = \Big(2\gamma -1)u\Big) \odot_\gamma h = (2\gamma-1)h = h \odot_\gamma \Big((2\gamma-1)u\Big)= h \odot_\gamma (h^{\odot_\gamma 2}).
\end{align}
Thus Jordan identity holds for all $\gamma$ when $n=2$. Now assume $n>2$. Let $x=e_1-e_2$ and $y=e_1$ so that
\begin{equation}
\overline{x}=0,
\qquad
\overline{y}=1,
\qquad
x\cdot x=2,
\qquad
x\cdot y=1,
\qquad
x\circ x=e_1+e_2,
\qquad
x\circ y=e_1.
\end{equation}
We then have
\begin{equation}
x^{\odot_\gamma 2}
=
\gamma(e_1+e_2)-2(1-\gamma)u/n,
\qquad
x\odot_{\gamma}y
=
\gamma e_1+(1-\gamma)(x-u)/n.
\end{equation}
Checking the Jordan identity gives
\begin{equation}
x^{\odot_\gamma 2} \odot_{\gamma}(x\odot_{\gamma}y)
-
x\odot_{\gamma}(x^{\odot_\gamma 2}\odot_{\gamma}y)
=
-2\gamma(1-\gamma)^2
\left(
(n-2)e_2-\sum_{k=3}^n e_k
\right)/n^2.
\end{equation}
Since the parenthesized vector is nonzero, the Jordan
identity forces
\begin{equation}
\gamma(1-\gamma)^2=0,
\end{equation}
so $\gamma=0$ or $\gamma=1$. These two values were already shown to
satisfy the Jordan identity, and this completes the proof.
\end{proof}

We leave it to future work to more fully explore the interplay between the ``simplest modification of the classical product on valuations'' and restriction to $\col(P)$. As we saw, the latter can be derived from the fundamental nonclassical coherence relation itself. At the same time, it is only a meaningful restriction because the representation of quantum mechanics afforded by a 3-design must be overcomplete. We hope that further study of the role played by this ``gauge freedom'' in representation will ultimately allow for an ever more transparent reconstruction of quantum mechanics, which centers the dual freedom of agent and world, mediated by a well-chosen reference measurement. 

%% file: sections/backmatter.tex
%%%%%%%%%%%%%%%%%%%%%%%%%%%%%%%%%%%%%%%%%%%%%%%%%%%%%%%%%

\SetUpBibliography
\bibliographystyle{unsrt}
\bibliography{dissertation.bib}

%%%%%%%%%%%%%%%%%%%%%%%%%%%%%%%%%%%%%%%%%%%%%%%%%%%%%%%%%

%% file: dissertation.bbl
\begin{thebibliography}{100}

\bibitem{fuchs2019qbismquantumtheoryheros}
Christopher~A. Fuchs and Blake~C. Stacey.
\newblock {{QBism}}: {{Quantum Theory}} as a {{Hero}}'s {{Handbook}}.
\newblock {\em Proceedings of the International School of Physics "Enrico Fermi"}, 197(Foundations of Quantum Theory):133--202, 2019.

\bibitem{fuchsQbismWhereNext2023}
Chris Fuchs.
\newblock {{QBism}}, {{Where Next}}?
\newblock In Philipp Berghofer and Harald~A. Wiltsche, editors, {\em Phenomenology and {{Qbism}}: {{New Approaches}} to {{Quantum Mechanics}}}. Routledge, 2023.

\bibitem{debrotaQBismsAccountQuantum2024}
John~B. DeBrota, Christopher~A. Fuchs, and R{\"u}diger Schack.
\newblock {{QBism}}'s account of quantum dynamics and decoherence.
\newblock {\em Physical Review A}, 110(5):052205, November 2024.

\bibitem{mullerProbabilisticTheoriesReconstructions2021}
Markus M{\"u}ller.
\newblock Probabilistic theories and reconstructions of quantum theory.
\newblock {\em SciPost Physics Lecture Notes}, page~28, March 2021.

\bibitem{Appleby_2017}
Marcus Appleby, Christopher~A. Fuchs, Blake~C. Stacey, and Huangjun Zhu.
\newblock Introducing the {{Qplex}}: A novel arena for quantum theory.
\newblock {\em The European Physical Journal D}, 71(7), July 2017.

\bibitem{slomczynskiMorphophoricPOVMsGeneralised2020}
Wojciech S{\l}omczy{\'n}ski and Anna Szymusiak.
\newblock Morphophoric {{POVMs}}, generalised qplexes, and 2-designs.
\newblock {\em Quantum}, 4:338, September 2020.

\bibitem{Huang_2020}
Hsin-Yuan Huang, Richard Kueng, and John Preskill.
\newblock Predicting many properties of a quantum system from very few measurements.
\newblock {\em Nature Physics}, 16(10):1050--1057, June 2020.

\bibitem{mao2024magicquditshadowestimation}
Chengsi Mao, Changhao Yi, and Huangjun Zhu.
\newblock The {{Magic}} in {{Qudit Shadow Estimation}} based on the {{Clifford Group}}, 2024.

\bibitem{Kliesch_2021}
Martin Kliesch and Ingo Roth.
\newblock Theory of {{Quantum System Certification}}.
\newblock {\em PRX Quantum}, 2(1), January 2021.

\bibitem{chen2024nonstabilizernessenhancesthriftyshadow}
Datong Chen and Huangjun Zhu.
\newblock Nonstabilizerness enhances thrifty shadow estimation, 2024.

\bibitem{Berghofer2024}
Philipp Berghofer.
\newblock Defending the quantum reconstruction program.
\newblock {\em European Journal for Philosophy of Science}, 14(3), September 2024.

\bibitem{Brassard2005}
Gilles Brassard.
\newblock Is information the key?
\newblock {\em Nature Physics}, 1(1):2--4, October 2005.

\bibitem{fuchs2002quantummechanicsquantuminformation}
Christopher~A. Fuchs.
\newblock Quantum mechanics as quantum information (and only a little more), 2002.

\bibitem{grinbaum2005notionreconstructionquantumtheory}
Alexei Grinbaum.
\newblock On the notion of reconstruction of quantum theory, 2005.

\bibitem{Chiribella2011}
Giulio Chiribella, Giacomo~Mauro D'Ariano, and Paolo Perinotti.
\newblock Informational derivation of quantum theory.
\newblock {\em Physical Review A}, 84(1), July 2011.

\bibitem{hardy2001quantumtheoryreasonableaxioms}
Lucien Hardy.
\newblock Quantum theory from five reasonable axioms, 2001.

\bibitem{Hardy2016}
Lucien Hardy.
\newblock Reconstructing quantum theory.
\newblock In Giulio Chiribella and Robert~W. Spekkens, editors, {\em Quantum Theory: {{Informational}} Foundations and Foils}, pages 223--248. Springer Netherlands, Dordrecht, 2016.

\bibitem{Hhn2017}
Philipp~Andres H{\"o}hn.
\newblock Toolbox for reconstructing quantum theory from rules on information acquisition.
\newblock {\em Quantum}, 1:38, December 2017.

\bibitem{Masanes_2011}
Llu{\'i}s Masanes and Markus~P M{\"u}ller.
\newblock A derivation of quantum theory from physical requirements.
\newblock {\em New Journal of Physics}, 13(6):063001, June 2011.

\bibitem{Selby2021}
John~H. Selby, Carlo~Maria Scandolo, and Bob Coecke.
\newblock Reconstructing quantum theory from diagrammatic postulates.
\newblock {\em Quantum}, 5:445, April 2021.

\bibitem{vandeWetering:2018kcp}
John {van de Wetering}.
\newblock An effect-theoretic reconstruction of quantum theory.
\newblock January 2018.

\bibitem{wilceRoyalRoadQuantum2018}
Alexander Wilce.
\newblock A {{Royal Road}} to {{Quantum Theory}} (or {{Thereabouts}}).
\newblock {\em Entropy}, 20(4):227, March 2018.

\bibitem{Nielsen_Chuang_2010}
Michael~A. Nielsen and Isaac~L. Chuang.
\newblock {\em Quantum Computation and Quantum Information: 10th Anniversary Edition}.
\newblock Cambridge University Press, Cambridge, 2010.

\bibitem{staceyQuantumTheorySymmetry2019}
Blake~C. Stacey.
\newblock Quantum {{Theory}} as {{Symmetry Broken}} by {{Vitality}}, 2019.

\bibitem{Drr1992}
Detlef D{\"u}rr, Sheldon Goldstein, and Nino Zangh{\'i}.
\newblock Quantum equilibrium and the origin of absolute uncertainty.
\newblock {\em Journal of Statistical Physics}, 67(5--6):843--907, June 1992.

\bibitem{Vaidman_2022}
Lev Vaidman.
\newblock Why the many-worlds interpretation?
\newblock {\em Quantum Reports}, 4(3):264--271, August 2022.

\bibitem{farautAnalysisSymmetricCones1994}
Jacques Faraut and Adam Kor{\'a}nyi.
\newblock {\em Analysis on {{Symmetric Cones}}}.
\newblock Oxford University PressOxford, December 1994.

\bibitem{Jordan1934-cm}
P~Jordan, J~{v. Neumann}, and E~Wigner.
\newblock On an algebraic generalization of the quantum mechanical formalism.
\newblock {\em Annals of Mathematics}, 35(1):29, January 1934.

\bibitem{Birkhoff1936}
Garrett Birkhoff and John~Von Neumann.
\newblock The logic of quantum mechanics.
\newblock {\em The Annals of Mathematics}, 37(4):823, October 1936.

\bibitem{hartkamperFoundationsQuantumMechanics1974}
A.~Hartk{\"a}mper, H.~Neumann, J.~Ehlers, K.~Hepp, H.~A. Weidenm{\"u}ller, and W.~Beiglb{\"o}ck, editors.
\newblock {\em Foundations of {{Quantum Mechanics}} and {{Ordered Linear Spaces}}: {{Advanced Study Institute Marburg}} 1973}, volume~29 of {\em Lecture {{Notes}} in {{Physics}}}.
\newblock Springer Berlin Heidelberg, Berlin, Heidelberg, 1974.

\bibitem{Davies1970}
E.~B. Davies and J.~T. Lewis.
\newblock An operational approach to quantum probability.
\newblock {\em Communications in Mathematical Physics}, 17(3):239--260, September 1970.

\bibitem{ludwigAxiomaticBasisQuantum1985}
G{\"u}nther Ludwig.
\newblock {\em An {{Axiomatic Basis}} for {{Quantum Mechanics}}}.
\newblock Springer Berlin Heidelberg, Berlin, Heidelberg, 1985.

\bibitem{mackey2004mathematical}
G.W. Mackey.
\newblock {\em Mathematical Foundations of Quantum Mechanics}.
\newblock Dover Books on Mathematics. Dover Publications, 2004.

\bibitem{Randall1978}
C.~H. Randall and D.~J. Foulis.
\newblock The operational approach to quantum mechanics.
\newblock In {\em Physical Theory as Logico-Operational Structure}, pages 167--201. Springer Netherlands, 1978.

\bibitem{BARNUM20113}
Howard Barnum and Alexander Wilce.
\newblock Information processing in convex operational theories.
\newblock {\em Electronic Notes in Theoretical Computer Science}, 270(1):3--15, 2011.

\bibitem{barrettInformationProcessingGeneralized2007}
Jonathan Barrett.
\newblock Information processing in generalized probabilistic theories.
\newblock {\em Physical Review A}, 75(3):032304, March 2007.

\bibitem{Janotta_2014}
Peter Janotta and Haye Hinrichsen.
\newblock Generalized probability theories: What determines the structure of quantum theory?
\newblock {\em Journal of Physics A: Mathematical and Theoretical}, 47(32):323001, July 2014.

\bibitem{lami2018nonclassicalcorrelationsquantummechanics}
Ludovico Lami.
\newblock Non-classical correlations in quantum mechanics and beyond, 2018.

\bibitem{PLAVALA20231}
Martin Pl{\'a}vala.
\newblock General probabilistic theories: {{An}} introduction.
\newblock {\em Physics Reports}, 1033:1--64, 2023.

\bibitem{Aubrun2022}
Guillaume Aubrun, Ludovico Lami, Carlos Palazuelos, and Martin Pl{\'a}vala.
\newblock Entanglement and superposition are equivalent concepts in any physical theory.
\newblock {\em Physical Review Letters}, 128(16), April 2022.

\bibitem{barnumCloningBroadcastingGeneric2006}
Howard Barnum, Jonathan Barrett, Matthew Leifer, and Alexander Wilce.
\newblock Cloning and {{Broadcasting}} in {{Generic Probabilistic Theories}}, 2006.

\bibitem{barnum2008teleportationgeneralprobabilistictheories}
Howard Barnum, Jonathan Barrett, Matthew Leifer, and Alexander Wilce.
\newblock Teleportation in general probabilistic theories, 2008.

\bibitem{Garner_2017}
Andrew J.~P. Garner, Markus~P. M{\"u}ller, and Oscar C.~O. Dahlsten.
\newblock The complex and quaternionic quantum bit from relativity of simultaneity on an interferometer.
\newblock {\em Proceedings of the Royal Society A: Mathematical, Physical and Engineering Sciences}, 473(2208):20170596, December 2017.

\bibitem{Plvala2016}
Martin Pl{\'a}vala.
\newblock All measurements in a probabilistic theory are compatible if and only if the state space is a simplex.
\newblock {\em Physical Review A}, 94(4), October 2016.

\bibitem{Barnum2014}
Howard Barnum, Markus~P M{\"u}ller, and Cozmin Ududec.
\newblock Higher-order interference and single-system postulates characterizing quantum theory.
\newblock {\em New Journal of Physics}, 16(12):123029, December 2014.

\bibitem{Heinosaari2019}
Teiko Heinosaari, Leevi Lepp{\"a}j{\"a}rvi, and Martin Pl{\'a}vala.
\newblock No-free-information principle in general probabilistic theories.
\newblock {\em Quantum}, 3:157, July 2019.

\bibitem{Krumm2019}
Marius Krumm and Markus~P. M{\"u}ller.
\newblock Quantum computation is the unique reversible circuit model for which bits are balls.
\newblock {\em npj Quantum Information}, 5(1), January 2019.

\bibitem{PhysRevLett.108.130401}
Markus~P. M{\"u}ller and Cozmin Ududec.
\newblock Structure of reversible computation determines the self-duality of quantum theory.
\newblock {\em Physical Review Letters}, 108(13):130401, March 2012.

\bibitem{Wright2021}
Victoria~J Wright and Stefan Weigert.
\newblock General probabilistic theories with a {{Gleason-type}} theorem.
\newblock {\em Quantum}, 5:588, November 2021.

\bibitem{Kolangatt2025}
Mayalakshmi Kolangatt, Thigazholi Muruganandan, Sahil~Gopalkrishna Naik, Tamal Guha, Manik Banik, and Sutapa Saha.
\newblock Bipartite polygon models: Entanglement classes and their nonlocal behaviour.
\newblock {\em Quantum}, 9:1599, January 2025.

\bibitem{Scheidl2010}
Thomas Scheidl, Rupert Ursin, Johannes Kofler, Sven Ramelow, Xiao-Song Ma, Thomas Herbst, Lothar Ratschbacher, Alessandro Fedrizzi, Nathan~K. Langford, Thomas Jennewein, and Anton Zeilinger.
\newblock Violation of local realism with freedom of choice.
\newblock {\em Proceedings of the National Academy of Sciences}, 107(46):19708--19713, November 2010.

\bibitem{PhysRevLett.23.880}
John~F. Clauser, Michael~A. Horne, Abner Shimony, and Richard~A. Holt.
\newblock Proposed experiment to test local hidden-variable theories.
\newblock {\em Physical Review Letters}, 23(15):880--884, October 1969.

\bibitem{cat}
John~D. Trimmer.
\newblock The present situation in quantum mechanics: A translation of schr\"odinger's "cat paradox" paper.
\newblock {\em Proceedings of the American Philosophical Society}, 124(5):323--338, 1980.

\bibitem{Hardy2011}
Lucien Hardy and William~K. Wootters.
\newblock Limited holism and real-vector-space quantum theory.
\newblock {\em Foundations of Physics}, 42(3):454--473, December 2011.

\bibitem{Daki2014}
B~Daki{\'c}, T~Paterek, and {\v C}~Brukner.
\newblock Density cubes and higher-order interference theories.
\newblock {\em New Journal of Physics}, 16(2):023028, February 2014.

\bibitem{hackingEmergenceProbabilityPhilosophical2006}
Ian Hacking.
\newblock {\em The {{Emergence}} of {{Probability}}: {{A Philosophical Study}} of {{Early Ideas}} about {{Probability}}, {{Induction}} and {{Statistical Inference}}}.
\newblock Cambridge University Press, 2 edition, July 2006.

\bibitem{PhysRevA.104.022207}
John~B. DeBrota, Christopher~A. Fuchs, Jacques~L. Pienaar, and Blake~C. Stacey.
\newblock Born's rule as a quantum extension of {{Bayesian}} coherence.
\newblock {\em Physical Review A: Atomic, Molecular, and Optical Physics}, 104(2):022207, August 2021.

\bibitem{RevModPhys.85.1693}
Christopher~A. Fuchs and R{\"u}diger Schack.
\newblock Quantum-{{Bayesian}} coherence.
\newblock {\em Reviews of Modern Physics}, 85(4):1693--1715, December 2013.

\bibitem{berghofer2023phenomenology}
P.~Berghofer and H.A. Wiltsche.
\newblock {\em Phenomenology and Qbism: {{New}} Approaches to Quantum Mechanics}.
\newblock Routledge Studies in the Philosophy of Mathematics and Physics. Taylor \& Francis, 2023.

\bibitem{Gefter:2024jbq}
Amanda Gefter.
\newblock Enaction for qbists.
\newblock November 2024.

\bibitem{Rowlands2010-pz}
Mark Rowlands.
\newblock {\em The New Science of the Mind}.
\newblock A Bradford Book. Bradford Books, Cambridge, MA, September 2010.

\bibitem{waldronIntroductionFiniteTight2018}
Shayne F.~D. Waldron.
\newblock {\em An {{Introduction}} to {{Finite Tight Frames}}}.
\newblock Applied and {{Numerical Harmonic Analysis}}. Springer New York, New York, NY, 2018.

\bibitem{obst2024wignerstheoremstabilizerstates}
Valentin Obst, Arne Heimendahl, Tanmay Singal, and David Gross.
\newblock Wigner's {{Theorem}} for stabilizer states and quantum designs.
\newblock {\em Journal of Mathematical Physics}, 65(11), November 2024.

\bibitem{kriegMinnesotaNotesJordan1999}
Aloys Krieg.
\newblock {\em The {{Minnesota Notes}} on {{Jordan Algebras}} and {{Their Applications}}}.
\newblock Number v.1710 in Lecture {{Notes}} in {{Mathematics Ser}}. Springer Berlin / Heidelberg, Berlin, Heidelberg, 1999.

\bibitem{alizadehIntroductionFormallyReal2012}
F.~Alizadeh.
\newblock An {{Introduction}} to {{Formally Real Jordan Algebras}} and {{Their Applications}} in {{Optimization}}.
\newblock In Miguel~F. Anjos and Jean~B. Lasserre, editors, {\em Handbook on {{Semidefinite}}, {{Conic}} and {{Polynomial Optimization}}}, volume 166, pages 297--337. Springer US, New York, NY, 2012.

\bibitem{barnumStronglySymmetricSpectral2019}
Howard Barnum and Joachim Hilgert.
\newblock Strongly symmetric spectral convex bodies are {{Jordan}} algebra state spaces, 2019.

\bibitem{ProbabilitiesBettingOdds}
Carlton Caves.
\newblock Probabilities as betting odds and the {{Dutch}} book.

\bibitem{vanfraassenBeliefWill1984}
C.~Van~Fraassen.
\newblock Belief and the {{Will}}.
\newblock {\em The Journal of Philosophy}, 81(5):235, May 1984.

\bibitem{definettiTheoryProbabilityCritical2017}
Bruno De~Finetti.
\newblock {\em Theory of Probability: A Critical Introductory Treatment}.
\newblock John Wiley \& Sons, Chichester, UK Hoboken, NJ, 2017.

\bibitem{definettiForesightItsLogical1992}
Bruno {de Finetti}.
\newblock Foresight: {{Its Logical Laws}}, {{Its Subjective Sources}}.
\newblock In Samuel Kotz and Norman~L. Johnson, editors, {\em Breakthroughs in {{Statistics}}: {{Foundations}} and {{Basic Theory}}}, pages 134--174. Springer, New York, NY, 1992.

\bibitem{GeneralizedInverses2003a}
Adi {Ben-Israel} and Thomas N.~E. Greville.
\newblock {\em Generalized {{Inverses}}}.
\newblock {{CMS Books}} in {{Mathematics}}. Springer-Verlag, New York, 2003.

\bibitem{hudsonLocallyNormalSymmetric1976}
R.~L. Hudson and G.~R. Moody.
\newblock Locally normal symmetric states and an analogue of de {{Finetti}}'s theorem.
\newblock {\em Zeitschrift f\"ur Wahrscheinlichkeitstheorie und Verwandte Gebiete}, 33(4):343--351, December 1976.

\bibitem{cavesUnknownQuantumStates2002}
Carlton~M. Caves, Christopher~A. Fuchs, and R{\"u}diger Schack.
\newblock Unknown quantum states: {{The}} quantum de {{Finetti}} representation.
\newblock {\em Journal of Mathematical Physics}, 43(9):4537--4559, September 2002.

\bibitem{lamiNonclassicalCorrelationsQuantum2018}
Ludovico Lami.
\newblock Non-classical correlations in quantum mechanics and beyond, March 2018.

\bibitem{weiss2024depolarizingreferencedevicesgeneralized}
Matthew~B. Weiss.
\newblock Depolarizing reference devices in generalized probabilistic theories, 2024.

\bibitem{szymusiakCanQBismExist2025}
Anna Szymusiak and Wojciech S{\l}omczy{\'n}ski.
\newblock Can {{QBism}} exist without {{Q}}? {{Morphophoric}} measurements in generalised probabilistic theories.
\newblock {\em Quantum}, 9:1598, January 2025.

\bibitem{scottTightInformationallyComplete2006}
A~J Scott.
\newblock Tight informationally complete quantum measurements.
\newblock {\em Journal of Physics A: Mathematical and General}, 39(43):13507--13530, October 2006.

\bibitem{debrotaSymmetricInformationallyComplete2020}
John~B. DeBrota, Christopher~A. Fuchs, and Blake~C. Stacey.
\newblock Symmetric informationally complete measurements identify the irreducible difference between classical and quantum systems.
\newblock {\em Physical Review Research}, 2(1):013074, January 2020.

\bibitem{schmidCharacterizationNoncontextualityFramework2021}
David Schmid, John~H. Selby, Elie Wolfe, Ravi Kunjwal, and Robert~W. Spekkens.
\newblock Characterization of {{Noncontextuality}} in the {{Framework}} of {{Generalized Probabilistic Theories}}.
\newblock {\em PRX Quantum}, 2(1):010331, February 2021.

\bibitem{schmidStructureTheoremGeneralizednoncontextual2024}
David Schmid, John~H. Selby, Matthew~F. Pusey, and Robert~W. Spekkens.
\newblock A structure theorem for generalized-noncontextual ontological models.
\newblock {\em Quantum}, 8:1283, March 2024.

\bibitem{shahandehUnifiedLinearAlgebraic2025}
Farid Shahandeh, Theodoros Yianni, and Mina Doosti.
\newblock A {{Unified Linear Algebraic Framework}} for {{Physical Models}} and {{Generalized Contextuality}}, December 2025.

\bibitem{selbyLinearProgramTesting2024}
John~H. Selby, Elie Wolfe, David Schmid, Ana~Bel{\'e}n Sainz, and Vinicius~P. Rossi.
\newblock Linear {{Program}} for {{Testing Nonclassicality}} and an {{Open-Source Implementation}}.
\newblock {\em Physical Review Letters}, 132(5):050202, January 2024.

\bibitem{tamGeometricTreatmentGeneralized1981}
Bit-Shun Tam.
\newblock A geometric treatment of generalized inverses and semigroups of nonnegative matrices.
\newblock {\em Linear Algebra and its Applications}, 41:225--272, December 1981.

\bibitem{weissCharacterizingQuantumStatespace2025}
Matthew~B. Weiss.
\newblock Characterizing quantum state-space with a single quantum measurement.
\newblock {\em Physical Review A}, 111(5):052205, May 2025.

\bibitem{matthewweissHeyredhatRedesigningV12026}
Matthew Weiss.
\newblock Heyredhat/redesigning: V1.
\newblock Zenodo, July 2026.

\bibitem{platoCreatingModernProbability1994}
Jan~Von Plato.
\newblock {\em Creating {{Modern Probability}}: {{Its Mathematics}}, {{Physics}} and {{Philosophy}} in {{Historical Perspective}}}.
\newblock Cambridge University Press, 1 edition, January 1994.

\bibitem{jeffreySubjectiveProbabilityReal2004}
Richard~C. Jeffrey.
\newblock {\em Subjective Probability: The Real Thing}.
\newblock Cambridge University Press, Cambridge, U.K New York, 2004.

\bibitem{hackingSlightlyMoreRealistic1967}
Ian Hacking.
\newblock Slightly {{More Realistic Personal Probability}}.
\newblock {\em Philosophy of Science}, 34(4):311--325, December 1967.

\bibitem{debrota2024quantumdynamicshappenspaper}
John~B. DeBrota, Christopher~A. Fuchs, and R{\"u}diger Schack.
\newblock Quantum {{Dynamics Happens Only}} on {{Paper}}: {{QBism}}'s {{Account}} of {{Decoherence}}.
\newblock {\em Physical Review A: Atomic, Molecular, and Optical Physics}, 110(5):052205, November 2024.

\bibitem{goldsteinPrevisionPrevision1983}
Michael Goldstein.
\newblock The {{Prevision}} of a {{Prevision}}.
\newblock {\em Journal of the American Statistical Association}, 78(384):817--819, December 1983.

\bibitem{shaferSubjectiveInterpretationConditional1983}
Glenn Shafer.
\newblock A {{Subjective Interpretation}} of {{Conditional Probability}}.
\newblock {\em Journal of Philosophical Logic}, 12(4):453--466, 1983.

\bibitem{fuchsBayesianConditioningReflection2012}
Christopher~A. Fuchs and R{\"u}diger Schack.
\newblock Bayesian {{Conditioning}}, the {{Reflection Principle}}, and {{Quantum Decoherence}}.
\newblock In Yemima {Ben-Menahem} and Meir Hemmo, editors, {\em Probability in {{Physics}}}, pages 233--247. Springer Berlin Heidelberg, Berlin, Heidelberg, 2012.

\bibitem{kolmogorovFoundationsTheoryProbability2018}
Andrej~Nikolaevi{\v c} Kolmogorov and A.~T. {Bharucha-Reid}.
\newblock {\em Foundations of the Theory of Probability}.
\newblock Dover Books on Mathematics. Dover Publications, Mineola, New York, second english edition, dover edition, republication of the 1956 second edition of the work originally published in 1950 by chelsea publishing, new york edition, 2018.

\bibitem{rockafellarConvexAnalysis1970}
R.~Tyrrell Rockafellar.
\newblock {\em Convex Analysis}.
\newblock Number~28 in Princeton Mathematical Series. Princeton University Press, Princeton, N.J, 1970.

\bibitem{macauleyNovelProofHeineBorel2008}
Matthew Macauley, Brian Rabern, and Landon Rabern.
\newblock A {{Novel Proof}} of the {{Heine-Borel Theorem}}, August 2008.

\bibitem{rudinPrinciplesMathematicalAnalysis1976}
Walter Rudin.
\newblock {\em Principles of Mathematical Analysis}.
\newblock International Series in Pure and Applied Mathematics. McGraw-Hill, New York, 3d ed edition, 1976.

\bibitem{rudinFunctionalAnalysis1991}
Walter Rudin.
\newblock {\em Functional Analysis}.
\newblock International Series in Pure and Applied Mathematics. McGraw-Hill, New York, 2nd ed edition, 1991.

\bibitem{buschQuantumStatesGeneralized2003}
P.~Busch.
\newblock Quantum {{States}} and {{Generalized Observables}}: {{A Simple Proof}} of {{Gleason}}'s {{Theorem}}.
\newblock {\em Physical Review Letters}, 91(12):120403, September 2003.

\bibitem{Caves2003GleasonTypeDO}
Carlton~M. Caves, Christopher~A. Fuchs, Kiran~K. Manne, and Joseph~M. Renes.
\newblock Gleason-type derivations of the quantum probability rule for generalized measurements.
\newblock {\em Foundations of Physics}, 34:193--209, 2003.

\bibitem{campbellGeneralizedInversesLinear2009}
Stephen~L. Campbell and Carl~D. Meyer.
\newblock {\em Generalized {{Inverses}} of {{Linear Transformations}}}.
\newblock {Society for Industrial and Applied Mathematics}, January 2009.

\bibitem{liGeneralFrameDecompositions1995}
Shidong Li.
\newblock On general frame decompositions.
\newblock {\em Numerical Functional Analysis and Optimization}, 16(9-10):1181--1191, January 1995.

\bibitem{piziakMatrixTheory2007}
Robert Piziak and P.L. Odell.
\newblock {\em Matrix {{Theory}}}.
\newblock {Chapman and Hall/CRC}, 0 edition, February 2007.

\bibitem{piziakFullRankFactorization1999}
R.~Piziak and P.~L. Odell.
\newblock Full {{Rank Factorization}} of {{Matrices}}.
\newblock {\em Mathematics Magazine}, 72(3):193--201, 1999.

\bibitem{ludovicolamiGeneralProbabilisticTheories2018}
{Lami, Ludovico}.
\newblock General probabilistic theories and meta-theoretical knowledge, December 2018.

\bibitem{axlerLinearAlgebraDone2024}
Sheldon~Jay Axler.
\newblock {\em Linear {{Algebra Done Right}}}.
\newblock Undergraduate {{Texts}} in {{Mathematics}}. Springer Nature, Cham, 2024.

\bibitem{boydConvexOptimization2004}
Stephen~P. Boyd and Lieven Vandenberghe.
\newblock {\em Convex Optimization}.
\newblock Cambridge University Press, Cambridge, 2004.

\bibitem{barrettFinettiTheoremTest2009}
Jonathan Barrett and Matthew Leifer.
\newblock The de {{Finetti}} theorem for test spaces.
\newblock {\em New Journal of Physics}, 11(3):033024, March 2009.

\bibitem{fuchsFinettiRepresentationTheorem2004}
Christopher~A. Fuchs, R{\"u}diger Schack, and Petra~F. Scudo.
\newblock De {{Finetti}} representation theorem for quantum-process tomography.
\newblock {\em Physical Review A}, 69(6):062305, June 2004.

\bibitem{chiribellaProbabilisticTheoriesPurification2010}
Giulio Chiribella, Giacomo~Mauro D'Ariano, and Paolo Perinotti.
\newblock Probabilistic theories with purification.
\newblock {\em Physical Review A}, 81(6):062348, June 2010.

\bibitem{scandoloInformationtheoreticFoundationsThermodynamics2019}
Carlo~Maria Scandolo.
\newblock Information-theoretic foundations of thermodynamics in general probabilistic theories, 2019.

\bibitem{hornMatrixAnalysis1985}
Roger~A. Horn and Charles~R. Johnson.
\newblock {\em Matrix {{Analysis}}}.
\newblock Cambridge University Press, 1 edition, December 1985.

\bibitem{debrotaLudersChannelsExistence2019}
John~B. DeBrota and Blake~C. Stacey.
\newblock L\textbackslash "uders channels and the existence of symmetric-informationally-complete measurements.
\newblock {\em Physical Review A}, 100(6):062327, December 2019.

\bibitem{rungtaUniversalStateInversion2001}
Pranaw Rungta, V.~Bu{\v z}ek, Carlton~M. Caves, M.~Hillery, and G.~J. Milburn.
\newblock Universal state inversion and concurrence in arbitrary dimensions.
\newblock {\em Physical Review A}, 64(4):042315, September 2001.

\bibitem{renesSymmetricInformationallyComplete2004}
Joseph~M. Renes, Robin {Blume-Kohout}, A.~J. Scott, and Carlton~M. Caves.
\newblock Symmetric informationally complete quantum measurements.
\newblock {\em Journal of Mathematical Physics}, 45(6):2171--2180, June 2004.

\bibitem{applebySICsAlgebraicNumber2017}
Marcus Appleby, Steven Flammia, Gary McConnell, and Jon Yard.
\newblock {{SICs}} and {{Algebraic Number Theory}}.
\newblock {\em Foundations of Physics}, 47(8):1042--1059, August 2017.

\bibitem{bengtssonSICsExplanations2020}
Ingemar Bengtsson.
\newblock {{SICs}}: {{Some Explanations}}.
\newblock {\em Foundations of Physics}, 50(12):1794--1808, December 2020.

\bibitem{fuchsSICQuestionHistory2017}
Christopher Fuchs, Michael Hoang, and Blake Stacey.
\newblock The {{SIC Question}}: {{History}} and {{State}} of {{Play}}.
\newblock {\em Axioms}, 6(3):21, July 2017.

\bibitem{cohnOptimalSimplicesCodes2016}
Henry Cohn, Abhinav Kumar, and Gregory Minton.
\newblock Optimal simplices and codes in projective spaces.
\newblock {\em Geometry \& Topology}, 20(3):1289--1357, July 2016.

\bibitem{harrowChurchSymmetricSubspace2013}
Aram~W. Harrow.
\newblock The {{Church}} of the {{Symmetric Subspace}}, 2013.

\bibitem{Weiss2021}
Christopher~A. Fuchs, Maxim Olshanii, and Matthew~Benjamin Weiss.
\newblock Quantum mechanics? {{It}}'s all fun and games until someone loses an i.
\newblock {\em Asian Journal of Physics}, 30(12):1707--1726, 2021.

\bibitem{hughesSphericalTtdesignsSmall2021}
Daniel Hughes and Shayne Waldron.
\newblock Spherical (t,t)-designs with a small number of vectors.
\newblock {\em Linear Algebra and its Applications}, 608:84--106, January 2021.

\bibitem{schrijver1986theory}
A.~Schrijver.
\newblock {\em Theory of Linear and Integer Programming}.
\newblock A {{Wiley-Interscience}} Publication. Wiley, 1986.

\bibitem{mullerTestingQuantumTheory2023}
Markus~P. M{\"u}ller and Andrew J.~P. Garner.
\newblock Testing {{Quantum Theory}} by {{Generalizing Noncontextuality}}.
\newblock {\em Physical Review X}, 13(4):041001, October 2023.

\bibitem{schmidShadowsSubsystemsGeneralized2025}
David Schmid, John~H. Selby, Vinicius~P. Rossi, Roberto~D. Baldij{\~a}o, and Ana~Bel{\'e}n Sainz.
\newblock Shadows and subsystems of generalized probabilistic theories: When tomographic incompleteness is not a loophole for contextuality proofs.
\newblock {\em Quantum}, 9:1880, October 2025.

\bibitem{markovsky2011low}
I.~Markovsky.
\newblock {\em Low Rank Approximation: {{Algorithms}}, Implementation, Applications}.
\newblock Communications and Control Engineering. Springer London, 2011.

\bibitem{diamondCVXPYPythonEmbeddedModeling2016}
Steven Diamond and Stephen Boyd.
\newblock {{CVXPY}}: {{A Python-Embedded Modeling Language}} for {{Convex Optimization}}, 2016.

\bibitem{ziegler2012lectures}
G.M. Ziegler.
\newblock {\em Lectures on Polytopes}.
\newblock Graduate Texts in Mathematics. Springer New York, 2012.

\bibitem{PycddlibPythonWrapper}
Pycddlib: {{A Python}} wrapper for cddlib.

\bibitem{dingTeachingTipWhen2013}
J.~Ding and N.~H. Rhee.
\newblock Teaching {{Tip}}: {{When}} a {{Matrix}} and {{Its Inverse Are Stochastic}}.
\newblock {\em The College Mathematics Journal}, 44(2):108--109, March 2013.

\bibitem{ferrieFrameRepresentationsQuantum2008}
Christopher Ferrie and Joseph Emerson.
\newblock Frame representations of quantum mechanics and the necessity of negativity in quasi-probability representations.
\newblock {\em Journal of Physics A: Mathematical and Theoretical}, 41(35):352001, September 2008.

\bibitem{ferrieFramedHilbertSpace2009}
Christopher Ferrie and Joseph Emerson.
\newblock Framed {{Hilbert}} space: Hanging the quasi-probability pictures of quantum theory.
\newblock {\em New Journal of Physics}, 11(6):063040, June 2009.

\bibitem{ferrieNecessityNegativityQuantum2010}
Christopher Ferrie, Ryan Morris, and Joseph Emerson.
\newblock Necessity of negativity in quantum theory.
\newblock {\em Physical Review A}, 82(4):044103, October 2010.

\bibitem{ferrieQuasiprobabilityRepresentationsQuantum2011}
Christopher Ferrie.
\newblock Quasi-probability representations of quantum theory with applications to quantum information science.
\newblock {\em Reports on Progress in Physics}, 74(11):116001, November 2011.

\bibitem{wernerAllmultipartiteBellcorrelationInequalities2001}
R.~F. Werner and M.~M. Wolf.
\newblock All-multipartite {{Bell-correlation}} inequalities for two dichotomic observables per site.
\newblock {\em Physical Review A}, 64(3):032112, August 2001.

\bibitem{pironioAllClauserHorne2014}
Stefano Pironio.
\newblock All {{Clauser}}--{{Horne}}--{{Shimony}}--{{Holt}} polytopes.
\newblock {\em Journal of Physics A: Mathematical and Theoretical}, 47(42):424020, October 2014.

\bibitem{fuchsQBismPolishingPoints2025}
Christopher~A. Fuchs and Blake~C. Stacey.
\newblock {{QBism}}, {{Polishing Some Points}}, 2025.

\bibitem{brunnerBellNonlocality2014}
Nicolas Brunner, Daniel Cavalcanti, Stefano Pironio, Valerio Scarani, and Stephanie Wehner.
\newblock Bell nonlocality.
\newblock {\em Reviews of Modern Physics}, 86(2):419--478, April 2014.

\bibitem{kunjwalMinimalStatedependentProof2014}
Ravi Kunjwal and Sibasish Ghosh.
\newblock Minimal state-dependent proof of measurement contextuality for a qubit.
\newblock {\em Physical Review A}, 89(4):042118, April 2014.

\bibitem{Bohr1928-vp}
Niels Bohr.
\newblock The quantum postulate and the recent development of atomic {{Theory}}.
\newblock {\em Nature}, 121(3050):580--590, April 1928.

\bibitem{baggott2011quantum}
J.~Baggott.
\newblock {\em The {{Quantum Story}}: {{A}} History in 40 Moments}.
\newblock OUP Oxford, 2011.

\bibitem{hou2016uncertaintyrelationsmultiobservables}
Jinchuan Hou and Kan He.
\newblock Uncertainty relations for any multi observables.
\newblock 2016.

\bibitem{appleby2011propertiesqbiststatespaces}
D.~M. Appleby, Asa Ericsson, and Christopher~A. Fuchs.
\newblock Properties of {{QBist State Spaces}}.
\newblock {\em Foundations of Physics}, 41(3):564--579, April 2010.

\bibitem{mazurekExperimentallyBoundingDeviations2021}
Michael~D. Mazurek, Matthew~F. Pusey, Kevin~J. Resch, and Robert~W. Spekkens.
\newblock Experimentally {{Bounding Deviations From Quantum Theory}} in the {{Landscape}} of {{Generalized Probabilistic Theories}}.
\newblock {\em PRX Quantum}, 2(2):020302, April 2021.

\bibitem{graboweckyExperimentallyBoundingDeviations2022}
Michael~J. Grabowecky, Christopher A.~J. Pollack, Andrew~R. Cameron, Robert~W. Spekkens, and Kevin~J. Resch.
\newblock Experimentally bounding deviations from quantum theory for a photonic three-level system using theory-agnostic tomography.
\newblock {\em Physical Review A}, 105(3):032204, March 2022.

\bibitem{barnum2013postclassicalprobabilitytheory}
Howard Barnum and Alexander Wilce.
\newblock Post-classical probability theory, 2013.

\bibitem{PhysRevLett.60.1103}
Hans Maassen and J.~B.~M. Uffink.
\newblock Generalized entropic uncertainty relations.
\newblock {\em Physical Review Letters}, 60(12):1103--1106, March 1988.

\bibitem{Heinrich_2019}
Markus Heinrich and David Gross.
\newblock Robustness of {{Magic}} and {{Symmetries}} of the {{Stabiliser Polytope}}.
\newblock {\em Quantum}, 3:132, April 2019.

\bibitem{e16031484}
D.~Marcus Appleby, Hoan~Bui Dang, and Christopher~A. Fuchs.
\newblock Symmetric informationally-complete quantum states as analogues to orthonormal bases and minimum-uncertainty states.
\newblock {\em Entropy. An International and Interdisciplinary Journal of Entropy and Information Studies}, 16(3):1484--1492, 2014.

\bibitem{cuffaro2024quantumstatesmaximalmagic}
Gianluca Cuffaro and Christopher~A. Fuchs.
\newblock Quantum states with maximal magic, 2024.

\bibitem{debrota2020varietiesminimaltomographicallycomplete}
John~B. DeBrota, Christopher~A. Fuchs, and Blake~C. Stacey.
\newblock The {{Varieties}} of {{Minimal Tomographically Complete Measurements}}.
\newblock {\em International Journal of Quantum Information}, 19(07):2040005, 2021.

\bibitem{bacon2005quantumschurtransformi}
Dave Bacon, Isaac~L. Chuang, and Aram~W. Harrow.
\newblock The {{Quantum Schur Transform}}: {{I}}. {{Efficient Qudit Circuits}}, 2005.

\bibitem{SEYMOUR1984213}
P.D Seymour and Thomas Zaslavsky.
\newblock Averaging sets: {{A}} generalization of mean values and spherical designs.
\newblock {\em Advances in Mathematics}, 52(3):213--240, 1984.

\bibitem{bengtsson2017discretestructuresfinitehilbert}
Ingemar Bengtsson and Karol Zyczkowski.
\newblock On discrete structures in finite {{Hilbert}} spaces, 2017.

\bibitem{Zhu_2011}
Huangjun Zhu and Berthold-Georg Englert.
\newblock Quantum state tomography with fully symmetric measurements and product measurements.
\newblock {\em Physical Review A}, 84(2), August 2011.

\bibitem{Gross_2021}
David Gross, Sepehr Nezami, and Michael Walter.
\newblock Schur--{{Weyl Duality}} for the {{Clifford Group}} with {{Applications}}: {{Property Testing}}, a {{Robust Hudson Theorem}}, and de {{Finetti Representations}}.
\newblock {\em Communications in Mathematical Physics}, 385(3):1325--1393, June 2021.

\bibitem{zhu2024momentsquditcliffordorbits}
Huangjun Zhu, Chengsi Mao, and Changhao Yi.
\newblock Third moments of qudit {{Clifford}} orbits and 3-designs based on magic orbits, 2024.

\bibitem{Zhu_2017}
Huangjun Zhu.
\newblock Multiqubit {{Clifford}} groups are unitary 3-designs.
\newblock {\em Physical Review A}, 96(6), December 2017.

\bibitem{Jones_2005}
Nick~S. Jones and Noah Linden.
\newblock Parts of quantum states.
\newblock {\em Physical Review A}, 71(1), January 2005.

\bibitem{fuchs2015strugglesblockuniverse}
Christopher~A. Fuchs, Maximilian Schlosshauer, and Blake~C. Stacey.
\newblock My {{Struggles}} with the {{Block Universe}}, 2015.

\bibitem{oszmaniec2021measuringrelationalinformationquantum}
Micha{\l} Oszmaniec, Daniel~J. Brod, and Ernesto~F. Galv{\~a}o.
\newblock Measuring relational information between quantum states, and applications.
\newblock {\em New Journal of Physics}, (1):013053, January 2024.

\bibitem{townsendJordanFormulationQuantum2016}
Paul~K. Townsend.
\newblock The {{Jordan}} formulation of {{Quantum Mechanics}}: A review, 2016.

\bibitem{akhiezerCommutatorMapReal2015}
Dmitri Akhiezer.
\newblock On the {{Commutator Map}} for {{Real Semisimple Lie Algebras}}.
\newblock {\em Moscow Mathematical Journal}, 15(4):609--613, 2015.

\bibitem{NCategoryCafe}
John Baez.
\newblock Dynamics in {{Jordan Algebras}}.

\bibitem{McCrimmon2003-kp}
Kevin McCrimmon.
\newblock {\em A {{Taste}} of {{Jordan Algebras}}}.
\newblock Universitext. Springer, New York, NY, 2004 edition, 2003.

\bibitem{jacobson1968structure}
N.~Jacobson.
\newblock {\em Structure and Representations of {{Jordan}} Algebras}.
\newblock Colloquium Publications. American Mathematical Society, 1968.

\bibitem{supicSelftestingQuantumSystems2020}
Ivan {\v S}upi{\'c} and Joseph Bowles.
\newblock Self-testing of quantum systems: A review.
\newblock {\em Quantum}, 4:337, September 2020.

\bibitem{jordan1933verallgemeinerungsmöglichkeiten}
P.~Jordan.
\newblock {\em \"Uber Verallgemeinerungsm\"oglichkeiten Des Formalismus Der Quantenmechanik}.
\newblock Sonderdrucke Aus Den Nachrichten von Der Gesellschaft Der Wissenschaften Zu G\"ottingen : {{Mathematisch-physikalische}} Klasse. Weidmann, 1933.

\bibitem{schafer1995introduction}
Richard~D. Schafer.
\newblock {\em An Introduction to Nonassociative Algebras}.
\newblock Dover Publications, 1995.

\bibitem{barnum2013localtomographyjordanstructure}
Howard Barnum and Alexander Wilce.
\newblock Local tomography and the {{Jordan}} structure of quantum theory, 2013.

\bibitem{orlitzkyRankComputationEuclidean2022}
Michael Orlitzky.
\newblock Rank computation in {{Euclidean Jordan}} algebras.
\newblock {\em Journal of Symbolic Computation}, 113:181--192, November 2022.

\bibitem{fanizzaSwapTestOptimal2020}
M.~Fanizza, M.~Rosati, M.~Skotiniotis, J.~Calsamiglia, and V.~Giovannetti.
\newblock Beyond the {{Swap Test}}: {{Optimal Estimation}} of {{Quantum State Overlap}}.
\newblock {\em Physical Review Letters}, 124(6):060503, February 2020.

\bibitem{gitiauxSWAPTestArbitrary2022}
Xavier Gitiaux, Ian Morris, Maria Emelianenko, and Mingzhen Tian.
\newblock {{SWAP}} test for an arbitrary number of quantum states.
\newblock {\em Quantum Information Processing}, 21(10):344, October 2022.

\bibitem{nakataComputationalComplexityUnitary2025}
Yoshifumi Nakata, Yuki Takeuchi, Martin Kliesch, and Andrew Darmawan.
\newblock Computational {{Complexity}} of {{Unitary}} and {{State Design Properties}}.
\newblock {\em PRX Quantum}, 6(3):030345, September 2025.

\bibitem{mcginleyPostselectionFreeLearningMeasurementInduced2024}
Max McGinley.
\newblock Postselection-{{Free Learning}} of {{Measurement-Induced Quantum Dynamics}}.
\newblock {\em PRX Quantum}, 5(2):020347, May 2024.

\bibitem{MarkusGrasslComputing}
{Markus Grassl: Computing Numerical and Exact SIC-POVMs -- nisq.pl}.

\bibitem{applebyConstructiveApproachZauners2025}
Marcus Appleby, Steven~T Flammia, and Gene~S Kopp.
\newblock A {{Constructive Approach}} to {{Zauner}}'s {{Conjecture}} via the {{Stark Conjectures}}, 2025.

\bibitem{royUnitaryDesignsCodes2009}
Aidan Roy and A.~J. Scott.
\newblock Unitary designs and codes.
\newblock {\em Designs, Codes and Cryptography}, 53(1):13--31, October 2009.

\bibitem{grossEvenlyDistributedUnitaries2007}
D.~Gross, K.~Audenaert, and J.~Eisert.
\newblock Evenly distributed unitaries: {{On}} the structure of unitary designs.
\newblock {\em Journal of Mathematical Physics}, 48(5):052104, May 2007.

\bibitem{bannaiExplicitConstructionExact2022}
Eiichi Bannai, Yoshifumi Nakata, Takayuki Okuda, and Da~Zhao.
\newblock Explicit construction of exact unitary designs.
\newblock {\em Advances in Mathematics}, 405:108457, August 2022.

\bibitem{barnumSelfdualityJordanStructure2023}
Howard Barnum, Cozmin Ududec, and John {van de Wetering}.
\newblock Self-duality and {{Jordan}} structure of quantum theory follow from homogeneity and pure transitivity, 2023.

\bibitem{Vinzant2014WhatIA}
Cynthia Vinzant.
\newblock What is... a spectrahedron.
\newblock {\em Notices of the American Mathematical Society}, 61:492--494, 2014.

\end{thebibliography}
